\documentclass[prx,amsmath,amssymb, notitlepage, onecolumn,
nofootinbib,
superscriptaddress,
longbibliography
]{revtex4-1}

\usepackage{amsmath}
\usepackage{tabularx,graphicx}
\usepackage{epstopdf}
\usepackage{makecell}
\usepackage{graphicx}
\usepackage{latexsym}
\usepackage{amssymb}
\usepackage{amsthm}
\usepackage{color, colortbl}
\usepackage{psfrag}
\usepackage{bbm}
\usepackage{bm}
\usepackage{titlesec}
\usepackage{dsfont}
\usepackage{feynmp}
\usepackage{slashed}
\usepackage[utf8]{inputenc}
\usepackage[symbols,nogroupskip,sort=none]{glossaries-extra}
\usepackage{nomencl}

\usepackage{multirow}
\usepackage{url}
\usepackage{xr}
\usepackage{xcite}
\usepackage{microtype}
\usepackage{tikz}
\usepackage{wasysym}
\usepackage{supertabular}
\usetikzlibrary{patterns}
\tikzset{every picture/.style={line width=0.75pt}} %set default line width to 0.75pt
\usepackage{float}

\usepackage{mathrsfs}
\usepackage[title]{appendix}

\usepackage{etoolbox}
\patchcmd{\appendices}{\quad}{: }{}{}
\usepackage{comment}

\makeatletter
\newcommand*{\addFileDependency}[1]{
  \typeout{(#1)}
  \@addtofilelist{#1}
  \IfFileExists{#1}{}{\typeout{No file #1.}}
}
\makeatother

\newcommand*{\myexternaldocument}[1]{%
    \externaldocument{#1}%
    \addFileDependency{#1.tex}%
    \addFileDependency{#1.aux}%
}

\myexternaldocument{main_05_12}

\newcommand{\mc}[1]{\mathcal{#1}}
\newcommand{\ew}[1]{\langle #1 \rangle}
\newcommand{\beq}{\begin{eqnarray}}
\newcommand{\eeq}{\end{eqnarray}}

\newcommand{\bsp}{\begin{aligned}}
\newcommand{\esp}{\end{aligned}}

\newcommand{\ie}{{i.e., }}
\newcommand{\eg}{{e.g., }}

\newcommand{\ket}[1]{|#1\rangle}
\newcommand{\bra}[1]{\langle#1|}

\newcommand{\lsupsc}[1]{^{#1}\hspace{-1pt}}
\newcommand{\bpm}{\begin{pmatrix}}
\newcommand{\epm}{\end{pmatrix}}
\newcommand{\bmm}{\begin{matrix}}
\newcommand{\emm}{\end{matrix}}
\newcommand{\Blangle}{\Biggl\langle\bmm} %the big \langle in the bra
\newcommand{\BRvert}{\emm\Biggr\vert} %the big \vert on the right
\newcommand{\BLvert}{\Biggl\vert\bmm} %the big \vert on the right
\newcommand{\Bvert}{\emm\Biggr\vert\bmm} % the big \vert in the middle of a bra-ket inner product.
\newcommand{\Brangle}{\emm\Biggr\rangle}

\usepackage{xcolor}
\definecolor{darkblue}{rgb}{0.,0.,0.4}
\definecolor{darkred}{rgb}{0.5,0.,0.}
\definecolor{BlueViolet}{RGB}{138,43,226}
\definecolor{SkyBlue}{RGB}{30,144,255}
\definecolor{DarkGreen}{RGB}{0,100,0}

\usepackage[colorlinks=true,linkcolor=darkblue,citecolor=blue,urlcolor=darkred]{hyperref}
\usepackage[normalem]{ulem}

\usepackage{mathrsfs}

\usepackage{comment}

\newcommand{\z}{\mathbb{Z}}

\makeatletter
\newcommand{\colim@}[2]{%
  \vtop{\m@th\ialign{##\cr
    \hfil$#1\operator@font colim$\hfil\cr
    \noalign{\nointerlineskip\kern1.5\ex@}#2\cr
    \noalign{\nointerlineskip\kern-\ex@}\cr}}%
}
\newcommand{\colim}{%
  \mathop{\mathpalette\colim@{\rightarrowfill@\textstyle}}\nmlimits@
}
\makeatother

\newtheorem{theorem}{Theorem}
\newtheorem{lemma}{Lemma}

\makenomenclature

\numberwithin{corollary}{section}
\numberwithin{theorem}{section}
\numberwithin{lemma}{section}
\numberwithin{definition}{section}
\numberwithin{example}{section}
\numberwithin{proposition}{section}
\numberwithin{remark}{section}

\usepackage{pst-all}

\usepackage{leftidx}

\usepackage[off]{auto-pst-pdf} 
\usepackage{booktabs}
\usepackage{yfonts}
\makeatletter

\usepackage[caption=false]{subfig}
\usepackage{enumitem}

\input{Figures}

\begin{document}

\title{Microscopic universal theory\\ of symmetry-enriched topological quantum spin liquids}

\author{Yingcheng Li}
\affiliation{Department of Physics, National University of Singapore, 117551, Singapore}

\author{Liujun Zou}
\affiliation{Department of Physics, National University of Singapore, 117551, Singapore}

\begin{abstract}
    An ultimate theory of a phase of matter should describe all its universal properties via quantities that are measurable numerically and experimentally. In this work, we present a microscopic universal theory of symmetry-enriched topological quantum spin liquids (TQSLs) in two spatial dimensions, which directly utilizes microscopically measurable quantities to describe the universal properties. This theory applies to generic TQSLs, which can be Abelian or non-Abelian, chiral or non-chiral. The symmetries are also general, which can include both internal and lattice symmetries, unitary and anti-unitary symmetries, and discrete and continuous symmetries. There can be spin-orbit coupling, the microscopic degrees of freedom may transform linearly or projectively under the symmetries, and the symmetries can permute anyons. The input of the theory is some microscopic states with anyons, operators that control the dynamics of anyons, and symmetry actions in the TQSL, and its output is a set of data characterizing the universal properties, whose underlying mathematical structure is a generalization of category theory. Based on this theory, we find an explicit bijective map between the universal data characterizing a TQSL with a symmetry described by a group $G$, where the symmetry actions may include both lattice and internal symmetries, and the corresponding universal data for a TQSL with only an internal symmetry group $G$, and thus establish a precise crystalline equivalence principle. We demonstrate our theory in symmetry-enriched TQSLs realized on quantum processors based on superconducting qubits, trapped ions, and Rydberg atoms, and in each example we verify the Lieb-Schultz-Mattis anomaly matching condition. Our theory provides a solid basis for identifying and manipulating symmetry-enriched TQSLs, which further paves the way for fault-tolerant quantum computation based on these systems.
\end{abstract}

\maketitle

\tableofcontents\

\section{Introduction}

Topological quantum spin liquids (TQSLs), or more generally, bosonic topological orders, are interesting gapped quantum phases of matter in spin or bosonic systems that exhibit long-range entanglement beyond the traditional Landau-Ginzburg paradigm. In two spatial dimensions, these exotic phases host unconventional particle-like excitations known as anyons, which can be neither boson nor fermion  \cite{wen2004quantum}. Besides being fundamentally interesting, such phases are also candidate hardware for fault-tolerant quantum computation \cite{Kitaev1997, Freedman2002, Nayak2007}. Although there is no conclusive evidence of them in solid-state materials yet \cite{Broholm2019}, TQSLs are identified in many model Hamiltonians \cite{Kitaev1997, Levin2004, Kitaev2006, Gong2013, Gong2014, Heinrich2016, Cheng2016, Gong2017, Verresen2020, Huang2021, Luo2022, Zhang2024} and realized in various intermediate-scale quantum processors \cite{Satzinger2021, Semeghini2021, Iqbal2023, Foss-Feig2023, Iqbal2023a}.

The topological properties of a TQSL, such as the statistics of the anyons, are well defined even without any symmetry{\footnote{Unless otherwise stated, all symmetries in this paper are ordinary 0-form invertible symmetries, which are the symmetries most relevant to numerical and experimental studies.}}. However, symmetries lead to richer physics including symmetry fractionalization as in quantum Hall effects \cite{Chen2016}, and two TQSLs with identical topological properties may be in distinct symmetry-enriched topological (SET) phases, \ie they cannot (can) be adiabatically connected if the symmetries are preserved (broken). Similar to the symmetry-independent topological properties, symmetries in TQSLs are not only of fundamental interest but also practically useful, as symmetry defects there create new routes to quantum computation \cite{Bombin2010, Lindner2012, Clarke2012, Alicea2015}.

Theoretically characterizing SET phases is both important from the conceptual perspective and necessary for experimentally identifying and manipulating topological phases, which are key operations in quantum computation based on these phases. In this regard, two fundamental questions are as follows.

\begin{itemize}
    
    \item What data characterizes the universal properties of symmetry-enriched TQSLs and what is the mathematical structure governing this data?

    \item In a symmetry-enriched TQSL, how can this data be extracted from physical quantities that can at least in principle be measured numerically and/or experimentally? 
    
\end{itemize}

In the first question, universal properties refer to common properties shared by all systems in the entire SET phase, which by definition are robust against perturbations to the Hamiltonian. It is these properties that endow topological phases with the potential for fault-tolerant quantum computation. For $(2+1)$D TQSLs without symmetries and with only internal symmetries, these properties are believed to be captured by, roughly speaking, the statistics and fusion of anyons and how symmetries act on anyons, and such data can be packaged into a mathematical structure, which is known as a category theory, and its extension with the symmetry group \cite{Kitaev2006, barkeshli2014, Tarantino2015, Lan2023}. However, lattice symmetries are also important, but a comprehensive theory for the universal properties of general TQSLs with generic lattice symmetries are lacking, although progress in special cases with specific classes of TQSLs and/or particular types of symmetries has been made \cite{Essin2012, Qi2015c, Zaletel2015, Barkeshli2016, Qi2017, Ding2025}.

The second question is also of paramount importance, as it concerns the bridge between the physically measurable quantities and the abstract mathematical structure governing the universal properties of symmetry-enriched TQSLs. A theory for the universal properties is testable only if this question is answered. Moreover, the answer to this question serves as the basis to detect and control symmetry-enriched TQSLs, which are indispensable in fault-tolerant quantum computation. For TQSLs without symmetries and with only internal symmetries, this question has been largely addressed in Refs. \cite{Kawagoe2019} and \cite{barkeshli2014}, respectively. However, for TQSLs with general symmetries that contain both internal and lattice symmetries, since even the first question is largely open, this second question is also poorly understood. As a result, although a mathematical classification of SET phases of TQSLs with general symmetries was derived in Ref. \cite{Ye2023}, for a given setup that realizes a TQSL with such a symmetry, there is no systematic way to identify which phase in the classification this setup belongs to. 

In this paper, we complete the framework for $(2+1)$D symmetry-enriched TQSLs and provide a comprehensive solution to both questions above. Our results apply to generic TQSLs, which may be Abelian or non-Abelian, chiral or non-chiral. The symmetries under consideration are also general and can include both internal and lattice symmetries, unitary and anti-unitary symmetries, and discrete and continuous symmetries. There can be spin-orbit coupling, the microscopic degrees of freedom may transform in linear or projective representations under the symmetries, and the symmetries can permute anyons. Our key insight is that the universal properties of an SET phase should be encoded in the low-energy, long-distance dynamics and symmetry properties of the anyons. The important dynamics of anyons are how they move, fuse and split, which can be described by certain {\it microscopically defined} moving and splitting operators \cite{Kawagoe2019}. Therefore, by studying these dynamics and their interplay with symmetries microscopically, we achieve a {\it microscopic universal theory} of symmetry-enriched TQSLs, which answers both questions above in one shot. In essence, this theory takes as input some microscopic anyon states, moving and splitting operators, and symmetry actions, and it outputs the universal properties of the symmetry-enriched TQSLs. The underlying mathematical structure of the universal data is a generalization of category theory, with its contents summarized in Sec. \ref{sec:outline} and details presented in Sec. \ref{sec:framework}.

Our microscopic universal theory is very powerful. In particular, using it we find an {\it explicit} bijective map between the data characterizing the universal properties of a TQSL with a general symmetry group $G$, which may include both internal and lattice symmetries, and the corresponding data of a TQSL with only an internal symmetry described by the same group $G$. The existence of such a map, known as the crystalline equivalence principle, was proposed before \cite{Song2016, Thorngren2016}, and it was a key ingredient in the mathematical classification of symmetry-enriched TQSLs \cite{Ye2023}. However, this proposal is made precise only in the current work because a general theory for TQSLs with lattice symmetries was previously lacking. Given a symmetric TQSL, our explicit crystalline equivalence principle then allows us to identify which SET phase in the classification of Ref. \cite{Ye2023} it belongs to.

To demonstrate our microscopic universal theory via examples, we apply it to various symmetry-enriched TQSLs already realized in some quantum processors \cite{Satzinger2021, Semeghini2021, Iqbal2023, Foss-Feig2023}. We show how to extract the universal data of these symmetry-enriched TQSLs from microscopic quantities. As a highly nontrivial check of our theory, we show that the Lieb-Schultz-Mattis anomaly matching conditions hold in all these examples. In essence, from the microscopic perspective the anomalies capture the interplay between the microscopic degrees of freedom and the symmetries, and from the perspective of the low-energy quantum phase they can be derived using the universal data of the phase \cite{Zou2026}. The matching between these two perspectives gives stringent constraints on which quantum phases can emerge in a system with a given symmetry setting, and it plays a crucial role in the classification of symmetry-enriched TQSLs \cite{Ye2023}. The successful matching of anomalies based on our theory demonstrates its consistency with various previous work from very different considerations \cite{Ye2021a, Ye2022, Ye2023}.

Before ending the Introduction, we highlight the methodological novelty of our approach (on top of the significance of our results). Previously, to establish such a general theory of topological phases, one is often assisted by some mathematical hints about the universal properties, such as the category theory of anyons. However, such a mathematical crutch is unavailable here due to the complexity of incorporating lattice symmetries into a category theory. Therefore, we use a physics-based approach, and our result is still a mathematically precise theory, \ie we directly take microscopically defined quantities and extract from them the mathematical structure governing the universal properties, which enables us to answer both fundamental questions at the beginning of the Introduction in one shot. We expect this philosophy to be also useful in the future study of other quantum phases.

\section{Outline and summary} \label{sec:outline}

The outline of the rest of the paper and the summary of the results are as follows.
\begin{itemize}

    \item In Sec. \ref{sec:setup}, we introduce the our setup and discuss the basic physics of TQSLs, including the concepts of anyons, the operators controlling their dynamics, and symmetry actions on them. We also sharpen the notion of SET phases of TQSLs.
    
    \item In Sec. \ref{sec:framework}, we present our microscopic universal theory. In more detail, we construct microscopically defined quantities to capture the physics of fusion, braiding and symmetry actions of anyons. These quantities are universal, \ie they are common across all TQSLs in the same SET phase. We discuss how these quantities are organized into a generalization of category theory. A summary of this theory is in Sec. \ref{subsec: summarizing universal data}.
    
    \item In Sec. \ref{sec:equivalence}, we establish a precise crystalline equivalence principle, given by Eq. \eqref{Eq:correspondence}, which is a bijective map between the universal data characterizing an SET phase with a general symmetry $G$ and the universal data for an SET phase with an internal symmetry $G$.
    
    \item In Sec. \ref{sec:application}, we apply the microscopic universal theory to three examples: the toric code, the ``flipped" toric code, and a $\z_2$ TQSL on ruby lattice. We demonstrate how to use our theory to identify the universal properties of the SET phases these example belong to, and verify the Lieb-Schultz-Mattis anomaly matching condition in each example.

    \item We conclude our paper in Sec. \ref{sec:discussion} with interesting future directions.

    \item Various appendices contain additional details.
    
\end{itemize}

\section{Setup, background and notations} \label{sec:setup}

In this section, we introduce our setup, discuss some basic theoretical background and specify some notations to be used later.

\subsection{Microscopic setup}

Our setup is a large but finite two dimensional lattice system with periodic boundary conditions, where all microscopic degrees of freedom (DOF) are bosonic. These DOF can be spins, pseudo-spins, mobile bosons, quantum rotors, etc. Such a system may arise from a fermionic system where all fermionic excitations have high energies, so they can be ignored in the low-energy physics, which is our focus. Mathematically, states in our system can be described by vectors in a Hilbert space that is a tensor product of local Hilbert spaces. The local Hilbert spaces may be finite or infinite dimensional, and their dimensions can be different. The system has a local Hamiltonian, and it may enjoy some symmetries, which can include both internal and lattice symmetries, unitary and anti-unitary symmetries, and discrete and continuous symmetries. There can be spin-orbit coupling, and the microscopic DOF may transform in linear or projective representations under the symmetries. In all cases, the symmetries are assumed to transform local operators to local operators and preserve the metric in space (see Appendix \ref{app: locality} for more discussions on the notion of locality in finite systems).

There are two reasons for us to consider large but finite lattices, instead of infinite lattices. First and foremost, all lattices in experiments are finite, no matter how large they are. Second and more technically, infinite lattice systems do not have the structure of a tensor product of local Hilbert spaces, which then requires the relatively unfamiliar tools of operator algebra to deal with (see, \eg Ref. \cite{Liu2024a} for more details). Note that the translation symmetry group in our finite lattice can only be $\z_{L_1}\times\z_{L_2}$, with $L_{1, 2}$ the size along the two directions. However, as long as the system is large enough, we expect that all physical results will reflect the physics in the thermodynamic limit, which can have a $\z\times\z$ translation symmetry.

\subsection{Basic physics of symmetry-enriched TQSLs}

With the above setup in mind, now we briefly discuss the basic physics of symmetry-enriched TQSLs. More details can be found in, \eg Refs. \cite{Kawagoe2019, barkeshli2014, Essin2012}.

\subsubsection{Anyon states} \label{subsec: anyon states}

A TQSL is a gapped system where the low-energy excitations are anyons, \ie quasi-particles that may be neither bosons nor fermions. More precisely, each anyon characterizes a collection of states with a quasi-particle that are in the same superselection sector, \ie these states can be converted to each other by local operators. In particular, nontrivial (trivial) anyons cannot (can) be created from a ground state by local operators \cite{Kitaev2006, Kawagoe2019}. Therefore, we denote the simplest states with nontrivial anyons by $|a_{1}, \bar a_{2}\rangle$, where $a$ and $\bar a$ are nontrivial anyons that are anti-particles of each other, \ie they are created together (but not individually) by a local operator from a ground state, and the subscript $1 (2)$ means that $a$ ($\bar a$) is located at position $x_1$ ($x_2$). We emphasize that, like all other quasi-particles, an anyon is usually not a point geometrically, and it can have a size, shape, and even some internal DOF. In our notation, $|a_{1}, \bar a_{2}\rangle$ represents a state where all local information of these anyons, \ie their sizes, shapes and internal DOF, is suppressed but implicitly specified{\footnote{Under periodic boundary conditions, a TQSL generally has multiple ground states, so when we discuss the state $|a_{1}, \bar a_{2}\rangle$, in principle, we should also specify from which ground state $a$ and $\bar a$ are created by a local operator. However, as we will see in Sec. \ref{sec:framework}, all universal data of a symmetry-enriched TQSL can be extracted from a large but finite region of the system, and because all these ground states look identical in such regions, for our purpose there is no need to specify the original ground state when we discuss anyon states.}}. 

Later we will also consider states with more than 2 anyons. To introduce the notations for such states, we note there are 2 types of anyons, Abelian and non-Abelian anyons. Fusing multiple Abelian anyons together results in a single type of anyon. However, when non-Abelian anyons $a$ and $b$ are fused, the resulting anyon $c$ may be of multiple types, and for a given type of $c$, there can also be multiple ``fusion vertex basis states", \ie a topologically protected degenerate subspace of locally indistinguishable states that is useful for fault-tolerant quantum computation \cite{Freedman2002, Nayak2007}. These fusion properties can be summarized as
\beq \label{eq: fusion}
a\times b=\sum_cN_{ab}^cc.
\eeq
Eq. \eqref{eq: fusion} means that the fusion outcome of anyons $a$ and $b$ can be any anyon $c$ in the right hand side, and $N_{ab}^c$ is a positive integer called fusion multiplicity, which is the dimension of the space of fusion vertex basis states for anyons $a$, $b$ and $c$.

With this in mind, as in Ref. \cite{Kawagoe2019}, we denote a 3-anyon state by $|a_{1}, b_{2}, c_{3}; \mu\rangle$, where anyons $a$, $b$ and $c$, at positions $x_1$, $x_2$ and $x_3$, respectively, are created locally from a ground state together, and $\mu$ indexes the fusion vertex basis state, \ie orthonormal basis states in the degenerate subspace of states with anyons $a$, $b$ and $c$. As before, in $|a_{1}, b_{2}, c_{3}; \mu\rangle$ all local information of the anyons is implicitly speficied. The above condition of local indistinguishability means that
\beq \label{eq: local indistinguishablity}
\langle a_{1}, b_{2}, c_{3}; \mu |O| a_{1}, b_{2}, c_{3}; \nu\rangle = ({\rm const.})\cdot\delta_{\mu\nu}
\eeq
for any operator $O$ supported in a region that does not contain more than one anyons, whenever $|a_1, b_2, c_3;\mu\rangle$ and $|a_1, b_2, c_3;\nu\rangle$ are within the same set of orthonormal fusion vertex basis states.

Before finishing discussing anyon states, we remark that we only consider short-range correlated anyon states. Namely, for any local region with at most one anyon and another region separated from it by at least the distance between the anyons, the mutual information between these regions is negligible.{\footnote{The mutual information between two regions $A$ and $C$ is defined as $I(A: C)=S(\rho_A)+S(\rho_C)-S(\rho_{AC})$, where $S(\rho)=-{\rm Tr}(\rho\ln\rho)$ is the von Neumann entropy of the density matrix $\rho$, and $\rho_{R}$ is the reduced density matrix of the system in the region $R$.}} We always take anyons to be far away from each other, and the states $|a_1, \bar a_2\rangle$ and $|a_1, b_2, c_3; \mu\rangle$ can always be chosen to be short-range correlated, although there are also other anyon states violating this condition (see Appendix \ref{app: short-range correlation}). This condition will be important in the following.

\subsubsection{Low-energy, long-distance dynamics of anyons} \label{subsubsec: dynamics of anyons}

As discussed in Introduction, the key insight to develop the microscopic universal theory of symmetry-enriched TQSLs is to examine the low-energy, long-distance dynamics of anyons and the symmetry actions. Here we discuss the low-energy, long-distance dynamics of anyons, and we will turn to the symmetry actions next.

Broadly speaking, anyons exhibit three important types of dynamics, \ie motion, splitting (\ie the opposite process of fusion), and braiding. As a braiding process, say, between anyons $a$ and $b$, can be viewed as splitting a trivial anyon into $a$ and $\bar a$, moving $a$ around $b$, and fusing $a$ and $\bar a$ to a trivial anyon, the braiding properties should be encoded in the motion and splitting (and fusion) properties. So here we focus on the motion and splitting of anyons.

An important lesson from the previous studies of topological orders is that the motion of nontrivial anyons can be realized by applying a string-like moving operator to the anyons \cite{Levin2004, Kawagoe2019}. Concretely, the moving operator $M^a_{21}$ is an operator supported in a string-like region connecting positions $x_1$ and $x_2$, such that
\beq \label{eq: moving operator}
M_{21}^a|a_{1}, \bar a_{1'}\rangle\propto|a_{2}, \bar a_{1'}\rangle.
\eeq
Here and below ``$\propto$" means that its two sides can differ by a phase factor. Eq. \eqref{eq: moving operator} means that $M^a_{21}$ moves the anyon $a$ from position $x_1$ to position $x_2$, and we do not fix the phase factor between the two sides because it is often difficult to determine experimentally.

Moreover, if the anyon $a$ is a fusion product of anyons $b$ and $c$, $a$ can split into $b$ and $c$ via a splitting operator, which can be viewed as a junction of the moving operators of $a$, $b$ and $c$. Concretely, the splitting operator $(S_{a,\mu}^{bc})_{21}$ is supported in the region around positions $x_1$ and $x_2$, such that
\beq \label{eq: splitting operator}
(S^{bc}_{a, \mu})_{21}|a_{1}, \bar a_{1'}\rangle\propto|b_{1}, c_{2}, \bar a_{1'}; \mu\rangle.
\eeq
Namely, $(S^{bc}_{a, \mu})_{21}$ splits an anyon $a$ at position $x_1$ into an anyon $b$ at $x_1$ and an anyon $c$ at $x_2$, so that the fusion vertex basis state has index $\mu$. Note that this definition of the splitting operator is slightly simpler but different than that in Ref. \cite{Kawagoe2019}.

A few remarks are in order before ending this part.

\begin{itemize}
	
	\item We emphasize that $M^a_{21}$ and $(S_{a,\mu}^{bc})_{21}$ are chosen after the anyon states $|a_{1}, \bar a_{1'}\rangle$, $|a_{2}, \bar a_{1'}\rangle$ and $|b_{1}, c_{2}, \bar a_{1'}; \mu\rangle$ are given, and these operators are not unique.
	
	\item In a generic TQSL with a local Hamiltonian, operators like $M^a_{21}$ and $(S_{a,\mu}^{bc})_{21}$ are supported in the entire space. However, they are concentrated in regions around $x_1$ and $x_2$ with exponentially decaying tails (see Appendix \ref{subapp: exponentially decaying tails} for more detail). In this work, we always take anyons to be far away from each other and ignore these tails, \ie we view these operators as fully supported in the region around $x_1$ and $x_2$.
	
	\item To satisfy Eqs. \eqref{eq: moving operator} and \eqref{eq: splitting operator}, the moving and splitting operators do not have to be invertible, even for Abelian anyons. However, we always assume that they are bounded{\footnote{Namely, these operators have finite operator norms, \ie their largest singular values are finite.}}. Then for short-range correlated states with Abelian or non-Ablian anyons, these operators can be approximated by unitary operators (see Appendix \ref{subapp: unitary approximation}). In the following, we always take these operators to be unitary and ignore the errors in this approximation.
\end{itemize}

\subsubsection{Symmetry actions in a TQSL}\label{subsubsec:symmetrysetup}

Now we turn to symmetries. As discussed in Sec. \ref{subsec: anyon states}, an anyon state contains global information, such as the anyon types and fusion vertex basis states, and local information, such as the sizes, shapes and internal DOF of anyons. When a symmetry acts on an anyon state, say, $|a_{1}, b_{2}, \bar c_{3}; \mu\rangle$, it can have multiple effects: 1) changing the types and positions of the anyons, 2) changing the fusion vertex basis states , and  3) changing the local information. Therefore, the following symmetry localization property is expected \cite{Essin2012, barkeshli2014}:
\beq \label{eq: symmetry localization}
R_g|a_{1}, b_{2}, \bar c_{3}; \mu\rangle
\propto V_g^a(^gx_1)V_g^b(^gx_2)V_g^{\bar c}(^gx_3)
\cdot \sum_{\mu'}[U_g(^ga, ^gb; ^gc)]_{\mu\mu'}|^ga_{^g1}, ^gb_{^g2}, ^g\bar c_{^g3}; \mu'\rangle,
\eeq
where $g\in G$ is an element in the symmetry group $G$, $R_g$ is the action of the symmetry $g$ in the microscopic system, anyon $a$ ($b$ and $\bar c$) at $x_1$ ($x_2$ and $x_3$) are transformed into anyon $^g a$ ($^g b$ and $^g \bar c$) at $^gx_1$ ($^gx_2$ and $^gx_3$), the unitary matrix $U_g(^ga, ^gb; ^gc)$ describes the change of fusion vertex basis state, and $V_g^a$ ($V_g^b$ and $V_g^{\bar c}$) are local unitary operators supported around $^gx_1$ ($^gx_2$ and $^gx_3$), which accounts for the change of local information.

A few remarks are in order.

\begin{itemize}
	
	\item No matter whether $R_g$ is a unitary or anti-unitary symmetry, Eq. \eqref{eq: symmetry localization} is expected to hold. But, in general, it is only expected to hold approximately, with errors exponentially small in the distances between the anyons and the sizes of the supports of $V_g^{a,b,\bar c}$. We will ignore these errors.
	
	\item As shown in Appendix \ref{subapp: general symmetry localization}, when $|a_1, b_2, \bar c_3; \mu\rangle$ and $|^ga_{^g1}, ^gb_{^g2}, ^g\bar c_{^g3}; \mu'\rangle$ are all short-range correlated and all anyons are far away from each other, given $R_g$, the position $x_1$ and the local information around the anyon $a$ and $^ga$, $V_g^a(^gx_1)$ is essentially fixed, no matter what anyons $b$ and $c$ are and what local information $b$ and $c$ carry. Similarly, $V_g^b(^gx_2)$ ($V_g^{\bar c}(^gx_3)$) is insensitive to anyons other than $b$ and $^gb$ ($c$ and $^g\bar c$). Consequently, when $c$ is a trivial anyon, which means $b=\bar a$, Eq. \eqref{eq: symmetry localization} reduces to $R_g|a_1, \bar a_2\rangle\propto V_g^a(^gx_1)V_g^{\bar a}(^gx_2)|^ga_{^g1}, ^g\bar a_{^g2}\rangle$ with $V_g^a(^gx_1)$ the same as that in Eq. \eqref{eq: symmetry localization}, which describes symmetry actions on 2-anyon states.
    
    \item On top of $|a_{1}, b_{2}, \bar c_{3}; \mu\rangle$, if finitely many other anyons are created in locations far away from $x_{1,2,3}$, Eq. \eqref{eq: symmetry localization} implies a multi-anyon generalization of itself, given by Eq. \eqref{eq: symmetry localization for n-anyon states}. In particular, the same $V$ operators in Eq. \eqref{eq: symmetry localization} still describe the localized symmetry actions around $^ga$, $^gb$ and $^g\bar c$.
	
\end{itemize}

\subsection{SET phases of TQSLs}\label{subsec: SETphases}

In the above, we have discussed our microscopic setup and the basic physics of symmetric TQSLs. Our major assumption about the physical properties of a TQSL is that its low-energy excitations are anyons, \ie mobile quasi-particles falling into different superselection sectors, which can fuse and split in the manner described above. Unlike many previous studies, we do not assume that a topological phase is described by a category theory or its variants. In fact, at this point, it is not even clear what it means for a topological phase to be described by a category-like theory, as the connection between the physical observables in a TQSL and concepts in the category-like theory has not been made precise. In Sec. \ref{sec:framework}, we will extract universal properties from microscopically defined physical observables, and thus illustrate how the category-like mathematical structure emerges.

To this end, we need to first refine the notion of SET phases of TQSLs, so that we can determine which properties are universal, \ie common among all systems in an entire SET phase. As discussed in Introduction, two symmetric TQSLs are in different SET phases if they cannot (can) be adiabatically connected in the presence (absence) of the symmetry. However, it is often useful to examine if two systems with different types of microscopic DOF, \eg a spin-1/2 system and a spin-1 system, are in the same SET phase. Therefore, we will use the following definition of SET phases.

First, consider two systems with gapped local Hamiltonians $H_1$ and $H_2$, respectively. They are viewed as in the same topological phase if by stacking each of them with an invertible phase, which can be thought of as a gapped phase without anyons, the combined systems are adiabatically connected. Namely, the two systems are in the same topological phase if and only if there are two invertible phases with local Hamiltonians $H_1'$ and $H_2'$, respectively, and a family of gapped local Hamiltonians that continuously interpolate between $H_1+H_1'$ and $H_2+H_2'$. Allowing stacking additional DOF allows us to examine whether two systems with different microscopic DOF are in the same phase, and including these DOF in the consideration is also physically motivated, since most real systems are accompanied by such DOF, which can be, \eg nuclear spins.

In the above, no symmetry is involved. Suppose that the two systems with Hamiltonians $H_{1,2}$ are in the same topological phase, then they are in the same SET phase if and only if these two systems, the above invertible phases with Hamiltonians $H'_{1,2}$, and the entire continuous interpolating family of gapped local Hamiltonians all have the same symmetry group.

We remark that our definition may differ from some other studies, where the stacked phases are required to be (symmetric) product states. Here we allow the stacked phases to be symmetric invertible phases for two reasons. First, in the context of TQSLs, we would like to focus on properties related to anyons, so all intertible phases are regarded as a single trivial phase in this perspective. Second, in the presence of lattice symmetries, sometimes no symmetric product state exists, while symmetric invertible states exist \cite{Li2024}.

A useful fact about the SET phases is the quasi-adiabatic continuation theorem \cite{Hastings2005, Hastings2010, Bachmann2012automorphic, Kapustin2022Noether}. Namely, suppose $|\psi_{1,2}\rangle$ is a ground state of $H_{1,2}+H_{1,2}'$, and $M_{1,2}$ ($S_{1,2}$) is some moving (splitting) operator in the first (second) system, then there exists a symmetric unitary locality-preserving operator $W$, such that
\beq \label{eq: quasi-adiabatic continuation 1}
W|\psi_1\rangle=|\psi_2\rangle,
WM_1W^\dag=M_2,
WS_1W^\dag=S_2.
\eeq
As all anyon states can be obtained by applying a set of splitting and moving operators to a ground state, Eq. \eqref{eq: quasi-adiabatic continuation 1} implies that for an anyon state $|\psi_{1}'\rangle$ in the first system, there is a corresponding anyon state $|\psi_2'\rangle=W|\psi_1'\rangle$ in the second system with the same types of anyons and fusion vertex basis states. Substituting this relation into Eq. \eqref{eq: symmetry localization}, we see that, given a localized symmetry operator $V_1$ for the first system, 
\beq \label{eq: quasi-adiabatic continuation 2}
V_2=WV_1W^\dag
\eeq
is a localized symmetry operator for the second system. In Sec. \ref{sec:framework}, we will use Eqs. \eqref{eq: quasi-adiabatic continuation 1} and \eqref{eq: quasi-adiabatic continuation 2} to show that all universal data we identify from the microscopically defined quantities are invariant under the action of $W$, which means that such data indeed characterizes universal properties of the entire SET phase. 

\section{Microscopic universal theory}\label{sec:framework}

Having introduced our setup and the relevant background, in this section we present the microscopic universal theory of symmetry-enriched TQSLs, which characterizes the universal properties of TQSLs via micrscopically defined quantities. The key features of the theory are as follows.

\begin{itemize}
	
	\item The input of the theory is the microscopically defined anyon states, moving and splitting operators, and the $V$ operators in the symmetry localization property (Eq. \eqref{eq: symmetry localization}), which are introduced in Sec. \ref{sec:setup}. Such microscopic information is often explicitly available in solvable models and experimental realizations in quantum processors \cite{Kitaev1997, Levin2004, Kitaev2006, Heinrich2016, Cheng2016, Verresen2020, Satzinger2021, Semeghini2021, Iqbal2023, Iqbal2023a, Foss-Feig2023}. For generic models, methods for extracting such information were also proposed \cite{Shi2019, Cian2022}.
	
	\item From the above microscopic information, a set of data will be identified, which is manifestly invariant under the action of quasi-adiabatic continuation in Eqs. \eqref{eq: quasi-adiabatic continuation 1} and \eqref{eq: quasi-adiabatic continuation 2}. Such data will be a representative of the ultimate universal data.
	
	\item Given a microscopic realization of a TQSL, as discussed in Sec. \ref{sec:setup}, there are different choices for the microscopic input, and thus different values for the above representative data. Different sets of representative data are related by some ``gauge transformations". The general mathematical form of the gauge transformations defines an equivalence relation between the representative data, \ie two sets of representative data are equivalent if they are related by such gauge transformations. The universal data is the equivalence class which the representative data belongs to, and two TQSLs are in the same SET phase if they have identical universal data.

    \item {\it Only part of} the universal data can be organized as some generalization of category theory, at least in the standard sense, and the full universal data contains additional information beyond the category theory. Nevertheless, this additional data does not affect the classification of SET phases, \ie the classification of the SET phases is still given by the classification of some generalization of categories. The demonstration of this additional data clarifies what it means for a TQSL to be described by a category-like theory.
	
\end{itemize}

In the rest of this section, we will discuss how to obtain the representative data from the microscopic input and how to characterize the equivalence relation between the representatives. Due to the complexity of the theory, we divide our discussion into multiple parts. First, in Sec. \ref{subsec: grading parameters}, we discuss how to organize the microscopic data. With this organization, the structure of the universal data becomes more clear. Second, in Sec. \ref{subsec:FR}, we discuss the representative data and their equivalence relation for TQSLs without symmetries, or with their symmetries omitted in the consideration. Then in Sec. \ref{subsec: including symmetries}, we discuss the representative data and their equivalence relation for TQSLs with general symmetries, which can include both internal and lattice symmetries. Finally, in \ref{subsec: summarizing universal data}, we summarize the structure of the universal data we derive.

Our discussions generalize the considerations in Refs. \cite{Kawagoe2019, barkeshli2014, Cheng2020} in multiple directions. In particular, our philosophy is to take a physics-based approach without assuming that the universal properties of TQSLs are described by certain category-like theory, while Refs. \cite{Kawagoe2019, barkeshli2014, Cheng2020} based their discussions on such assumptions. In fact, we identify universal properties not captured by the category-like theory in the standard sense. Also, unlike the present paper, Refs. \cite{Kawagoe2019, barkeshli2014, Cheng2020} do not discuss general lattice symmetries.

\subsection{Organizing the input microscopic data} \label{subsec: grading parameters}

In our microscopic universal theory, the input data consists of 1) a set of positions to put anyons, 2) a set of states with anyons, and 3) a set of moving and splitting operators and a set of $V$ operators as in Eq. \eqref{eq: symmetry localization}. According to the discussion in Sec. \ref{sec:setup}, this input data is not unique. Not surprisingly, when we use different input data to calculate the universal data, we will get different results, and the ultimate universal data contains all these different results packaged together as a whole. To better illustrate the structure of the universal properties of SET phases of TQSLs, here we discuss how we organize our microscopic input data. In particular, we will introduce the important parameters $\gamma$, $\Gamma$ and $\omega$ in Eqs. \eqref{eq: gamma main}, \eqref{eq: Gamma main} and \eqref{eq: omega main}, respectively.

To start, note that from the definition of the moving operators, we have
\beq \label{eq: gamma main}
M_{32}^a M_{21}^a|a_{1}, \bar a_0\rangle &= \gamma^a_{321}M^a_{31}|a_{1}, \bar a_0\rangle,
\eeq
where the position $x_3$ may or may not be the same as the position $x_1$, and $\gamma^a_{321}$ is a phase factor. Roughly speaking, these $\gamma$ phase factors describe whether different moving operators can be seamlessly connected.

Also, from the definition of the moving and splitting operators, there is a unitary matrix $(\Gamma^{ab}_c)_{2'1'21}$, such that
\beq \label{eq: Gamma main}
(S^{ab}_{c;\mu})_{21}\ket{c_1,\bar c_0}
=\sum_{\mu'}[(\Gamma^{ab}_c)_{2'1'21}]_{\mu\mu'}M^b_{22'}M^a_{11'}(S^{ab}_{c;\mu'})_{2'1'}M^c_{1'1}\ket{c_{1},\bar c_0}.
\eeq
Here the subscript ``$2'1'21$" in $(\Gamma^{ab}_c)_{2'1'21}$ keeps track of the supports of the splitting operators on the two sides of this equation, and this equation holds because both sides are states with anyons $a$, $b$ and $\bar c$ at positions $x_1$, $x_2$ and $x_0$, respectively, with the same local information around these anyons. Note that we allow the fusion vertex basis states indexed by $\mu$ and $\mu'$ to be different sets of orthonormal basis states. Roughly speaking, these $\Gamma$ matrices describe how to transport a splitting operator from one location to another location by moving operators.

Moreover, as shown in Appendix \ref{app: approximate factorization}, there is a phase factor $(\omega^a_g)_{12}$, such that the following equation holds when all anyons are far apart:
\beq \label{eq: omega main}
R_g M^{\lsupsc{\bar g}a}_{\lsupsc{\bar g}1\lsupsc{\bar g}2} R_g^{-1}V^{\lsupsc{\bar g}a}_g(x_2)\ket{a_{2},\bar a_0}
=(\omega^a_{g})_{12} V^{\lsupsc{\bar g}a}_g(x_1)M^{a}_{12}\ket{a_{2},\bar a_0}.
\eeq
Here the subscript ``12" keeps track of the supports of the $V$ operators on the two sides of this equation. Roughly speaking, these $\omega$ phase factors describes how to transport the $V$ operators from one location to another location by moving operators.

By choosing the moving, splitting and $V$ operators appropriately, the above parameters $\gamma$ and $\omega$ can be arbitrary phase factors, and $\Gamma$ can be arbitrary unitary matrices (with the rank determined by the relevant fusion multiplicity). When we construct the representatives of universal data later, we will fix a set of positions to put anyons and fix these $\gamma$, $\Gamma$ and $\omega$ when we choose the moving, splitting and $V$ operators. Note that although the moving, splitting and $V$ operators in one TQSL are generally not valid moving, splitting and $V$ operators in another TQSL, even if they are in the same SET phase, any given set of $\gamma$, $\Gamma$ and $\omega$ can be defined for all TQSLs in an SET phase. After fixing $\gamma$, $\Gamma$ and $\omega$, the moving, splitting and $V$ operators are still not completely fixed. In the main text, we will mostly focus on the case where all $\gamma$, $\Gamma$ and $\omega$ are $\mathbbm{1}$. Namely, the moving, splitting and $V$ operators are chosen so that the following equations hold:
\beq \label{eq: seamless}
\begin{split}
	M_{32}^a M_{21}^a|a_{1}, \bar a_{1'}\rangle &= M^a_{31}|a_{1}, \bar a_{1'}\rangle,
\end{split}
\eeq
\beq \label{eq: simplifying condition 1}
(S^{ab}_{c;\mu})_{21}\ket{c_1,\bar c_{1'}}
=M^b_{22'}M^a_{11'}(S^{ab}_{c;\mu})_{2'1'}M^c_{1'1}\ket{c_{1},\bar c_{1'}},
\eeq
\beq \label{eq: simplifying condition 2}
R_g M^{\lsupsc{\bar g}a}_{\lsupsc{\bar g}1\lsupsc{\bar g}2} R_g^{-1}V^{\lsupsc{\bar g}a}_g(x_2)\ket{a_{2},\bar a_{1'}}
=V^{\lsupsc{\bar g}a}_g(x_1)M^{a}_{12}\ket{a_{2},\bar a_{1'}}.
\eeq
We remark that this choice of $\gamma$, $\Gamma$ and $\omega$ is by no means more special than other choices, from a microscopic perspective. At this point, it stands out merely because of its apparent simplicity. Later, we will see that the universal data obtained with this choice can be neatly organized as some generalization of category theory. However, in principle, representatives of the universal data can also be calculated using other choices of $\gamma$, $\Gamma$ and $\omega$, and the full structure of all representative universal data for all choices of $\gamma$, $\Gamma$ and $\omega$ will be described in Appendix \ref{app: full structure}.

\begin{figure}
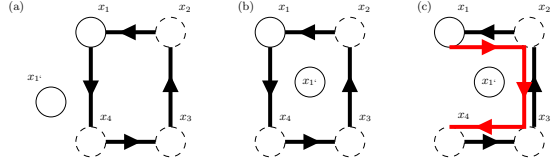

\centering
\scalebox{0.5}{\beyondA}
\caption{The solid circles represent the anyons in states $|a_{1}, \bar a_{1'}\rangle$ and $|a_{1}, \bar a_{1''}\rangle$. If moving operators $M^a_{21}$, $M^a_{32}$, $M^a_{43}$ and $M^a_{14}$ are supported around the black lines in panels (a) and (b) and if they satisfy Eq. \eqref{eq: seamless}, \ie $M^a_{14}M^a_{43}M^a_{32}M^a_{21}|a_{1}, \bar a_{1'}\rangle=M^a_{14}M^a_{41}|a_{1}, \bar a_{1'}\rangle=|a_{1}, \bar a_{1'}\rangle$, then the analogous equation $M^a_{14}M^a_{43}M^a_{32}M^a_{21}|a_{1}, \bar a_{1''}\rangle=|a_{1}, \bar a_{1''}\rangle$ is not expected to hold, as there can be an additional mutual braiding phase factor between $a$ and $\bar a$ when $M^a_{14}M^a_{43}M^a_{32}M^a_{21}$ acts on $|a_{1}, \bar a_{1''}\rangle$. To make both Eq. \eqref{eq: seamless} and this analogous equation hold, once $M^a_{21}$, $M^a_{32}$ and $M^a_{43}$ are taken to be supported around the black lines in panels (a) and (b), $M^a_{41}$ can be taken to be supported around the red line in panel (c).}
\label{fig:beyondA1A2}
\end{figure}

Before moving on, we note that if some moving operators satisfy Eq. \eqref{eq: seamless}, they may not satisfy the analogous equations with $|a_{1}, \bar a_{1'}\rangle$ replaced by $|a_{1}, \bar a_{1''}\rangle$ (see Fig. \ref{fig:beyondA1A2} (a, b)). However, when we consider anyon states and moving operators later, we will always choose the moving operators so that no nontrivial anyon is enclosed by connecting the moving operators (see, \eg Fig. \ref{fig:beyondA1A2} (c)). Then not only Eq. \eqref{eq: seamless}, but also its analogs with $|a_{1}, \bar a_{1'}\rangle$ replaced by another state with an anyon $a$ at $x_1$ (and some other anyons at other positions), hold.

\subsection{Omitting symmetries}\label{subsec:FR}

With the above setup, now we discuss the universal microscopic theory of TQSLs where symmetries are omitted. Suppose that the microscopic information, including the positions to put anyons, some anyon states, and some moving and splitting operators in Sec. \ref{sec:setup}, is given. Our goal is to identify the universal properties of the TQSL phase from this information. The quasi-adiabatic continuation theorem suggests that the types of anyons and the fusion multiplicity $N_{ab}^c$ in Eq. \eqref{eq: fusion} are universal \cite{Hastings2005}, so our aim is to identify other universal properties. As discussed in Introduction and Sec. \ref{sec:setup}, such universal properties should be encoded in the motion and splitting processes of anyons, which contain the anyons' fusion and braiding properties.

It is natural that some properties associated with braiding, such as the self-statistics of the anyons, are universal. But besides $N_{ab}^c$, what other fusion properties are universal? Note that $N_{ab}^c$ is associated with a single fusion process, and we need to consider multiple successive fusion processes to find other universal properties related to fusion. For example, if anyons $a$, $b$ and $c$ can fuse to $d$, one may first fuse $a$ and $b$ to get anyon $e$ and then fuse $e$ and $c$ to get $d$, or first fuse $b$ and $c$ to get anyon $f$ and then fuse $a$ and $f$ to get $d$. There must be nontrivial relations between the intermediate states in these two processes, and some universal properties can be encoded in such relations.

Below we make these ideas concrete.

\subsubsection{Representative of universal data} \label{subsubsec: representative FR}

{\it\quad\quad\quad\quad\quad\quad\quad\quad\quad\quad\quad\quad\quad\quad\quad\quad\quad\quad\underline{Fusion and splitting}}\\

We begin with the consideration of the two processes of fusing anyons $a$, $b$ and $c$ into $d$. It turns out to be more convenient to think of the opposite processes, where anyon $d$  splits into $a$, $b$ and $c$. Concretely, we start with a state $|d_{0}, \bar d_{0'}\rangle$ and apply some moving and splitting operators to it, which leads to the two states below and the intermediate steps are precisely the above two splitting processes:

\begin{equation} \label{eq: general F states main}
	\bsp
	&\ket{e, \mu, \nu}=M^{b}_{64}M^{a}_{52}(S^{ab}_{e,\mu})_{42}M^e_{21}M^{c}_{73}(S^{ec}_{d,\nu})_{31}M^d_{10}\ket{d_0, \bar d_{0'}},\\
	&\ket{f, \rho, \sigma}=M^{c}_{74'}M^{b}_{63'}(S^{bc}_{f,\rho})_{4'3'}M^f_{3'2'}M^{a}_{51'}(S^{af}_{d,\sigma})_{2'1'}M^d_{1'0}\ket{d_0, \bar d_{0'}}.
	\esp
\end{equation}
Here $x_0, x_1, x_2, x_3, x_4, x'_1, x'_2, x'_3, x'_4, x_5, x_6, x_7$ are a set of 12 positions to put anyons, some of which can be identical, as long as in all intermediate states no two anyons are close to each other. These two states are expressed in terms of graphics as follows:
\begin{equation} \label{eq: F states graph}
	\scalebox{0.7}{\generalF}.
\end{equation}
These graphs represent the spacetime trajectories corresponding to the processes that create the two states, each red wavy line represents a splitting process, each tilted black solid line represents a moving process, each vertical black solid line can be viewed as identity operator, and each dashed line indicates a position.

The processes considered here belong to the most general class of processes where anyon $d$ splits in the two different ways described above, in the sense that the splitting processes can happen at different locations, and the anyons can move freely at every stage before they reach the final state. At the same time, these processes are the simplest ones in these most general processes, in the sense that the motion of anyons before and after each splitting process is achieved by applying a single moving operator, rather than multiple moving operators. This simplicity plays no role in the main text because Eq. \eqref{eq: seamless} is assumed here, but it simplifies the analysis in Appendix \ref{app: full structure} (see Appendix \ref{app: review Kawagoe-Levin} for a brief review). When some of the 12 positions we choose coincide with each other, the states in Eq. \eqref{eq: general F states main} reduce to their analogs in Ref. \cite{Kawagoe2019}. In this case, all splitting processes occur at the same location. The motivation for Ref. \cite{Kawagoe2019} to define two such states is to simplify the procedure to identify a presumed category theory that describes the topological phase. However, we do not assume that the topological phase is described by any category-like theory. Instead, we look for universal properties encoded in microscopic quantities, so we consider these general processes.

The relation between the states in Eq. \eqref{eq: general F states main} is captured by a matrix $[F^{abc}_d]_{\{e, \mu, \nu\}, \{f, \rho, \sigma\}}$, where $\{e, \mu, \nu\}$ is the row index and $\{f, \rho, \sigma\}$ is the column index:
\begin{equation}\label{Eq:generalF main}
	[F^{abc}_d]_{\{e, \mu, \nu\}, \{f, \rho, \sigma\}}=\langle f, \rho, \sigma| e, \mu, \nu\rangle.
\end{equation} 
In Eq. \eqref{eq: general F states main}, we choose the moving and splitting operators so that both $\ket{e, \mu, \nu}$ and $\ket{f, \rho, \sigma}$ have anyon $a$ at $x_5$, $b$ at $x_6$ and $c$ at $x_7$, and both states have the same local information around the anyons, so by varying $e, \mu, \nu$ the set of states $\{|e, \mu, \nu\rangle\}$ form an orthonormal basis of the degenerate subspace of states with anyons $a$, $b$ and $c$ (which fuse to $d$), and the same applies to the set of states $\{\ket{f, \rho, \sigma}\}$. Therefore, $F^{abc}_d$ is unitary. Moreover, as shown in Appendix \ref{subsubapp: F relation}, when $\gamma=\Gamma=\omega=\mathbbm{1}$, the $F$-matrices satisfy the so-called pentagon equation:
\beq \label{eq: pentagon equation}
\sum_\delta[F^{fcd}_e]_{\{g, \beta, \gamma\}, \{l, \nu, \delta\}}[F^{abl}_e]_{\{f, \alpha, \delta\}, \{k, \mu, \lambda\}}=\sum_{h, \sigma, \psi, \rho}[F^{abc}_g]_{\{f, \alpha, \beta\}, \{h, \psi, \sigma\}}[F^{ahd}_e]_{\{g, \sigma, \gamma\}, \{k, \rho, \lambda\}}[F^{bcd}_k]_{\{h, \psi, \rho\}, \{l, \nu, \mu\}}.
\eeq

Importantly, under the quasi-adiabatic continuation in Eq. \eqref{eq: quasi-adiabatic continuation 1}, $|e, \mu, \nu\rangle\rightarrow W|e, \mu, \nu\rangle$ and $|f, \rho, \sigma\rangle\rightarrow W|f, \rho, \sigma\rangle$, so the $F$-matrix in Eq. \eqref{Eq:generalF main} is invariant. One might therefore identify the $F$-matrices as the universal data that characterizes a topological phase. However, this is only partially appropriate, since the precise value of the $F$-matrix depends on many subjective choices, including the positions we choose to put the anyons, the local information around anyons $d$ and $\bar d$ in the state $\ket{d_{0}, \bar d_{0'}}$ and the specific choices of the moving and splitting operators in Eq. \eqref{eq: general F states main}. In Sec. \ref{subsubsec: equivalence of FR}, we will discuss how different sets of $F$-matrices in a given TQSL are related by ``gauge transformations", \ie how the $F$-matrices transform when the subjective choices change. These gauge transformations organize the possible $F$-matrices into equivalence classes, such that two sets of $F$-matrices are equivalent if they are related by such gauge transformations. The equivalence class that the $F$-matrices belong to should be identified as the targeted universal data, and a particular set of $F$-matrices is just a representative in the equivalence class.

One may ask whether we can also take, say, equivalence classes of $\langle d_{2}, \bar d_{0'}|M^d_{21}|d_{1}, \bar d_{0'}\rangle$ as universal data of a topological phase, since it is also invariant under the quasi-adiabatic continuation in Eq. \eqref{eq: quasi-adiabatic continuation 1}. Note that given states $|d_{1}, \bar d_{0'}\rangle$ and $|d_{2}, \bar  d_{0'}\rangle$, we can always choose the phase of $M^d_{21}$ so that $\langle d_{2}, \bar d_{0'}|M^d_{21}|d_{1}, \bar d_{0'}\rangle=1$. So for all TQSLs and all anyon $d$, there is only one such equivalence class, which can be represented by 1. Therefore, this data is universal in a trivial way, but it is not useful for characterizing topological phases. Similarly, if $O$ is a generic microscopic operator that is unrelated to the moving and splitting operators (\eg $O$ may be a Pauli operator on a qubit), $O_W=WOW^\dagger$, and $|\psi_W\rangle=W|\psi\rangle$ with $|\psi\rangle$ a ground state of a system in the topological phase, then although $\langle\psi_W|O_W|\psi_W\rangle$ is invariant under the quasi-adiabatic continuation in Eq. \eqref{eq: quasi-adiabatic continuation 1}, no nontrivial universal property can be extracted from it (see Appendix \ref{app: universal}). In the following, we will not consider such trivially universal properties. In contrast, the equivalence class of the $F$-matrices can be different for different TQSLs, so it is useful universal data that characterizes the phases of TQSLs.

One may also ask whether there are other universal properties related to fusion and splitting, in addition to the equivalence class of $F$-matrices (and the fusion rules in Eq. \eqref{eq: fusion}). For example, one may consider splitting an anyon to more than three anyons and look for universal properties associated with such processes. However, we believe that all such properties are already included in the equivalence class of $F$-matrices, since the processes leading to the $F$-matrices, \ie splitting an anyon into three anyons, are the building blocks of these more complicated processes.\\

{\it\quad\quad\quad\quad\quad\quad\quad\quad\quad\quad\quad\quad\quad\quad\quad\quad\quad\quad\quad\quad\quad\underline{Braiding}}\\

Next, we turn to braiding. It is expected that both the half-braiding exchange statistics of two anyons of the same type and the full braiding statistic between two anyons of different types contain universal properties. To capture both kinds of processes in a unified manner, we look for universal properties encoded in half-braiding exchange processes, no matter whether the two anyons are of the same type or not.

\begin{figure*}[h!]
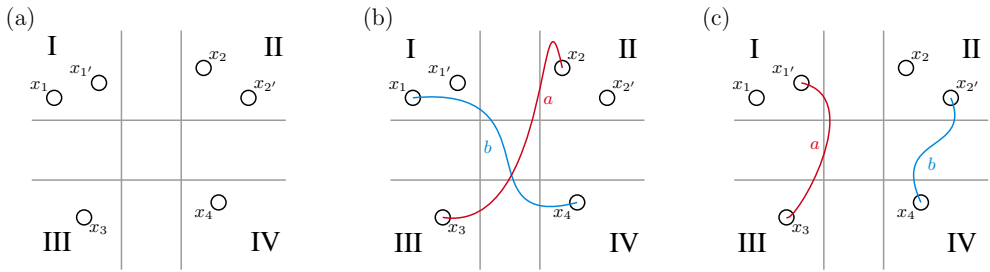

	\centering
	\scalebox{0.75}{\generalRillustrate}
	\caption{(a) The $4$ regions I, II, III, and IV are all outside of the two strips, such that moving along  {II}-{I}-{III}-{IV} should wind around the overlapping region of the strips in a counter-clockwise manner.
		(b) and (c): The two processes differ by a braiding.}
	\label{fig:generalRillustrate main}
\end{figure*}

For such exchange processes of anyons $a$ and $b$ in real experiments, $a$ ($b$) may not move to exactly the original position of $b$ ($a$), but some position close to it. To describe such realistic exchange processes, we first use two orthogonal strips to divide the space into multiple regions (see Fig. \ref{fig:generalRillustrate main}(a)). We require the strips to be wide enough to contain the supports of the moving and splitting operators of anyons. We then choose the following positions to put the anyons: two positions $x_1$ and $x_{1'}$ in region {I}, two positions $x_2$ and $x_{2'}$ in region {II}, one position $x_3$ in region {III}, one position $x_4$ in region {IV}, and one position $x_0$ that can be anywhere. There is no constraint on the relative position of $x_1$ and $x_{1'}$ ($x_2$ and $x_{2'}$).

Next, we consider the following two states{\footnote{Here if we start with a pair of states $|b_1, a_2, \bar c_{0'}; \mu\rangle$ and $|a_1, b_2, \bar c_{0'}; \nu\rangle$, instead of a common initial state $|c_0, \bar c_{0'}\rangle$, we will get some trivially universal data that cannot be used to characterize the underlying topological phase.}}:
\begin{equation} \label{eq: R states general}
	\bsp
	&\ket{\mu}=M^a_{32}M^b_{41}(S^{ba}_{c,\mu})_{21}M^c_{10}\ket{c_0,\bar c_{0'}},\\
	&\ket{\nu}=M^b_{42'}M^a_{31'}(S^{ab}_{c,\nu})_{2'1'}M^c_{1'0}\ket{c_0,\bar c_{0'}},
	\esp
\end{equation}
where the moving and splitting operators above satisfy the following conditions. The supports of the splitting operator $(S^{ba}_{c, \mu})_{21}$ and $(S^{ab}_{c, \nu})_{2'1'}$ should not overlap with regions {III} and {IV}, the support of the moving operator $M^b_{41}$ ($M^a_{32}$) is required to not overlap with regions {II} and {III} (regions {I} and {IV}), and the support of the moving operator $M^b_{42'}$ ($M^a_{31'}$) should not overlap with regions {I} and {III} (regions {II} and {IV}). As such, in the two processes encoded in Eq. \eqref{eq: R states general}, the anyon $c$ splits into $a$ and $b$ in two different ways, and, relative to the processes in $|\nu\rangle$, in the processes in $|\mu\rangle$ the anyons $a$ and $b$ exchange their positions in a counter-clockwise manner. Using graphic presentations similar to Eq. \eqref{eq: F states graph}, this position exchange is more obvious:
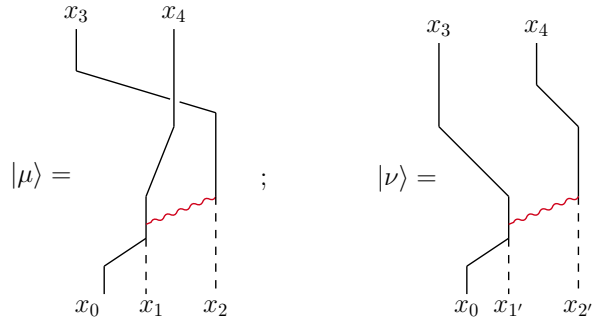
\begin{equation*}
	\scalebox{0.7}{\generalR}
\end{equation*}

The relation between the two states in Eq. \eqref{eq: R states general} is captured by a matrix $[R^{ab}_c]_{\mu\nu}$, where the $\mu$ is the row index and $\nu$ is the column index:
\begin{equation}\label{eq: general R main}
	[R^{ab}_c]_{\mu\nu}=\langle\nu|\mu\rangle.
\end{equation}
Similar to the $F$-matrices, our $R$-matrices are again different from the $R$-matrices in Ref. \cite{Kawagoe2019} (see Appendix \ref{app: review Kawagoe-Levin} for a brief review). In particular, in the definition of the $R$-matrices in Ref. \cite{Kawagoe2019}, all splitting processes occur at the same location, which is not required here.

Again, in Eq. \eqref{eq: R states general} the moving and splitting operators are chosen so that both $|\mu\rangle$ and $|\nu\rangle$ have anyon $a$ at $x_3$ and anyon $b$ at $x_4$, and both states have the same local information around the anyons, so by varying $\mu$ the set of states $\{\ket{\mu}\}$ form an orthonormal basis of the degenerate subspace of states with anyons $a$ and $b$ (which fuse to $c$), and the same applies to the set of states $\{\ket{\nu}\}$. Therefore, the $R$-matrix in Eq. \eqref{eq: general R main} is unitary. Moreover, as shown in Appendix \ref{subsubapp: R relations}, when $\gamma=\Gamma=\omega=\mathbbm{1}$, our $F$-matrices defined in Eq. \eqref{Eq:generalF main} and $R$-matrices defined in Eq. \eqref{eq: general R main} satisfy the so-called hexagon equations:
\beq \label{eq: hexagon equations}
\begin{split}
	\sum_{\lambda, \gamma}[R^{ac}_e]_{\alpha\lambda}[F^{acb}_d]_{\{e, \lambda, \beta\}, \{g,\gamma,\nu\}}[R^{bc}_g]_{\gamma\mu}&=\sum_{f, \sigma, \delta, \psi}[F^{cab}_d]_{\{e, \alpha, \beta\}, \{f, \delta, \sigma\}}[R^{fc}_d]_{\sigma\psi}[F^{abc}_d]_{\{f, \delta, \psi\},\{g, \mu, \nu\}},\\
	\sum_{\lambda, \gamma}[(R^{ca}_e)^{-1}]_{\alpha\gamma}[F^{acb}_d]_{\{e, \lambda, \beta\},\{g, \gamma, \nu\}} [(R^{cb}_g)^{-1}]_{\gamma\mu}&=\sum_{f, \sigma, \delta, \mu}[F^{cab}_d]_{\{e,\alpha,\beta\},\{f,\delta, \sigma\}}[(R^{cf}_d)^{-1}]_{\sigma\psi}[F^{abc}_d]_{\{f,\delta,\psi\},\{g,\mu, \nu\}}.
\end{split}	
\eeq

Importantly, under the quasi-adiabatic continuation in Eq. \eqref{eq: quasi-adiabatic continuation 1}, $|\mu\rangle\rightarrow W|\mu\rangle$ and $|\nu\rangle\rightarrow W|\nu\rangle$, so the $R$-matrices in Eq. \eqref{eq: general R main} are invariant, just as the $F$-matrices in Eq. \eqref{Eq:generalF main}. Similarly, the $R$-matrix depends on many subjective choices, \ie the positions to put the anyons, the local information around anyons $c$ and $\bar c$ in the state $\ket{c_{1}, \bar c_{0'}}$, and the specific choices of the moving and splitting operators in Eq. \eqref{eq: R states general}. In Sec. \ref{subsubsec: equivalence of FR}, we will discuss how different sets of $R$-matrices in a given TQSL are related by gauge transformations, and how these gauge transformations define equivalence classes of $R$-matrices. The equivalence class that the $R$-matrices belong to is the targeted universal data related to braiding, and a particular set of $R$-matrices is a representative in this equivalence class.

\subsubsection{Equivalence relation between representative data} \label{subsubsec: equivalence of FR}

Now we discuss the equivalence relations between the representatives of the universal data, \ie the $F$- and $R$-matrices defined in Eqs. \eqref{Eq:generalF main} and \eqref{eq: general R main}, respectively. As discussed in Sec. \ref{subsubsec: representative FR}, these equivalence relations originate from the intrinsic non-uniqueness of the positions and local information of the anyons, and the moving and splitting operators involved in Eqs. \eqref{eq: general F states main} and \eqref{eq: R states general}. We will see that these equivalence relations can be organized using some ``gauge transformations", and the equivalence class that a set of $F$- and $R$-matrices belong to will be universal data that characterizes the quantum phase of TQSLs. Moreover, this universal data has the mathematical structure of a unitary modular tensor category.

Here we focus on the simplest case, where the positions and local information of the anyons are fixed. We will show that, when the moving and splitting operators in Eqs. \eqref{eq: general F states main} and \eqref{eq: R states general} are changed, the $F$- and $R$-matrices will transform according to Eq. \eqref{eq: vertex basis gauge transformation}. In Appendix \ref{app: full structure}, we show that if not only the moving and splitting operators, but also the positions and local information of the anyons are changed, the transformations of the $F$- and $R$-matrices are still given by Eq. \eqref{eq: vertex basis gauge transformation}.

To start, note that fixing the positions and local information of the anyons puts severe constraints on the possible moving and splitting operators, \ie if $M^{a}_{21}$ and $\tilde M^{a}_{21}$ are two moving operators for the anyon state $\ket{a_{1},\bar a_{0'}}$, and $(S^{ab}_{c, \mu})_{2, 1}$ and $(\tilde S^{bc}_{a, \mu'})_{2, 1}$ are two splitting operators for this state, then
\begin{equation} \label{eq: new moving and splitting operators}
	\bsp
	&\tilde M^a_{21}\ket{a_{1},\bar a_{0'}}= e^{i\phi_{21}^a} M^a_{21}\ket{a_{1},\bar a_{0'}},\\
	&(\tilde S^{bc}_{a,\mu'})_{21}\ket{a_{1},\bar a_{0'}}
	=\sum_{\mu}[(\Omega_0(b,c;a))_{21}]_{\mu'\mu}(S^{bc}_{a,\mu})_{21}\ket{a_{1},\bar a_{0'}}.
	\esp
\end{equation}
Namely, choosing a different moving operator induces a phase $\phi_{21}^a$, and choosing a different splitting operator induces a unitary matrix $\Omega_0(b, c; a)$. When $\Omega_0(b, c; a)$ is non-diagonal, it induces a change of the fusion vertex basis states. Note that, in Eqs. \eqref{eq: general F states main} and \eqref{eq: R states general}, the moving and splitting operators are applied to states with anyons other than their targeted anyons. For example, in the first equation of Eq. \eqref{eq: general F states main}, although the targeted anyon of the moving operator $M^c_{73}$ is $c$, $M^c_{73}$ is applied to a state with anyons $c$, $e$ and $\bar d$. However, as long as the other anyons are far away from the targeted anyons, the effects induced by different choices of the moving and splitting operators only depend on the targeted anyons of these operators, but not the other anyons. For example, if $\tilde M^c_{73}|c_{3}, \bar c_{0'}\rangle=e^{i\phi_{73}^c}M_{73}^c|c_{3}, \bar c_{0'}\rangle$, then $\tilde M^c_{73}(S^{ec}_{d, \nu})_{31}|d_{1}, \bar d_{0'}\rangle=e^{i\phi_{32}^c}M^c_{73}(S^{ec}_{d, \nu})_{31}|d_{1}, \bar d_{0'}\rangle$.{\footnote{To see it, note that $\langle d_1, \bar d_{0'}|(S^{ec}_{d,\nu})_{31}^\dagger \tilde M^c_{73} M^c_{73}(S^{ec}_{d, \nu})_{31}|d_1, \bar d_{0'}\rangle=\langle e_1, c_3, \bar d_{0'}; \nu|\tilde M^c_{73} M^c_{73}|e_1, c_3, \bar d_{0'}; \nu\rangle=\langle c_3, \bar c_{0'}|(S^{\bar de}_{\bar c,\nu})_{10'}^\dagger \tilde M^c_{73} M^c_{73}(S^{\bar de}_{\bar c,\nu})_{10'}|c_3, \bar c_{0'}\rangle\approx \langle c_3, \bar c_{0'}|\tilde M^c_{73} M^c_{73}|c_3, \bar c_{0'}\rangle\langle c_3, \bar c_{0'}|(S^{\bar de}_{\bar c,\nu})_{10'}^\dagger (S^{\bar de}_{\bar c,\nu})_{10'}|c_3, \bar c_{0'}\rangle=\langle c_3, \bar c_{0'}|\tilde M^c_{73} M^c_{73}|c_3, \bar c_{0'}\rangle$, where the condition of short-range correlation has been used.}} Moreover, Eqs. \eqref{eq: seamless} and \eqref{eq: simplifying condition 1} put constraints on the phases $\phi_{21}^a$ and matrices $\Omega_0(b, c; a)$ at different locations:
\begin{equation}\label{eq: constraining transformation}
\bsp
    &e^{i\phi^a_{32}}e^{i\phi^a_{21}}=e^{i\phi^a_{31}},\quad e^{i\phi^a_{21}}=e^{-i\phi^a_{12}},\\
    &[(\Omega_0(a,b;c))_{21}]_{\mu\mu'}=[(\Omega_0(a,b;c))_{2'1'}]_{\mu\mu'}\frac{e^{i\phi^a_{11'}}e^{i\phi^b_{22'}}}{e^{i\phi^c_{11'}}}.
\esp
\end{equation}

Therefore, replacing the moving and splitting operators in Eqs. \eqref{eq: general F states main} and \eqref{eq: R states general} by new ones, the relevant states transform as
\begin{widetext}
	\beq
	\bsp
	|e, \mu, \nu\rangle
    &\rightarrow e^{i(\phi^b_{64}+\phi^a_{52}+\phi^e_{21}+\phi^c_{73}+\phi^d_{10})}\sum_{\mu'\nu'}[(\Omega_0(a, b; e))_{42}]_{\mu\mu'}[(\Omega_0(e, c; a))_{31}]_{\nu\nu'}|e, \mu', \nu'\rangle\\
    &=e^{i(\phi^b_{62}+\phi^a_{51}+\phi^c_{72}+\phi^d_{10})}\sum_{\mu'\nu'}e^{\phi^b_{12}}[(\Omega_0(a, b; e))_{21}]_{\mu\mu'}e^{\phi^c_{12}}[(\Omega_0(e, c; a))_{21}]_{\nu\nu'}|e, \mu', \nu'\rangle,\\
	|f, \rho, \sigma\rangle
    &\rightarrow e^{i(\phi^c_{74'}+\phi^b_{63'}+\phi^f_{3'2'}+\phi^a_{51'}+\phi^d_{1'0})}\sum_{\rho'\sigma'}[(\Omega_0(b, c; f))_{4'3'}]_{\rho\rho'}[(\Omega_0(a, f; d))_{2'1'}]_{\sigma\sigma'}|f, \rho', \sigma'\rangle\\
    &=e^{i(\phi^b_{62}+\phi^a_{51}+\phi^c_{72}+\phi^d_{10})}\sum_{\rho'\sigma'}e^{\phi^c_{12}}[(\Omega_0(b, c; f))_{21}]_{\rho\rho'}e^{\phi^f_{12}}[(\Omega_0(a, f; d))_{21}]_{\sigma\sigma'}|f, \rho', \sigma'\rangle,\\
	|\mu\rangle
    &\rightarrow e^{i(\phi^a_{32}+\phi^b_{41}+\phi^c_{10})}\sum_{\mu'}[(\Omega_0(b, a; c))_{21}]_{\mu\mu'}|\mu'\rangle\\
    &=e^{i(\phi^a_{32}+\phi^b_{41}+\phi^c_{10})}\sum_{\mu'}e^{i\phi^a_{12}}[(\Omega_0(b, a; c))_{21}]_{\mu\mu'}|\mu'\rangle,\\
	|\nu\rangle
    &\rightarrow e^{i(\phi^b_{42'}+\phi^a_{31'}+\phi^c_{1'0})}\sum_{\nu'}[(\Omega_0(a, b; c))_{2'1'}]_{\nu\nu'}|\nu'\rangle\\
    &=e^{i(\phi^a_{32}+\phi^b_{41}+\phi^c_{10})}\sum_{\nu'}e^{i\phi^b_{12}}[(\Omega_0(a, b; c))_{21}]_{\nu\nu'}|\nu'\rangle,
	\esp
	\eeq
\end{widetext}
where Eq. \eqref{eq: constraining transformation} is used.

Consequently, there exist some unitary matrices $\Omega^{ab}_c$, such that the $F$- and $R$-matrices transform as
\begin{equation} \label{eq: vertex basis gauge transformation}
	\bsp
	&[F^{abc}_d]_{\{e,\mu,\nu\},\{f,\rho,\sigma\})}\to\sum_{\mu',\nu',\rho',\sigma'}[F^{abc}_d]_{\{e,\mu',\nu'\},\{f,\rho',\sigma'\}}\times[\Omega^{ab}_e]_{\mu\mu'}[\Omega^{ec}_d]_{\nu\nu'}[(\Omega^{bc}_f)^{-1}]_{\sigma'\sigma}[(\Omega^{af}_d)^{-1}]_{\rho'\rho},\\
	&[R^{ab}_c]_{\mu\nu}\to \sum_{\mu'\nu'}[R^{ab}_c]_{\mu'\nu'}[\Omega^{ba}_c]_{\mu\mu'}[(\Omega^{ab}_c)^{-1}]_{\nu'\nu},
	\esp
\end{equation}
In the present case, $\Omega^{ab}_c=e^{i\phi^b_{12}}(\Omega_0(a, b; c))_{21}$.

We use Eq. \eqref{eq: vertex basis gauge transformation} to define the equivalence classes of the $F$- and $R$-matrices. Concretely, two sets of $F$- and $R$-matrices are considered equivalent if they are related by a transformation of the form of Eq. \eqref{eq: vertex basis gauge transformation}. So each TQSL is characterized by an equivalence class of $F$- and $R$-matrices, and this equivalence class is universal in the entire topological phase because their representatives, \ie the $F$- and $R$-matrices, are invariant under quasi-adiabatic continuation, as discussed in Sec. \ref{subsubsec: representative FR}.

The universal data of TQSLs discussed so far can be organized into a mathematical structure known as a unitary modular tensor category (UMTC) \cite{Kitaev2006}, where the anyon types are the objects of the category, and the UMTC also has $F$- and $R$-matrices satisfying the same pentagon and hexagon equations as in Eqs. \eqref{eq: pentagon equation} and \eqref{eq: hexagon equations}, and subject to the same equivalence relation as in Eq. \eqref{eq: vertex basis gauge transformation}. In the UMTC literature, the $F$- and $R$-matrices are also called the $F$- and $R$-symbols, and Eq. \eqref{eq: vertex basis gauge transformation} is called a vertex basis gauge transformation. We will also use these terminologies.

We stress again that we did not start with the assumption that topological phases are described by UMTCs. In fact, before connecting physical observables in a TQSL with concepts in a UMTC as in Eqs. \eqref{Eq:generalF main} and \eqref{eq: general R main}, it is not even clear what it means for a topological phase to be described by a UMTC. Instead, we started with the physical assumption that the low-energy excitations of a TQSL are anyons, \ie mobile quasi-particles falling into different superselection sectors, which can fuse and split as discussed in Sec. \ref{sec:setup}, and we have demonstrated the emergence of a UMTC based on this assumption. 

One notable feature of this discussion is that, although the $F$- and $R$-matrices are ``gauge non-invariant" in UMTC because they change under Eq. \eqref{eq: vertex basis gauge transformation}, they still correspond to physical observables given by Eqs. \eqref{Eq:generalF main} and \eqref{eq: general R main}. This is different from, say, gauge connections in a gauge theory, which are deemed to be not observable since they are gauge non-invariant. Of course, from the $F$- and $R$-matrices, one can construct quantities invariant under Eq. \eqref{eq: vertex basis gauge transformation}, such as the self-statistics of anyons, which are also physically observable.

\subsection{Including symmetries} \label{subsec: including symmetries}

In Secs. \ref{subsec:FR}, we have presented the microscopic universal theory of TQSLs with symmetry omitted, where the universal properties of the topological phase are expressed using microscopically defined quantities. It remains that, from the same principles, we now present the microscopic universal theory of TQSLs with symmetries, which can contain both internal and lattice symmetries that may not commute with each other. To this end, on top of the theory in Sec. \ref{subsec:FR}, we need to capture the effects of symmetries.

As discussed in Sec. \ref{sec:setup}, the effects of a symmetry on an anyon state can be organized into three parts: changing the anyon types, changing the fusion vertex basis states, and changing the local information around the anyons. The first two effects are non-local, and universal physics independent of microscopic details is expected. Even for the last effect, as we will see, there will be universal physics associated with symmetry fractionalization, \ie some anyons can carry fractional quantum numbers of the symmetries, which generalizes the concept of fractional charge in quantum Hall physics. Our task is to describe these properties using microscopic quantities.

\subsubsection{Representative of universal data} \label{subsubsec: representative internal}

As before, the universal data is the equivalence class of some representative data. Here we discuss the representative data, and we turn to the equivalence class in Sec. \ref{subsubsec: equivalence internal}.

To obtain the universal data, we examine the three effects of the symmetries, whose corresponding symmetry group is denoted by $G$. First, the change of the types of anyons under symmetry $g\in G$ is described by a bijective map $\rho_g$ between the anyon types, \ie
\beq \label{eq: anyon permutation}
\rho_g(a)= ^ga,
\eeq
where $^ga$ represents the anyon obtained by acting the symmetry $g$ on anyon $a$.

\begin{figure}[h!]
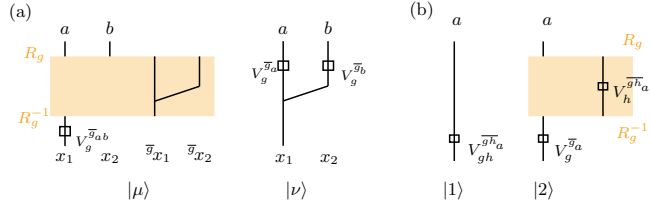

	\centering
	\scalebox{0.75}{\latticeUeta}
	\caption{(a) Graphic representations of the states in Eq. \eqref{eq: U-symbol states}. (b) Graphic representations of the states in Eq. \eqref{eq: eta states}. The shaded region means the actions of $R_g$ and $R_g^{-1}$, where the positions of the anyons can be changed.}
	\label{fig:latticeUeta}
\end{figure}

Next, we turn to the change of fusion vertex basis states encoded in the unitary matrix $U_g(^ga,^gb; ^gc)$ in Eq. \eqref{eq: symmetry localization}. In principle, Eq. \eqref{eq: symmetry localization} can already be viewed as a microscopic definition of $U_g(^ga,^gb; ^gc)$. However, just from Eq. \eqref{eq: symmetry localization} it is unclear how $U_g(^ga,^gb; ^gc)$ can be related to the $F$- and $R$-matrices defined using moving and splitting operators in Eqs. \eqref{Eq:generalF main} and \eqref{eq: general R main}. To capture their relations, an expression of $U_g(a,b;c)$ involving the moving and splitting operators is desired. To get such an expression, we write $R_g(S^{^{\bar g}a^{\bar g}b}_{^{\bar g}c,\mu})_{^{\bar g}2^{\bar g}1}|^{\bar g}c_{^{\bar g}1}, ^{\bar g}\bar c_{^{\bar g}0'}\rangle$ in two different ways, where $\bar g=g^{-1}$:
\begin{equation*}
\bsp
&R_g(S^{^{\bar g}a^{\bar g}b}_{^{\bar g}c,\mu})_{^{\bar g}2^{\bar g}1}|^{\bar g}c_{^{\bar g}1}, ^{\bar g}\bar c_{^{\bar g}0'}\rangle
\propto  R_g|^{\bar g}a_{^{\bar g}1}, ^{\bar g}b_{^{\bar g}2}, ^{\bar g}\bar c_{^{\bar g}0'}; \mu\rangle\\
\propto&V_g^{^{\bar g}a}(x_1)V_g^{^{\bar g}b}(x_2)V_g^{^{\bar g}\bar c}(x_0')\sum_{\nu}[U_{g}(a, b; c)]_{\mu\nu}|a_1, b_2, \bar c_{0'}; \nu\rangle\\
\propto& V_g^{^{\bar g}a}(x_1)V_g^{^{\bar g}b}(x_2)V_g^{^{\bar g}\bar c}(x_0')\sum_{\nu}[U_{g}(a, b; c)]_{\mu\nu}(S^{ab}_{c, \nu})_{21}|c_1, \bar c_{0'}\rangle,\\
&R_g(S^{^{\bar g}a^{\bar g}b}_{^{\bar g}c,\mu})_{^{\bar g}2^{\bar g}1}|^{\bar g}c_{^{\bar g}1}, ^{\bar g}\bar c_{^{\bar g}0'}\rangle
=R_g(S^{^{\bar g}a^{\bar g}b}_{^{\bar g}c,\mu})_{^{\bar g}2^{\bar g}1}R_g^{-1}R_g|^{\bar g}c_{^{\bar g}1}, ^{\bar g}\bar c_{^{\bar g}0'}\rangle\\
\propto &R_g(S^{^{\bar g}a^{\bar g}b}_{^{\bar g}c,\mu})_{^{\bar g}2^{\bar g}1}R_g^{-1}V_g^{^{\bar g}c}(x_1)V_g^{^{\bar g}\bar c}(x_0')|c_1, \bar c_{0'}\rangle.
\esp
\end{equation*}
Since $V_g^{^{\bar g}a}(x_1)V_g^{^{\bar g}b}(x_2)(S^{ab}_{c, \nu})_{21}$ and $R_g(S^{^{\bar g}a^{\bar g}b}_{^{\bar g}c,\mu})_{^{\bar g}2^{\bar g}1}R_g^{-1}V_g^{^{\bar g}c}(x_1)$ are supported around $x_{1,2}$, while $V_g^{^{\bar g}\bar c}(x_0')$ is supported around $x_0'$ which is far from $x_{1, 2}$, according to Appendix \ref{app: approximate factorization}, $V_g^{^{\bar g}a}(x_1)V_g^{^{\bar g}b}(x_2)\sum_{\nu}[U_{g}(a, b; c)]_{\mu\nu}(S^{ab}_{c, \nu})_{21}|c_1, \bar c_{0'}\rangle$ and $R_g(S^{^{\bar g}a^{\bar g}b}_{^{\bar g}c,\mu})_{^{\bar g}2^{\bar g}1}R_g^{-1}V_g^{^{\bar g}c}(x_1)|c_1, \bar c_{0'}\rangle$ only differ by a phase factor. Therefore, we can write
\beq \label{eq: U-matrix}
[U(a, b; c)]_{\mu\nu}=\langle\nu|\mu\rangle,
\eeq
with
\beq \label{eq: U-symbol states}
\bsp
&\ket{\mu}=R_g (S^{\lsupsc{\bar g}a \lsupsc{\bar g}b}_{\lsupsc{\bar g}c,\mu})_{^{\bar g}2^{\bar g}1}R_g^{-1}V^{\lsupsc{\bar g}c}_g(x_1)\ket{c_1,\overline{c}_{0'}},\\
&\ket{\nu}=V^{\lsupsc{\bar g}a}_g(x_1)V^{\lsupsc{\bar g}b}_g(x_2)(S^{ab}_{c,\nu})_{21}\ket{c_1,\overline{c}_{0'}}.
\esp
\eeq
Graphic representations of these two states are in Fig. \ref{fig:latticeUeta} (a). We remark that Eq. \eqref{eq: U-matrix} follows from Eq. \eqref{eq: symmetry localization} and captures the change of the fusion vertex basis states, and its advantage compared to Eq. \eqref{eq: symmetry localization} is that it allows us to find the relation between the $U$-matrix and the other universal data. Also, readers should not confuse the states in Eq. \eqref{eq: U-symbol states} with those in Eq. \eqref{eq: R states general}.

Lastly, we discuss the change of local information around the anyons, which encodes the physics of symmetry fractionalization, \ie the anyons can carry fractional quantum numbers under the symmetries. Intuitively, symmetry fractionalization is similar to projective representations in quantum mechanics, which reflects a mismatch between the {\it local} actions of $R_{gh}$ and $R_gR_h$ on anyons. With this motivation, we compare
\begin{equation*}
R_{gh}|^{\overline{gh}}a_{^{\overline{gh}}1}, ^{\overline{gh}}\bar a_{^{\overline{gh}}0'}\rangle\propto V_{gh}^{^{\overline{gh}}a}(x_1)V_{gh}^{^{\overline{gh}}\bar a}(x_0')|a_1, \bar a_{0'}\rangle
\end{equation*}
and
\begin{equation*}
\bsp
&R_gR_h|^{\overline{gh}}a_{^{\overline{gh}}1}, ^{\overline{gh}}\bar a_{^{\overline{gh}}0'}\rangle
\propto R_gV_h^{^{\overline{gh}}a}(^{\bar g}x_1)V_h^{^{\overline{gh}}\bar a}(^{\bar g}x_0')|^{\bar g}a_{^{\bar g}1}, ^{\bar g}\bar a_{^{\bar g}0'}\rangle\\
=&(R_gV_h^{^{\overline{gh}}a}(^{\bar g}x_1)R_g^{-1})(R_gV_h^{^{\overline{gh}}\bar a}(^{\bar g}x_0')R_g^{-1})R_g|^{\bar g}a_{^{\bar g}1}, ^{\bar g}\bar a_{^{\bar g}0'}\rangle\\
\propto& (R_gV_h^{^{\overline{gh}}a}(^{\bar g}x_1)R_g^{-1})(R_gV_h^{^{\overline{gh}}\bar a}(^{\bar g}x_0')R_g^{-1})\cdot V_g^{^{\bar g}a}(x_1)V_g^{^{\bar g}a}(x_0')|a_1, \bar a_{0'}\rangle.
\esp
\end{equation*}
Because $R_{gh}\propto R_gR_h$ (no matter whether the microscopic degrees of freedom transform in linear or projective representations of $G$), $V_{gh}^{^{\overline{gh}}a}(x_1)$ and $R_gV_h^{^{\overline{gh}}a}(^{\bar g}x_1)R_g^{-1}V_g^{^{\bar g}a}(x_1)$ are supported around $x_1$, while $V_{gh}^{^{\overline{gh}}\bar a}(x_0')$ and $R_gV_h^{^{\overline{gh}}\bar a}(^{\bar g}x_0')R_g^{-1}V_g^{^{\bar g}a}(x_0')$ are supported around $x_0'$ that is far away from $x_1$, according to Appendix \ref{app: approximate factorization}, the following two states (see Fig. \ref{fig:latticeUeta} (b) for their graphic representations)
\beq \label{eq: eta states}
\bsp
&\ket{1}=V^{\lsupsc{\overline{gh}}a}_{gh}(x_1)\ket{a_1,\bar a_{0'}},\\
&\ket{2}=R_g V^{\lsupsc{\overline{gh}}a}_h(^{\bar g}x_1)R_g^{-1}V^{\lsupsc{\bar g}a}_g(x_1)\ket{a_1,\bar a_{0'}}
\esp
\eeq
differ by a phase factor, and we define $\eta_a(g, h)$ as this phase factor
\beq \label{eq: eta}
\eta_a(g, h)=\langle 1|2\rangle.
\eeq
This $\eta_a(g, h)$ measures the mismatch between the {\it local} actions of $R_{gh}$ and $R_gR_h$ on anyon $a$, and it is the analog of the phase factors that define projective representations, so it captures symmetry fractionalization.

Put together, the maps $\rho_g: a\rightarrow ^ga$ in Eq. \eqref{eq: anyon permutation}, the $U$-matrices defined in Eq. \eqref{eq: U-matrix} and the phase factors $\eta$ defined in Eq. \eqref{eq: eta} capture the three effects that the symmetries can have on the anyons. These pieces of data are not independent. In fact, just like the $F$- and $R$-matrices, which satisfy some nontrivial relations in Eqs. \eqref{eq: pentagon equation} and \eqref{eq: hexagon equations}, our new data also satisfy some nontrivial relations:

\begin{align}
    &K^{q(g)}[F^{\lsupsc{\overline{g}}a\lsupsc{\overline{g}}b\lsupsc{\overline{g}}c}_{\lsupsc{\overline{g}}d}]_{\{\lsupsc{\overline{g}}e,\mu',\nu'\},\{\lsupsc{\overline{g}}f,\rho',\sigma'\}}K^{q(g)}\notag\\
    &=\sum_{\mu,\nu,\rho,\sigma}[U_g(a,b;e)]_{\mu'\mu}[U_g(e,c;d)]_{\nu'\nu}[F^{abc}_{d}]_{\{e,\mu,\nu\},\{f,\rho,\sigma\}}[(U_g(b,c;f)^{-1})]_{\rho\rho'}[(U_g(a,f;d))^{-1}]_{\sigma\sigma'},\label{Eq:internalFU}\\
    &K^{q(g)}[J_g(R^{\lsupsc{\overline{g}}a\lsupsc{\overline{g}}b}_{\lsupsc{\overline{g}}c})]_{\mu'\nu'}K^{q(g)}= \sum_{\mu\nu}[U_g(b,a;c)]_{\mu'\mu}[R^{ab}_c]_{\mu\nu}[(U_g(a,b;c))^{-1}]_{\nu\nu'},\label{Eq:internalRU}\\
    &\sum_{\nu}K^{q(g)}[U_h(\lsupsc{\overline{g}}a,\lsupsc{\overline{g}}b;\lsupsc{\overline{g}}c)]_{\mu\nu}K^{q(g)}[U_g(a,b;c)]_{\nu\rho}=\frac{\eta_c(g,h)}{\eta_a(g,h)\eta_b(g,h)}[U_{gh}(a,b;c)]_{\mu\rho},\label{Eq:internalUeta}\\
    &K^{q(g)}\eta_{\lsupsc{\overline{g}}a}(h,k)K^{q(g)}\eta_a(g,hk)=\eta_a(gh,k)\eta_a(g,h),\label{Eq:internaletaeta}
\end{align}
with $K$ the complex conjugation operation, $q(g)=0$ ($q(g)=1$) if the symmetry $g$ is unitary (anti-unitary), and $J_g (R^{^{\bar g}a^{\bar g}b}_{^{\bar g}c}) = R^{^{\bar g}a^{\bar g}b}_{^{\bar g}c}$ ($J_g(R^{^{\bar g}a^{\bar g}b}_{^{\bar g}c})=(R^{^{\bar g}b^{\bar g}a}_{^{\bar g}c})^{-1}$) if the symmetry $g$ preserves (reverses) the spatial orientation. We emphasize that $q(g)$ is purely determined by whether $g$ is anti-unitary, and it does not depend on whether $g$ reserves the spatial orientation or whether $g^2$ is the identity operation. Similarly, the effect of $J_g$ is purely determined by whether $g$ reverses the spatial orientation, and it does not depend on whether $g$ is anti-unitary or whether $g^2$ is identity.

Here we briefly explain the meanings of Eqs. \eqref{Eq:internalFU}-\eqref{Eq:internaletaeta}, while leaving their derivation to Appendix \ref{app: full structure}. Eqs. \eqref{Eq:internalFU} and \eqref{Eq:internalRU} describe how symmetries can change the fusion vertex basis states in a way consistent with the fusion and braiding properties of the TQSL. These two equations are very similar to Eq. \eqref{eq: vertex basis gauge transformation}, if the symmetry is a unitary internal symmetry that does not change the anyon types, and if $U_g$ here is replaced by $\Omega$ there. This similarity is not a coincidence, because both these two equations and Eq. \eqref{eq: vertex basis gauge transformation} characterize how the $F$- and $R$-matrices should change when the fusion vertex basis states change. When $g$ is anti-unitary, the appearance of complex conjugation should be expected. For example, because the self-statistics of the anyons should be conjugated under time reversal, the $R$-matrix, which is related to braiding, should also be conjugated. When $g$ reverses the spatial orientation, it turns counter-clockwise braiding into clockwise braiding, which is the origin of the nontrivial effect of $J_g$ here. Eq. \eqref{Eq:internalUeta} describes the ``composition rule" of the symmetry actions on the fusion vertex basis states. Eq. \eqref{Eq:internaletaeta} comes from the ``associativity" of the local operators $V$ in Eq. \eqref{eq: symmetry localization}, and it is very similar to the corresponding equation of the phase factors that define projective representations (see, \eg Appendix C of Ref. \cite{Chen2010}).

Importantly, the maps $\rho_g$, the $U$-matrices and the phase factors $\eta$ are all manifestly invariant under quasi-adiabatic continuation in Eqs. \eqref{eq: quasi-adiabatic continuation 1} and \eqref{eq: quasi-adiabatic continuation 2}. Moreover, since the maps $\rho_g$ do not depend on any non-universal choice, which are similar to the anyon types, they are universal data of an entire SET phase of symmetric TQSLs. On the other hand, just like the $F$- and $R$-matrices in Eqs. \eqref{Eq:generalF main} and \eqref{eq: general R main}, the $U$-matrices and phase factors $\eta$ are representatives of the universal data of the SET phase, because they depend on many subjective choices. The equivalence relations between these representatives will be discussed in Sec. \ref{subsubsec: equivalence internal}.

\subsubsection{Equivalence relation between representative data} \label{subsubsec: equivalence internal}

Now we discuss how the $U$-matrix and the phase factor $\eta$ defined in Eqs. \eqref{eq: U-matrix} and \eqref{eq: eta} depend on the subjective choices, and the resulting equivalence relation. Similar to the $F$- and $R$-matrices in Eqs. \eqref{Eq:generalF main} and \eqref{eq: general R main}, according to the definitions in Eqs. \eqref{eq: U-matrix} and \eqref{eq: eta}, the $U$-matrix and phase factor $\eta$ also depend on the positions and local information of the anyons, and the choices of the moving and splitting operators. A new ingredient is that they also depend on the $V$ operators in the symmetry localization property Eq. \eqref{eq: symmetry localization}. We will see that when the these subjective choices change, not only that the $F$- and $R$-matrices transform according to Eq. \eqref{eq: vertex basis gauge transformation}, the $U$-matrix and phase factor $\eta$ will also undergo ``gauge transformations" according to \begin{equation}\label{eq: general U eta transformation}
	\bsp
	&U_g(a,b;c)\to \frac{\delta^{a}_g\delta^{b}_g}{\delta^{c}_g}\cdot K^{q(g)}(\Omega^{\lsupsc{\overline{g}}a\lsupsc{\overline{g}}b}_{\lsupsc{\overline{g}}c})K^{q(g)} U_g(a,b;c)(\Omega^{ab}_{c})^{-1},\\
	&\eta_a(g,h) \to \frac{\delta^{a}_{gh}}{K^{q(g)}\delta^{\lsupsc{\overline{g}}a}_hK^{q(g)}\delta^{a}_{g}}\eta_a(g,h),
	\esp
\end{equation}
where $\delta_g^a$ is a set of phase factors, and $\Omega^{ab}_{c}$ is the same set of unitary matrices as those appearing in Eq. \eqref{eq: vertex basis gauge transformation}.

Below we focus on the simplest case where the positions and local information of anyons are fixed, and show that the $U$-matrix and phase factor $\eta$ will change as in Eq. \eqref{eq: general U eta transformation} when the moving, splitting and $V$ operators change. In Appendix \ref{app: full structure}, we show that even if the positions and local information of anyons are also changed, the $U$-matrix and phase factor $\eta$ will still change as in Eq. \eqref{eq: general U eta transformation}.

It turns out that the effects of changing the splitting operators and changing the $V$ operators can be separated. We first consider the effect of changing the $V$ operators while fixing the splitting operators. Suppose $R_g|a_1, \bar a_0\rangle\propto V_g^a(x_1)V_g^{\bar a}(x_0)|^g a_1, ^g {\bar a}_0\rangle\propto \tilde V_g^a(x_1)\tilde V_g^{\bar a}|^g a_1, ^g{\bar a}_{0}\rangle$, \ie $V$ and $\tilde V$ are two sets of localized symmetry actions. Because $V_g^a(x_1)$ and $\tilde V_g^a(x_1)$ are supported around $x_1$, which is far from the supports of $V_g^{\bar a}(x_0)$ and $\tilde V_g^{\bar a}(x_0)$, according to Appendix \ref{app: approximate factorization}, $\tilde V_g^a(x_1)|^g a_1, ^g{\bar a}_0\rangle=(\delta^{^ga}_g(x_1))^{-1}V_g^a(x_1)|^g a_1, ^g{\bar a}_0\rangle$ holds, with $\delta^{^ga}_g(x_1)$ a phase factor that can potentially depend on the position $x_1$. However, it is straightforward to check that Eq. \eqref{eq: simplifying condition 2} forces $\delta^{^ga}_g(x_1)$ to be independent of $x_1$, so we simply denote this phase by $\delta^{^ga}_g$. That is, the effects of two sets of $V$ operators only differ by a position-independent phase factor. Substituting this relation into Eqs. \eqref{eq: U-symbol states} and \eqref{eq: eta states} gives the transformations of the relevant states:
\beq \label{eq: symmetry action gauge transformation of states}
\bsp
&|\mu\rangle\rightarrow (\delta^{c}_g)^{-1}|\mu\rangle,
|\nu\rangle\rightarrow(\delta^{a}_g\delta^b_g)^{-1}|\nu\rangle,\\
&|1\rangle\rightarrow (\delta^a_{gh})^{-1}|1\rangle, |2\rangle\rightarrow (K^{q(g)}\delta_h^{^{\bar g}a}K^{q(g)}\delta_g^{a})^{-1}|2\rangle.
\esp
\eeq
Then from Eqs. \eqref{eq: U-matrix} and \eqref{eq: eta}, the $U$-matrix and phase factor $\eta$ precisely transform as in Eq. \eqref{eq: general U eta transformation}, with $\Omega^{ab}_c=1$ for all anyons $a$, $b$ and $c$.

Next, we consider the effect of changing the splitting operator while fixing the $V$ operators. Note that when the $V$ operators are fixed, Eq. \eqref{eq: simplifying condition 2} imposes constraints on the phases $\phi^a_{21}$ and $\phi^{\lsupsc{g}a}_{\lsupsc{g}2\lsupsc{g}1}$,
\begin{equation}\label{eq: constraining transformation 2}
    e^{i\phi^a_{12}}=K^{q(g)}e^{i\phi^{\lsupsc{\bar g}a}_{\lsupsc{\bar g}1\lsupsc{\bar g}2}}K^{q(g)}.
\end{equation}
Keeping this constraint mind, substituting Eq. \eqref{eq: new moving and splitting operators} into Eqs. \eqref{eq: U-symbol states} and \eqref{eq: eta states} yields
\begin{equation}
    \bsp
|\mu\rangle &\rightarrow \sum_{\mu'}K^{q(g)}[(\Omega_0(\lsupsc{\bar g}a,\lsupsc{\bar g}b;\lsupsc{\bar g}c)_{\lsupsc{\bar g}2\lsupsc{\bar g}1}]_{\mu\mu'}K^{q(g)}|\mu'\rangle\\
&=K^{q(g)}e^{i\phi^{\lsupsc{\bar g}b}_{\lsupsc{\bar g}2\lsupsc{\bar g}1}}K^{q(g)}K^{q(g)}e^{i\phi^{\lsupsc{\bar g}a}_{\lsupsc{\bar g}11}+i\phi^{\lsupsc{\bar g}b}_{\lsupsc{\bar g}11}-i\phi^{\lsupsc{\bar g}c}_{\lsupsc{\bar g}11}}\sum_{\mu'}e^{i\phi^{\lsupsc{\bar g}b}_{12}}[(\Omega_0(\lsupsc{\bar g}a,\lsupsc{\bar g}b;\lsupsc{\bar g}c)_{21}]_{\mu\mu'}K^{q(g)}|\mu'\rangle\\
&=e^{i\phi^{b}_{21}}K^{q(g)}e^{i\phi^{\lsupsc{\bar g}a}_{\lsupsc{\bar g}11}+i\phi^{\lsupsc{\bar g}b}_{\lsupsc{\bar g}11}-i\phi^{\lsupsc{\bar g}c}_{\lsupsc{\bar g}11}}\sum_{\mu'}e^{i\phi^{\lsupsc{\bar g}b}_{12}}[(\Omega_0(\lsupsc{\bar g}a,\lsupsc{\bar g}b;\lsupsc{\bar g}c)_{21}]_{\mu\mu'}K^{q(g)}|\mu'\rangle,\\
|\nu\rangle &\rightarrow \sum_{\nu'}[(\Omega_0(a,b;c))_{21}]_{\nu\nu'}|\nu'\rangle\\
&=e^{i\phi^{b}_{21}}\sum_{\nu'}e^{i\phi^{b}_{12}}[(\Omega_0(a,b;c))_{21}]_{\nu\nu'}|\nu'\rangle,\\
|1\rangle&\rightarrow |1\rangle, \quad |2\rangle\rightarrow |2\rangle,
\esp
\end{equation}
where we have used Eqs. \eqref{eq: constraining transformation} and \eqref{eq: constraining transformation 2}. Then from Eqs. \eqref{eq: U-matrix} and \eqref{eq: eta}, the $U$-matrix and phase factor $\eta$ precisely transform as in Eq. \eqref{eq: general U eta transformation}, with $\Omega^{ab}_c=e^{i\phi^b_{12}}\Omega_0(a, b; c)_{21}$ and $\delta^a_g=K^{q(g)}e^{i\phi^{\lsupsc{\bar g}a}_{\lsupsc{\bar g}11}}K^{q(g)}$ for all anyons $a,b,c$ and $g\in G$. Put together, changing the splitting and $V$ operators will induce a transformation of the $U$-matrix and phase factor $\eta$, which is given by Eq. \eqref{eq: general U eta transformation}.

In the special case where all symmetries are internal, together with the universal data described in Sec. \ref{subsec:FR}, which has the structure of a unitary modular tensor category, the new data, which includes the maps $\rho_g$ in Eq. \eqref{eq: anyon permutation}, the $U$-matrices and phase factors $\eta$ that satisfy Eqs. (\ref{Eq:internalFU}-\ref{Eq:internaletaeta}) and are subject to the equivalence relation defined in Eq. \eqref{eq: general U eta transformation}, forms a $G$-crossed braided tensor category ($G$-BTC) \cite{barkeshli2014}. There the $U$-matrix and phase factor $\eta$ are often referred to as the $U$-symbol and $\eta$-symbol, respectively, and the transformations in Eq. \eqref{eq: general U eta transformation} with $\delta_g^a=1$ are referred to as vertex basis gauge transformations of the $U$-symbols, while those with $\Omega^{ab}_c=\mathbbm{1}$ are referred to as symmetry action gauge transformations. We will adopt these terminologies below. Again, we stress that we did not start by assuming that the SET phase is described by a $G$-BTC. Instead, our analysis demonstrates its emergence.

\subsection{Summarizing the microscopic universal theory} \label{subsec: summarizing universal data}

Now we summarize Sec. \ref{sec:framework}. In Sec. \ref{sec:framework}, we present a microscopic universal theory of TQSLs with a general symmetry $G$, which can contain potentially non-commuting internal and lattice symmetries. Using the microscopically defined anyon states, and the moving, splitting and $V$ operators introduced in Sec. \ref{sec:setup}, we can characterize the universal properties of the SET phases of these TQSLs. In the main text, we focus on the case where Eqs. \eqref{eq: seamless}, \eqref{eq: simplifying condition 1} and \eqref{eq: simplifying condition 2} hold. In Appendix \ref{app: full structure}, we discuss the general case without assuming Eqs. \eqref{eq: seamless}, \eqref{eq: simplifying condition 1} and \eqref{eq: simplifying condition 2}.

The universal data from our theory can be organized as follows.

\begin{enumerate}
	
	\item The anyon types, the fusion rules and fusion multiplicities $N_{ab}^c$ in Eq. \eqref{eq: fusion}.

    This piece of data is contained in the microscopic anyon states in Sec. \ref{sec:setup}, so it is part of the input of our theory.
	
	\item The equivalence class of $F$- and $R$-matrices.
	
	This piece of data characterizes universal properties related to fusion and braiding of anyons. 
	
	Representative $F$-matrices and $R$-matrices are defined by Eqs. \eqref{Eq:generalF main} and \eqref{eq: general R main}, respectively, for any choice of the parameters $\gamma$, $\Gamma$ and $\omega$ in Eqs. \eqref{eq: gamma main}, \eqref{eq: Gamma main} and \eqref{eq: omega main}.
    
    When Eqs. \eqref{eq: seamless}, \eqref{eq: simplifying condition 1} and \eqref{eq: simplifying condition 2} are assumed, the $F$-matrices and $R$-matrices satisfy the pentagon and hexagon equations, Eqs. \eqref{eq: pentagon equation} and \eqref{eq: hexagon equations}, and they are subject to equivalence relation defined in Eq. \eqref{eq: vertex basis gauge transformation}.
    
    When the parameters $\gamma$, $\Gamma$ and $\omega$ in Eqs. \eqref{eq: gamma main}, \eqref{eq: Gamma main} and \eqref{eq: omega main} are not all $\mathbbm{1}$, the pentagon and hexagon equations will be replaced by Eqs. \eqref{eq: general pentagon}, \eqref{eq: general hexagon 1} and \eqref{eq: general hexagon 2}, respectively. The equivalence relation of the $F$-matrices is given by Eqs. \eqref{Eq:generalFVBGT} and \eqref{Eq:generalFtransformation}, and the equivalence relation of the $R$-matrices is given by Eqs. \eqref{Eq:generalRVBGT}, \eqref{Eq:generalRtransformation1} and \eqref{Eq:generalRtransformation2}.
	
	\item The maps $\rho_g$ in Eq. \eqref{eq: anyon permutation}. 
	
	This data characterizes how the symmetry $g\in G$ changes the types of anyons.
	
	\item The equivalence class of $U$- and $\eta$-symbols. 
	
	This piece of data characterizes how the symmetries change the fusion vertex basis states and local information of the anyons, which encode the physics of symmetry fractionalization.
	
	Representative $U$-symbol and $\eta$-symbol are defined by Eqs. \eqref{eq: U-matrix} and \eqref{eq: eta}, respectively, for any choice of the parameters $\gamma$, $\Gamma$ and $\omega$ in Eqs. \eqref{eq: gamma main}, \eqref{eq: Gamma main} and \eqref{eq: omega main}.

    When Eqs. \eqref{eq: seamless}, \eqref{eq: simplifying condition 1} and \eqref{eq: simplifying condition 2} are assumed, in addition to Eqs. \eqref{eq: pentagon equation}, \eqref{eq: hexagon equations}, the $F$-, $R$-, $U$- and $\eta$-symbols satisfy also satisfy Eqs. \eqref{Eq:internalFU}, \eqref{Eq:internalRU}, \eqref{Eq:internalUeta} and \eqref{Eq:internaletaeta}. The $U$- and $\eta$-symbols are subject to the equivalence relation defined by Eq. \eqref{eq: general U eta transformation}.

    When the parameters $\gamma$, $\Gamma$ and $\omega$ in Eqs. \eqref{eq: gamma main} are not all $\mathbbm{1}$, Eqs. \eqref{Eq:internalFU}, \eqref{Eq:internalRU}, \eqref{Eq:internalUeta} and \eqref{Eq:internaletaeta} are generalized to Eqs. \eqref{eq: general FU equation}, \eqref{eq: general RU equation}, \eqref{Eq:relationUeta} and \eqref{eq: general eta eta equation}, respectively. The equivalence relation defined by Eq. \eqref{eq: general U eta transformation} is generalized to Eqs. \eqref{Eq:generalUVBGT}, \eqref{Eq:generalUSAGT}, \eqref{Eq:generalUtransformation}, \eqref{Eq:generaletaSAGT} and \eqref{Eq:generaletatransformation}.

\end{enumerate}

We can schematically write the universal data of an SET phase of TQSLs as
\beq
\mathcal{C}=\bigoplus_{\gamma,\Gamma,\omega}\mathcal{C}_{\gamma,\Gamma,\omega},
\eeq
where each $\mathcal{C}_{\gamma,\Gamma,\omega}$ represents the above set of data obtained using our microscopic universal theory, with a set of fixed positions to put anyons and a set of parameters $\gamma$, $\Gamma$ and $\omega$ in Eqs. \eqref{eq: gamma main}, \eqref{eq: Gamma main} and \eqref{eq: omega main}. The symbol ``$\oplus$" just means that each set of $\mathcal{C}_{\gamma,\Gamma,\omega}$ should be viewed as different, and the full universal data includes all $\mathcal{C}_{\gamma,\Gamma,\omega}$'s.

The simplest part of $\mc{C}$, which is also the case discussed in the main text, is $\mc{C}_{1, \mathbbm{1}, 1}$, which is obtained by demanding Eqs. \eqref{eq: seamless}, \eqref{eq: simplifying condition 1} and \eqref{eq: simplifying condition 2}. If there is no symmetry, then the data $\mc{C}_{1, \mathbbm{1}, 1}$ can be organized as a unitary modular tensor category. If there is only internal symmetry, then $\mc{C}_{1, \mathbbm{1}, 1}$ can be organized as a $G$-crossed braided tensor category ($G$-BTC). If there is lattice symmetry, we dub the data $\mc{C}_{1, \mathbbm{1}, 1}$ a {\it generalized $G$-crossed braided tensor category} (generalized $G$-BTC).

We stress that, from a microscopic perspective, $\mc{C}_{1, \mathbbm{1}, 1}$ is not more special than other $\mathcal{C}_{\gamma,\Gamma,\omega}$, and all of them are universal data of the SET phase that can be measured numerically and experimentally, at least in principle. However, unlike $\mc{C}_{1, \mathbbm{1}, 1}$, the data in general $\mathcal{C}_{\gamma,\Gamma,\omega}$ cannot be organized as a category theory, at least not in the standard sense. One way to see it is to note that the $F$-matrices in general $\mathcal{C}_{\gamma,\Gamma,\omega}$ do not satisfy the pentagon equation, but Eq. \eqref{eq: general pentagon}.

Now an important question arises: Is the classification of SET phases still given by the classification of certain categories? After all, the presence of general $\mathcal{C}_{\gamma,\Gamma,\omega}$ may enrich the classification. Our answer to this question is no. To see it, note that the data in different $\mathcal{C}_{\gamma,\Gamma,\omega}$'s can be obtained from each other by simply changing the parameters $\gamma$, $\Gamma$ and $\omega$. However, given the anyon types, fusion rule, fusion multiplicities and the maps $\rho_g$, the possible choices of $\gamma$, $\Gamma$ and $\omega$ are the same for all SET phases, which means that the relation between different $\mathcal{C}_{\gamma,\Gamma,\omega}$'s is independent of the SET phase. Therefore, each $\mathcal{C}_{\gamma,\Gamma,\omega}$ actually contains the information of the full data $\mc{C}$, and the classification of these SET phases is still given by the classification of $\mc{C}_{1, \mathbbm{1}, 1}$, the generalized $G$-BTCs.

\section{Crystalline equivalence principle}\label{sec:equivalence}

In Sec. \ref{sec:framework}, we have established the microscopic universal theory of $(2+1)$-dimensional TQSLs with a general symmetry $G$, which can contain potentially non-commuting internal and lattice symmetries. In particular, we have found that the classification of SET phases of such TQSLs is given by the classification of generalized $G$-crossed braided tensor categories (generalized $G$-BTCs). In this section, we present a remarkable explicit bijective map between generalized $G$-BTCs and the standard $G$-BTCs in Ref. \cite{barkeshli2014}, which is summarized in Eq. \eqref{Eq:correspondence} and leads to a precise crystalline equivalence principle (CEP). The existence of this map shows that not only the classification of SET phases of TQSLs with a general symmetry $G$ agrees with the classification of SET phases of TQSLs with a purely internal symmetry $G$, but the two sets of data that characterize these two types of SET phases also have an explicit correspondence.

The CEP was first noted in Ref. \cite{Song2016} and then substantiated in Ref. \cite{Thorngren2016}. Its original statement is that the classification of symmetry-protected topological (SPT) phases with a lattice symmetry group $G$ coincides with the classification of SPT phases with an internal symmetry group $G$, as long as all reflection symmetries in the former are viewed as anti-unitary symmetries in the latter. Supported by many case studies \cite{Isobe2015, Barkeshli2016, Huang2017, Cheng2017a, Zou2017a, Else2018a, Else2018, Else2019, Jiang2019}, the CEP further extends to declare that not only the two classifications agree, but the precise characterizations of the two types of topological phases also have correspondence, and, moreover, such correspondence is not restricted to SPT phases and can be extended to SET phases and even symmetry-enriched gapless phases. The CEP then becomes a key assumption in many studies of quantum matter with lattice symmetries, because it allows one to convert the problem to another one about quantum matter with internal symmetries, which is much better understood. For example, it plays an important role in Refs. \cite{Ye2021a} and \cite{Ye2023}, which classifies topological and critical quantum spin liquids with both internal and lattice symmetries. However, the previous understanding of CEP is incomplete, because of the lack of a general theory for quantum matter with lattice symmetries that does not itself rely on CEP. Consequently, given a system realizing a TQSL with lattice symmetries, previously there was no systematic approach to identify which SET phase in the classification of Ref. \cite{Ye2023} it belongs to. Now our explicit map in Eq. \eqref{Eq:correspondence} makes the CEP precise and allow us to address these issues.

To obtain a precise CEP, we just need to establish an invertible map between the $\rho_g$ maps, $U$- and $\eta$-symbols in a generalized $G$-BTC and those in a standard $G$-BTC. Note that if there is no spacetime orientation reversal symmetries, \eg reflection and time reversal, the equations that the $U$- and $\eta$-symbols should satisfy and the equivalence relations they are subject to are identical for generalized $G$-BTCs and standard $G$-BTCs, given by Eqs. (\ref{Eq:internalFU}-\ref{Eq:internaletaeta}) and \eqref{eq: general U eta transformation}. So in this simple case the targeted map is just the identity map, \ie $\rho_g=\rho'_g$, $U_g(a, b; c)=U'_{g}(a, b; c)$ and $\eta_a(g, h)=\eta'_a(g, h)$ for any $g, h\in G$ and anyons $a, b, c$, with the unprimed (primed) objects from the generalized (standard) $G$-BTC.

The more challenging case is when some symmetries involve spacetime orientation reversal. The intuition to understand this case is as follows. Topological phases should be described by some topological quantum field theories at low energies, which have a CRT symmetry. If a topological phase is compatible with a reflection symmetry, then the CRT symmetry implies that it is also compatible with a CT symmetry. So if a microscopic realization of this topological phase has an exact reflection symmetry, there should also be a realization of it with a CT symmetry, which is an anti-unitary order-2 internal symmetry and is identified as a microscopic time reversal symmetry. Taking this intuition forward, in Appendix \ref{appd:CEP}, we show that the following is the targeted map in the most general case:
\begin{widetext}
	\begin{equation}\label{Eq:correspondence}
		\bsp
		&\rho'_g(a)=\left\{ \begin{array}{ll}
			\rho_g(a)=^ga, &\textrm{if $g$ preserves spatial orientation,}\\
			\rho_g(\bar a)=^g\bar{a}, &\textrm{if $g$ reverses spatial orientation.}
		\end{array}\right.\\
		&U'_{g}(a,b;c)= \left\{ \begin{array}{ll}
			U_g(a,b;c),&\textrm{if $g$ preserves spatial orientation,}\\
			K^{q(g)}(A^{\lsupsc{\overline g}b\lsupsc{\overline g}a}_{\lsupsc{\overline g}c}R^{\lsupsc{\overline g}a\lsupsc{\overline g}b}_{\lsupsc{\overline g}c})K^{q(g)}U_g(a,b;c),&\textrm{if $g$ reverses spatial orientation.}
		\end{array}\right.\\
		& \eta'_a(g_1,g_2)=\left\{\begin{array}{ll}
			\eta_a(g_1,g_2),&\textrm{if $g_1$ preserves spatial orientation,}\\
			\eta_a(g_1,g_2)K^{q(g_1)}U_{g_2}(\lsupsc{\overline{g_1}} a,\lsupsc{\overline{g_1}}\bar a;0)K^{q(g_1)},&\textrm{if $g_1$ reverses spatial orientation but $g_2$ preserves it,}\\
			\eta_a(g_1,g_2)K^{q(g_1)}U_{g_2}(\lsupsc{\overline{g_1}}\bar a,\lsupsc{\overline{g_1}}a;0)\theta_{\lsupsc{\overline{g_1}}a}\varkappa^{*}_{\lsupsc{\overline{g_1}}a}K^{q(g_1)},&\textrm{if both $g_1$ and $g_2$ reverse spatial orientation.}
		\end{array}\right.
		\esp
	\end{equation}
\end{widetext}
where the unprimed (primed) objects are from the generalized $G$-BTC (standard $G$-BTC), and 0 represents the trivial anyon. Under this map, a unitary (anti-unitary) spatial reflection symmetry in a generalized $G$-BTC becomes an anti-unitary (unitary) internal symmetry in the standard $G$-BTC. The fact that $\rho'_g(a)=\rho_g(\bar a)$ is because the composition of the R and CRT symmetries in the effective topological field theory is the CT symmetry. The matrix $A$ is unitary and satisfies Theorem \ref{thm: CRT}, which is constructed in Appendix \ref{appd:CEP} and physically implements the CRT operation in the effective topological field theory. The phase factor $\varkappa_a$ is the Frobenius-Schur indicator $\varkappa_a=d_a[F^{a\bar aa}_a]_{00}$, where $d_a$ is called the quantum dimension of the anyon $a$ (see, for example, equation (8.10) in Ref. \cite{Simonbook2023} for its definition). The inverse map of Eq. \eqref{Eq:correspondence} is given by Eq. \eqref{Eq:inverse correspondence app}.

In Appendix \ref{appd:CEP}, the correctness of this map is established by showing that as long as the maps $\rho_g$, the $U$- and $\eta$-symbols satisfy Eqs. (\eqref{Eq:internalFU}-\ref{Eq:internaletaeta}) and are subject to the equivalence relation in Eq. \eqref{eq: general U eta transformation}, then the maps $\rho'_g$, the $U'$- and $\eta'$-symbols in Eq. \eqref{Eq:correspondence} satisfy the standard $G$-BTC version of Eqs. (\ref{Eq:internalFU}-\ref{Eq:internaletaeta}), which are given by Eqs. (\ref{eq: internal F Uprime}-\ref{eq: internal etaprime etaprime}), and are subject to the standard $G$-BTC version of equivalence relation in Eq. \eqref{eq: general U eta transformation}, which is given by Eqs. \eqref{Eq:gaugetransformationUp} and \eqref{Eq:gaugetransformationetap}. In the main text, we proceed by applying this precise CEP in Sec. \ref{sec:application}.

In passing, we note that a topological quantum field theory also has an RT symmetry in addition to the CRT symmetry, so one may wonder if one can use the RT symmetry to construct the map in CEP. In Appendix \ref{appd:CEP}, we argue that this is invalid.

\section{Applications}\label{sec:application}

After describing our microscopic universal theory in Sec. \ref{sec:framework} and crystalline equivalence principle in Sec. \ref{sec:equivalence}, in this section, we demonstrate our theoretical framework using three symmetry-enriched TQSLs, some of which have been realized in quantum processors based on superconducting qubits, trapped ions and Rydberg atoms \cite{Satzinger2021, Semeghini2021, Iqbal2023, Foss-Feig2023}.
For each example, we will explain in detail the procedure of extracting a set of $F$-, $R$-, $U$- and $\eta$-symbols, whose equivalence class is universal across all systems in the same SET phase. We further show that the Lieb-Schultz-Mattis (LSM) anomaly matching condition holds in each example, which presents a highly nontrivial check of our theory that shows its consistency with various previous work from very different angles \cite{Ye2021a, Ye2022, Ye2023}.

\subsection{Toric code} \label{subsec: toric code}

\begin{figure*}
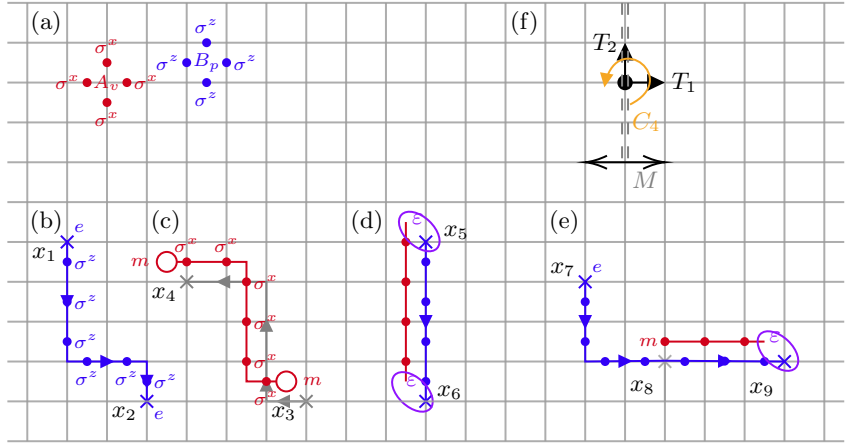

    \centering
    \toriccode
    \caption{(a) The $A_v$ operators and $B_p$ operators. (b) From a ground state, the state $|e_{1} ,e_{2}\rangle$ can be created by applying the blue $e$ string operator, which is realized by applying $\sigma^z$ operators to a string connecting the two vertices $x_1$ and $x_2$. (c) From a ground state, the state $|m_{3}, m_{4}\rangle$ can be created by applying the red $m$ string operator, which is realized by applying $\sigma^x$ operators to a string connecting the two vertices. When moving or splitting an anyon $m$ from $x_3$ to $x_4$, we always choose the support of the string operator to be at the right side of the gray arrowed line pointing from $x_3$ to $x_4$. (d) From a ground state, the state $|\varepsilon_{5}, \varepsilon_{6}\rangle$ can be created by first applying the blue $e$ string operator and then applying the red $m$ string operator. (e) From a ground state, the state $|e_{5}, m_{6}, \varepsilon_{7}\rangle$ can be created by first applying the blue $e$ string operator and then applying the red $m$ string operator. (f) The generators of the $p4m$ symmetry, where $T_{1,2}$ are translations by 1 unit cell along the two perpendicular directions, $C_4$ is a 4-fold rotation around a site with $C_4^2$ denoted as $C_2$, and $M$ is a reflection that flips the horizontal direction.}
    \label{fig:toriccode}
\end{figure*}

Our first example is a $p4m\times \z_2^T$ symmetric $\z_2$ TQSL described by the celebrated toric code model \cite{Kitaev1997} and realized using superconducting qubits and trapped ions \cite{Satzinger2021, Iqbal2023, Foss-Feig2023}. The model is defined on a square lattice, with a qubit residing on each edge of the lattice (see Fig. \ref{fig:toriccode}). The Hamiltonian of the model is
\begin{equation}\label{Eq:toriccode}
    H_{\rm{TC}}=-\sum_{v}A_v-\sum_{p}B_p,
\end{equation}
where an $A_v$ operator acts on the 4 qubits on links connected with the vertex $v$ by applying $\sigma^x$ to each of these qubits, and a $B_p$ operator acts on the 4 qubits surrounding the plaquette $p$ by applying $\sigma^z$ each of these qubits. The Hamiltonian Eq. \eqref{Eq:toriccode} has a $p4m\times \mathbb{Z}_2^T$ symmetry, where $p4m$ is the standard symmetry of a square lattice defined in Fig. \ref{fig:toriccode}, and the $\mathbb{Z}_2^T$ symmetry is a time reversal that flips the sign of all Pauli operators, \ie $\mc{T}\sigma^i\mc{T}=-\sigma^i$ for all $i=1,2,3$ for all qubits. The ground states are simultaneous eigenstates of all $A_v$'s and $B_p$'s with eigenvalues all being 1. There are three nontrivial anyons in this TQSL: the $e$ anyon signaled by having $A_v=-1$ at a vertex $v$, the $m$ anyon signaled by having $B_p=-1$ at a plaquette $p$, and the $\varepsilon$ anyon, which is the composition of $e$ and $m$ anyons. Their fusion rules are captured by
\beq \label{eq: toric code fusion}
e\times e=m\times m=0,
e\times m=\varepsilon,
\eeq
with 0 the trivial anyon. The anyon types are not changed by any symmetry, \ie $\rho_g(a)=a$ for all $g\in p4m\times\z_2^T$ and $a\in\{0, e, m, \varepsilon\}$.

\subsubsection{Extracting data} \label{subsubsec: toric code data}

Now we extract a set of $F$-, $R$-, $U$- and $\eta$-symbols in this symmetry-enriched TQSL, whose equivalence class is a universal property of the entire SET phase. Following the recipe in Sec. \ref{sec:framework}, to do this, we need to specify the relevant anyon states, the moving, splitting and $V$ operators in Eqs. \eqref{eq: general F states main}, \eqref{eq: R states general}, \eqref{eq: U-symbol states} and \eqref{eq: eta states}, and use Eqs. \eqref{Eq:generalF main}, \eqref{eq: general R main}, \eqref{eq: U-matrix} and \eqref{eq: eta} to calculate the $F$-, $R$-, $U$- and $\eta$-symbols. In particular, we will choose the microscopic input data such that Eqs. \eqref{eq: seamless}, \eqref{eq: simplifying condition 1} and \eqref{eq: simplifying condition 2} are satisfied. Note that in this example, according to Eq. \eqref{eq: toric code fusion}, the fusion outcome of any two anyons is unique and the fusion multiplicity is always 1, so we will simply denote a 3-anyon state by $|a_1, b_2, c_3\rangle$, with the index $\mu$ suppressed. Similarly, we denote the splitting operator, say, $(S^{ab}_{c, \mu})_{21}$, by $(S^{ab})_{21}$. The $F$-, $R$- and $U$-symbols will be denoted by $F^{abc}$, $R^{ab}$ and $U_g(a, b)$, respectively. In fact, the examples in Secs. \ref{subsec: flipped toric code} and \ref{subsec: ruby lattice} are also $\z_2$ TQSLs, and we will use the same notations there.

We first set up the conventions of anyon states in Sec. \ref{subsec: anyon states}. We use the positions of vertices to label the positions of all types of anyons, and, for notational convenience, we denote by $|0\rangle$ the simultaneous eigenstate of all $\sigma^z$ operators with eigenvalue 1.
\begin{itemize}
    \item An anyon state $|e_{1}, e_{2}\rangle$ is defined by applying the following projectors to $\ket{0}$ and then normalizing the state: $\frac{1-A_v}{2}$ for vertices at $x_1$ and $x_2$, $\frac{1+A_v}{2}$ for all other vertices, and $\frac{1+B_p}{2}$ for all plaquettes (see Fig. \ref{fig:toriccode} (b)). This state is a simultaneous eigenstate of all $A_v$ and $B_p$ operators, with the eigenvalues being $-1$ for the $A_v$ operators at $x_1$ and $x_2$, and all other eigenvalues being 1.

    \item An anyon state $|m_{3}, m_{4}\rangle$ is defined by applying the following projectors to $\ket{0}$ and then normalizing the state: $\frac{1-B_p}{2}$ for the {\it upper-left} plaquettes of $x_3$ and $x_4$, $\frac{1+B_p}{2}$ for all other plaquettes, and $\frac{1+A_v}{2}$ for all vertices (see Fig. \ref{fig:toriccode} (c)). This state is a simultaneous eigenstate of all $A_v$ and $B_p$ operators, with the eigenvalues being $-1$ for the $B_p$ operators for the upper-left plaquettes of $x_3$ and $x_4$, and all other eigenvalues being 1.

    \item An anyon state $|\varepsilon_{5}, \varepsilon_{6}\rangle$ is defined by applying the following projectors to $\ket{0}$ and then normalizing the state: $\frac{1-A_v}{2}$ for vertices $x_5$ and $x_6$, $\frac{1-B_p}{2}$ for the upper-left plaquettes of $x_5$ and $x_6$, and $\frac{1+A_v}{2}$ and $\frac{1+B_p}{2}$ for all other vertices and plaquettes (see Fig. \ref{fig:toriccode} (d)). This state is a simultaneous eigenstate of all $A_v$ and $B_p$ operators, with the eigenvalues for $A_v$ operators at $x_5$ and $x_6$ being $-1$, the eigenvalues for the $B_p$ operators for the upper-left plaquettes of $x_5$ and $x_6$ being $-1$, and all other eigenvalues being 1.

    \item A 3-anyon state $|e_{7}, m_{8}, \varepsilon_{9}\rangle$ is similarly defined by applying the following projectors to $\ket{0}$ and then normalizing the state: $\frac{1-B_p}{2}$ for the upper-left plaquettes of $x_8$ and $x_9$, $\frac{1-A_v}{2}$ for the vertex at $x_7$ and $x_9$, and $\frac{1+B_p}{2}$ and $\frac{1+A_v}{2}$ for all other plaquettes and vertices (see Fig. \ref{fig:toriccode} (e)). This state is also a simultaneous eigenstate of all $A_v$ and $B_p$ operators, and the eigenvalues can be similarly read off as before.
\end{itemize}
To summarize our convention, an $e_1$ anyon resides at the vertex $x_1$, an $m_2$ anyon resides in the upper-left plaquette of the vertex $x_2$, and an $\varepsilon_3$ anyon is a composite of $e_3$ and $m_3$.

With the above definitions of anyon states in Sec. \ref{subsec: anyon states}, we can define the moving and splitting operators in Sec. \ref{subsubsec: dynamics of anyons} using Eqs. \eqref{eq: moving operator} and \eqref{eq: splitting operator}. In this model, the moving and splitting operators can be defined using the string operators introduced in Fig. \ref{fig:toriccode}. Concretely, for any nontrivial anyon $b$, we find an operator $O^b_{21}$ such that $|b_2, b_0\rangle=O^b_{21}|b_1, b_0\rangle$, where $O^b_{21}$ is a string operator of $b$ that connects $x_{1,2}$. Then, a string operator $O^b_{21}$ for any anyon $b$ supported between $x_1$ and $x_2$ also serves as $(S^{ab})_{21}=M^b_{21}=O^b_{21}$, independent of what the anyon $a$ in $(S^{ab})_{21}$ is. The string operators $O^b_{21}$ are defined as follows.
\begin{itemize}
    \item For $e$ anyons, we first specify the support of a string operator $O^e_{21}$ by choosing a string that goes through the lattice edges and connects $x_1$ and $x_2$. Then we apply $\sigma^z$ operators to all qubits on this string (see Fig. \ref{fig:toriccode} (b) for an example).
    
    \item For $m$ anyons, we first specify the support of a string operator $O^m_{21}$. We start by choosing a string that goes through the lattice edges and connects $x_1$ and $x_2$. Then we translate this string by half-plaquette upward and half-plaquette leftward, rendering its end points to be in the upper-left plaquettes of $x_1$ and $x_2$. 
    \begin{equation*}
        \TCmstringchoice
    \end{equation*}
    This new string connects the upper-left plaquettes of $x_1$ and $x_2$ and cuts the edges on the path. The string operator $O^m_{21}$ is the product of $\sigma^x$ operators acting on the qubits cut by the above red string (see Fig. \ref{fig:toriccode} (c) for another example). 

    \item For $O^\varepsilon_{21}$ operators, we first define them for the cases where the two anyons are on the same horizontal or vertical line. When $x_1$ and $x_2$ are on the same horizontal (vertical) line, we directly connect them with a straight segment, and apply $\sigma^z$ operators to all the spins along this segment. Then, we translate this segment by half-edge upwards and half-edge leftwards, and apply $\sigma^x$ operators to all the spins cut by this translated segment. This operator is denoted as $\hat O^\varepsilon_{21}$.
    \begin{equation}
        \scalebox{0.9}{\TCepsilonstringchoicea}
    \end{equation}
    A generic $O^\varepsilon_{43}$ is obtained by multiplying together all the straight segments. We first specify the support of the string operator by choosing a string connecting $x_3$ and $x_4$. Suppose this string consists of $N$ straight segments. We label each corner point as $x_{3_i}$, $i=1,2...N-1$, and the segments are denoted by $3_{i}3_{i-1}$. Then, 
    \begin{equation}
        O^\varepsilon_{43}=\hat O^\varepsilon_{43_{N-1}}\hat O^\varepsilon_{3_{N-1}3_{N-2}}\cdots \hat O^\varepsilon_{3_{2}3_1}\hat O^\varepsilon_{3_13},
    \end{equation}
    where $O^\varepsilon_{43_{N-1}}$, $O^\varepsilon_{3_{i}3_{i-1}}$, and $O^\varepsilon_{3_13}$ are defined above for straight segments. See below for an example drawn in figure. 
        \begin{equation}
        \scalebox{0.8}{\TCepsilonstringchoiceb}
    \end{equation}
    When $x_3$ and $x_4$ are on the same horizontal or vertical line, $O^\varepsilon_{43}=\hat O^\varepsilon_{43}$. See Fig. \ref{fig:toriccode} (d) for an example. It is straight forward to see that $O^\varepsilon_{43}=(O^\varepsilon_{34})^{\dagger}$.
\end{itemize}
One can verify that these moving and splitting operators indeed satisfy their defining equations in Eqs. \eqref{eq: moving operator} and \eqref{eq: splitting operator}, as well as Eqs. \eqref{eq: seamless} and \eqref{eq: simplifying condition 1}. 

The last piece of microscopic input is the $V$ operators in Eq. \eqref{eq: symmetry localization}. To obtain them, it will be useful to set up a coordinate system where the origin is the $C_4$ rotation center, and the two orthogonal axes are given by the directions of translations $T_{1,2}$, so that each vertex is labeled by a pair of integers.

For the translation symmetry, using the anyon states defined above and Eq. \eqref{eq: symmetry localization}, we can set
\begin{equation}
    V_{T_1^{n_1}}^a(x)=V_{T_2^{n_2}}^a(x)=1
\end{equation}
for any $n_{1,2}\in\z$, any anyon $a$ and any position $x$. For the $C_4$ rotation symmetry, using the anyon states with two $e$ anyons defined above and Eq. \eqref{eq: symmetry localization}, we can set
\begin{equation}
    V^e_{C_4}(x)=1
\end{equation}
for any position $x$. For $m$ and $\varepsilon$ anyons, however, the $C_4$ rotation will map the anyons as
\begin{equation}
\bsp
    &R_{C_4}\BLvert\scalebox{0.6}{\manyonposition{x_1}}\Brangle=\BLvert\scalebox{0.6}{\manyonrotation{\lsupsc{C_4}x_1}}\Brangle\\
    =&\sigma^x_l(\lsupsc{C_4}x_1)\BLvert\scalebox{0.6}{\manyonposition{\lsupsc{C_4}x_1}}\Brangle,
\esp
\end{equation}
where we have used the notation $\sigma^x_{i}(x_1),\;i=l,r,u,b$ to represent the $\sigma^x$ operator acting on the left, right, upper, and bottom qubit of the vertex $x_1$, respectively, as indicated below.
\begin{equation} \label{eq: nearby Pauli X operators}
    \paulix
\end{equation}
Using the previous definitions of two-$m$ and two-$\varepsilon$ states and Eq. \eqref{eq: symmetry localization}, we can set
\begin{equation}
    V^m_{C_4}(x)=V^{\varepsilon}_{C_4}(x)=\sigma^x_l(x).
\end{equation}
By inspecting $C_2$ and $(C_4)^{-1}$ similarly, we can set, for any position $x$, 
\begin{equation}
    \bsp
    &V^m_{C_2}(x)=V^{\varepsilon}_{C_2}(x)=\sigma^x_l(x)\sigma^x_b(x),\\
    &V^m_{(C_4)^{-1}}(x)=V^{\varepsilon}_{(C_4)^{-1}}(x)=\sigma^x_u(x),
    \esp
\end{equation}
A choice of the $V$ operators for the reflection symmetry $M$ can be similarly derived:
\begin{equation}
    V_M^e(x)=1, V_{M}^m(x)=V^{\varepsilon}_M(x)=\sigma^x_{u}(x),
\end{equation}
for any position $x$. 

For the time-reversal symmetry $\mathcal{T}$, note that it flips sign for all the Pauli operators, so an anyon state acquires a $-1$ if it is created by applying an odd number of Pauli operators to the ground state. Since an $\varepsilon$ anyon is the composite of $e$ and $m$ anyons, our definition of two-$\varepsilon$ states ensures that these states are always created by an even number of Pauli operators, and hence
\begin{equation}
    V_{T}^{\varepsilon}(x)=1
\end{equation}
for any $x$. For $e$ and $m$ anyons, to be compatible with our definitions of two-anyon states and Eq. \eqref{eq: simplifying condition 2}, we can set the $V$ operators as
\begin{equation}
    V^e_{\mathcal{T}}(x)=V^{m}_{\mathcal{T}}(x)=(-1)^{l_1+l_2},
\end{equation}
where $l_1$ and $l_2$ are the coordinates of the vertex $x=(l_1,l_2)$. This choice can be depicted by
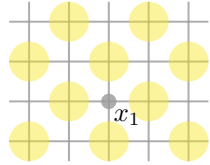
\begin{equation*}
    \toriccodeTRV
\end{equation*}
and the $V^{e(m)}_{\mathcal{T}}$ operators are $-1$ (1) at the shaded (unshaded) vertices.

Given the $V$ operators for the generators, the $V$ operators for a generic symmetry action $(T_1)^{n_1}(T_2)^{n_2}(C_4)^{n_3}M^{n_4}\mathcal{T}^{n_5}$ can be set as
\begin{equation}\label{Eq:TCgeneralV}
    \bsp
    &V_{(T_1)^{n_1}(T_2)^{n_2}(C_4)^{n_3}M^{n_4}\mathcal{T}^{n_5}}^a(x)=\\
    &V_{\mathcal{T}^{n_5}}^a(\lsupsc{M^{-n_4}(C_4)^{-n_3}(T_2)^{-n_2}(T_1)^{-n_1}}x)\times\\&
    V_{M^{n_4}}^a(\lsupsc{(C_4)^{-n_3}(T_2)^{-n_2}(T_1)^{-n_1}}x)\times\\
    &V_{(C_4)^{n_3}}^a(\lsupsc{(T_2)^{-n_2}(T_1)^{-n_1}}x)V_{(T_2)^{n_2}}^a(\lsupsc{(T_1)^{-n_1}}x)V_{(T_1)^{n_1}}^a(x).
    \esp
\end{equation}
One can check that the above definition of $V$ operators satisfies Eq. \eqref{eq: simplifying condition 2}. 

Using the setup above, we find that all the $F$-matrices are $1$ and the $R$-matrices are given by $R^{ab}=e^{i\pi a_eb_m}$, where if the anyon $a=0$ then $(a_e, a_m)=(0, 0)$, if the anyon $a=e$ then $(a_e, a_m)=(1, 0)$, if the anyon $a=m$ then $(a_e, a_m)=(0, 1)$, and if the anyon $a=\varepsilon$ then $(a_e, a_m)=(1, 1)$. Given the $V$ operators, using Eq. \eqref{eq: eta}, we can explicitly calculate the $\eta$-phases via
\begin{equation}\label{Eq:TCderiveeta}
    \eta_a(g,h)=\text{sign}(V_{gh}^a(x)R_g V_h^a(\lsupsc{\bar g}x)R_g^{\dagger}V_g^a(x)).
\end{equation}
Note that when $g$ contains time-reversal action, conjugating $V_h^a(\lsupsc{\bar g}x)$ with $R_g$ flips the sign of all Pauli matrices in between. It is straightforward to show that the $\eta$-phases satisfy Eq. \eqref{Eq:internaletaeta}. For $U$-matrices, since the fusion vertices are all $1$ dimensional, the $U$-matrices are all phases in this example. It is simple to find that the $U$-phases for translation, rotation, and time reversal symmetries are trivial, \ie
\begin{equation}\label{Eq:TCUother}
    \bsp
    &U_{T_1}(a,b)=U_{T_2}(a,b)=U_{C_4}(a,b)=U_{(C_4)^{-1}}(a,b)=U_{C_2}(a,b)=U_{\mathcal{T}}(a,b)=1.
    \esp
\end{equation}
The $U$-phases are no longer trivial for the reflection symmetry $M$. Let us consider $U_{M}(\varepsilon,\varepsilon)$ for example. According to the definition Eq. \eqref{eq: U-matrix}, $U_M(\varepsilon, \varepsilon)$ is 
\begin{equation*}
    \bsp
    &\Blangle\scalebox{0.7}{\etaanyonUa}\BRvert (V_M^{\epsilon}(x_1))^\dagger (V_M^{\epsilon}(x_2))^\dagger\BLvert\scalebox{0.7}{\etaanyonUb}\Brangle\\
    =&\Blangle\scalebox{0.7}{\etaanyonUa}\Bvert\scalebox{0.7}{\etaanyonUc}\Brangle=-1.
    \esp
\end{equation*}
The last equality comes from the anti-commutativity of $\sigma^x$ and $\sigma^z$ and the fact that the ground state is a $+1$ eigenstate for all contractible closed string operators. The physical picture of this $-1$ is also evident: compared to the original splitting operator at $x_1x_2$, the mirror-reflected splitting operator has an additional effect coming from an $m$ anyon winding around an $e$ anyon, as indicated in the equation. Therefore, this $-1$ can be attributed to the nontrivial mutual statistics of $e$ and $m$. We can similarly derive all the $U_M$-phases. The nontrivial $U_M$-phases are
\begin{equation}\label{Eq:TCUM}
    U_{M}(\varepsilon,\varepsilon)=U_{M}(m,e)=U_M(m,\varepsilon)=U_M(\varepsilon,e)=-1.
\end{equation}
Given the $U$-phases for all the symmetry generators, we can find the $U$-phases for any symmetry action using Eq. \eqref{Eq:internalUeta}. From our choice of $V$ operators, we find that, for $c=a\times b$, we always have
\begin{equation}
\bsp
    &\text{sign}(V^a_g(x))\cdot \text{sign}(V^b_g(x))=\text{sign}(V^c_g(x)),\\
    &\text{sign}(R_gV^a_g(x)R_g^{-1})\cdot \text{sign}(R_gV^b_g(x)R_g^{-1})
    =\text{sign}(R_gV^c_g(x)R_g^{-1}).
\esp
\end{equation}
Hence, within our choice, we have $\eta_c(g,h)=\eta_a(g,h)\eta_b(g,h)$ for $c=a\times b$. The explicit expression of $U$-phases for generic symmetry actions in this case is therefore simplified to 
\begin{equation}\label{Eq:TCderiveU}
    U_{gh}(a,b;c)=U_h(a,b;c)U_g(a,b;c).
\end{equation}
The explicit values of the $U$- and $\eta$-symbols can be calculated by combining the above equations.

\subsubsection{Identifying SET phase and matching anomaly} \label{subsubsec: SET toric code}

Having extracted the universal data, we now identify which SET phase obtained in Ref. \cite{Ye2023} the toric code belongs to, and calculate its anomaly. Note that in Ref. \cite{Ye2023}, all SET phases are mapped to their corresponding versions with purely internal symmetries using a presumed crystalline equivalence principle.

According to Ref. \cite{Ye2023}, in the present setting, the SET phases are characterized by the fractionalization pattern of the $p4m\times\z_2^T$ symmetry on the $e$ and $m$ anyons. For example, the two translations $T_1$ and $T_2$ may ``anti-commute" on $e$ and/or $m$. More precisely, this fractionalization pattern is characterized by an element in $\mathcal{H}^2(p4m\times \z_2^T, \mathbb{Z}_2\times \mathbb{Z}_2)$, where $\mathbb{Z}_2\times \mathbb{Z}_2$ is from the fusion rules in this TQSL. So identifying the SET phase amounts to identifying the element in $\mathcal{H}^2(p4m\times Z_2^T, \mathbb{Z}_2\times \mathbb{Z}_2)$ that the $\eta'$-phases of the toric code correspond to, where $\eta'$ can be obtained from $\eta$ via Eq. \eqref{Eq:correspondence}. Concretely, we can define a 10-entry vector (see Table XIII in Ref. \cite{Ye2023})
\beq \label{Eq:p4mbasis}
\bsp
\vec v_{p4m\times\z_2^T}
=
(&B_{xy}, B_{c^2}, A_{x+y}^2, A_{x+y}A_m, A_c^2,
A_m^2, A_{x+y}t, A_ct, A_mt, t^2)^T,
\esp
\eeq
with each entry being a 2-cocycle defined in Appendix \ref{app: SET Z2}. Then there are 2 other 10-entry vectors
$\vec\chi^e$ and $\vec\chi^m$, such that (up to coboundary transformations)
\beq \label{eq: expansion toric code}
\bsp
&\eta'_e(g_1, g_2)=(-1)^{\vec\chi^e\cdot\vec v_{p4m\times\z_2^T}},\\
&\eta'_m(g_1, g_2)=(-1)^{\vec\chi^m\cdot\vec v_{p4m\times\z_2^T}}.
\esp
\eeq
Each pair of $\vec\chi^e$ and $\vec\chi^m$ specifies an element in $\mathcal{H}^2(p4m\times \z_2^T, \mathbb{Z}_2\times \mathbb{Z}_2)$, and thus an SET phase in Ref. \cite{Ye2023}. In Appendix \ref{app: SET Z2}, we explain the physical meanings of these $\eta'_e(g_1, g_2)$ and $\eta'_m(g_1, g_2)$, in terms of the fractionalization patterns of the $p4m\times\z_2^T$ symmetry on the $e$ and $m$ anyons.

To identify which element in $\mathcal{H}^2(p4m\times Z_2^T, \mathbb{Z}_2\times \mathbb{Z}_2)$ the toric code corresponds to, we use our precise crystalline equivalence principle Eq. \eqref{Eq:correspondence} to map the data obtained in Sec. \ref{subsubsec: toric code data} to get the universal data for a corresponding SET phase with a purely internal $p4m\times\z_2^T$ symmetry. For the $\z_2$ TQSL, taking all $A$-matrices in Eq. \eqref{Eq:correspondence} to be $\mathbbm{1}$ always satisfies Eqs. \eqref{Eq:CEPAAcondition}-\eqref{Eq:CEPAUcondition}. Applying Eq. \eqref{Eq:correspondence}, the obtained $\eta'$-phases can be written as Eq. \eqref{eq: expansion toric code}, with (see Appendix \ref{app: SET Z2} for the derivation)
\beq \label{eq: coefficients toric code}
\bsp
&\vec\chi^e=(0, 0, 0, 0, 0, 0, 1, 0, 0, 0)^T,\\
&\vec\chi^m=(0, 0, 0, 0, 0, 0, 1, 1, 1, 0)^T.
\esp
\eeq
This pair of $\chi^e$ and $\chi^m$ specifies an SET phase in Ref. \cite{Ye2023}.

Next, we turn to the anomaly of the toric code. According to Eq. (D9) in Ref. \cite{Ye2023}, the anomaly of a $p4m\times\z_2^T$ symmetric $\z_2$ TQSL can be specified by the following 20 anomaly indicators:
\begin{equation} \label{eq: p4m indicators}
    \bsp
    &I_0=\mathcal{I}_0,\\
    &I_1=\mathcal{I}_1(\mathcal{T}),\quad I_2=\mathcal{I}_1(M),\quad I_3=\mathcal{I}_1(C_2\mathcal{T}),\\
    &I_4=\mathcal{I}_1(C_4M),\quad I_5=\mathcal{I}_2(\mathcal{T},C_2\mathcal{T}),\quad I_6=\mathcal{I}_2(\mathcal{T},M),\\
    &I_7=\mathcal{I}_2(\mathcal{T},C_4M),\quad I_8=\mathcal{I}_2(C_2\mathcal{T},M),\\
    &I_9=\mathcal{I}_2(C_2\mathcal{T},C_4M),\quad I_{10}=\mathcal{I}_1(T_1M),\quad I_{11}=\mathcal{I}_1(T_1C_2\mathcal{T}),\\
    &I_{12}=\mathcal{I}_1(T_1T_2C_2\mathcal{T}),\quad I_{13}=\mathcal{I}_2(\mathcal{T},T_1M),\\
    &I_{14}=\mathcal{I}_2(\mathcal{T},T_1C_2\mathcal{T}),\quad I_{15}=\mathcal{I}_2(\mathcal{T},T_1T_2C_2\mathcal{T}),\\
    &I_{16}=\mathcal{I}_2(T_2C_2\mathcal{T},M),\quad I_{17}=\mathcal{I}_2(M,T_2C_2M),\\
    &I_{18}=\mathcal{I}_2(T_1T_2^{-1}C_2\mathcal{T},C_4M),\quad I_{19}=\mathcal{I}_2(T_1T_2C_2\mathcal{T},T_1M),
    \esp
\end{equation}
where
\begin{equation}\label{Eq:baseindicators}
    \bsp
    &\mathcal{I}_0=\frac{1}{D}\sum_a\theta_a=1;\\
    &\mathcal{I}_1(\mathcal{T}_1)=\frac{1}{D}\sum_{a}\theta_a\eta'_a(\mathcal{T}_1,\mathcal{T}_1);\\
    &\mathcal{I}_2(\mathcal{T}_1,\mathcal{T}_2)=\frac{1}{D}\sum_{a,b,c}\frac{\theta_{a\times b}}{\theta_a\theta_b}R(c,c)U'_{\mathcal{T}_1}(a,c)U'_{\mathcal{T}_1}(a\times c,c)\\
    &\quad U'_{\mathcal{T}_2}(c,b\times c)U'_{\mathcal{T}_2}(c,b) \eta'_{a}(\mathcal{T}_1,\mathcal{T}_1)\eta'_b(\mathcal{T}_2,\mathcal{T}_2)\frac{\eta'_c(\mathcal{T}_2,\mathcal{T}_1)}{\eta'_c(\mathcal{T}_1,\mathcal{T}_2)},
    \esp
\end{equation}
where $\theta_{a}=(R^{aa})^2$ is the topological spin of anyon $a$, and $\mc{T}_1$ and $\mc{T}_2$ are two anti-unitary order-2 generators of the internal $p4m\times\z_2^T$ symmetry. For example, the argument of the fifth anomaly indicator $I_5=\mathcal{I}_2(\mathcal{T},C_2\mathcal{T})$ is from a $\mathbb{Z}_2^T\times\mathbb{Z}_2^T$ subgroup of $p4m\times\mathbb{Z}_2^T$, where the first $\mathbb{Z}_2^T$ is generated by $\mathcal{T}$ and the second $\mathbb{Z}_2^T$ is generated by $C_2\mathcal{T}$. In particular, the original mirror symmetry $M$ is regarded as an anti-unitary symmetry under the map Eq. \eqref{Eq:correspondence}. These anomaly indicators were originally derived in Eqs. (46), (50), and (55) of Ref. \cite{Ye2022}, and here we are only using their simplified form for $\z_2$ TQSL with $F^{abc}=1$.

Using the $R$-, $U'$- and $\eta'$-symbols obtained by applying Eq. \eqref{Eq:correspondence} to the results in Sec. \ref{subsubsec: toric code data}, the anomaly indicators can be calculated using Eq. \eqref{eq: p4m indicators}, and the results are
\begin{equation}
    \begin{array}{lll}
    I_0=1,&I_1=1,&I_2=1\\
    I_3=1,&I_4=1,&I_5=1\\
    I_6=1,&I_7=1,&I_8=1\\
    I_9=1,&I_{10}=1,&I_{11}=-1\\
    I_{12}=1,&I_{13}=1,&I_{14}=-1\\
    I_{15}=1,&I_{16}=1,&I_{17}=1\\
    I_{18}=1,&I_{19}=1.&\\
    \end{array}
\end{equation}
These results exactly match the Lieb-Schultz-Mattis anomaly of such lattice systems, which is given by column $c$ of Table XXV in Ref. \cite{Ye2023}. The success of this anomaly matching is a highly nontrivial check of our theory.

\subsection{Flipped toric code} \label{subsec: flipped toric code}

The second example is a variation of the toric code, where the degrees of freedom and their locations are the same as the toric code, but the Hamiltonian is
\begin{equation}\label{Eq: flipped toric code}
    H_{\rm{fTC}}=-\sum_{v}A_v+\sum_{p}B_p.
\end{equation}
This ``flipped" toric code differs from the original toric code in Sec. \ref{subsec: toric code} by merely the $+$ sign in front of the $B_p$ terms, and we assume that the system has an even number of plaquettes in total. This system has the same $p4m\times\mathbb{Z}_2^T$ symmetry and anomaly as the toric code. The $+$ sign in front of the $B_p$ terms indicates that the ground state now has a background $m$ anyon in each plaquette. An interesting feature of this model is that the existence of the background $m$ anyons leads to the translation fractionalization of $e$ anyons. Namely, the effective actions of $T_1$ and $T_2$ on a single $e$ anyon anti-commute \cite{Essin2014}. In the language of our microscopic universal theory, this translation fractionalization means, when Eqs. \eqref{eq: seamless}, \eqref{eq: simplifying condition 1} and \eqref{eq: simplifying condition 2} are satisfied, we have
\begin{equation} \label{eq: translation fractionalization}
    \frac{\eta_e(T_1,T_2)}{\eta_e(T_2,T_1)}=-1.
\end{equation}
Below we will verify this feature explicitly.

To the best of our knowledge, this TQSL has not been experimentally realized. We discuss it here as a warmup that connects the discussion on toric code in Sec. \ref{subsec: toric code} to the discussion on the $\z_2$ TQSL in Sec. \ref{subsec: ruby lattice}, which also host background anyons, displays translation fractionalization, but has been experimentally realized \cite{Semeghini2021}.

\subsubsection{Extracting data}

We now extract the universal data for the flipped toric code and show that the translation fractionalization indeed happens.

We first specify the anyon states. As before, we use the positions of vertices to label the positions of all types of anyons. 
\begin{itemize}
    \item An anyon state $|e_{1}, e_{2}\rangle$ is defined by applying the following projectors to $\ket{0}$ and then normalizing the state: $\frac{1-A_v}{2}$ for vertices at $x_1$ and $x_2$, $\frac{1+A_v}{2}$ for all other vertices, and $\frac{1-B_p}{2}$ for all plaquettes. 

    \item An anyon state $|m_{3}, m_{4}\rangle$ is defined by applying the following projectors to $\ket{0}$ and then normalizing the state: $\frac{1+B_p}{2}$ for the upper-left plaquettes of $x_3$ and $x_4$, $\frac{1-B_p}{2}$ for all other plaquettes, and $\frac{1+A_v}{2}$ for all vertices. 

    \item An anyon state $|\varepsilon_{5}, \varepsilon_{6}\rangle$ is defined by applying the following projectors to $\ket{0}$ and then normalizing the state: $\frac{1-A_v}{2}$ for vertices $x_5$ and $x_6$, $\frac{1+B_p}{2}$ for the upper-left plaquettes of $x_5$ and $x_6$, and $\frac{1+A_v}{2}$ and $\frac{1-B_p}{2}$ for all other vertices and plaquettes.

    \item A 3-anyon state $|e_{7}, m_{8}, \varepsilon_{9}\rangle$ is similarly defined by applying the following projectors to $\ket{0}$ and then normalizing the state: $\frac{1+B_p}{2}$ for the upper-left plaquettes of $x_8$ and $x_9$, $\frac{1-A_v}{2}$ for the vertex at $x_7$ and $x_9$, and $\frac{1-B_p}{2}$ and $\frac{1+A_v}{2}$ for all other plaquettes and vertices. 
\end{itemize}
All these states are simultaneous eigenstates of all $A_v$ and $B_p$ operators, and their eigenvalues can be straightforwardly read off. Again, an $e_1$ anyon resides at the vertex $x_1$, an $m_2$ anyon resides in the upper-left plaquette of the vertex $x_2$, and an $\varepsilon_3$ anyon is a composite of $e_3$ and $m_3$.

Next, we turn to the moving and splitting operators. A direct consequence of the background $m$ anyons is that the string operators of $e$ and $\varepsilon$ anyons defined in Sec. \ref{subsubsec: toric code data} no longer satisfy Eq. \eqref{eq: seamless}. For example, the $-$ sign in the following equation means that Eq. \eqref{eq: seamless} is violated:
\begin{equation}
    \BLvert\scalebox{0.7}{\flipTCdeformationa}\Brangle=-\BLvert\scalebox{0.7}{\flipTCdeformationb}\Brangle,
\end{equation}
where each red circle represents a background $m$ anyon and the blue segments represent the string operators of $e$ in Sec. \ref{subsubsec: toric code data}. The reason for this $-$ sign is evident: the two sides of the above equation differ by a full braiding of $e$ and $m$ anyons, and hence the $-$ sign is the result of the braiding.

The strategy to find moving, splitting and $V$ operators that satisfy Eqs. \eqref{eq: seamless}, \eqref{eq: simplifying condition 1} and \eqref{eq: simplifying condition 2} is to find the connection between this flipped toric code and the original toric code. Note that the ground state of the flipped toric code (denoted as $\ket{\tilde \psi}$) is connected to the ground state of the original toric code model (denoted as $\ket{\psi}$) via a product of commuting string operators $\sigma^x$ applying alternatively as the following,
\begin{equation} \label{Eqfig:TCtoflipTC}
    \ket{\tilde \psi}=\prod ' \sigma^x\ket{\psi}=\BLvert\scalebox{0.7}{\flipTCgsb}\Brangle=\BLvert\scalebox{0.7}{\flipTCgsc}\Brangle,
\end{equation}
where $\prod'$ means that the product is only over the qubits colored in red above. So the appropriate string operator for anyon $a$ in the flipped toric code, denoted as $\tilde O^a_{21}$, can be achieved by dressing the string operators in the original toric code with the conjugation of the product of $\sigma^x$ operators organized as in Eq. \eqref{Eqfig:TCtoflipTC}, \ie $\tilde O^a_{21}=\prod'\sigma^xO^a_{21}\prod' \sigma^x$. For example, if an $e$ string operator $\tilde O^e_{21}$ is supported on the following blue curve,
\begin{equation}
    \bmm\flipTCstring\emm,
\end{equation}
then the corresponding moving operator should be $M^e_{21}=\tilde O^e_{21}=-O^e_{21}$. As such, when we choose $(S^{ab})_{21}=M^b_{21}=\tilde O^b_{21}$ for anyons $a$ and $b$, the moving operators clearly satisfy Eq. \eqref{eq: seamless} because 
\begin{equation*}
    \bsp
    &M^b_{32}M^b_{21}=\tilde O^b_{32}\tilde O^b_{21}\\
    =&(\prod' \sigma^x)O^b_{32}(\prod' \sigma^x)\times (\prod' \sigma^x)O^b_{21}(\prod' \sigma^x)\\
    =&(\prod' \sigma^x)O^b_{31}(\prod' \sigma^x)\\
    =&M^b_{31}.
    \esp
\end{equation*}
As a result, the moving operators for $m$ anyons are the same as the previous model $\tilde O^m_{21}=O^m_{21}$, while those for the $e$ and $\varepsilon$ anyons should have an additional $-1$ sign when an odd number of $\sigma^z$ operators are conjugated by the $\prod'\sigma^x$ operator in Eq. \eqref{Eqfig:TCtoflipTC}. 

Finally, we need the $V$ operators. To find these operators, we set a coordinate as in Sec. \ref{subsubsec: toric code data}. Without loss of generality, we assume that the origin, which is also the $C_4$ rotation center, locates at the gray solid circle below:
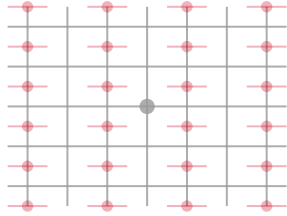
\begin{equation*}
    \flipTCorigin.
\end{equation*}
This figure sets the relative position between the origin and the qubits being acted by $\prod'\sigma^x$ in Eq. \eqref{Eqfig:TCtoflipTC}. Given the definitions of anyon states, moving and splitting operators, it is straightforward to find the $V$ operators. Nontrivial $V$ operators for symmetry generators are summarized as follows. Denote the coordinates of the position $x$ by $(l_1,l_2)$. Then
\begin{equation}\label{Eq:flipTCVs}
    \bsp
    &V^e_{T_1}(x)=V^{\varepsilon}_{T_1}(x)=(-1)^{l_2},\\
    &V^e_{C_4}(x)=(-1)^{l_1l_2},\; V^m_{C^4}(x)=\sigma^x_l,\; V^{\varepsilon}_{C_4}(x)=(-1)^{l_1l_2}\sigma^x_l,\\
    &V^m_{C_2}(x)=V^{\varepsilon}_{C_2}(x)=\sigma^x_l\sigma^x_b,\\
    &V^e_{C_4}(x)=(-1)^{l_1l_2},\; V^m_{(C_4)^{-1}}(x)=\sigma^x_u,\\
    &V^{\varepsilon}_{(C_4)^{-1}}(x)=(-1)^{l_1l_2}\sigma^x_u,\\
    &V_{M}^m(x)=V^{\varepsilon}_M(x)=\sigma^x_{u},\\
    &V^e_{\mathcal{T}}(x)=V^{m}_{\mathcal{T}}(x)=(-1)^{l_1+l_2},
    \esp
\end{equation}
where $\sigma^x_{l,r,u,d}$ has the same meaning as in Eq. \eqref{eq: nearby Pauli X operators}. The $V$ operator for a generic symmetry action is again given by Eq. \eqref{Eq:TCgeneralV}.

Given the definition of anyon states, the moving and splitting operators, and the $V$ operators, it is again straightforward to show that all the $F$-matrices are $1$ and the $R$-matrices are given by $R(a,b)=e^{i\pi a_eb_m}$, just as the toric code. The $U$-phases are still given by Eqs. \eqref{Eq:TCUother}, \eqref{Eq:TCUM}, and \eqref{Eq:TCderiveU}. The $\eta$-phases are derived by combining Eqs. \eqref{Eq:flipTCVs}, \eqref{Eq:TCgeneralV}, and \eqref{Eq:TCderiveeta}. The explicit values of $\eta$-phases are different from the those in toric code. For example, we can show that
\begin{equation}
    \bsp
    &\eta_e(T_1,T_2)=V_{T_1T_2}^e(x_1)V^e_{T_2}(\lsupsc{\bar T_1}x_1)V^e_{T_1}(x_1)=1;\\
    &\eta_e(T_2,T_1)=V_{T_1T_2}^e(x_1)V^e_{T_1}(\lsupsc{\bar T_2}x_1)V^e_{T_2}(x_1)=-1,
    \esp
\end{equation}
Hence Eq. \eqref{eq: translation fractionalization} holds, indicating the expected translation fractionalization.

\subsubsection{Identifying SET phase and matching anomaly}

With the data obtained above, using the same approach as in Sec. \ref{subsubsec: SET toric code}, we find that for the flipped toric code
\beq \label{eq: coefficients flipped toric code}
\bsp
&\vec\chi^e=(1, 1, 1, 0, 0, 0, 1, 0, 0, 0)^T,\\
&\vec\chi^m=(0, 0, 0, 0, 0, 0, 1, 1, 1, 0)^T.
\esp
\eeq
These two vectors specify an SET phase in Ref. \cite{Ye2023}.

Moreover, using the same method as in Sec. \ref{subsubsec: SET toric code}, we can verify that the flipped toric code indeed has the same Lieb-Schultz-Mattis anomaly as the toric code, which is expected.

\subsection{$\z_2$ TQSL on ruby lattice} \label{subsec: ruby lattice}

\begin{figure*}
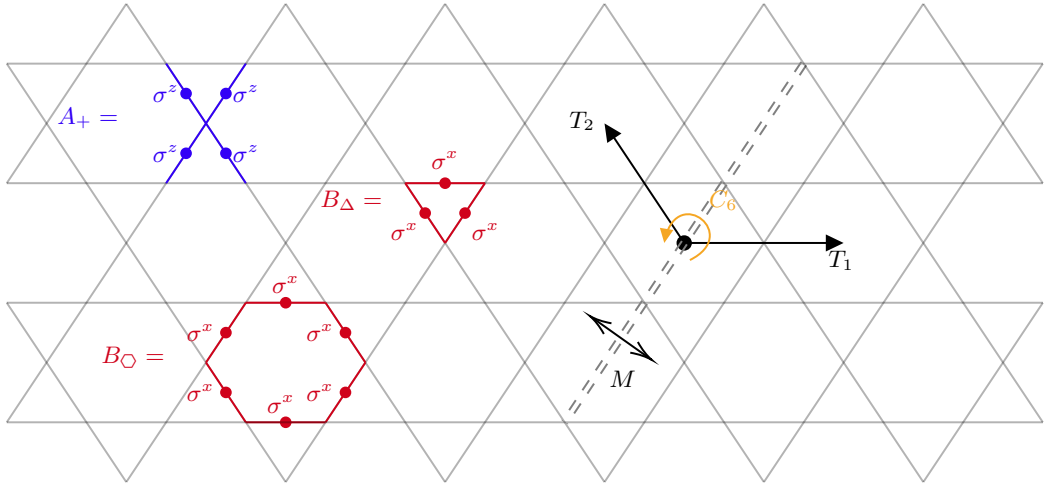

    \centering
    \rubyTC
    \caption{The qubits are on the links of the kagome lattice, which are also sites of the ruby lattice. For the generators of the $p6m$ symmetry, $T_{1,2}$ are translations by 1 unit cell along the two directions as indicated, $C_6$ is a 6-fold rotation around a plaquette with $C_6^2$ denoted by $C_3$ and $C_6^3$ denoted by $C_2$, and $M$ is a reflection whose reflection axis if the double dashed line.}
    \label{fig:rubyTC}
\end{figure*}

The last example is the $\z_2$ TQSL on a ruby lattice, which has been realized using Rydberg atoms \cite{Semeghini2021}. The degrees of freedom in this TQSL resides on the sites of a ruby lattice, which can also be viewed as edges of a Kagome lattice. The system has a $p6m\times\z_2^T$ symmetry, where the generators $p6m$ are defined in Fig. \ref{fig:rubyTC}, and the $\z_2^T$ acts by complex conjugation, so that all qubits are Kramers singlets under $\z_2^T$. This symmetry is anomaly free, which can be easily seen from the fact that it is compatible with a product state $|0\rangle$, the simultaneous eigenstate of all $\sigma^z$ operators with eigenvalue 1.

The Hamiltonian for this TQSL is complicated to analyze, but it turns out that the following exactly solvable Hamiltonian describes a TQSL in the same SET phase as the original one \cite{Tarabunga2022} (see other related Hamiltonians in Refs. \cite{Samajdar2022} and \cite{Verresen2022}):
\begin{equation} \label{eq: ruby Hamiltonian}
    H_{\textrm{rTC}}=W\sum_{+}A_+-J_1\sum_{\Delta}B_{\Delta}-J_2\sum_{\hexagon}B_{\hexagon},
\end{equation}
where the coefficients $W, J_1, J_2$ are all positive. An $A_+$ operator acts on a vertex and applies $\sigma^z$ to qubits on all four edges connecting the vertex (see Fig. \ref{fig:rubyTC}), a $B_{\hexagon}$ ($B_{\Delta}$) operator acts on a hexagon (triangle) plaquette and applies $\sigma^x$ to the qubits on the six (four) edges enclosing the plaquette. We will use Eq. \eqref{eq: ruby Hamiltonian} to analyze the universal properties of this SET phase.

Eq. \eqref{eq: ruby Hamiltonian} actually realizes the same $\z_2$ topological phase as the one in Secs. \ref{subsec: toric code} and \ref{subsec: flipped toric code} \cite{Tarabunga2022}. Moreover, the ground state is an eigenstate of all $A_+$, $B_{\hexagon}$ and $B_{\Delta}$ operators, with eigenvalues for all $A_+$ operators being $-1$, and the eigenvalues for all $B_{\hexagon}$ and $B_{\Delta}$ operators being 1, which indicates the presence of a background anyon at each vertex, similar to the flipped toric code in Sec. \ref{subsec: flipped toric code}. Using the convention in Ref. \cite{Tarabunga2022}, this background anyon is denoted by $e$.

\subsubsection{Extracting data} \label{subsubsec: ruby data}

Now we apply similar procedures as in Secs. \ref{subsec: toric code} and \ref{subsec: flipped toric code} to extract a representative set of universal data that characterizes the SET phase Eq. \eqref{eq: ruby Hamiltonian} belongs to.

We first set up the convention for anyon states. We use the position of the centers of hexagon plaquettes to label the positions of anyons. 
\begin{itemize}
    \item An anyon state $\ket{m_1,m_2}$ is defined by applying the following projectors to $\ket{0}$ and then normalizing the state: $\frac{1-B_{\hexagon}}{2}$ for hexagon plaquettes at $x_1$ and $x_2$, $\frac{1+B_{\hexagon}}{2}$ for all other hexagon plaquettes, $\frac{1+B_\Delta}{2}$ for all triangle plaquettes, and $\frac{1-A_{+}}{2}$ for all vertices. 

    \item An anyon state $\ket{e_3,e_4}$ is defined by applying the following projectors to $\ket{0}$ and then normalizing the state: $\frac{1+A_{+}}{2}$ for the {\it bottom-left} vertices of the hexagon plaquettes at $x_3$ and $x_4$, $\frac{1+B_{\hexagon}}{2}$ for all hexagon plaquettes, $\frac{1+B_\Delta}{2}$ for all triangle plaquettes, and $\frac{1-A_{+}}{2}$ for all other vertices. 

    \item An anyon state $\ket{\varepsilon_5,\varepsilon_6}$ is defined by applying the following projectors to $\ket{0}$ and then normalizing the state: $\frac{1-B_{\hexagon}}{2}$ for the hexagon plaquette at $x_5$ and $x_6$, $\frac{1+A_{+}}{2}$ for the bottom-left vertices of the hexagon plaquettes at $x_5$ and $x_6$, $\frac{1+B_{\hexagon}}{2}$ for all other hexagon plaquettes, $\frac{1-A_{+}}{2}$ for all other vertices, and $\frac{1+B_\Delta}{2}$ for all triangle plaquettes. 

    \item An anyon state $\ket{e_7,m_8,\varepsilon_9}$ is defined by applying the following projectors to $\ket{0}$ and then normalizing the state: $\frac{1-B_{\hexagon}}{2}$ for the hexagon plaquette at $x_8$ and $x_9$, $\frac{1+A_{+}}{2}$ for the bottom-left vertices of the hexagon plaquettes at $x_7$ and $x_9$, $\frac{1+B_{\hexagon}}{2}$ for all other hexagon plaquettes, $\frac{1-A_{+}}{2}$ for all other vertices, and $\frac{1+B_\Delta}{2}$ for all triangle plaquettes. 
\end{itemize}
All these states are simultaneous eigenstates of all $A_+$, $B_{\hexagon}$ and $B_\Delta$ operators, and the corresponding eigenvalues can be easily read off. To summarize our convention, an $m_1$ anyon resides in the hexagon plaquette at $x_1$, an $e_2$ anyon resides at the bottom-left vertex of the hexagon plaquette at $x_2$, and an $\varepsilon_3$ anyon is the combination of $m_3$ and $e_3$.

After specifying the anyon states, we now set up the convention for moving and splitting operators. Again, we specify the string operators $O^a_{21}$ for each nontrivial anyon $a$ in this TQSL, and the moving and splitting operators are chosen to be $S^{ab}_{21}=M^b_{21}=O^b_{21}$. Below we take the direction of a string operator $O^a_{21}$ to be starting from $x_1$ and pointing to $x_2$. 
\begin{itemize}
    \item For $O^m_{21}$ operator, we choose a string connecting the hexagon plaquettes $x_1$ and $x_2$, such that, along the direction of $O^m_{21}$, the $m$ anyon always goes through the triangle on the right side with respect to the moving direction. Then we may attempt to apply $\sigma^z$ operators to all qubits cut by the string, as depicted in the following.
    \begin{equation*}
        \scalebox{0.7}{\rubyTCmstring}.
    \end{equation*}
    Notably, this naive combination of $\sigma^z$ operators does not satisfy Eq. \eqref{eq: seamless}. For example, as shown in the picture below, if we move an $m$ anyon to the next plaquette and then move it back using this naive string operator, this smallest closed loop operator of $m$ anyon is identical to an $A_+$ operator and has $-1$ eigenvalue on the ground state, implying the violation of Eq. \eqref{eq: seamless}:
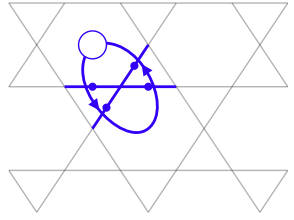
    \begin{equation*}
        \scalebox{0.7}{\rubyTCmloop}.
    \end{equation*}
    So we use the same strategy as in Sec. \ref{subsec: flipped toric code} to find the appropriate string operators, \ie we first annihilate all the background $e$ anyons by applying a certain configuration of $\sigma^x$ operators, such that that each vertex is connected by an odd number of links on which there is a qubit to be acted by the $\sigma^x$ operator(s), and then we conjugate the naive $m$ string operators with these $\sigma^x$ operators. We choose the configuration of $\sigma^x$ operators as indicated by the red lines below.
    \begin{equation*}
    \scalebox{0.6}{\rubyTCdimera}
    \end{equation*}
    The $m$ string operators are then defined as $O^m_{21}=(-1)^{\#}\prod \sigma^z$, where $\prod \sigma^z$ are the Pauli $z$ operators acting on the chosen string, and $\#$ is the number of red lines the chosen string intersects with. 
    
    \item For $O^e_{21}$ operator, we find the shortest line that connects the bottom-left vertex of $x_1$ and $x_2$, as shown in the below (the red line). The $O^e_{21}$ operator defined by applying Pauli-$x$ operators to all the spins on the line.    

    \begin{equation*}
        \scalebox{0.7}{\rubyTCestring}.
    \end{equation*}

    \item The string operators for $\varepsilon$ anyons are defined in a similar manner as the toric code model. we first define $\hat O^\varepsilon_{21}$ operators for the cases where $x_1$ and $x_2$ can translate to each other by applying $(T_1)^{\pm1}$ or $(T_2)^{\pm1}$. In such cases, we directly connect them with a straight segment, and apply $O^e_{21}$ using this segment as the string connecting $x_1$ and $x_2$. Then, we translate this segment to let it connect the bottom-left vertex of $x_1$ and $x_2$ and apply $\sigma^x$ operators to all the spins on this moved string.
    \begin{equation*}
        \scalebox{0.6}{\rubyTCepsilonstring}
    \end{equation*}
    
    A generic $O^\varepsilon_{43}$ is obtained by multiplying together all the straight segments. We first specify the support support of the string operator by choosing a string connecting $x_3$ and $x_4$, whose straight segments only go in $(T_1)^{\pm1}$ or $(T_2)^{\pm1}$ directions, but not $(T_1T_2)^{\pm1}$. Suppose this string consists of $N$ straight segments. We label each corner point as $x_{3_i}$, $i=1,2...N-1$, and the segments are denoted by $3_{i}3_{i-1}$. Then, 
    \begin{equation}
        O^\varepsilon_{43}=\hat O^\varepsilon_{43_{N-1}}\hat O^\varepsilon_{3_{N-1}3_{N-2}}\cdots \hat O^\varepsilon_{3_{2}3_1}\hat O^\varepsilon_{3_13},
    \end{equation}
    where $O^\varepsilon_{43_{N-1}}$, $O^\varepsilon_{3_{i}3_{i-1}}$, and $O^\varepsilon_{3_13}$ are defined above for straight segments. see below for an example.
    \begin{equation*}
        \scalebox{0.6}{\rubyTCepsilonstringb}
    \end{equation*}

\end{itemize}

Having specified the moving and splitting operators, we now find the $V$ operators. We again set a coordinate with the origin at the $C_6$ rotation center (\ie the solid gray dot below) and label each plaquette with an integer pair $x=(l_1,l_2)$, where increasing $l_1$ ($l_2$) goes along the direction of $T_1$ ($T_2$).
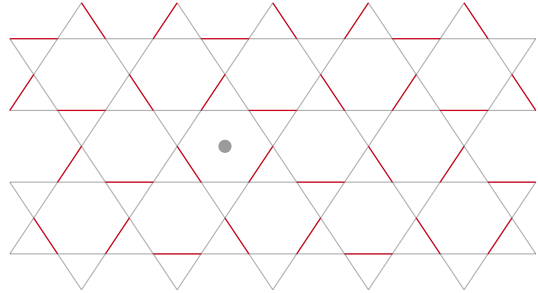
\begin{equation}
    \scalebox{0.6}{\rubyTCdimerb}
\end{equation}
The $V$ operators for $e$ anyons are summarized as follows.
\begin{equation}
    \bsp
    &V^e_{T_1}(x)=V^e_{T_2}(x)=1;\\
    &V^e_{C_6}(x)=\bmm
    \scalebox{0.5}{\rubyTCVea}\emm,\;V^e_{C_3}(x)=\bmm
    \scalebox{0.5}{\rubyTCVeb}\emm,\; V^e_{C_2}(x)=\bmm
    \scalebox{0.5}{\rubyTCVec}\emm;\\
    &V^e_{(C_3)^-1}(x)=\bmm
    \scalebox{0.5}{\rubyTCVed}\emm,\;V^e_{(C_6)^-1}(x)=\bmm
    \scalebox{0.5}{\rubyTCVee}\emm;\\
    &V^e_{M}(x)=V^e_{\mathcal{T}}(x)=1.
    \esp
\end{equation}
Each red line means applying $\sigma^x$ to the qubits on it. 

For $m$ anyons, the $V$ operators are summarized as follows. For the translation generators, reflection, and the time-reversal generator, the $V$ operators are
\begin{equation}
    V^m_{T_1}(x)=V^m_{T_2}(x)=(-1)^{l_2},V^m_{M}=V^m_{\mathcal{T}}(x)=1.   
\end{equation}
For the $C_6$ rotation symmetry, the $V$ operators barely have a closed form. We summarize the $V_{C_6}^m$ operators around the rotation center in the following.
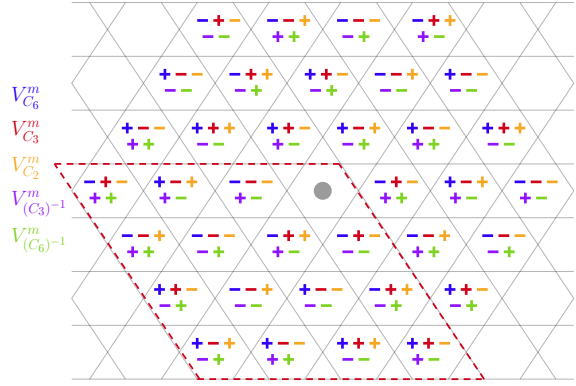
\begin{equation*}
    \scalebox{0.45}{\rubyTCVmall}
\end{equation*}
Here, the $+$ ($-$) sign in each plaquette means that the $V^m$ operator at this position is $+1$ ($-1$), different colors represent different group elements of $C_6$ as indicated. The gray dot in the center is the rotation center, where the $V^m$ operators are $+1$. The region enclosed by red dashed line is the ``unit cell" of $V^m_{C_6^k}$ operators; translating this `unit cell' along $(T_1)^{\pm1}$ or $(T_2)^{\pm1}$ gives the $V^m_{C_6^k}$ operators for the entire space.

The $V^{\varepsilon}_g$ are chosen to be $V^{\varepsilon}_g(x)=V^e_g(x)V^m_g(x)$ for any $x$ and $g$, and for generic symmetry action $g$, the $V_g$ operators are given by 
\begin{equation}\label{Eq:TCgeneralVruby}
    \bsp
    &V_{(T_1)^{n_1}(T_2)^{n_2}(C_6)^{n_3}M^{n_4}\mathcal{T}^{n_5}}^a(x)=\\
    &V_{\mathcal{T}^{n_5}}^a(\lsupsc{M^{-n_4}(C_6)^{-n_3}(T_2)^{-n_2}(T_1)^{-n_1}}x)\times\\&
    V_{M^{n_4}}^a(\lsupsc{(C_6)^{-n_3}(T_2)^{-n_2}(T_1)^{-n_1}}x)\times\\
    &V_{(C_6)^{n_3}}^a(\lsupsc{(T_2)^{-n_2}(T_1)^{-n_1}}x)V_{(T_2)^{n_2}}^a(\lsupsc{(T_1)^{-n_1}}x)V_{(T_1)^{n_1}}^a(x).
    \esp
\end{equation}

It is then straightforward to find that the $F$ symbols are still $1$ and the $R$ symbols are $R(a,b)=e^{i\pi a_eb_m}$. The $U$-phases of symmetry generators are given by 
\begin{equation}\label{Eq:TCUotherruby}
    \bsp
    &U_{T_1}(a,b)=U_{T_2}(a,b)=U_{C_6^k}(a,b)=U_{\mathcal{T}}(a,b)=1,\\
    &U_{M}(\varepsilon,\varepsilon)=U_{M}(m,e)=U_M(m,\varepsilon)=U_M(\varepsilon,e)=-1.
    \esp
\end{equation}
All other $U$-phases can be calculated via
\begin{equation}
    U_{gh}(a,b;c)=U_h(a,b;c)U_g(a,b;c).
\end{equation}
The $\eta$-phases are calculated by
\begin{equation}
    \eta_a(g,h)=\text{sign}(V_{gh}^a(x)R_g V_h^a(\lsupsc{\bar g}x)R_g^{\dagger}V_g^a(x)).
\end{equation}
Notably, we have $\eta_m(T_1,T_2)/\eta_m(T_2,T_1)=-1$, indicating that the $m$ anyons have translation fractionalization.

\subsubsection{Identifying SET phase and matching anomaly}

With the data obtained above, we now identify the SET phase in Ref. \cite{Ye2023} that Eq. \eqref{eq: ruby Hamiltonian} belongs to, and calculate its anomaly. 

Similar to the previous cases, each SET phase in this case is characterized by the fractionalization pattern of the $p6m\times\z_2^T$ symmetry on the $e$ and $m$ anyons, which is mathematically described by an element in $\mathcal{H}^2(p6m\times Z_2^T, \mathbb{Z}_2\times \mathbb{Z}_2)$. So identifying the SET phase amounts to identifying the element in $\mathcal{H}^2(p6m\times Z_2^T, \mathbb{Z}_2\times \mathbb{Z}_2)$ that the $\eta'$-phases of Eq. \eqref{eq: ruby Hamiltonian} correspond to. Concretely, we define a 7-entry vector (see Table XIII in Ref. \cite{Ye2023}):
\beq \label{eq: vector basis p6m}
\bsp
\vec v_{p6m\times\z_2^T}=(B_{xy}, A_{c^2}, A_cA_m, A_m^2, A_ct, A_mt, t^2)^T,
\esp
\eeq
with each entry being a 2-cocycle. Note that these 2-cocycles are different from those in Eq. \eqref{eq: p4m basis cocycles}, although we use similar symbols in these two cases. The precise definitions of these 2-cocycles are given in Appendix \ref{app: SET Z2}. Then there are 2 other 7-entry vectors $\vec\chi^e$ and $\vec\chi^m$, such that (up to coboundary transformations)
\beq \label{eq: expansion p6m}
\bsp
&\eta'_e(g_1, g_2)=(-1)^{\vec\chi^e\cdot\vec v_{p6m\times\z_2^T}},\\
&\eta'_m(g_1, g_2)=(-1)^{\vec\chi^m\cdot\vec v_{p6m\times\z_2^T}}.
\esp
\eeq
Each pair of $\vec\chi^e$ and $\vec\chi^m$ specifies an element in $\mc{H}^2(p6m\times\z_2^T, \z_2\times\z_2)$, and thus an SET phase in Ref. \cite{Ye2023}. In Appendix \ref{app: SET Z2}, we explain the physical meanings of these $\eta'_e(g_1, g_2)$ and $\eta'_m(g_1, g_2)$, in terms of the fractionalization patterns of the $p4m\times\z_2^T$ symmetry on the $e$ and $m$ anyons.

To identify which element in $\mc{H}^2(p6m\times\z_2^T, \z_2\times\z_2)$ the TQSL described by Eq. \eqref{eq: ruby Hamiltonian} corresponds to, we use our precise crystalline equivalence principle Eq. \eqref{Eq:correspondence} to map the data obtained in Sec. \ref{subsubsec: ruby data} to get the universal data for a corresponding SET phase with a purely internal $p6m\times\z_2^T$ symmetry. The obtained $\eta'$-phases can be written as Eq. \eqref{eq: expansion p6m}, with (see Appendix \ref{app: SET Z2} for the derivation):
\beq \label{eq: coefficients ruby}
\vec\chi^e=\vec\chi^m=(0, 0, 0, 0, 0, 0, 0)^T.
\eeq
These two vectors specify an SET phase in Ref. \cite{Ye2023}. Note that the apparent triviality of these two vectors is due to the choice of the ``basis" in Eq. \eqref{eq: basis p6m}, and it does not mean that the information of symmetry fractionalization is gone. For example, we still have $\eta'_m(T_1, T_2)/\eta'_m(T_2, T_1)=-1$, showing the translation fractionalization.

Finally, we turn to the anomaly. According to Eq. (D8) in Ref. \cite{Ye2023}, the anomaly of a $p6m\times\z_2^T$ symmetric $\z_2$ TQSL can be specified by the following 14 anomaly indicators (again, we are using similar symbols as in Eq. \eqref{eq: p4m indicators}, but they have different meanings):
\begin{equation}
    \bsp
    &I_0=\mathcal{I}_0\\
    &I_1=\mathcal{I}_1(\mathcal{T})\quad I_2=\mathcal{I}_1(M)\quad I_3=\mathcal{I}_1(C_2\mathcal{T})\\
    &I_4=\mathcal{I}_1(C_2M)\quad I_5=\mathcal{I}_2(\mathcal{T},C_2\mathcal{T})\quad I_6=\mathcal{I}_2(\mathcal{T},M)\\
    &I_7=\mathcal{I}_2(C_2\mathcal{T},M)\quad I_8=\mathcal{I}_2(C_2\mathcal{T},C_2M)\\
    &I_9=\mathcal{I}_2(M,C_2M)\quad I_{10}=\mathcal{I}_1(T_1T_2C_2\mathcal{T})\\ &I_{11}=\mathcal{I}_2(M,T_1T_2C_2\mathcal{T})\quad
    I_{12}=\mathcal{I}_2(\mathcal{T},T_1T_2C_2\mathcal{T})\\ &I_{13}=\mathcal{I}_2(M,T_1T_2C_2M),
    \esp
\end{equation}
where $\mathcal{I}_0$, $\mathcal{I}_1$, and $\mathcal{I}_2$ are given in Eq. \eqref{Eq:baseindicators}. It is straightforward to check that all anomaly indicators are 1 in this case, indicating the absence of any nontrivial anomaly, as expected.

\section{Discussion}\label{sec:discussion}

In this paper, we have presented a microscopic universal theory for generic symmetry-enriched TQSLs in two spatial dimensions. For a given TQSL, this theory takes as input some microscopic anyon states and some operators that control the motion, splitting and symmetry actions of the anyons, and it outputs a set of data that characterizes the universal properties of the entire SET phase this TQSL belongs to. The mathematical structure underlying this universal data is some generalization of the category theory. Our theory offers a theoretical foundation for identifying and manipulating topological phases numerically and experimentally, at least in principle.

Based on this theory, we have established a precise crystalline equivalence principle, which is an explicit bijective map between the universal data that characterizes an SET phase with a symmetry group $G$, which can include both lattice and internal symmetries, and an SET phase with a purely internal symmetry $G$. We have demonstrated our theory in various examples, and verified that the Lieb-Schultz-Mattis anomaly matching conditions hold in all these examples.

We finish this paper by listing some important future directions.

\begin{enumerate}
    
    \item Our theory is physics-based and it leads to mathematically precise results. Although we believe in the correctness of our theory, some of our working assumptions have not been rigorously proved. For example, we have not shown from the first principles that in a TQSL there must be finitely many types of deconfined anyons, nor have we proved the symmetry localization property Eq. \eqref{eq: symmetry localization}. Also, we have not proved that there is no universal data other than what we identity. For mathematical rigor, it is necessary to prove these assumptions, ideally based solely on the fundamental principles of quantum mechanics.

    Some previous work in this direction has been put forward, which focuses on infinite-size systems \cite{Cha2018, Ogata2021a, Kawagoe2024}. Although it is useful to understand the thermodynamic limit that is defined in infinite size, we emphasize that, ultimately, it is important to rigorously formulate and prove these results in {\it finite} systems, because all experimentally relevant lattice systems are finite, no matter how large they are. In this regard, one should estimate all finite-size errors, as done in some of our appendices.

    \item The input of our microscopic universal theory is some anyon states and operators that control the dynamics and symmetry actions of the anyons. In a realistic system where a TQSL emerges, it is often challenging to find these states and operators. Therefore, it is useful to develop numerical and experimental methods to identify them. In this direction, Refs. \cite{Shi2019, Cian2022, Morampudi2016, Kirchner2025} are potentially useful. In particular, besides other numerical and experimental difficulties, in our formulation we need to specify the parameters $\gamma$, $\Gamma$ and $\omega$ in Eqs. \eqref{eq: gamma main}, \eqref{eq: Gamma main} and \eqref{eq: omega main} in order to obtain a representative set of universal data. Determining these parameters numerically is possible, but it is rather challenging experimentally, which often requires difficult interference experiments. Interestingly, the scheme designed in Ref. \cite{Kawagoe2019} to extract the $F$- and $R$-symbols is independent of these parameters (at the expense of introducing some other experimental challenges). It is useful to design a scheme to extract the $U$- and $\eta$-symbols that does not depend on these parameters.

    \item Our paper focuses on spin and bosonic systems, but the best experimentally studied topological phases are fractional quantum Hall systems, which are fermionic systems \cite{Stormer1999}. Some previous theories of such systems, which have not been made microscopically concrete and have not been developed for general symmetries, can be found in Refs. \cite{Aasen2021, Bulmash2021, Bulmash2021b, Barkeshli2021, Lan2023}. It is important to generalize our framework to these systems. We expect that most of our formalism can be applied there.

    \item In a symmetric TQSL, the symmetry defects, such as lattice dislocations and magnetic domain walls, are expected to have interesting properties \cite{barkeshli2014}. Our theory captures the interplay between the anyons and symmetry, so we expect that all universal properties of the symmetry defects should also be encoded in our theory. It is interesting to extract these properties from our theory. We note that a thorough microscopic understanding of these properties is needed to implement quantum computation using symmetry defects.

    \item Currently, a widely used method to study TQSLs is based on some gauge theories or their variants, which are often obtained through some parton constructions or their analogs \cite{Wen2001}. As discussed by many previous papers, although this method is useful, it is neither systematic nor conceptually direct \cite{Kitaev2006, Essin2012, barkeshli2014, Ye2023}. It is useful to fit the language employed in this method into our microscopic universal theory.

    \item Some SET phases of TQSLs beyond the gauge theoretic method were predicted in Ref. \cite{Ye2023}. It is interesting to use our microscopic universal theory as the basis to construct models to realize these phases.

    \item Recently, the studies on mixed states and on systems with average symmetries have gained much attention \cite{Ma2022a, Ma2023}. It is interesting to generalize our theory to these settings.
    
    \item Some phenomena related to symmetry fractionalization can be re-interpreted using generalized symmetries \cite{Delmastro2022, Brennan2022}. It is interesting to reformulate our theory in that language.
    
\end{enumerate}

\begin{acknowledgements}

We thank Wen Wei Ho, Michael Levin, Chong Wang, Weicheng Ye, Jinmin Yi for helpful discussions. LZ is supported by the National University of Singapore start-up grants A-0009991-00-00 and A-0009991-01-00.

\end{acknowledgements}

\begin{appendices}

\section{Locality in finite systems} \label{app: locality}

As discussed in Sec. \ref{sec:setup}, in this paper we consider large but finite lattice systems, and we aim to find universal low-energy, long-distance properties of symmetry-enriched TQSLs. Studying finite systems is important, because all experimental systems are finite, no matter how large they are. However, the notion of locality, which is the basis of many-body physics, is most cleanly defined in infinite-size systems. In this appendix, we clarify the notion of locality in finite systems. For most considerations in this paper, there may not be any confusion even without the discussion in this appendix, and this appendix is presented for completeness and further clarity.

Suppose that the size of the system is $L_1\times L_2$, with $L_{1,2}$ the linear size of the system along each direction. We call operators whose supports can be fully covered by a disk with diameter $D$ {\it size-$D$ operators}. We assume there is a $D$ such that we only have access to size-$D$ operators, \ie the Hamiltonian of the system can always be written as a sum of such operators. Note that expectation values of size-$D$ operators are generically insufficient for specifying the state of our system, and for this purpose we also need correlation functions of multiple size-$D$ operators separated by distances potentially larger than $D$. These correlation functions can be measured via linear and non-linear responses, even if we can only access size-$D$ operators \cite{Morampudi2016, Kirchner2025}.

The assumption of a maximal range $D$ for the size of our accessible operators is experimentally motivated. Conceptually, $D$ sets the length scale of local operators, \ie size-$D$ operators are viewed as local operators and operators with larger sizes will be considered non-local. Also, $D$ is roughly the length scale of the ``coarse grained sites" or ``spatial resolution of measurements" in the real-space renormalization group picture.

The above definition of local operators has implications on anyon states. In general, the linear sizes of the anyons are at least of the order of the correlation length, $\xi$. Therefore, we demand that $D\geqslant c_1\xi$, with $c_1$ some order-1 constant. Because local operators are just size-$D$ operators and an anyon by definition represents a collection of states that can be converted into each other by local operators, the distances between the anyons must be larger than $D$ (and, of course, smaller than the system size). In other words, if two anyons were to be put together so that their distance is smaller than $D$, they would automatically be viewed as a single anyon corresponding to one of their fusion outcomes, rather than two individual anyons, since we do not resolve the physics at length scales below $D$.

The value of $D$ has no universal meaning. The universal long-distance physics we aim to describe should only include physical quantities that converge for large enough $D$ and large enough system size. More precisely, for {\it any} fixed finite length scale $D>D_0$, where $D_0$ is some other fixed length scale, consider a sequence of systems such that the $i$-th system has size $L_1^{(i)}\times L_2^{(i)}$, with $L_1^{(i)}\geqslant L_1^{(i-1)}$ and $L_2^{(i)}\geqslant L_2^{(i-1)}$, for $i=1, 2, \cdots$, where $L_{1, 2}^{(0)}$ is some fixed length scale. We assume that both $L_1^{(i)}$ and $L_2^{(i)}$ increase indefinitely as $i$ increases. Suppose a physical quantity can be defined for all these systems. As $i$ increases, if this quantity calculated in the $i$-th system converges to a value that is {\it independent of $D$}, then we say that this physical quantity represents a long-distance property. Here $D_0$ and $L_{1,2}^{(0)}$ can be viewed as the length scales above which the universal long-distance physics starts to emerges, and these scales are generally much larger than the correlation length.

We illustrate the notion of long-distance properties using examples discussed in this paper. Here we focus on the $U$-matrix defined in Eq. \eqref{eq: U-matrix}, while the discussions can be straightforwardly generalized to the $F$-, $R$- and $\eta$-symbols defined in Eqs. \eqref{Eq:generalF main}, \eqref{eq: general R main} and \eqref{eq: eta}, respectively. As in Eqs. \eqref{eq: U-matrix} and \eqref{eq: U-symbol states}, the $U$-matrix is defined microscopically in terms of the anyon states, splitting operators, and the $V$ operators in Eq. \eqref{eq: symmetry localization}. For any fixed $D>D_0$, since the distances between anyons are larger than $D$, the distance between $x_1$ and $x_2$, denoted by $|\vec x_2-\vec x_1|$, in the splitting operator $(S^{ab}_{c, \mu})_{21}$ is larger than $D$. So it is not always valid to fix $|\vec x_2-\vec x_1|$ and change $D$. Instead, we will always take $|\vec x_2-\vec x_1|=c_2D$ with $c_2>1$ a constant independent of $D$ and the system size. Then in the $i$-th system with size $L_1^{(i)}\times L_2^{(i)}$, as long as this size is large enough to contain the anyons and operators in Eq. \eqref{eq: U-symbol states}, Eqs. \eqref{eq: U-matrix} and \eqref{eq: U-symbol states} make sense, and we denote the $U$-matrix obtained in this case by $U_g(a, b; c; D, L_1^{(i)}, L_2^{(i)}, c_2)$. For a given $c_2$, if the equivalence class (as defined in detail in Appendix \ref{app: full structure}) of $\lim_{i\rightarrow\infty}U_g(a, b; c; D, L_1^{(i)}, L_2^{(i)}, c_2)$ exists and is independent of $D$, this equivalence class represents a long-distance property. In this paper, we assume, but do not attempt to prove, that this limiting equivalence class exists and is independent of $D$, since the existence of a thermodynamic limit is another highly nontrivial subject beyond our current scope.

\section{Short-range correlations of anyon states} \label{app: short-range correlation}

In Sec. \ref{subsec: anyon states}, we have remarked that we only consider short-range correlated anyon states, \ie anyon states where the mutual information between any local region with at most one anyon and any other region is negligible, as long as these regions are separated by a distance no less than the distances between the anyons. In this appendix, we first show that in any topological phase there are always anyon states that violate the condition of short-range correlation, and then argue that there are also short-range correlated anyon states in all topological phases.

Suppose that all anyon states are short-range correlated, then there is a contradiction, because, given a short-range correlated anyon state $|a_1, \bar a_{1'}\rangle$, one can find an anyon state violating the condition of short-range correlation in the same topological phase. To do so, on top of the original system, we first add two more qubits that are in a pure product state and that are located around 1 and $1'$, respectively. By definition, the new system is in the same topological phase as the original system. Now we consider a new state $|a_1, \bar a_{1'}\rangle\otimes|{\rm Bell}_{11'}\rangle$, which is obtained by making the two new qubits form a Bell state without changing the original system. For local regions around 1 and $1'$, the mutual information increases by $2\ln 2$. Therefore, this new anyon state violates the condition of short-range correlation. It is straightforward to generalize this construction to show that there must be states with more than 2 anyons that violate the condition of short-range correlation.

Note that in the above construction, the mutual information between two regions with a distance much larger than the distance between 1 and $1'$ is still negligible, and the condition of short-range correlation is violated just because some operation around 1 and $1'$ is performed, which does not change the correlation between other regions.

\begin{figure}
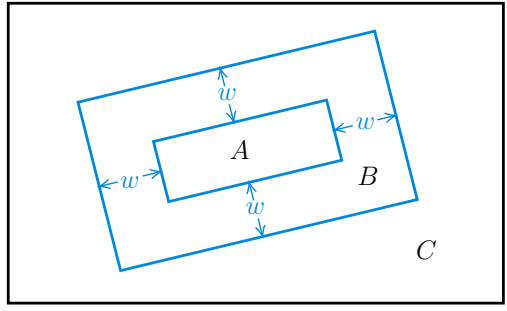

    \centering
    \scalebox{0.7}{\ABCpartition}
    \caption{The finite rectangular region $A$ is shielded from $C$ by the annulus-type rectangular region $B$. The widths of all sides of $B$ are $w$, and the size of $C$ can be comparable to the total system size. Together, regions $A$, $B$ and $C$ partition the entire lattice, which is under periodic boundary conditions. Theorem 1 of Ref. \cite{Yi2025} states that if two states are connected by a quasi-adiabatic continuation $W$, and if one of them has exponentially decaying mutual information between regions $A$ and $C$ (\ie $I(A: C)$, the mutual information between $A$ and $C$, satisfies that $I(A: C)\leqslant I_0e^{-w/\xi}$, where $I_0$ does not depend on the size of $C$ (but can depend on the size of $A$ polynomially) and $\xi$ is independent of the partition), then the other state also has exponentially decaying mutual information between $A$ and $C$. The most remarkable aspect of this theorem is that $I_0$ does not depend on the size of $C$, which is important in the application of Theorem \ref{thm: unitary approximation 1}.}
    \label{fig: partition}
\end{figure}

Next, we argue that short-range correlated anyon states exist in all topological phases, which takes 3 steps:

\begin{enumerate}
	
	\item For TQSLs with a gappable boundary, which can be described by a solvable Levin-Wen model \cite{Levin2004}, under an assumption, we prove that there are anyon states where the mutual information between two regions $A$ and $C$ as in Fig. \ref{fig: partition} is identically zero, \ie $I(A:C)=0$, as long as $A$ contains at most one anyon, and $A$ and $C$ are far enough.
	
	\item For any TQSL in the same phase as the solvable model above, when the region $A$ in Fig. \ref{fig: partition} contains at most one anyon, we use the result in step 1 to prove the existence of anyon states where $I(A: C)$ decays exponentially in $w$, the distance between $A$ and $C$.
	
	\item For a TQSL without a gappable boundary, such as a chiral phase, we again use the result of step 1 to find anyon states where $I(A: C)$ in Fig. \ref{fig: partition} decays exponentially as $w$ increases, if $A$ contains at most one anyon.
	
\end{enumerate}

The above argument is similar to Ref. \cite{Yi2025}, which proves the decay behavior of the mutual information in the ground states of a $(2+1)$D TQSL. Here we extend that result to states with anyons, where the region $A$ contains at most one anyon. As a result, as long as the anyons are far enough, there are short-range correlated anyon states. Below we carry out the three steps above.

\subsection{Vanishing mutual information in solvable models}

First, we argue that in the Levin-Wen model \cite{Levin2004}, there are anyon states where the mutual information between two regions $A$ and $C$ as in Fig. \ref{fig: partition} is identically zero, as long as $A$ and $C$ are far enough and region $A$ contains at most one anyon.

Our argument relies on the following assumption, \ie the relevant anyon state can be realized as a ground state of a Hamiltonian $H=-\sum_jP_j$, where $\{P_j\}$ is a set of mutually commuting local projectors, such that for any operator $O$ supported in any contractible region that contains at most one anyon and is far away from other anyons, $P OP=C_OP$, where $P=\Pi_jP_j$ is the projector onto the ground state subspace of $H$ and $C_O$ is a c-number. This assumption can be explicitly verified for all Levin-Wen models with only Abelian anyons. In particular, $H$ is obtained from the original Levin-Wen Hamiltonian by flipping the signs of various terms. On the other hand, for non-Abelian topological phases that can be realized by the Levin-Wen models, so far this assumption has only been proved for states with ``flux-type" anyons, but not for states with ``charge-type" or ``dyon-type" anyons. However, we believe that this assumption is generally true and will base the following discussion on it. In passing, note that if the support of an operator $O$ contains more than one anyons, the condition $P OP=C_OP$ is generally violated.

To show that the mutual information $I(A: C)=0$, we consider any operator $O_1$ supported on region $A$ in Fig. \ref{fig: partition} and another operator $O_2$ supported on region $C$. The meaning that $A$ and $C$ are far enough is that 1) no projector $P_j$ acts on both $A$ and $C$, and 2) all operators $O_1$ supported in region $A$ satisfy that $PO_1P=C_{O_1}P$. Suppose that $\rho$ is the density matrix of any ground state of $H$ or any classical mixture of the ground states of $H$. We define $\tilde P_{1, 2}=\Pi_{{\rm supp}(P_j)\cap{\rm supp}(O_{1,2})=\emptyset}P_j$, and find that
\beq
\begin{split}
&{\rm Tr}(\rho O_1O_2)
={\rm Tr}(P^2\rho O_1O_2)={\rm Tr}(P\rho P O_1O_2)\\
=&{\rm Tr}(\rho P O_1O_2P)={\rm Tr}(\rho P\tilde P_1 O_1O_2\tilde P_2P)\\
=&{\rm Tr}(\rho PO_1\tilde P_1\tilde P_2O_2P)={\rm Tr}(\rho PO_1PO_2P)\\
=&{\rm Tr}(\rho O_1){\rm Tr}(\rho O_2),
\end{split}
\eeq
where the condition $PO_1P=C_{O_1}P$ is used in the last step. This relation implies that $I(A: C)=0$.

\subsection{Exponentially decaying mutual information in topological phases with gappable boundaries}

Next, we show that in the entire topological phase which the solvable model discussed above belongs to, there are anyon states where $I(A: C)$ decays exponentially as the distance between $A$ and $C$ increases, \ie $I(A: C)\leqslant I_0e^{-w/\xi}$ with $I_0$ and $\xi$ independent of the size of $C$.

To this end, consider a generic system in the same topological phase as the above solvable model, and suppose that $|\psi_0\rangle$ is an anyon state in the above solvable model that has $I(A: C)=0$ when $A$ and $C$ are far enough and $A$ contains at most one anyon. Suppose that the quasi-adiabatic continuation operator that converts the ground state of the solvable model to the ground state of this generic system is $W$ \cite{Hastings2005, Hastings2010, Bachmann2012automorphic, Kapustin2022Noether}. For the anyon state $|\psi_0\rangle$ in the solvable model, $|\psi\rangle = W|\psi_0\rangle$ is an anyon state in the generic system, again with at most one anyon in region $A$. Then Theorem 1 of Ref. \cite{Yi2025} can be invoked to conclude that $I(A: C)\leqslant I_0e^{-w/\xi}$, where $I_0$ does not depend on the size of $C$ (but can depend on the size of $A$ polynomially), and $\xi$ is the correlation length, which is determined by $W$ and does not depend on the partition.

\subsection{Exponentially decaying mutual information in topological phases without gappable boundaries}

Finally, we turn to topological phases without gappable boundaries, such as a chiral phase, which cannot be represented by solvable models of the above form.

In this case, we can stack the time reversal partner of this topological phase with it, so that the combined system can be represented by a solvable model of the above form \cite{Muger2003, Kirillov2010, Davydov2010, Balsam2010, Balsam2010a, Kirillov2011}. Denote the anyon state with zero mutual information in this solvable model by $|\psi_0\rangle$, then there is a quasi-adiabatic continuation operator $W$, such that $W|\psi_0\rangle=|\psi\rangle\otimes|^\mc{T}\psi\rangle$, where $|\psi\rangle$ is a corresponding anyon state in the topological phase of interest here, which has no gappable boundaries, and $|^\mc{T}\psi\rangle$ is the time reversal partner of $|\psi\rangle$. Note that although $|^{\mc{T}}\psi\rangle$ contains the time reversal partners of the anyons in $|\psi\rangle$, because each single anyon in $|\psi_0\rangle$ is labeled by an anyon from $|\psi\rangle$ together with another anyon in $|^\mc{T}\psi\rangle$, the number of anyons in a given region is still the same for all three states $|\psi_0\rangle$, $|\psi\rangle$ and $|^\mc{T}\psi\rangle$. Now Theorem 1 of Ref. \cite{Yi2025} implies that $I(A: C)\leqslant I_0e^{-w/\xi}$ for $|\psi\rangle\otimes|^\mc{T}\psi\rangle$. Because $I(A: C)$ for $|\psi\rangle\otimes|^\mc{T}\psi\rangle$ is twice of $I(A: C)$ for $|\psi\rangle$, $I(A: C)\leqslant I_0e^{-w/\xi}/2$ for $|\psi\rangle$, which is indeed exponentially small in $w$.

\section{More on moving and splitting operators, and symmetry localization} \label{app: approximating operators}

In Sec. \ref{sec:setup}, we have stated that in general the moving and splitting operators are supported in the entire lattice, but they are concentrated in a local region with exponentially decaying tails. We have also claimed that although the moving and splitting operators do not have to be unitary or even invertible, they can be approximated by unitary operators. Moreover, we have mentioned that the symmetry localization property for 3-anyon states in Eq. \eqref{eq: symmetry localization} implies its own generalizations. In this appendix, we sharpen these statements.

\subsection{Exponentially decaying tails of the moving and splitting operators} \label{subapp: exponentially decaying tails}

First, we discuss the ``spatial profiles" of the moving and splitting operators. These results are well known among the experts, and our purpose here is to provide a self-contained rigorous explanation of them.

Using arguments similar to the ones in Appendix \ref{app: short-range correlation}, we see that the moving and splitting operators in a generic $(2+1)$D TQSL can be related to a corresponding operator in a solvable model via a quasi-adiabatic continuation operator $W$, at least after the TQSL is stacked with its time reversal partner{\footnote{When the TQSL is stacked with its time reversal partner, the relevant moving and splitting operators are also stacked with their time reversal partners. This does not change the following result that the spatial profile of these operators are concentrated in a finite region, with only exponentially decaying tails.}}. For concreteness, consider a moving operator denoted by $M$ and its corresponding operator $M_0$ in the solvable model, \ie $M=WM_0W^\dagger$. In general, $M$ is supported in the entire system. However, using the fact that $M_0$ is supported in a local region, we will see that $M$ can be approximated by another operator $\tilde M$ that is supported in the finite region, with an error that is exponentially small in the size of this region. The same analysis applies to the splitting operator. So both the moving and splitting operators are concentrated in a local region, with some exponentially small tails leaking out.

To obtain the above result, denote the support of $M_0$ by $A_0$, denote by $A$ another region that contains $A_0$, and denote by $A^c$ the complement of $A$ in the entire lattice. Then define
\beq
\tilde M=\int dW_0 W_0MW_0^\dagger,
\eeq
where $W_0$ is a unitary operator supported in $A^c$, and $\int dW_0=1$ is the normalized Haar measure of these unitary operators{\footnote{Strictly speaking, here we need to assume that the local Hilbert spaces are finite dimensional, otherwise the normalized Haar measure of unitary operators does not exist. However, we expect that for TQSLs realized in systems where some local Hilbert spaces are infinite dimensional, the moving and splitting operators are still concentrated in a local region.}}. $\tilde M$ is known as the conditional expectation of $M$, and it is straightforward to show that $\tilde M$ is only supported in $A$. Next, we bound the norm difference between $M$ and $\tilde M$:
\beq
\begin{split}
    &||M-\tilde M||
    =||M-\int dW_0 W_0MW_0^\dagger||\\
    =&||\int dW_0M-\int dW_0W_0MW_0^\dagger||\\
    =&||\int dW_0[M, W_0]||
    \leqslant\int dW_0||[M, W_0]||\\
    =&\int dW_0||[WM_0W^\dagger, W_0]||\\
    \leqslant& \int dW_0Ce^{-w/\xi}||M_0||=Ce^{-w/\xi}||M||,
\end{split}
\eeq
where the Lieb-Robinson bound is used in the second last step \cite{Lieb:1972wy, Hastings2010a}, $w$ is the distance between $A^c$ and $A_0$, and $C$ and $\xi$ are constants determined by $W$. Therefore, $M$ is concentrated in a local region with some exponentially decaying tails leaking out, and the same result applies to the splitting operators.

\subsection{Approximating local operators by local unitary operators} \label{subapp: unitary approximation}

Next, we show that, on short-range correlated states, the effects of an operator can be approximated by the effect of a unitary operator with a slightly enlarged support, and the error in this approximation is controlled by the decay behavior of the mutual information of the state. Our discussion is based on the following general theorem{\footnote{We thank Jinmin Yi for formulating and proving this theorem.}}, which will be proved at the end of this subsection.

\begin{theorem} \label{thm: unitary approximation 1}
	Let $|\psi\rangle$ and $|\phi\rangle=O|\psi\rangle$ be two normalized states, where $O$ is a bounded operator supported in region $A$ as in Fig. \ref{fig: partition}. Then there is a unitary $U$ supported in region $AB$, such that
	\beq \label{eq: unitary approximation}
	|\langle\phi|U|\psi\rangle|\geqslant 1-\sqrt{\ln 2\cdot I_{|\psi\rangle}(A: C)/2}(1+||O||^2),
	\eeq
	where $I_{|\psi\rangle}(A: C)$ is the mutual information between $A$ and $C$ in Fig. \ref{fig: partition} for the state $|\psi\rangle$ and $||O||$ is the operator norm of $O$.
\end{theorem}

Let us explain the intuition behind this theorem. When the bounded operator $O$ acts on $|\psi\rangle$, its effects on a region $C$ away from its support $A$ are controlled by $||O||$ and the correlation in the state $|\psi\rangle$, and this correlation can be characterized by the mutual information $I_{|\psi\rangle}(A: C)$. If $I_{|\psi\rangle}(A: C)$ is small enough {\it even if $C$ has a size comparable to the system size}, which is satisfied by our short-range correlated anyon states discussed in Appendix \ref{app: short-range correlation}, $O$ essentially has no effect on $C$. In other words, all effects of $O$ are contained in $AB$, which can then be achieved by a unitary operator $U$ supported in $AB$. Note that although $U$ and $O$ have similar effects on $|\psi\rangle$, their effects on other states can be very different.

As an implication of this theorem, given a moving or splitting operator that is supported in a region $A$, the effects of this operator on a short-range correlated anyon state can approximated by a unitary operator with an enlarged support, $AB$, and the error in this approximation is controlled by Eq. \eqref{eq: unitary approximation}, which is small as long as $B$ is reasonably wide. We note that for the special case where the state has zero mutual information, the unitary approximation for the moving operators was discussed in Refs. \cite{Shi2018, Shi2019}.

Finally, we prove Theorem \ref{thm: unitary approximation 1}. 

\begin{proof} \label{proof: unitary approximation 1}
	We denote by $\psi_A$ the reduced density matrix of $|\psi\rangle$ on region $A$, $\phi_{AC}$ the reduced density matrix of $|\phi\rangle$ on region $AC$, etc (see Fig. \ref{fig: partition}). We can bound the trace difference between $\psi_C$ and $\phi_C$ by $||O||$ and $I_{|\psi\rangle}(A: C)$:
	\beq
	\begin{split}
		&||\psi_C-\phi_C||_1\\
		=&||{\rm Tr}_A(\psi_{AC}-O\psi_{AC}O^\dagger)||_1\\
		=&||{\rm Tr}_A[(\psi_{AC}-\psi_A\otimes\psi_C)\\
		&-O(\psi_{AC}-\psi_A\otimes\psi_C)O^\dagger||_1\\
		\leqslant&||\psi_{AC}-\psi_A\otimes\psi_C||_1(1+||O||^2)\\
		\leqslant&\sqrt{2\ln 2\cdot I_{|\psi\rangle}(A: C)}(1+||O||^2)
	\end{split}
	\eeq
	where in the second equality we have used the fact that $O$ is only supported in $A$ and $|\phi\rangle$ is normalized, in the first inequality we have used the monotonicity of the trace difference under partial trace, and in the last inequality we have used the Pinsker inequality.
	
	Next, we use the Fuchs-van de Graaf inequality to bound $F(\psi_C, \phi_C)$, the fidelity between $\psi_C$ and $\phi_C$,
	\beq
	\begin{split}
		&F(\psi_C, \phi_C)\\
		\geqslant& \left(1-\frac{1}{2}||\psi_C-\phi_C||_1\right)^2\\
		\geqslant& \left(1-\sqrt{\ln 2\cdot I_{|\psi\rangle}(A: C)/2}(1+||O||^2)\right)^2.
	\end{split}
	\eeq
	Therefore, Theorem \ref{thm: unitary approximation 1} is proved by using the Uhlmann's theorem, which states there must be a unitary operator $U$ supported in $AB$, such that
	\beq
	\begin{split}
		&|\langle\phi|U|\psi\rangle|
		=\sqrt{F(\psi_C, \phi_C)}\\
		\geqslant&1-\sqrt{\ln 2\cdot I_{|\psi\rangle}(A: C)/2}(1+||O||^2).
	\end{split}
	\eeq
\end{proof}

A straightforward generalization of Theorem \ref{thm: unitary approximation 1} is the following theorem.

\begin{theorem} \label{thm: unitary approximation 2}
	Let $|\psi\rangle$, $|\phi_1\rangle=O_1|\psi\rangle$ and $|\phi_2\rangle=O_2|\psi\rangle$ be three normalized states, where $O_1$ and $O_2$ are both bounded operators supported in region $A$ as in Fig. \ref{fig: partition}. Then there is a unitary $U$ supported in region $AB$, such that
	\beq \label{eq: unitary approximation 2}
	\bsp
	&|\langle\phi_2|U|\phi_1\rangle|\\
	&\ \geqslant 1-\sqrt{\ln 2\cdot I_{|\psi\rangle}(A: C)/2}(2+||O_1||^2+||O_2||^2),
	\esp
	\eeq
	where $I_{|\psi\rangle}(A: C)$ is the mutual information between $A$ and $C$ of the state $|\psi\rangle$ in Fig. \ref{fig: partition} and $||O_{1,2}||$ is the operator norm of $O_{1,2}$.
\end{theorem}

\subsection{Implications of symmetry localization} \label{subapp: general symmetry localization}

In the last part of this appendix, we discuss some implications of the symmetry localization property of 3-anyon states in Eq. \eqref{eq: symmetry localization}. First, we show that the operator $V_g^a(^gx_1)$ ($V_g^b(^gx_2)$ and $V_g^{\bar c}(^gx_3)$) in Eq. \eqref{eq: symmetry localization} is insensitive to anyons other than $a$ and $^ga$ ($b$ and $^gb$, and $\bar c$ and $^g\bar c$). For example, $V_g^a(^gx_1)$ only depends on the symmetry action $R_g$, the position of the anyon $a$, and the local information around the anyons $a$ and $^ga$. In particular, no matter what anyons $b$ and $c$ are and what local information they carry, $V_g^a(^gx_1)$ can be chosen to be the same. A corollary of this result is the symmetry localization property of 2-anyon states, given by Eq. \eqref{eq: symmetry localization 2-anyon}. Second, Eq. \eqref{eq: symmetry localization} implies its own generalization to multi-anyon states with finitely many anyons, given by Eq. \eqref{eq: symmetry localization for n-anyon states}. Note that all these implications of Eq. \eqref{eq: symmetry localization} hold approximately in short-range correlated anyon states, with errors exponentially small in the distances between the anyons. We will discuss the magnitudes of these errors. Throughout this paper we assume that Eq. \eqref{eq: symmetry localization} holds without any error, although there should be some error that is exponentially small in the distances between the anyons in reality, which has not been rigorously derived in the literature. When such errors are included, it is expected that the actual errors of the implications of Eq. \eqref{eq: symmetry localization} discussed in this appendix are larger than the errors derived below, but only by an amount that decays exponentially as the distances of the anyons increase.

\subsubsection{Insensitivity of the $V$ operators}

We start by elaborating on the insensitivity of the $V$ operators in Eq. \eqref{eq: symmetry localization}, \ie $V_g^a(^g x_1)$ only depends on the symmetry action $R_g$, the position of the anyon $a$, and the local information around the anyons $a$ and $^ga$.

Our discussion is based on the following lemma.

\begin{lemma} \label{lemma: fixing V}
    Let $|\psi_1\rangle$, $|\psi_2\rangle$, $|\phi_1\rangle$ and $|\phi_2\rangle$ be four normalized states, and $R_gPR_g^{-1}$ and $Q$ be unitary operators supported in region $C$ in Fig. \ref{fig: partition}, with $R_g$ the symmetry action of $g\in G$. Suppose that $R_g|\psi_1\rangle\propto U_{A1}U_{C1}|\phi_1\rangle$ and $R_g|\psi_2\rangle\propto U_{A2}U_{C2}|\phi_2\rangle$, where $U_{A1}$ and $U_{A2}$ are unitary operators supported in region $A$, and $U_{C1}$ and $U_{C2}$ are unitary operators supported in region $C$. Then the fidelity $F(R_g|\psi_2\rangle, U_{A1}U_{C2}|\phi_2\rangle)=|\langle \phi_2|U_{C2}^\dagger U_{A1}^\dagger R_g|\psi_2\rangle|^2$ satisfies that
    \beq \label{eq: fixing V lemma}
    \bsp
        &\sqrt{F(R_g|\psi_2\rangle, U_{A1}U_{C2}|\phi_2\rangle)}\\
        \geqslant& 1-\sqrt{1-F(|\psi_1\rangle, P|\psi_2\rangle)}-\sqrt{1-F(|\phi_1\rangle, Q|\phi_2\rangle)}
        -\sqrt{2I_{|\phi_2\rangle}(A: C)},
    \esp
    \eeq
    with $I_{|\phi_2\rangle}(A: C)$ the mutual information between regions $A$ and $C$ in state $|\phi_2\rangle$.
\end{lemma}

\begin{proof}
    Note that
    \beq \label{eq: convert fidelity}
    \bsp
    &F(R_g|\psi_2\rangle, U_{A1}U_{C2}|\phi_2\rangle)\\
    =&F(U_{A2}U_{C2}|\phi_2\rangle, U_{A1}U_{C2}|\phi_2\rangle)\\
    =&F(U_{A2}|\phi_2\rangle, U_{A1}|\phi_2\rangle),
    \esp
    \eeq
    so a lower bound of $F(R_g|\psi_2\rangle, U_{A1}U_{C2}|\phi_2)$ is just the lower bound of $F(U_{A2}|\phi_2\rangle, U_{A1}|\phi_2)$. To get such a lower bound, we will first derive a lower bound of $F(U_{A2}R_gPR_g^{-1}U_{C2}|\phi_2\rangle, U_{A1}U_{C1}Q|\phi_2\rangle)$ and then use Theorem \ref{thm: approximate operator factorization}. To get a lower bound of $F(U_{A2}R_gPR_g^{-1}U_{C2}|\phi_2\rangle, U_{A1}U_{C1}Q|\phi_2\rangle)$, we derive an upper bound of $||\rho_{U_{A2}R_gPR_g^{-1}U_{C2}|\phi_2\rangle}-\rho_{U_{A1}U_{C1}Q|\phi_2\rangle}||_1$ and use the Fuchs-van de Graaf inequality, where $\rho_{|\psi\rangle}=|\psi\rangle\langle\psi|$ is the density matrix of $|\psi\rangle$.

    Note that
    \beq
    \bsp
    &||\rho_{U_{A2}R_gPR_g^{-1}U_{C2}|\phi_2\rangle}-\rho_{U_{A1}U_{C1}Q|\phi_2\rangle}||_1\\
    \leqslant &||\rho_{U_{A2}R_gPR_g^{-1}U_{C2}|\phi_2\rangle}-\rho_{U_{A1}U_{C1}|\phi_1\rangle}||_1\\
    &+||\rho_{U_{A1}U_{C1}|\phi_1\rangle}-\rho_{U_{A1}U_{C1}Q|\phi_2\rangle}||_1\\
    =&||\rho_{U_{A2}R_gPR_g^{-1}U_{C2}|\phi_2\rangle}-\rho_{U_{A1}U_{C1}|\phi_1\rangle}||_1\\
    &+||\rho_{|\phi_1\rangle}-\rho_{Q|\phi_2\rangle}||_1\\
    \leqslant& 2\sqrt{1-F(U_{A2}R_gPR_g^{-1}U_{C2}|\phi_2\rangle, U_{A1}U_{C1}|\phi_1\rangle)}\\
    +&2\sqrt{1-F(|\phi_1\rangle, Q|\phi_2\rangle)}\\
    =&2\sqrt{1-F(R_gP|\psi_2\rangle, R_g|\psi_1\rangle)}+2\sqrt{1-F(|\phi_1\rangle, Q|\phi_2\rangle)}\\
    =&2\sqrt{1-F(|\psi_1\rangle, P|\psi_2\rangle)}+2\sqrt{1-F(|\phi_1\rangle, Q|\phi_2\rangle)},
    \esp
    \eeq
    where the first inequality is the triangle inequality and the second inequality is the Fuchs-van de Graaf inequality. Then we use the Fuchs-van de Graaf inequality again to get
    \beq
    \bsp
    &\sqrt{F(U_{A2}R_gPR_g^{-1}U_{C2}|\phi_2\rangle, U_{A1}U_{C1}Q|\phi_2\rangle)}\\
    \geqslant&1-\frac{1}{2}||\rho_{U_{A2}R_gPR_g^{-1}U_{C2}|\phi_2\rangle}-\rho_{U_{A1}U_{C1}Q|\phi_2\rangle}||_1\\
    \geqslant& 1-\sqrt{1-F(|\psi_1\rangle, P|\psi_2\rangle)}-\sqrt{1-F(|\phi_1\rangle, Q|\phi_2\rangle)}.
    \esp
    \eeq
    Because $R_gPR_g^{-1}U_{C2}$ and $U_{C1}Q$ are unitary operators supported in the region $C$, and $U_{A1}$ and $U_{A2}$ are unitary operators supported in the region $A$, Eq. \eqref{eq: fixing V lemma} is obtained by combining Eq. \eqref{eq: convert fidelity} and Theorem \ref{thm: approximate operator factorization}.
\end{proof}

Now we use Lemma \ref{lemma: fixing V} to deduce the insensitivity of the operator $V_g^a(^gx_1)$, by taking $|\psi_1\rangle=|a_1, b_2, \bar c_3; \mu_1\rangle$, $|\psi_2\rangle=|a_1, b'_2, \bar c'_3; \mu_2\rangle$, $|\phi_1\rangle=\sum_{\nu_1}[U_g(^g a, ^gb; ^gc)]_{\mu_1\nu_1}|^ga_{^g1}, ^gb_{^g2}, ^g\bar c_{^g3}; \nu_1\rangle$ and $|\phi_2\rangle=\sum_{\nu_2}[U_g(^g a, ^gb'; ^gc')]_{\mu_2\nu_2}|^ga_{^g1}, ^gb'_{^g2}, ^g\bar c'_{^g3}; \nu_2\rangle$. Here the anyon $b$ ($c$) can be different from the anyon $b'$ ($c'$), and even if they are the same, they can carry different local information. We assume that all these states are short-range correlated and the anyons are far away from each other. Moreover, we consider the case where the reduced density matrices of $|\psi_1\rangle$ and $|\psi_2\rangle$ around $x_1$ are identical, and the reduced density matrices of $|\phi_1\rangle$ and $|\phi_2\rangle$ around $^gx_1$ are nearly identical, which can be achieved, for example, by taking $|\psi_1\rangle=(S^{b\bar c}_{\bar a,\mu_1})_{32}|a_1, \bar a_2\rangle$ and $|\psi_2\rangle=(S^{b'\bar c'}_{\bar a,\mu_2})_{32}|a_1, \bar a_2\rangle$, $|\phi_1\rangle=\sum_{\nu_1}[U_g(^g a, ^gb; ^gc)]_{\mu_1\nu_1}(S^{^gb^g\bar c}_{^g\bar a, \nu_1})_{^g3^g2}|^ga_{^g1}, ^g\bar a_{^g2}\rangle$ and $|\phi_2\rangle=\sum_{\nu_2}[U_g(^g a, ^gb'; ^gc')]_{\mu_2\nu_2}(S^{^gb'^g\bar c'}_{^g\bar a, \nu_2})_{^g3^g2}|^ga_{^g1}, ^g\bar a_{^g2}\rangle$. Note that according to Theorem \ref{thm: unitary approximation 2}, in a short-range correlated anyon states with anyons far apart, there exists a unitary operator $Q$ supported around $^gx_2$ and $^gx_3$, such that $F(|\phi_1\rangle, Q|\phi_2\rangle)\geqslant (1-\sqrt{\ln 2\cdot I_{|^ga_{^g1}, ^g\bar a_{^g2}\rangle}(A': C')/2}(2+\sqrt{N_{b\bar c}^{\bar a}}+\sqrt{N_{b'\bar c'}^{\bar a}}))^2$, where $A'$ is a region that covers the supports of $(S^{^gb^g\bar c}_{^g\bar a, \nu_1})_{^g3^g2}$ and $(S^{^gb'^g\bar c'}_{^g\bar a, \nu_2})_{^g3^g2}$, and $C'$ is a subset in the complement of $A'$. In deriving this inequality, we have used that $||\sum_{\nu_1}[U_g(^g a, ^gb; ^gc)]_{\mu_1\nu_1}(S^{^gb^g\bar c}_{^g\bar a, \nu_1})_{^g3^g2}||\leqslant |\sum_{\nu_1}[U_g(^g a, ^gb; ^gc)]_{\mu_1\nu_1}|\leqslant\sqrt{N_{b\bar c}^{\bar a}}$ and $||\sum_{\nu_2}[U_g(^g a, ^gb'; ^gc')]_{\mu_2\nu_2}(S^{^gb'^g\bar c'}_{^g\bar a, \nu_2})_{^g3^g2}||\leqslant |\sum_{\nu_1}[U_g(^g a, ^gb'; ^gc')]_{\mu_1\nu_1}|\leqslant\sqrt{N_{b'\bar c'}^{\bar a}}$.

Eq. \eqref{eq: symmetry localization} means that 
\beq
\bsp
&|\psi_1\rangle\propto V_g^a(^gx_1)V_g^b(^gx_2)V_g^{\bar c}(^gx_3)|\phi_1\rangle,\\
&|\psi_2\rangle\propto \tilde V_g^a(^gx_1)V_g^{b'}(^gx_2)V_g^{\bar c'}(^gx_3)|\phi_2\rangle,
\esp
\eeq
with the $V$ operators some unitaries supported around their respective regions. Our goal is to show that we can always approximately take $\tilde V_g^a(^gx_1)=V_g^a(^gx_1)$.

To this end, we can set up the partition in Fig. \ref{fig: partition} such that the region $A$ covers the position $^gx_1$ and the region $C$ covers the positions $^gx_2$ and $^gx_3$. Then the conditions in Lemma \ref{lemma: fixing V} are satisfied, with $P=(S^{bc}_{\bar a, \mu_1})_{32}^{-1}(S^{b'c'}_{\bar a, \mu_2})_{32}$, $Q$ as described above, $U_{A1}=V_g^a(^gx_1)$, $U_{C1}=V_g^b(^gx_2)V_g^{\bar c}(^gx_3)$, $U_{A2}=\tilde V_g^a(^gx_1)$, and $U_{C2}=V_g^{b'}(^gx_2)V_g^{\bar c'}(^gx_3)$. Therefore, in a short-range correlated anyon state with the anyons are far apart, Lemma \ref{lemma: fixing V} shows that
\beq
\bsp
&\sqrt{F(R_g|\psi_2\rangle, U_{A1}U_{C2}|\phi_2\rangle)}\\
\geqslant &
1-\{1-[1-\sqrt{\ln 2\cdot I_{|^ga_{^g1}, ^g\bar a_{^g2}\rangle}(A': C')/2}\\
&\cdot(2+\sqrt{N_{b\bar c}^{\bar a}}+\sqrt{N_{b'\bar c'}^{\bar a}})]^2\}^{1/2}-\sqrt{2I_{|\phi_2\rangle}(A: C)}.
\esp
\eeq
This inequality shows the insensitivity of $V_g^a(^gx_1)$ to anyons other than $a$ and $^ga$, \ie we can indeed always take $\tilde V_g^a(^gx_1)=V_g^a(^gx_1)$, which only results in fidelity that deviates from 1 by an amount that is exponentially small in the distances between the anyons.

Similar analysis shows that $V_g^{b}(^gx_2)$ ($V_g^{\bar c}(^gx_3)$) is insensitive to anyons other than $b$ and $^gb$ ($c$ and $^gc$).

Applying this analysis to the special situation where $b=\bar a$ and $\bar c=0$, we get the symmetry localization property for 2-anyon states:
\beq \label{eq: symmetry localization 2-anyon}
R_g|a_1, \bar a_2\rangle\propto V_g^{a}(^gx_1)V_g^{\bar a}(^g x_2)|^ga_{^g1}, ^g\bar a_{^g2}\rangle,
\eeq
where $V_g^a(^gx_1)$ is the same as that in Eq. \eqref{eq: symmetry localization}, and the fidelity can be bounded as
\beq
\bsp
&\sqrt{F(R_g|a_1, \bar a_2\rangle,  V_g^{a}(^gx_1)V_g^{\bar a}(^g x_2)|^ga_{^g1}, ^g\bar a_{^g2}\rangle)}\\
\geqslant& 1-\sqrt{2I_{|^ga_{^g1}, ^g\bar a_{^g2}\rangle}(A: C)},
\esp
\eeq
which deviates from 1 by an amount exponentially small in the distance between the anyons for short-range correlated anyon states.

\subsubsection{From 3-anyon states to $n$-anyon states}

Next, we generalize Eq. \eqref{eq: symmetry localization} to multi-anyon states.

We first introduce our notation for a multi-anyon state. A multi-anyon state is specified by an order of fusing them, and the successive fusion outcomes and fusion vertex basis states. For example, a 5-anyon state with anyons $a^{(1,2,3,4,5)}$ at positions $x_{1,2,3,4,5}$ can be written as
\begin{equation}
    \ket{a^{(1)}_1,a^{(2)}_2,a^{(3)}_3,a^{(4)}_4,a^{(5)}_5;e^{(1)},e^{(2)};\mu_1,\mu_2,\mu_3},
\end{equation}
where $a^{(1)}$ and $a^{(2)}$ fuse to $e^{(1)}$ with fusion vertex index $\mu_1$, $e^{(1)}$ and $a^{(3)}$ fuse to $e^{(2)}$ with fusion vertex index $\mu_2$, $e^{(2)}$ and $a^{(4)}$ fuse to $\bar a^{(5)}$ with fusion vertex index $\mu_3$, and, finally, $a^{(5)}$ and $\bar a^{(5)}$ fuse to a trivial anyon 0 with a trivial fusion vertex. This state can be represented by the following fusion tree:
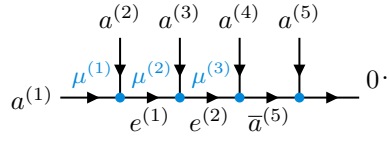
\begin{equation*}
    \bmm\fiveanyonfusiontree\emm.
\end{equation*}
In general, in our notation, we always fuse anyons from left to right, and the intermediate fusion outcomes (\ie internal anyons) are labeled correspondingly, just like this example of 5-anyon state.

The physical meaning of the internal anyons is that, for example, if one finds a subregion containing both $a^{(1)}$ and $a^{(2)}$ and measures the braiding between the anyons in this subregion and other anyons, which can be defined microscopically using our $R$-matrix in Eq. \eqref{eq: general R main}, the result is as if this subregion contains an anyon $e^{(1)}$; if the subregion contains $a^{(1)}$, $a^{(2)}$ and $a^{(3)}$, then the braiding between the anyons in this subregion and other anyons will agree with the braiding between $e^{(2)}$ and other anyons.\footnote{One may also ask what the result is if the subregion contains $a^{(2)}$ and $a^{(3)}$ for the five-anyon state. In our notation, the result is a certain superposition of the possible fusion outcome of $a^{(2)}$ and $a^{(3)}$, with the probability amplitude determined by the $F$-symbols.} Another way of thinking of this 5-anyon state is to first create from a ground state an anyon $a^{(5)}$ at $x_5$ and another anyon $\bar a^{(5)}$ at $x_4$, then split the anyon $\bar a^{(5)}$ at $x_4$ into an anyon $a^{(4)}$ at $x_4$ and another anyon $e^{(2)}$ at $x_3$ with fusion vertex basis state labeled by $\mu_3$, next split the anyon $e^{(2)}$ at $x_3$ into an anyon $a^{(3)}$ at $x_3$ and another anyon $e^{(1)}$ at $x_2$ with fusion vertex basis state labeled by $\mu_2$, and finally split the anyon $e^{(1)}$ at $x_2$ into an anyon $a^{(1)}$ at $x_1$ and another anyon $a^{(2)}$ at $x_2$ with fusion vertex basis state labeled by $\mu_1$. Namely,
\beq
\ket{a^{(1)}_1,a^{(2)}_2,a^{(3)}_3,a^{(4)}_4,a^{(5)}_5;e^{(1)},e^{(2)};\mu_1,\mu_2,\mu_3}\propto (S^{a^{(2)}a^{(1)}}_{e^{(1)}, \mu_1})_{12}(S^{a^{(3)}e^{(1)}}_{e^{(2)}, \mu_2})_{23}(S^{a^{(4)}e^{(2)}}_{\bar a^{(5)}, \mu_3})_{34}(S^{a^{(5)}\bar a^{(5)}}_{0, 1})_{45}|{\rm ground\ state}\rangle,
\eeq
where the $S$ operators are a set of splitting operators. This concept of multi-anyon state can be similarly generalized to states with any number of anyons.

For simplicity, we will write the notation for $n$-anyon states as
\begin{equation*}
    \ket{a^{(1)}_1,a^{(2)}_2,a^{(3)}_3,\cdots, a^{(n)}_n;{\{e\},\{\mu\}}},
\end{equation*}
where the anyon $a^{(i)}$ is at the position $x_i$ for $i=1, 2, \cdots, n$, $\{e\}$ labels all the internal anyons and $\{\mu\}$ labels all the fusion vertex basis states. The corresponding fusion tree is
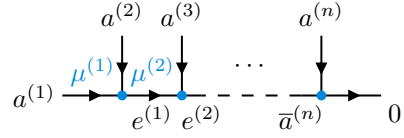
\begin{equation*}
    \bmm\generalfusiontree\emm.
\end{equation*}
For example, the fusion trees of the two-anyon state $|a_1, \bar a_2\rangle$ and three-anyon state $|a_1, b_2, c_3; \mu\rangle$ are
\begin{equation}
    \bmm\twoandthreeanyontrees\emm.
\end{equation}

We now prove that, given the symmetry localization property for 3-anyon states in Eq. \eqref{eq: symmetry localization}, we can generalize it to short-range correlated $n$-anyon states with finitely many anyons that are far away from each other. We will see that the following relation holds approximately:
\beq \label{eq: symmetry localization for n-anyon states}
\bsp
&R_g \ket{a^{(1)}_1,a^{(2)}_2,\cdots, \bar a^{(n)}_n;{\{e\}; \{\mu\}}} \propto\sum_{\{\mu'\}}[U_g(^ga^{(1)},^ga^{(2)},\cdots, ^ga^{(n)}; \{^ge\})]_{\{\mu\}\{\mu'\}}\\
&\cdot\prod_{i=1}^{n-1}\Big(V_g^{a^{(i)}}(^gx_i)\Big)V_g^{a^{(n)}}(^gx_n)
\ket{\lsupsc{g}a^{(1)}_{\lsupsc{g}1},\lsupsc{g}a^{(2)}_{\lsupsc{g}2},\cdots, ^g\bar a^{(n)}_{^gn};\{\lsupsc{g}e\}; \{\mu'\}}.
\esp
\eeq
In the above,
\beq
\ket{a^{(1)}_1,a^{(2)}_2,\cdots, \bar a^{(n)}_n;{\{e\}; \{\mu\}}}=\ket{a^{(1)}_1,a^{(2)}_2,\cdots, \bar a^{(n)}_n; e^{(1)}, e^{(2)}, \cdots, e^{(n-3)}; \mu_1, \mu_2, \cdots, \mu_{n-2}}
\eeq
and
\beq
\bsp
&[U_g(^ga^{(1)},^ga^{(2)},\cdots, ^ga^{(n)}; \{^ge\})]_{\{\mu\}\{\mu'\}}\\
=&[U_g(^ga^{(1)}, ^ga^{(2)}; ^ge^{(1)})]_{\mu_1\mu_1'}[U_g(^ge^{(1)}, ^ga^{(3)}; ^ge^{(2)})]_{\mu_2\mu_2'}\cdots [U_g(^ge^{(n-3)}, ^ga^{(n-1)}; ^ga^{(n)})]_{\mu_{n-2}\mu_{n-2}'}.
\esp
\eeq
As before, the $V$ operators above are the same as in Eq. \eqref{eq: symmetry localization}. This relation should be well expected, since it just means that the symmetry actions on a multi-anyon state have the three anticipated effects: changing the types and positions of the anyons, changing the fusion vertex basis states, and changing the local information. The nontrivial part of our analysis is to bound the error in this relation for states where the anyons are separated by finite distances. In the rest of this appendix, we derive this bound, which is given by Eq. \eqref{eq: error in symmetry localization for n anyons}.

For notational simplicity, denote the left and right hand sides of Eq. \eqref{eq: symmetry localization for n-anyon states} by $|\psi_1^{(n)}\rangle$ and $|\psi_2^{(n)}\rangle$, respectively. Also, denote the density matrix of any pure state $|\psi\rangle$ by $\rho_{|\psi\rangle}$. We will relate $||\rho_{|\psi_1^{(n+1)}\rangle}-\rho_{|\psi_2^{(n+1)}\rangle}||_1$ to $||\rho_{|\psi_1^{(n)}\rangle}-\rho_{|\psi_2^{(n)}\rangle}||_1$ or its analog.{\footnote{The trace norm of an operator $O$ is $||O||_1={\rm Tr}(\sqrt{O^\dagger O})$.}} Because $||\rho_{|\psi_1^{(3)}\rangle}-\rho_{|\psi_2^{(3)}\rangle}||_1=0$ according to Eq. \eqref{eq: symmetry localization}, such a relation allows us to bound $||\rho_{|\psi_1^{(n)}\rangle}-\rho_{|\psi_2^{(n)}\rangle}||_1$ for any $n$.

To proceed, note that
\beq
\bsp
|\psi_1^{(n+1)}\rangle
=&R_g\ket{a^{(1)}_1,a^{(2)}_2,\cdots, a^{(n)}_n, \bar a^{(n+1)}_{n+1}; e^{(1)}, e^{(2)}, \cdots, e^{(n-2)}; \mu_1, \mu_2, \cdots, \mu_{n-1}}\\
=&R_g(S^{a^{(n)}\bar a^{(n+1)}}_{\bar e^{(n-2)}, \mu_{n-1}})_{n+1, n}R_g^{-1}R_g\ket{a^{(1)}_1, a^{(2)}_2, \cdots, a^{(n-1)}_{n-1}, \bar e^{(n-2)}; e^{(1)}, e^{(2)}, \cdots, e^{(n-3)}; \mu_1, \mu_2, \cdots, \mu_{n-2}}\\
=&R_g(S^{a^{(n)}\bar a^{(n+1)}}_{\bar e^{(n-2)}, \mu_{n-1}})_{n+1, n}R_g^{-1}|\psi_1^{(n)'}\rangle
\esp
\eeq
and
\beq
\bsp
&|\psi_2^{(n+1)}\rangle\\
=&\sum_{\{\mu'\}}[U_g(^ga^{(1)},^ga^{(2)},\cdots, ^ga^{(n)}, ^ga^{(n+1)}; \{^ge\})]_{\{\mu\}\{\mu'\}}
\cdot\prod_{i=1}^{n}\Big(V_g^{a^{(i)}}(^gx_i)\Big)V_g^{\bar a^{(n+1)}}(^gx_{n+1})\\
&\ket{\lsupsc{g}a^{(1)}_{\lsupsc{g}1},\lsupsc{g}a^{(2)}_{\lsupsc{g}2},\cdots, ^ga^{(n)}_{^gn}, ^g\bar a^{(n+1)}_{^g(n+1)};\{\lsupsc{g}e\},\{\mu'\}}\\
=&\sum_{\{\mu'\}}[U_g(^ga^{(1)},^ga^{(2)},\cdots, ^ga^{(n)}, ^ga^{(n+1)}; \{^ge\})]_{\{\mu\}\{\mu'\}}
\cdot\prod_{i=1}^{n}\Big(V_g^{a^{(i)}}(^gx_i)\Big)V_g^{\bar a^{(n+1)}}(^gx_{n+1})\\
&(S^{^ga^{(n)}{^g\bar a^{(n+1)}}}_{^g\bar e^{(n-2)}, \mu'_{n-1}})_{^g(n+1)^gn}
\ket{^ga^{(1)}_{^g1}, ^ga^{(2)}_{^g2}, \cdots, ^ga^{(n-1)}_{^g(n-1)}, ^g\bar e^{(n-2)}_{^gn}; ^ge^{(1)}, ^ge^{(2)}, \cdots, ^ge^{(n-3)}; \mu_1', \mu_2', \cdots, \mu_{n-2}'},
\esp
\eeq
with
\beq
|\psi_1^{(n)'}\rangle=R_g\ket{a^{(1)}_1, a^{(2)}_2, \cdots, a^{(n-1)}_{n-1}, \bar e^{(n-2)}; e^{(1)}, e^{(2)}, \cdots, e^{(n-3)}; \mu_1, \mu_2, \cdots, \mu_{n-2}}.
\eeq
Then
\beq \label{eq: recurssion relation}
\bsp
&||\rho_{|\psi_1^{(n+1)}\rangle}-\rho_{|\psi_2^{(n+1)}\rangle}||_1\\
\leqslant& ||\rho_{|\psi_1^{(n+1)}\rangle}-\rho_{R_g(S^{a^{(n)}\bar a^{(n+1)}}_{\bar e^{(n-2)}, \mu_{n-1}})_{n+1, n}R_g^{-1}|\psi_2^{(n)'}\rangle}||_1+||\rho_{R_g(S^{a^{(n)}\bar a^{(n+1)}}_{\bar e^{(n-2)}, \mu_{n-1}})_{n+1, n}R_g^{-1}|\psi_2^{(n)'}\rangle}-\rho_{|\psi_2^{(n+1)}\rangle}||_1\\
=& ||\rho_{|\psi_1^{(n)'}\rangle}-\rho_{|\psi_2^{(n)'}\rangle}||_1+||\rho_{R_g(S^{a^{(n)}\bar a^{(n+1)}}_{\bar e^{(n-2)}, \mu_{n-1}})_{n+1, n}R_g^{-1}|\psi_2^{(n)'}\rangle}-\rho_{|\psi_2^{(n+1)}\rangle}||_1\\
\leqslant&||\rho_{|\psi_1^{(n)'}\rangle}-\rho_{|\psi_2^{(n)'}\rangle}||_1+2\sqrt{1-F(R_g(S^{a^{(n)}\bar a^{(n+1)}}_{\bar e^{(n-2)}, \mu_{n-1}})_{n+1, n}R_g^{-1}|\psi_2^{(n)'}\rangle, |\psi_2^{(n+1)}\rangle)},\\
\esp
\eeq
where the first inequality is the triangle inequality, the last inequality is the Fuchs-van de Graaf inequality, with the fidelity between two pure states $|\alpha\rangle$ and $|\beta\rangle$ defined as $F(|\alpha\rangle, |\beta\rangle)=|\langle\alpha|\beta\rangle|^2$, and
\beq
\bsp
&|\psi_2^{(n)'}\rangle\\
=&\sum_{\{\mu'\}}[U_g(^g a^{(1)}, ^ga^{(2)}, \cdots, ^ga^{(n-1)}, ^ge^{(n-2)}; ^ge^{(1)}, ^ge^{(2)}, \cdots, ^ge^{(n-3)})]_{\{\mu\}\{\mu'\}}\\
&\cdot\prod_{i=1}^{n-1}\Big(V_g^{a^{(i)}}(^gx_i)\Big)V_g^{\bar e^{(n-2)}}(^gx_n)\ket{^ga^{(1)}_{^g1}, ^ga^{(2)}_{^g2}, \cdots, ^ga^{(n-1)}_{^g(n-1)}, ^g\bar e^{(n-2)}_{^gn}; ^ge^{(1)}, ^ge^{(2)}, \cdots, ^ge^{(n-3)}; \mu'_1, \mu'_2, \cdots, \mu'_{n-2}}.
\esp
\eeq

Next, we need the fidelity in the second term of Eq. \eqref{eq: recurssion relation}, which can be obtained from
\beq
\bsp
&\langle\psi_2^{(n+1)}|R_g(S^{a^{(n)}\bar a^{(n+1)}}_{\bar e^{(n-2)}, \mu_{n-1}})_{n+1, n}R_g^{-1}|\psi_2^{(n)'}\rangle\\
=&
\sum_{\{\mu'\mu''\}}[U_g(^ga^{(1)},^ga^{(2)},\cdots, ^ga^{(n)}, ^ga^{(n+1)}; \{^ge\})]_{\{\mu\}\{\mu'\}}^*
[U_g(^ga^{(1)}, ^ga^{(2)}, \cdots, ^ga^{(n-1)}, ^ge^{(n-2)}; \{^ge\})]_{\{\mu\}\{\mu''\}}\\
&\cdot\bra{^ga^{(1)}_{^g1}, ^ga^{(2)}_{^g2}, \cdots, ^ga^{(n-1)}_{^g(n-1)}, ^g\bar e^{(n-2)}_{^gn}; ^ge^{(1)}, ^ge^{(2)}, \cdots, ^ge^{(n-3)}; \mu_1', \mu_2', \cdots, \mu_{n-2}'}
(S^{^ga^{(n)}{^g\bar a^{(n+1)}}}_{^g\bar e^{(n-2)}, \mu'_{n-1}})_{^g(n+1)^gn}^\dagger\\
&\cdot\prod_{i=1}^{n}\Big(V_g^{a^{(i)}}(^gx_i)\Big)^\dagger \Big(V_g^{\bar a^{(n+1)}}(^gx_{n+1})\Big)^\dagger
R_g(S^{a^{(n)}\bar a^{(n+1)}}_{\bar e^{(n-2)}, \mu_{n-1}})_{n+1, n}R_g^{-1}\prod_{j=1}^{n-1}\Big(V_g^{a^{(j)}}(^gx_j)\Big)V_g^{\bar e^{(n-2)}}(^gx_n)\\
&\cdot\ket{^ga^{(1)}_{^g1}, ^ga^{(2)}_{^g2}, \cdots, ^ga^{(n-1)}_{^g(n-1)}, ^g\bar e^{(n-2)}_{^gn}; ^ge^{(1)}, ^ge^{(2)}, \cdots, ^ge^{(n-3)}; \mu''_1, \mu''_2, \cdots, \mu''_{n-2}}\\
=&\sum_{\{\mu'\mu''\}}[U_g(^ga^{(1)},^ga^{(2)},\cdots, ^ga^{(n)}, ^ga^{(n+1)}; \{^ge\})]_{\{\mu\}\{\mu'\}}^*
[U_g(^ga^{(1)}, ^ga^{(2)}, \cdots, ^ga^{(n-1)}, ^ge^{(n-2)}; \{^ge\})]_{\{\mu\}\{\mu''\}}\\
&\cdot\bra{^ga^{(1)}_{^g1}, ^ga^{(2)}_{^g2}, \cdots, ^ga^{(n-1)}_{^g(n-1)}, ^g\bar e^{(n-2)}_{^gn}; ^ge^{(1)}, ^ge^{(2)}, \cdots, ^ge^{(n-3)}; \mu_1', \mu_2', \cdots, \mu_{n-2}'}
(S^{^ga^{(n)}{^g\bar a^{(n+1)}}}_{^g\bar e^{(n-2)}, \mu'_{n-1}})_{^g(n+1)^gn}^\dagger\\
&\cdot\Big(V_g^{a^{(n)}}(^gx_n)\Big)^\dagger \Big(V_g^{\bar a^{(n+1)}}(^gx_{n+1})\Big)^\dagger
R_g(S^{a^{(n)}\bar a^{(n+1)}}_{\bar e^{(n-2)}, \mu_{n-1}})_{n+1, n}R_g^{-1}V_g^{\bar e^{(n-2)}}(^gx_n)\\
&\cdot\ket{^ga^{(1)}_{^g1}, ^ga^{(2)}_{^g2}, \cdots, ^ga^{(n-1)}_{^g(n-1)}, ^g\bar e^{(n-2)}_{^gn}; ^ge^{(1)}, ^ge^{(2)}, \cdots, ^ge^{(n-3)}; \mu''_1, \mu''_2, \cdots, \mu''_{n-2}}\\
=&\sum_{\{\mu'\}}[U_g(^ga^{(1)},^ga^{(2)},\cdots, ^ga^{(n)}, ^ga^{(n+1)}; \{^ge\})]_{\{\mu\}\{\mu'\}}^*
[U_g(^ga^{(1)}, ^ga^{(2)}, \cdots, ^ga^{(n-1)}, ^ge^{(n-2)}; \{^ge\})]_{\{\mu\}\{\mu'\}}\\
&\cdot \bra{^ge_{^g1}^{(n-2)}, ^g\bar e_{^gn}^{(n-2)}}O^\dag \\
&\cdot(S^{^ga^{(n)}{^g\bar a^{(n+1)}}}_{^g\bar e^{(n-2)}, \mu'_{n-1}})_{^g(n+1)^gn}^\dagger \Big(V_g^{a^{(n)}}(^gx_n)\Big)^\dagger \Big(V_g^{\bar a^{(n+1)}}(^gx_{n+1})\Big)^\dagger
R_g(S^{a^{(n)}\bar a^{(n+1)}}_{\bar e^{(n-2)}, \mu_{n-1}})_{n+1, n}R_g^{-1}V_g^{\bar e^{(n-2)}}(^gx_n)\\
&\cdot O \ket{^ge_{^g1}^{(n-2)}, ^g\bar e_{^gn}^{(n-2)}},
\esp
\eeq
where $O$ is a product of some splitting and moving operators supported around $^gx_1, ^gx_2, \cdots, ^gx_{n-1}$, such that
\beq
O \ket{^ge_{^g1}^{(n-2)}, ^g\bar e_{^gn}^{(n-2)}}=\ket{^ga^{(1)}_{^g1}, ^ga^{(2)}_{^g2}, \cdots, ^ga^{(n-1)}_{^g(n-1)}, ^g\bar e^{(n-2)}_{^gn}; ^ge^{(1)}, ^ge^{(2)}, \cdots, ^ge^{(n-3)}; \mu''_1, \mu''_2, \cdots, \mu''_{n-2}}.
\eeq

According to Ref. \cite{Wolf2007}, the above equation implies that
\beq \label{eq: clustering bound}
\bsp
\Bigg|
&
\langle\psi_2^{(n+1)}|R_g(S^{a^{(n)}\bar a^{(n+1)}}_{\bar e^{(n-2)}, \mu_{n-1}})_{n+1, n}R_g^{-1}|\psi_2^{(n)'}\rangle\\
&-
\sum_{\{\mu'\}}[U_g(^ga^{(1)},^ga^{(2)},\cdots, ^ga^{(n)}, ^ga^{(n+1)}; \{^ge\})]_{\{\mu\}\{\mu'\}}^*
[U_g(^ga^{(1)}, ^ga^{(2)}, \cdots, ^ga^{(n-1)}, ^ge^{(n-2)}; \{^ge\})]_{\{\mu\}\{\mu'\}}\\
&\cdot \bra{^ge_{^g1}^{(n-2)}, ^g\bar e_{^gn}^{(n-2)}}O^\dag O \ket{^ge_{^g1}^{(n-2)}, ^g\bar e_{^gn}^{(n-2)}}\\
&\cdot \bra{^ge_{^g1}^{(n-2)}, ^g\bar e_{^gn}^{(n-2)}}(S^{^ga^{(n)}{^g\bar a^{(n+1)}}}_{^g\bar e^{(n-2)}, \mu'_{n-1}})_{^g(n+1)^gn}^\dagger \Big(V_g^{a^{(n)}}(^gx_n)\Big)^\dagger \Big(V_g^{\bar a^{(n+1)}}(^gx_{n+1})\Big)^\dagger\\
&R_g(S^{a^{(n)}\bar a^{(n+1)}}_{\bar e^{(n-2)}, \mu_{n-1}})_{n+1, n}R_g^{-1}V_g^{\bar e^{(n-2)}}(^gx_n)
\ket{^ge_{^g1}^{(n-2)}, ^g\bar e_{^gn}^{(n-2)}}
\Bigg|\\
\leqslant&\sqrt{2I_{\ket{^ge^{(n-2)}_{^g1}, ^g\bar e^{(n-2)}_{^gn}}}(A_{^g1,^g2,\cdots, ^g(n-1)}: A_{^gn, ^g(n+1)})}\\
&\Bigg|\sum_{\{\mu'\}}[U_g(^ga^{(1)},^ga^{(2)},\cdots, ^ga^{(n)}, ^ga^{(n+1)}; \{^ge\})]_{\{\mu\}\{\mu'\}}^*
[U_g(^ga^{(1)}, ^ga^{(2)}, \cdots, ^ga^{(n-1)}, ^ge^{(n-2)}; \{^ge\})]_{\{\mu\}\{\mu'\}}\Bigg|.
\esp
\eeq
Here $A_{^g1,^g2,\cdots, ^g(n-1)}$ and $A_{^gn, ^g(n+1)}$ are the smallest rectangular regions that cover $^gx_1, ^gx_2, \cdots, ^gx_{n-1}$ and $^gx_n, ^gx_{n+1}$, respectively.

Using Eq. \eqref{eq: U-matrix}, the left hand side of Eq. \eqref{eq: clustering bound} becomes
\beq
\bsp
\Bigg|
&\langle\psi_2^{(n+1)}|R_g(S^{a^{(n)}\bar a^{(n+1)}}_{\bar e^{(n-2)}, \mu_{n-1}})_{n+1, n}R_g^{-1}|\psi_2^{(n)'}\rangle\\
&-\sum_{\{\mu'\}}[U_g(^ga^{(1)},^ga^{(2)},\cdots, ^ga^{(n)}, ^ga^{(n+1)}; \{^ge\})]_{\{\mu\}\{\mu'\}}^*
[U_g(^ga^{(1)}, ^ga^{(2)}, \cdots, ^ga^{(n-1)}, ^ge^{(n-2)}; \{^ge\})]_{\{\mu\}\{\mu'\}}\\
&\cdot[U_g(^ga^{(n)}, ^g\bar a^{(n+1)}; ^g\bar e^{(n-2)})]_{\mu_{n-1}\mu'_{n-1}}
\Bigg|
\esp
\eeq
The second term in the above can be simplified using a cyclic property, $U_g(a, b; c)\propto U_g(b, \bar c; \bar a)$:
\beq
\bsp
&\sum_{\{\mu'\}}[U_g(^ga^{(1)},^ga^{(2)},\cdots, ^ga^{(n)}, ^ga^{(n+1)}; \{^ge\})]_{\{\mu\}\{\mu'\}}^*
[U_g(^ga^{(1)}, ^ga^{(2)}, \cdots, ^ga^{(n-1)}, ^ge^{(n-2)}; \{^ge\})]_{\{\mu\}\{\mu'\}}\\
&\cdot[U_g(^ga^{(n)}, ^g\bar a^{(n+1)}; ^g\bar e^{(n-2)})]_{\mu_{n-1}\mu'_{n-1}}\\
\propto&\sum_{\mu'_1, \mu'_2, \cdots, \mu'_{n-1}}
[U_g(^ga^{(1)}, ^ga^{(2)}; ^ge^{(1)})]^*_{\mu_1\mu'_1}
[U_g(^ge^{(1)}, ^ga^{(3)}; ^ge^{(2)})]^*_{\mu_2\mu'_2}
\cdots
[U_g(^ge^{(n-2)}, ^ga^{(n)}; ^ga^{(n+1)})]^*_{\mu_{n-1}\mu'_{n-1}}\\
&\cdot
[U_g(^ga^{(1)}, ^ga^{(2)}; ^ge^{(1)})]_{\mu_1\mu'_1}
[U_g(^ge^{(1)}, ^ga^{(3)}; ^ge^{(2)})]_{\mu_2\mu'_2}
\cdots
[U_g(^ge^{(n-3)}, ^ga^{(n-1)}; ^ge^{(n-2)})]_{\mu_{n-2}\mu'_{n-2}}
\\
&\cdot[U_g(^ga^{(n)}, ^g\bar a^{(n+1)}; ^g\bar e^{(n-2)})]_{\mu_{n-1}\mu'_{n-1}}\\
=&1.
\esp
\eeq
The cyclic property is a consequence of Eqs. \eqref{eq: symmetry localization} and \eqref{eq: U-matrix}:
\beq \label{eq: cyclic property}
\bsp
&[U_g(a,b;c)]_{\mu\nu}\\
\propto&\bra{a_1,b_2,\bar c_{0'};\nu}V^{\lsupsc{\bar g}b}_g(x_2)^\dagger V^{\lsupsc{\bar g}a}_g(x_1)^\dagger V^{\lsupsc{\bar g}\bar c}_g(x_{0'})^\dagger\cdot
R_g\ket{\lsupsc{\bar g}a_{\lsupsc{\bar g}1},\lsupsc{\bar g}b_{\lsupsc{\bar g}2},\lsupsc{\bar g}\bar c_{\lsupsc{\bar g}0'};\mu}\\
\propto&\bra{a_1,\bar a_{2}}(S^{b\bar c}_{\bar a,\nu})_{0'2}^\dagger V^{\lsupsc{\bar g}b}_g(x_2)^\dagger V^{\lsupsc{\bar g}a}_g(x_1)^\dagger V^{\lsupsc{\bar g}\bar c}_g(x_{0'})^\dagger\cdot
R_g(S^{\lsupsc{\bar g}b\lsupsc{\bar g}\bar c}_{\lsupsc{\bar g}\bar a,\nu})_{\lsupsc{\bar g}{0'} \lsupsc{\bar g}2}R^{-1}_g V^{\lsupsc{\bar g}a}_g(x_1) V^{\lsupsc{\bar g}\bar a}_g(x_2)\ket{a_1,\bar a_2}\\
=&U(b,\bar c;\bar a)_{\mu\nu}.
\esp
\eeq

The right hand side of Eq. \eqref{eq: clustering bound} can be bounded using the triangle inequality and Cauchy-Schwartz inequality:
\beq
\bsp
&\sqrt{2I_{\ket{^ge^{(n-2)}_{^g1}, ^g\bar e^{(n-2)}_{^gn}}}(A_{^g1,^g2,\cdots, ^g(n-1)}: A_{^gn, ^g(n+1)})}\\
&\Bigg|\sum_{\{\mu'\}}[U_g(^ga^{(1)},^ga^{(2)},\cdots, ^ga^{(n)}, ^ga^{(n+1)}; \{^ge\})]_{\{\mu\}\{\mu'\}}^*
[U_g(^ga^{(1)}, ^ga^{(2)}, \cdots, ^ga^{(n-1)}, ^ge^{(n-2)}; \{^ge\})]_{\{\mu\}\{\mu'\}}\Bigg|\\
=&\sqrt{2I_{\ket{^ge^{(n-2)}_{^g1}, ^g\bar e^{(n-2)}_{^gn}}}(A_{^g1,^g2,\cdots, ^g(n-1)}: A_{^gn, ^g(n+1)})}\sum_{\mu'_{n-1}}[U_g(^ge^{(n-2)}, ^ga^{(n)}; ^ga^{(n+1)})]^*_{\mu_{n-1}\mu'_{n-1}}\\
\leqslant &\sqrt{2N_{e^{(n-2)}a^{(n)}}^{a^{(n+1)}}I_{\ket{^ge^{(n-2)}_{^g1}, ^g\bar e^{(n-2)}_{^gn}}}(A_{^g1,^g2,\cdots, ^g(n-1)}: A_{^gn, ^g(n+1)})}.
\esp
\eeq

Therefore,
\beq
\bsp
&\sqrt{F(R_g(S^{a^{(n)}\bar a^{(n+1)}}_{\bar e^{(n-2)}, \mu_{n-1}})_{n+1, n}R_g^{-1}|\psi_2^{(n)'}\rangle, |\psi_2^{(n+1)}\rangle)}\\
\geqslant&
1-\sqrt{2N_{e^{(n-2)}a^{(n)}}^{a^{(n+1)}}I_{\ket{^ge^{(n-2)}_{^g1}, ^g\bar e^{(n-2)}_{^gn}}}(A_{^g1,^g2,\cdots, ^g(n-1)}: A_{^gn, ^g(n+1)})}.
\esp
\eeq
Or, equivalently,
\beq
\bsp
&F(R_g(S^{a^{(n)}\bar a^{(n+1)}}_{\bar e^{(n-2)}, \mu_{n-1}})_{n+1, n}R_g^{-1}|\psi_2^{(n)'}\rangle, |\psi_2^{(n+1)}\rangle)\\
\geqslant &
\Bigg[{\rm max}\Bigg\{1-\sqrt{2N_{e^{(n-2)}a^{(n)}}^{a^{(n+1)}}I_{\ket{^ge^{(n-2)}_{^g1}, ^g\bar e^{(n-2)}_{^gn}}}(A_{^g1,^g2,\cdots, ^g(n-1)}: A_{^gn, ^g(n+1)})}, 0\Bigg\}\Bigg]^2
\esp
\eeq
Then Eq. \eqref{eq: recurssion relation} becomes
\beq
\bsp
&||\rho_{|\psi_1^{(n+1)}\rangle}-\rho_{|\psi_2^{(n+1)}\rangle}||_1\\
\leqslant&
||\rho_{|\psi_1^{(n)'}\rangle}-\rho_{|\psi_2^{(n)'}\rangle}||_1\\
&+
2\Bigg\{
1-\Bigg[{\rm max}\Bigg\{1-\sqrt{2N_{e^{(n-2)}a^{(n)}}^{a^{(n+1)}}I_{\ket{^ge^{(n-2)}_{^g1}, ^g\bar e^{(n-2)}_{^gn}}}(A_{^g1,^g2,\cdots, ^g(n-1)}: A_{^gn, ^g(n+1)})}, 0\Bigg\}\Bigg]^2
\Bigg\}^{\frac{1}{2}}.
\esp
\eeq
Iterating the above relation, we get
\beq \label{eq: error in symmetry localization for n anyons}
\bsp
&||\rho_{|\psi_1^{(n)}\rangle}-\rho_{|\psi_2^{(n)}\rangle}||_1\\
\leqslant&
2\sum_{i=4}^n
\Bigg\{
1-\Bigg[{\rm max}\Bigg\{1-\sqrt{2N_{e^{(i-3)}a^{(i-1)}}^{a^{(i)}}I_{\ket{^ge^{(i-3)}_{^g1}, ^g\bar e^{(i-3)}_{^g(i-1)}}}(A_{^g1,^g2,\cdots, ^g(i-2)}: A_{^g(i-1), ^g(i)})}, 0\Bigg\}\Bigg]^2
\Bigg\}^{\frac{1}{2}}.
\esp
\eeq
If the anyon states under consideration are short-range correlated and all anyons are far apart, the above bound approaches zero exponentially as the distances between the anyons increase. It is interesting to note the dependence on the fusion multiplicity in this bound, which suggests that the error becomes larger when the dimension of the space of fusion vertex basis states becomes larger.

\section{Approximate opereator factorization} \label{app: approximate factorization}

In Sec. \ref{sec:setup}, we have claimed that Eq. \eqref{eq: simplifying condition 2} can always be achieved approximately by choosing the phases of the $V$ operators. Similar treatments are also used multiple times later. In this appendix, we explain why this is valid and discuss the error in this approximation.

We start by writing $R_g|^{\bar g}a_{^{\bar g}1}, ^{\bar g}\bar a_{^{\bar g}0}\rangle$ in two different ways:
\beq \label{eq: deriving MV condition}
\bsp
R_g|^{\bar g}a_{^{\bar g}1}, ^{\bar g}\bar a_{^{\bar g}0}\rangle
&\propto V_g^{^{\bar g}a}(x_1)V_g^{^{\bar g}\bar a}(x_0)|a_{1}, \bar a_{0}\rangle\\
&\propto V_g^{^{\bar g}a}(x_1)V_g^{^{\bar g}\bar a}(x_0)M^a_{12}|a_{2}, \bar a_{0}\rangle\\
&=V_g^{^{\bar g}a}(x_1)M^a_{12}V_g^{^{\bar g}\bar a}(x_0)|a_{2}, \bar a_{0}\rangle,\\
R_g|^{\bar g}a_{^{\bar g}1}, ^{\bar g}\bar a_{^{\bar g}0}\rangle
&\propto R_gM^{^{\bar g}a}_{^{\bar g}1^{\bar g}2}|^{\bar g}a_{^{\bar g}2}, ^{\bar g}\bar a_{^{\bar g}0}\rangle=R_gM^{^{\bar g}a}_{^{\bar g}1^{\bar g}2}R_g^{-1}R_g|^{\bar g}a_{^{\bar g}2}, ^{\bar g}\bar a_{^{\bar g}0}\rangle\\
&\propto R_gM^{^{\bar g}a}_{^{\bar g}1^{\bar g}2}R_g^{-1}V_g^{^{\bar g}a}(x_2)V_g^{^{\bar g}\bar a}(x_0)|a_{2}, \bar a_{0}\rangle,
\esp
\eeq
We see that $V_g^{^{\bar g}a}(x_1)M^a_{12}$ and $R_gM^{^{\bar g}a}_{^{\bar g}1^{\bar g}2}R_g^{-1}V_g^{^{\bar g}a}(x_2)$ are unitary operators supported around $x_{1,2}$, and $V_g^{^{\bar g}\bar a}(x_0)$ is a unitary operator supported around $x_0$ that is far away from $x_{1,2}$. The following theorem implies that the two sides of Eq. \eqref{eq: simplifying condition 2} only differ by a phase factor approximately, with an error exponentially small in the distance between $x_0$ and $x_{1, 2}$.

\begin{theorem} \label{thm: approximate operator factorization}
	Let $|\psi\rangle$ be a normalized state, $P_A$ and $Q_A$ be two bounded operators supported in region $A$ of Fig. \ref{fig: partition}, and $P_C$ and $Q_C$ be two bounded operators supported in region $C$ of Fig. \ref{fig: partition}. Then
	\beq \label{eq: approximate factorization}
	\bsp
	&{\rm min}\left\{\sqrt{F(P_A|\psi\rangle, Q_A|\psi\rangle)}, \sqrt{F(P_C|\psi\rangle, Q_C|\psi\rangle)}\right\}\\
    &\geqslant \sqrt{F(P_AP_C|\psi\rangle, Q_AQ_C|\psi\rangle)}-\sqrt{2I_{|\psi\rangle}(A: C)}\cdot||P_AQ_A^\dagger||\cdot||P_CQ_C^\dagger||,
	\esp
	\eeq
	where $F(|\psi_1\rangle, |\psi_2\rangle)=|\langle\psi_1|\psi_2\rangle|^2$ is the fidelity between the states  $|\psi_1\rangle$ and $|\psi_2\rangle$, and $I_{|\psi\rangle}(A: C)$ is the mutual information between regions $A$ and $C$ of the state $|\psi\rangle$ in Fig. \ref{fig: partition}.
\end{theorem}

\begin{proof} \label{proof: approximate operator factorization}
	According to Ref. \cite{Wolf2007},
	\beq
	\bsp
	&|\langle\psi|P_AQ_A^\dagger P_CQ_C^\dagger|\psi\rangle
	-
	\langle\psi|P_AQ_A^\dagger|\psi\rangle\langle\psi|P_CQ_C^\dagger|\psi\rangle
	|\leqslant \sqrt{2I_{|\psi\rangle}(A: C)}\cdot||P_AQ_A^\dagger||\cdot||P_CQ_C^\dagger||,
	\esp
	\eeq
	so
	\beq
	\bsp
	&|\sqrt{F(P_AP_C|\psi\rangle, Q_AQ_C|\psi\rangle)}-\sqrt{F(P_A|\psi\rangle, Q_A|\psi\rangle)F(P_C|\psi\rangle, Q_C|\psi\rangle)}|\\
	=&|\langle\psi|P_AQ_A^\dagger P_CQ_C^\dagger|\psi\rangle-\langle\psi|P_AQ_A^\dagger|\psi\rangle\langle\psi|P_CQ_C^\dagger|\psi\rangle|\\
	\leqslant& \sqrt{2I_{|\psi\rangle}(A: C)}\cdot||P_AQ_A^\dagger||\cdot||P_CQ_C^\dagger||,
	\esp
	\eeq
	Using that $0\leqslant F(P_{A, C}|\psi\rangle, Q_{A, C}|\psi\rangle)\leqslant 1$, we get Eq. \eqref{eq: approximate factorization}.
\end{proof}

As an application of Theorem \ref{thm: approximate operator factorization}, for a short-range correlated state $|\psi\rangle$, if $P_AP_C|\psi\rangle$ and $Q_AQ_C|\psi\rangle$ only differ by a phase and if $A$ and $C$ are far away from each other so that $\sqrt{2I_{|\psi\rangle}(A: C)}\cdot||P_AQ_A^\dagger||\cdot||P_CQ_C^\dagger||\leqslant \sqrt{F(P_AP_C|\psi\rangle, Q_AQ_C|\psi\rangle)}=1$, then 
\beq
F(P_A|\psi\rangle, Q_A|\psi\rangle)\geqslant\left(1-\sqrt{2I_{|\psi\rangle}(A: C)}\cdot||P_AQ_A^\dagger||\cdot||P_CQ_C^\dagger||\right)^2.
\eeq
Namely, $P_A|\psi\rangle$ and $Q_A|\psi\rangle$ are almost the same, up to an error that is exponentially small in the distance between $A$ and $C$. The same applies to $P_C|\psi\rangle$ and $Q_C|\psi\rangle$. Namely, the equivalence between the effect of the operator $P_AP_C$ on $|\psi\rangle$ and the effect of the operator $Q_AQ_C$ on $|\psi\rangle$ can be approximately factorized into the two regions $A$ and $C$. We will often use this factorization in this work, while ignoring the exponentially small errors when the regions $A$ and $C$ are far away from each other.

We emphasize that the condition of short-range correlation is important for this factorization. An example where the factorization fails can be seen from the following 2-qubit example. Suppose $|\psi\rangle=(|0_A1_C\rangle+|1_A0_C\rangle)/\sqrt{2}$. One can verify that $P_AP_C|\psi\rangle=Q_AQ_C|\psi\rangle$, where $P_A=|0_A\rangle\langle 1_A|+|1_A\rangle\langle 0_A|$, $P_C=|0_C\rangle\langle 1_C|+|1_C\rangle\langle 0_C|$, and $Q_A=Q_C$ are the identity operator. In this case, the fidelities$F(P_A|\psi\rangle, Q_A|\psi\rangle)=F(P_C|\psi\rangle, Q_C|\psi\rangle)=0$. Namely, $P_A|\psi\rangle$ and $Q_A|\psi\rangle$ are very different, and the reason is that the two qubits are strongly correlated.

Returning to Eq. \eqref{eq: deriving MV condition}, by taking $P_A=V_g^{^{\bar g}a}(x_1)M^a_{12}$, $Q_A=R_gM^{^{\bar g}a}_{^{\bar g}1^{\bar g}2}R_g^{-1}V_g^{^{\bar g}a}(x_2)$, $P_C=Q_C=V_g^{^{\bar g}\bar a}(x_0)$ and $|\psi\rangle=|^{\bar g}a_{^{\bar g}1}, ^{\bar g}\bar a_{^{\bar g}0}\rangle$, we conclude that the two sides of Eq. \eqref{eq: simplifying condition 2} only approximately differ by a phase factor, so by choosing the phases of the $V$ operators appropriately, Eq. \eqref{eq: simplifying condition 2} can be achieved, with an error exponentially small in the distance between $x_0$ and $x_{1,2}$, which will be ignored because these regions are far away from each other.

\section{More on trivially universal properties} \label{app: universal}

In Sec. \ref{subsubsec: representative FR}, we have mentioned that if $O$ is a generic microscopic operator that is unrelated to the moving and splitting operators (\eg $O$ may be a Pauli operator on a qubit), $O_W=WOW^\dagger$, and $|\psi_W\rangle=W|\psi\rangle$ with $|\psi\rangle$ a ground state of a system in the topological phase, then although $\langle\psi_W|O_W|\psi_W\rangle$ is invariant under the quasi-adiabatic continuation in Eq. \eqref{eq: quasi-adiabatic continuation 1}, no nontrivial universal properties can be extracted from it. In this appendix, we elaborate on this point. As a concrete example, we can assume that $O=Z_1$, the Pauli-$Z$ operator acting on the first qubit of the system.

First of all, because $O$ is assumed to be unrelated to the moving and splitting operators, \ie it is not a sum of products of some moving and splitting operators, we should not expect that $\langle\psi_W|O_W|\psi_W\rangle$ has any meaningful relation with anyons. This observation already suggests that the fact that $\langle\psi_W|O_W|\psi_W\rangle$ is invariant under the quasi-adiabatic continuation in Eq. \eqref{eq: quasi-adiabatic continuation 1} cannot help us characterize the underlying topological phase.

Next, let us examine which universal properties we can possibly extract from the fact that $\langle\psi_W|O_W|\psi_W\rangle$ is invariant under the quasi-adiabatic continuation in Eq. \eqref{eq: quasi-adiabatic continuation 1}. One attempt is to consider the statement that there exists a quasi-local unitary operator whose ground state expectation value in all systems of the topological phase takes a particular value. Here the quasi-local operator means an operator concentrated around a local region but can have some exponentially decaying tail. This statement is true, with the quasi-local operator being $O_W$ and the particular value of the ground state expectation value being $\langle\psi|O|\psi\rangle$. However, this statement trivially holds for all quantum states, no matter whether the state is a ground state of the topological phase of interest. More precisely, given any complex number $z$ with $|z|\leqslant 1$ and any state $|\phi\rangle$, it is straightforward to find a unitary operator $O_1$ acting only on the first qubit, such that the $\langle\phi|O_1|\phi\rangle=z$. Therefore, this universal statement is trivial and cannot help us characterize the underlying topological phase.

We should contrast the above trivial universal statement and the nontrivial universal properties we identify in the main text of the paper. The universal data discussed in the main text also takes the form of $\langle\psi_W|O_W|\psi_W\rangle$, but the operator $O$ there are not some arbitrary microscopic operator. Instead, there $O$ is made of some moving and splitting operators, as well as some symmetry action operators. These operators are very special and related to anyons, which is why they can imprint universal properties tied to the anyons. Moreover, quantities obtained from these operators satisfy some highly nontrivial relations, such as the pentagon and hexagon equations in Eqs. \eqref{eq: pentagon equation} and \eqref{eq: hexagon equations}, which are specific properties of topological phases not shared by generic phases.

\section{Review of Kawagoe-Levin's $F$- and $R$-symbols} \label{app: review Kawagoe-Levin}

In this appendix, we review the microscopic definitions of $F$- and $R$- matrices by Kawagoe and Levin, which can be viewed as the $F$- and $R$-symbols in a braided tensor category \cite{Kawagoe2019}.

\begin{figure*}[h!]
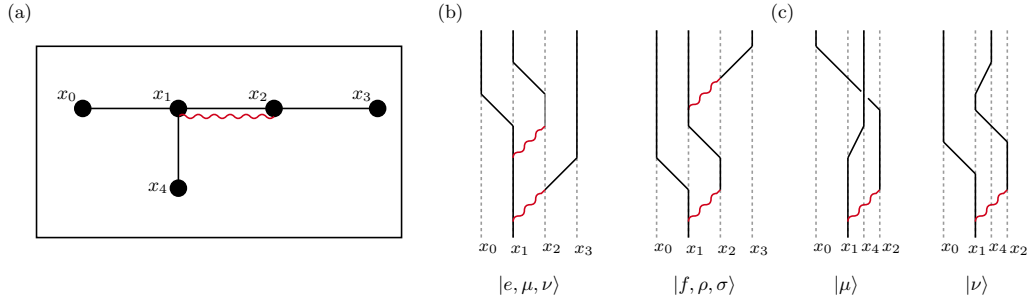

    \centering
    \scalebox{0.8}{\FRsymbols}
    \caption{(a) The 6 positions $x_{0,1,2,3,4}$ and $x_0'$ are far from each other, with $x_{0'}$ not shown in the figure. We only use splitting operators that split an anyon at $x_1$ to an anyon at $x_1$ and another anyon at $x_2$ (red curly line), and moving operators that move an anyon between $x_i$ and $x_{i-1}$ (black lines). (b) and (c) The graphic representations of the spacetime trajectories of the processes encoded in the states in Eqs. \eqref{eq: kawagoe levin F symbol states app} and \eqref{eq: kawagoe levin R symbol states app}, respectively.}
	\label{fig:FRsymbols}
\end{figure*}

We fix $6$ positions $x_0$, $x_1$, $x_2$, $x_3$, $x_4$, and $x_{0'}$ to host anyons (see Fig. \ref{fig:FRsymbols}). To define the $F$-symbols, we start with a state $|d_{1}, \bar d_{0'}\rangle$ and apply some moving and splitting operators to it, which leads to the two states below and the intermediate steps encode the processes in Fig. \ref{fig:FRsymbols} (b):
\begin{equation}\label{eq: kawagoe levin F symbol states app}
	\bsp
	&\ket{e,\mu,\nu}=M^{b}_{12}M^a_{01}(S^{ab}_{e,\mu})_{21}M^c_{32}(S^{ec}_{d,\nu})_{21}\ket{d_{1},\overline{d}_{0'}},\\
	&\ket{f,\rho,\sigma}=M^c_{32}(S^{bc}_{f,\rho})_{21}M^{f}_{12}M^a_{01}(S^{af}_{d,\sigma})_{21}\ket{d_{1},\overline{d}_{0'}}.
	\esp
\end{equation}
The Kawagoe-Levin $F$-matrix is defined as
\beq \label{eq: kawagoe levin F matrix app}
[F^{abc}_d]_{\{e, \mu, \nu\}, \{f, \rho, \sigma\}}=\langle f, \rho, \sigma| e, \mu, \nu\rangle.
\eeq
It is shown in Ref. \cite{Kawagoe2019} that, when changing the subjective choices in Eq. \eqref{eq: kawagoe levin F symbol states app}, the $F$-matrices Eq. \eqref{eq: kawagoe levin F matrix app} also transform in the same form as in Eq. \eqref{eq: vertex basis gauge transformation}. We repeat the transformation here.
\begin{equation}
    [F^{abc}_d]_{\{e,\mu,\nu\},\{f,\rho,\sigma\})}\to\sum_{\mu',\nu',\rho',\sigma'}[F^{abc}_d]_{\{e,\mu',\nu'\},\{f,\rho',\sigma'\}}\times[\Omega^{ab}_e]_{\mu\mu'}[\Omega^{ec}_d]_{\nu\nu'}[(\Omega^{bc}_f)^{-1}]_{\sigma'\sigma}[(\Omega^{af}_d)^{-1}]_{\rho'\rho}.
\end{equation}
Moreover, in appendix A of Ref. \cite{Kawagoe2019}, it is proved that the $F$-matrices Eq. \eqref{eq: kawagoe levin F matrix app} satisfy the pentagon equation \cite{Kawagoe2019}:
	\beq \label{eq: pentagon equation app}
	\sum_\delta[F^{fcd}_e]_{(g, \beta, \gamma), (l, \nu, \delta)}[F^{abl}_e]_{(f, \alpha, \delta), (k, \mu, \lambda)}=\sum_{h, \sigma, \psi, \rho}[F^{abc}_g]_{(f, \alpha, \beta), (h, \psi, \sigma)}[F^{ahd}_e]_{(g, \sigma, \gamma), (k, \rho, \lambda)}[F^{bcd}_k]_{(h, \psi, \rho), (l, \nu, \mu)}.
	\eeq

To define $R$-matrices, we start with a state $\ket{c_{1}, \bar c_{0'}}$ and apply some moving and splitting operators to it, which leads to the following two states with the intermediate steps shown in Fig. \ref{fig:FRsymbols} (c):
\beq \label{eq: kawagoe levin R symbol states app}
\begin{split}
	&|\mu\rangle=M^a_{01}M^a_{12}M^b_{41}(S^{ba}_{c, \mu})_{21}\ket{c_{1}, \bar c_{0'}},\\
	&|\nu\rangle=M^b_{41}M^b_{12}M^a_{01}(S^{ab}_{c, \nu})_{21}\ket{c_{1}, \bar c_{0'}}.
\end{split}
\eeq
That is, the anyon $c$ splits into $a$ and $b$ in two different ways, and, relative to the processes in $|\nu\rangle$, in the processes in $|\mu\rangle$ the anyons $a$ and $b$ exchange their positions in a counter-clockwise manner. The Kawagoe-Levin $R$-matrix is defined as
\beq \label{eq: kawagoe levin R matrix app}
[R^{ab}_c]_{\mu\nu}=\langle\nu|\mu\rangle.
\eeq

There are two points to be clarified here, which were not discussed in Ref. \cite{Kawagoe2019}.

\begin{figure*}[h!]
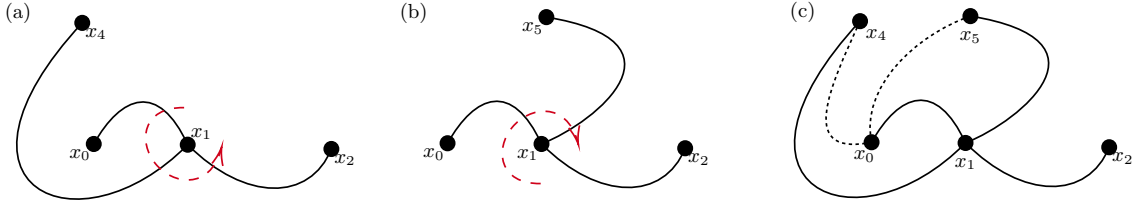

    \centering
    \scalebox{0.85}{\coneregion}
    \caption{(a) The supports of $M^{a}_{01}$, $M^{b}_{41}$ and $M^{a}_{12}$ ($M^{b}_{12}$) sweep around $x_1$ in a counter-clockwise manner. (b) The supports of $M^{a}_{01}$, $M^{b}_{51}$ and $M^{a}_{12}$ ($M^{b}_{12}$) sweep around $x_1$ in a clockwise manner. (c) We compare the $R$-matrices extracted with counter-clockwise sweeping moving operators and the $\hat R$-matrices extracted with clockwise sweeping moving operators.}
    \label{fig:coneregion}
\end{figure*}

First, to let the $R$-matrices Eq. \eqref{eq: kawagoe levin R matrix app} be compatible with the convention in category theory, that is, $R^{ab}_c$ results from the process where anyon $b$ winds around anyon $a$ for a half loop in a counter-clockwise direction, there are some constraints on the support of the moving operators. For each moving operator, there should exist a simply-connected region that 1) contains the support of this operator (after truncating its exponentially decaying tails discussed in Appendix \ref{subapp: exponentially decaying tails}), 2) does not contain anyon positions that are not connected by this operator, and 3) does not intersect with the supports of the other moving operators. As such, the moving operators $M_{01}^{a_1}$, $M_{12}^{a_2}$, and $M_{41}^{a_3}$ intersect only around $x_1$, for any anyons $a_1$, $a_2$ and $a_3$ in this TQSL. We then require that the supports of $M^{a}_{01}$, $M^{b}_{41}$ and $M^{a}_{12}$ ($M^{b}_{12}$) sweep around $x_1$ in a counter-clockwise manner, as shown in Fig. \ref{fig:coneregion} (a). In particular, $M^{a}_{12}$ and $M^{b}_{12}$ do not need to have the same support, but they should both satisfy the above requirement. In this way, it is guaranteed that the anyon $b$ winds around the anyon $a$ in a counter-clockwise direction during the process that is used to extract the $R$-matrix. In contrast, Fig. \ref{fig:coneregion} (b) shows an example where the supports of $M^{a}_{01}$, $M^{b}_{51}$ and $M^{a}_{12}$ ($M^{b}_{12}$) sweep around $x_1$ in a clockwise manner.

Second, if the supports of $M^{a}_{01}$, $M^{b}_{41}$ and $M^{a}_{12}$ ($M^{b}_{12}$) sweep around $x_1$ in a clockwise manner, then Eq. \eqref{eq: kawagoe levin R matrix app} gives $(R^{ba}_c)^{-1}$ instead of $R^{ab}_c$. To prove this, we still let the supports of $M^{a}_{01}$, $M^{b}_{41}$ and $M^{a}_{12}$ ($M^{b}_{12}$) sweep around $x_1$ in a clockwise manner, and introduce an $x_5$ with the supports of $M^{a}_{01}$, $M^{b}_{51}$ and $M^{a}_{12}$ ($M^{b}_{12}$) sweep around $x_1$ in a clockwise manner (see Fig. \ref{fig:coneregion} (c)). Consider the quantity extracted using $M^{a}_{01}$, $M^{b}_{51}$, $M^{a}_{12}$, and $M^{b}_{12}$ via definition Eq. \eqref{eq: kawagoe levin R matrix app}. We denote this quantity as $[\hat R^{ab}_c]_{\mu\nu}=_{5}\hspace{-3pt}\ew{\nu|\mu}_5$, where
    \begin{equation}
    \bsp
    \ket{\mu}_5=M^{a}_{01}M^a_{12}M^b_{51}(S^{ba}_{c,\mu})_{21}\ket{c_1,\bar c_{0'}},\\
    \ket{\nu}_5=M^{b}_{51}M^b_{12}M^a_{01}(S^{ab}_{c,\nu})_{21}\ket{c_1,\bar c_{0'}}.
    \esp
    \end{equation}

To find the relation between these $\hat R$-matrices and the $R$-matrices in Eq. \eqref{eq: kawagoe levin R matrix app}, we find two moving operators $M^b_{05}$ and $M^a_{40}$ that support in the dashed line in Fig. \ref{fig:coneregion} (c). These two moving operators are taken to satisfy Eq. \eqref{eq: seamless}. Then we find that 
\begin{equation*}
\bsp
    M^b_{05}M^a_{40}\ket{\mu}_5&=M^b_{05}M^a_{40}M^{a}_{01}M^a_{12}M^b_{51}(S^{ba}_{c,\mu})_{21}\ket{c_1,\bar c_{0'}}\\
    &=M^a_{41}M^a_{12}M^b_{01}(S^{ba}_{c,\mu})_{21}\ket{c_1,\bar c_{0'}}\equiv\ket{\mu'},\\
    M^b_{05}M^a_{40}\ket{\nu}_5&=M^b_{05}M^a_{40}M^{b}_{51}M^b_{12}M^a_{01}(S^{ab}_{c,\nu})_{21}\ket{c_1,\bar c_{0'}}\\
    &=M^b_{01}M^b_{12}M^a_{41}(S^{ab}_{c,\nu})_{21}\ket{c_1,\bar c_{0'}}\equiv\ket{\nu'}.\\
\esp
\end{equation*}
We have used the fact that $M^b_{05}$ commutes with $M^a_{41}$, and $M^a_{40}$ commutes with $M^{b}_{51}$ and $M^b_{12}$. On one hand, since the moving operators are chosen to be unitary, we have $_{5}\ew{\nu|(M^a_{40})^{\dagger}(M^b_{05})^{\dagger}M^b_{05}M^a_{40}|\mu}_5={_5}\ew{\nu|\mu}_5=[\hat R^{ab}_c]_{\mu\nu}$. On the other hand, we also notice that $\ew{\nu'|\mu'}=[R^{ba}_c]_{\nu\mu}^*$ according to Eq. \eqref{eq: kawagoe levin R matrix app}. Hence, we have $[\hat R^{ab}_c]_{\mu\nu}=[R^{ba}_c]_{\mu\nu}^{-1}$.

Changing the subjective choices in Eq. \eqref{eq: kawagoe levin R symbol states app} gives a similar transformation as in Eq. \eqref{eq: vertex basis gauge transformation}. Since there are constraints on the moving operators, we require that the new moving operators do not violate the constraints. As such, for example, when the moving and splitting operators change according to Eq. \eqref{eq: new moving and splitting operators}, the $R^{ab}_c$ transforms as
\begin{equation}
    [R^{ab}_c]_{\mu\nu}\to \sum_{\mu'\nu'}[R^{ab}_c]_{\mu'\nu'}[\Omega^{ba}_c]_{\mu\mu'}[(\Omega^{ab}_c)^{-1}]_{\nu'\nu},
\end{equation}
with $\Omega^{ab}_c=e^{i\phi^b_{12}}(\Omega_0(a, b; c))_{21}$. If one insists to make the supports of new $M^{a}_{01}$, $M^{b}_{41}$ and $M^{a}_{12}$ ($M^{b}_{12}$) operators sweep around $x_1$ in a clockwise manner, then
\begin{equation}
    [R^{ab}_c]_{\mu\nu}\to \sum_{\mu'\nu'}[R^{ba}_c]^{-1}_{\mu'\nu'}[\Omega^{ba}_c]_{\mu\mu'}[(\Omega^{ab}_c)^{-1}]_{\nu'\nu},
\end{equation}
with $\Omega^{ab}_c=e^{i\phi^b_{12}}(\Omega_0(a, b; c))_{21}$. Moreover, in the Appendix B of Ref. \cite{Kawagoe2019}, it is proved that the $R$-matrices Eq. \eqref{eq: kawagoe levin R matrix app} satisfy the hexagon equations:
\begin{widetext}
	\beq \label{eq: hexagon equations app}
	\begin{split}
		\sum_{\lambda, \gamma}[R^{ac}_e]_{\alpha\lambda}[F^{acb}_d]_{(e, \lambda, \beta)(g,\gamma,\nu)}(R^{bc}_g)_{\gamma\mu}&=\sum_{f, \sigma, \delta, \psi}[F^{cab}_d]_{\{e, \alpha, \beta\}, \{f, \delta, \sigma\}}[R^{fc}_d]_{\sigma\psi}[F^{abc}_d]_{\{f, \delta, \psi\}),\{g, \mu, \nu\}},\\
		\sum_{\lambda, \gamma}[(R^{ca}_e)^{-1}]_{\alpha\gamma}[F^{acb}_d]_{\{e, \lambda, \beta\},\{g, \gamma, \nu\}} [(R^{cb}_g)^{-1}]_{\gamma\mu}&=\sum_{f, \sigma, \delta, \mu}[F^{cab}_d]_{\{e,\alpha,\beta\},\{f,\delta, \sigma\}}[(R^{cf}_d)^{-1}]_{\sigma\psi}[F^{abc}_d]_{\{f,\delta,\psi)(g,\mu, \nu\}}.
	\end{split}	
	\eeq
\end{widetext}

From the above discussion, the consistency equations satisfied by these $F$- and $R$-matrices, and the equivalence relation they are subject to, make them valid $F$- and $R$-symbols in a braided tensor category.

\section{The full structure of the universal data of symmetry-enriched TQSLs} \label{app: full structure}

In Sec. \ref{sec:framework}, we have introduced the universal data of symmetry-enriched TQSLs. The starting point there is the physical properties of TQSLs introduced in Secs. \ref{sec:setup} and \ref{sec:framework}, based on which we have defined the data that characterizes the four features in TQSLs: 1) the associativity of fusion of anyons, 2) the braiding of anyons, 3) the change of fusion vertex basis states under the symmetry transformation, and 4) symmetry fractionalization. The general idea of defining this data is as follows. For each of the four features, we start with a certain 2-anyon state, perform two different physical processes that are related to this feature, and then take the inner product of the two final states. The two processes must consist of applications of only moving and splitting operators, symmetry transformations, and $V$ operators, so that the results are related to the physics of anyons, rather than including irrelevant high-energy physics. The two resulting final states are required to differ by at most a unitary operation in the degenerate subspace of the locally indistinguishable fusion vertex basis states, so that the results can reflect nontrivial universal properties of the SET phases, rather than being completely arbitrary. 

In Sec. \ref{sec:framework}, we focus on the data under the assumption that Eqs. \eqref{eq: seamless}, \eqref{eq: simplifying condition 1}, and \eqref{eq: simplifying condition 2} hold. However, as we have emphasized, Eqs. \eqref{eq: seamless}, \eqref{eq: simplifying condition 1}, and \eqref{eq: simplifying condition 2} are not necessary for defining the universal data. In this appendix, we discuss the full structure of the universal data. In particular, we introduce three parameters to quantify how Eqs. \eqref{eq: seamless}, \eqref{eq: simplifying condition 1}, and \eqref{eq: simplifying condition 2} are violated: a phase factor $\gamma$ resulting from the violation of Eq. \eqref{eq: seamless}; a unitary matrix $\Gamma$ comparing the fusion vertex basis states at different positions, which quantifies the violation of Eq. \eqref{eq: simplifying condition 1}; and a phase factor $\omega$ resulting from the violation of Eq. \eqref{eq: simplifying condition 2}. Concretely, given a set of anyon states, moving, splitting and $V$ operators, the following relations hold:
\begin{align}
&M^a_{31}\ket{a_1,\bar a}=\gamma^{a}_{321} M^a_{32}M^a_{21}\ket{a_{1},\bar a};\label{Eq:unseamless}\\
&(S^{ab}_{c;\mu})_{21}\ket{c_1,\bar c}=\sum_{\mu'}[(\Gamma^{ab}_c)_{2'1'21}]_{\mu\mu'}M^b_{22'}M^a_{11'}(S^{ab}_{c;\mu'})_{2'1'}M^c_{1'1}\ket{c_{1},\bar c};\label{Eq:generalMS}\\
&R_g M^{\lsupsc{\bar g}a}_{\lsupsc{\bar g}1\lsupsc{\bar g}2} R_g^{-1}V^{\lsupsc{\bar g}a}_g(x_2)\ket{a_{2},\bar a}=(\omega^a_{g})_{12} V^{\lsupsc{\bar g}a}_g(x_1)M^{a}_{12}\ket{a_{2},\bar a}\label{Eq:generalMV}.
\end{align}
The above relations are general. In particular, the two sides of Eq. \eqref{eq: seamless} can only differ by a phase factor, and the two sides of Eq. \eqref{eq: simplifying condition 1} can only differ by a unitary matrix, because of the definitions of anyon states and the moving and splitting operators. The reason why the two sides of Eq. \eqref{eq: simplifying condition 2} can only differ by a phase factor is explained in Appendix \ref{app: approximate factorization}.

Note the notational difference of the above anyon states compared to the main text. For notational simplicity, just as above, we may omit the position index for the $\bar a$ anyon in this appendix, because this anyon is far away from the region where we perform the various processes and is unimportant. Furthermore, in this appendix we will mostly use diagrammatic derivations. In particular, Eqs. \eqref{Eq:unseamless}, \eqref{Eq:generalMS} and \eqref{Eq:generalMV} can be expressed by the following diagrams.
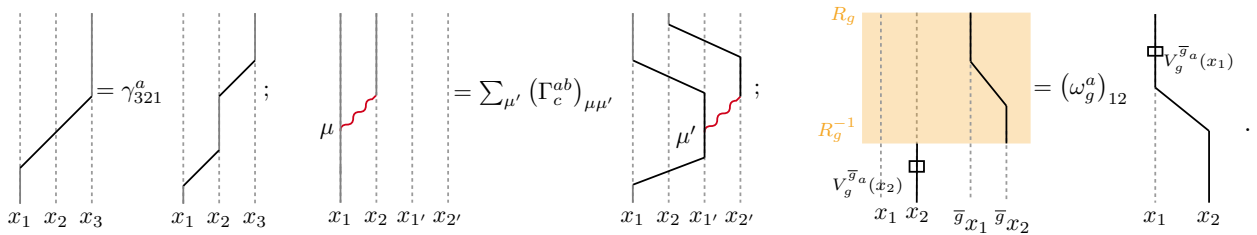
\begin{align*}
    \bmm\scalebox{0.9}{\generalconditions}\emm.
\end{align*}

Also note that the moving and splitting operators with different endpoints can be chosen to commute with each other in most cases. In the following, we will always make such a choice unless otherwise stated. 

In the following, from Appendix \ref{subapp: F} to Appendix \ref{subapp: eta}, we will consider a specific TQSL in the SET phase. In Appendix \ref{subapp: graded structure}, we will illustrate a graded structure of the universal data in the entire SET phase, where the three parameters $\gamma$, $\Gamma$, and $\omega$ are the grading parameters.

This appendix is written in a largely self-contained manner, and many definitions given in the main text are also given here. In the following, we will sometimes talk about the categorical $F$-, $R$-, $U$- and $\eta$-symbols. The categorical $F$- and $R$-symbols can be viewed as the $F$- and $R$-matrices defined in Ref. \cite{Kawagoe2019}, which is also reviewed in Appendix \ref{app: review Kawagoe-Levin}. These are also the $F-$ and $R$-matrices defined in Eqs. \eqref{Eq:generalF main} and \eqref{eq: general R main}, assuming Eqs. \eqref{eq: seamless} and \eqref{eq: simplifying condition 1}. The categorical $U$- and $\eta$-symbols can be viewed as the $U$-matrices and $\eta$-phases defined in Eqs. \eqref{eq: U-matrix} and \eqref{eq: eta}, assuming Eqs. \eqref{eq: seamless}, \eqref{eq: simplifying condition 1} and \eqref{eq: simplifying condition 2} and that the symmetry is purely internal.

\subsection{The associativity of fusion: $F$-matrices} \label{subapp: F}

We begin with the first feature, \ie the associativity of the fusion of anyons. The general process to capture this feature is as follows. We start from a certain anyon state $\ket{d_0,\bar d}$ with anyon $\bar d$ far away from the region where the relevant processes are performed, split the anyon $d$ into three anyons $a$, $b$, and $c$ in two different ways, and compare the difference between these two processes by taking the inner product of the two final states. The processes considered below are the most general class of processes, in the sense that the splitting processes can happen at different locations, and the anyons are allowed to move freely at every stage before they reach the final state. At the same time, these processes are the simplest ones in these most general processes, in the sense that the motion of anyons before and after each splitting process is achieved by applying a single moving operator, rather than multiple moving operators.

According to this general idea, the associativity of fusion is captured by
\begin{equation}\label{Eq:generalF}
    [(F^{abc}_d)_{012341'2'3'4'567}]_{\{e, \mu, \nu\}, \{f, \rho, \sigma\}}(\{\ket{\psi}\},\{M\},\{S\})=\langle f, \rho, \sigma| e, \mu, \nu\rangle,
\end{equation} 
where 
\begin{equation}
\bsp
    &\ket{e, \mu, \nu}=M^{b}_{64}M^{a}_{52}(S^{ab}_{e,\mu})_{42}M^e_{21}M^{c}_{73}(S^{ec}_{d,\nu})_{31}M^d_{10}\ket{d_0, \bar d_{0'}},\\
    &\ket{f, \rho, \sigma}=M^{c}_{74'}M^{b}_{63'}(S^{bc}_{f,\rho})_{4'3'}M^f_{3'2'}M^{a}_{51'}(S^{af}_{d,\sigma})_{2'1'}M^d_{1'0}\ket{d_0, \bar d_{0'}}.
\esp
\end{equation}
In terms of graphics, these two states are expressed by
\begin{equation}
    \scalebox{0.7}{\generalF}.
\end{equation}
As we can see from the definition, $F$-matrices have explicit dependence on the choice of anyon states collectively represented by $\{|\psi\rangle\}$, the choice of the moving and splitting operators collectively represented by $\{M\}$ and $\{S\}$, respectively, and 12 positions collectively represented by $012341'2'3'4'567$. Below we will sometimes write the position labels as $\{x\}$ for simplicity. All moving and splitting operators in the above processes can be chosen to commute with each other as long as they have different endpoints.

If $x_0$, $x_1$, $x_2$, $x_6$, $x_{1'}$ and $x_{3'}$ are all the same position, and $x_3$, $x_4$, $x_{2'}$ and $x_{4'}$ are also all the same position, then the processes that define the above $F$-matrix are precisely the same as the protocol defined by Kawagoe and Levin \cite{Kawagoe2019} (see Appendix \ref{app: review Kawagoe-Levin} for a review). In particular, when $x_0$, $x_1$, $x_2$, $x_6$, $x_{1'}$,  $x_{3'}$ are identified as a single position, the moving operators between them become phase factors that can be absorbed into some splitting operators. Similar statement is true when $x_3$, $x_4$, $x_{2'}$ and $x_{4'}$ are identified as a single position. Therefore, the Kawagoe-Levin definition of $F$-matrices, being the categorical $F$-symbols, are special cases of the microscopic $F$-matrices in Eq. \eqref{Eq:generalF}.

On the other hand, in general, the $F$-matrix defined in Eq. \eqref{Eq:generalF} is different from the Kawagoe-Levin $F$-matrix in Ref. \cite{Kawagoe2019}. Below we discuss the properties of the general microscopic $F$-matrices. There are two questions to be answered. First, what is the relation between the $F$-matrix in Eq. \eqref{Eq:generalF} and the categorical $F$-symbol? Given the pentagon equation Eq. \eqref{eq: pentagon equation} satisfied by the categorical $F$-symbol, this relation allows us to write down an analogous equation satisfied by the $F$-matrix in Eq. \eqref{Eq:generalF}. Second, since the $F$-matrix in Eq. \eqref{Eq:generalF} has explicit dependence on the choices of $\{\ket{\psi}\}$, $\{M\}$ and $\{S\}$, how does it transform when these choices change? The answer to this question allows us to understand the equivalence class of the $F$-matrices.

\subsubsection{The relation between microscopic $F$-matrices and categorical $F$-symbols} \label{subsubapp: F relation}

The strategy to answer the first question is to compare the above definition, Eq. \eqref{Eq:generalF}, with the $F$-matrices defined in Ref. \cite{Kawagoe2019} and Appendix \ref{app: review Kawagoe-Levin}. We pick up four positions $\tilde x_0$, $\tilde x_1$, $\tilde x_2$, and $\tilde x_3$ to extract the $F$-symbols using the Kawagoe-Levin formalism; these $F$-symbols are denoted by $[(\tilde F^{abc}_d)_{\tilde0\tilde1\tilde2\tilde3}]_{\{e,\mu',\nu'\},\{f,\rho',\sigma'\}}(\{\ket{\psi}\},\{M\},\{S\})=\widetilde{\langle f,\rho',\sigma'|}\widetilde{e,\mu',\nu'\rangle}$ in the following, where $\widetilde{|f,\rho,\sigma\rangle}$ and $\widetilde{|e,\mu,\nu\rangle}$ are defined via the process introduced in Eq. \eqref{eq: kawagoe levin F symbol states app}. We also find $12$ positions $x_0, x_1, x_2, x_3, x_4, x_{1'}, x_{2'}, x_{3'}, x_{4'}, x_5, x_6, x_7$ to extract the $F$-matrices defined in Eq. \eqref{Eq:generalF}. We can find the relation between $\ket{e,\mu,\nu}$ and $\widetilde{\ket{e,\mu,\nu}}$, and $\ket{f,\rho,\sigma}$ and $\widetilde{\ket{f,\rho,\sigma}}$, depicted as follows.
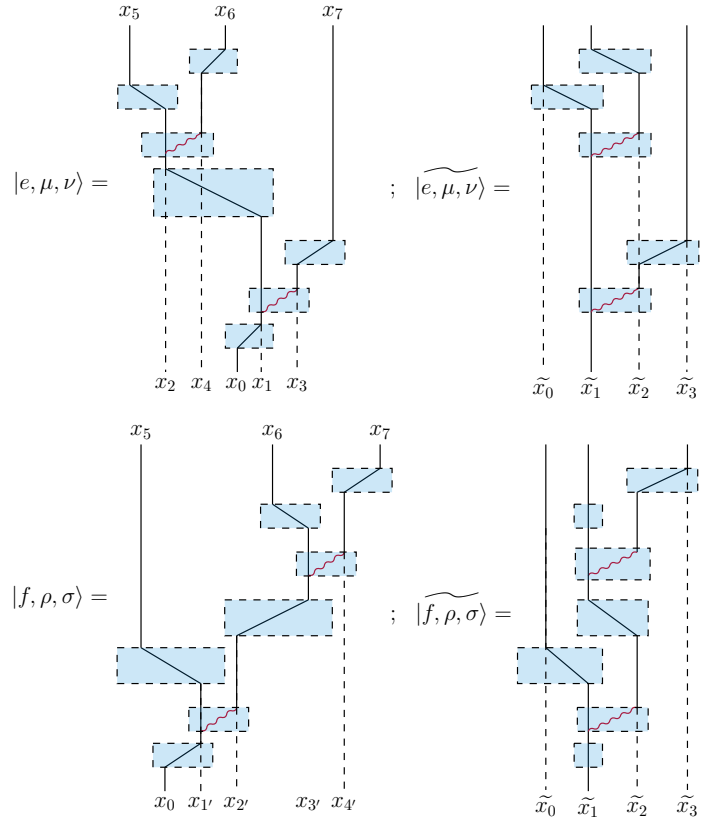
\begin{align*}
    \scalebox{0.6}{\compareFa}\\
    \scalebox{0.6}{\compareFb}
\end{align*}
All the moving and splitting operators are highlighted in boxes in the above figures. Note that the initial state used to define $F$ is $\ket{d_0,\bar d}$ with anyon $\bar d$ far away from the region of interest, and the initial state used to define $\tilde F$ using the Kawagoe-Levin protocol is $M^{d}_{\tilde00}\ket{d_0,\bar d}$. According to Eqs. \eqref{Eq:unseamless} and \eqref{Eq:generalMS}, the moving and splitting operators involving only $x$'s can be substituted using the moving and splitting operators involving only $\tilde x$'s and the moving operators between $x$'s and $\tilde x$'s. For example,
\begin{equation*}
    \bsp
    &(S^{ab}_{c;\mu})_{21}\ket{c_1,\bar c_{0'}}=\sum_{\mu'}[(\Gamma^{ab}_c)_{\tilde2\tilde121}]_{\mu\mu'}M^a_{1\tilde1}M^b_{2\tilde2}(S^{ab}_{c;\mu'})_{\tilde2\tilde1}M^c_{\tilde11}\ket{c_{1},\bar c_{0'}},\\
    &M^a_{21}\ket{c_1,\bar c_{0'}}=\gamma^{a}_{2\tilde2\tilde11} M^a_{2\tilde2}M^a_{\tilde2\tilde1}M^a_{\tilde11}\ket{a_{1},\bar a_{0'}},
    \esp
\end{equation*}
where $\gamma^{a}_{2\tilde2\tilde11}$ is the abbreviation of $\gamma^a_{2\tilde2\tilde 1}\gamma^a_{\tilde2\tilde11}$. As such, the state $\ket{e,\mu,\nu}$ can be expressed using the moving and splitting operators involving only $\tilde x$'s and the moving operators between $x$'s and $\tilde x$'s:
\begin{equation} \label{eq: relating F states temp}
    \bsp
    &\ket{e, \mu, \nu}=M^{a}_{52}M^{b}_{64}M^{c}_{73}(S^{ab}_{e,\mu})_{42}M^e_{21}(S^{ec}_{d,\nu})_{31}M^d_{10}\ket{d_0,d_{0'}}\\
    =&\left[\gamma^a_{5\tilde0\tilde12}M^{a}_{5\tilde0}M^{a}_{\tilde0\tilde1}M^{a}_{\tilde12}\right]\left[\gamma^b_{6\tilde1\tilde24}M^{b}_{6\tilde1}M^b_{\tilde1\tilde2}M^b_{\tilde24}\right]\left[\sum_{\mu'}[(\Gamma^{ab}_e)_{\tilde2\tilde142}]_{\mu\mu'}M^a_{2\tilde1}M^b_{4\tilde2}(S^{ab}_{e,\mu})_{\tilde2\tilde1}M^e_{\tilde12}\right]\left[\gamma^e_{2\tilde11}M^e_{2\tilde1}M^e_{\tilde11}\right]\times\\
    &\left[\gamma^c_{7\tilde3\tilde23}M^{c}_{7\tilde3}M^c_{\tilde3\tilde2}M^c_{\tilde23}\right]\left[\sum_{\nu'}[(\Gamma^{ec}_d)_{\tilde2\tilde131}]_{\nu\nu'}M^e_{1\tilde1}M^c_{3\tilde2}(S^{ec}_{d,\nu'})_{\tilde2\tilde1}M^{d}_{\tilde11}\right]\left[\gamma^d_{1\tilde10}M^d_{1\tilde1}M^d_{\tilde10}\right]\ket{d_0,d_{0'}}.\\
    \esp
\end{equation}
This relation is graphically shown as follows.
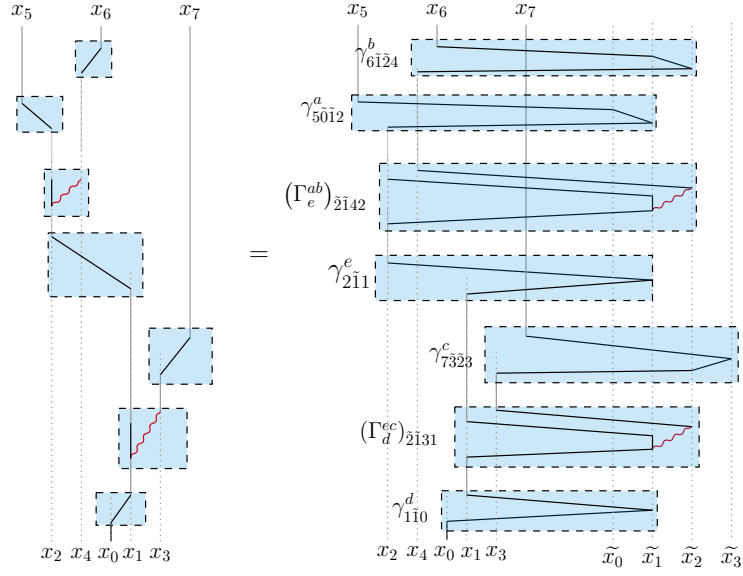
\begin{equation*}
    \scalebox{0.6}{\compareFc}
\end{equation*}
Moreover, each term on the right hand side of Eq. \eqref{eq: relating F states temp} can be written as follows.
\begin{equation*}
    \bsp
    &\left[M^{a}_{5\tilde0}M^{a}_{\tilde0\tilde1}M^{a}_{\tilde12}\right]\left[M^{b}_{6\tilde1}M^b_{\tilde1\tilde2}M^b_{\tilde24}\right]\left[M^a_{2\tilde1}M^b_{4\tilde2}(S^{ab}_{e,\mu})_{\tilde2\tilde1}M^e_{\tilde12}\right]\\
    &\cdot\left[M^e_{2\tilde1}M^e_{\tilde11}\right]\left[M^{c}_{7\tilde3}M^c_{\tilde3\tilde2}M^c_{\tilde23}\right]\left[M^e_{1\tilde1}M^c_{3\tilde2}(S^{ec}_{d,\nu'})_{\tilde2\tilde1}M^{d}_{\tilde11}\right]\left[M^d_{1\tilde1}M^d_{\tilde10}\right]\ket{d_0,d_{0'}}\\=
    &( \gamma _{\tilde{2} 4\tilde{2}}^{b})^{-1}( \gamma _{\tilde{1} 2\tilde{1}}^{a})^{-1}( \gamma _{\tilde{1} 2\tilde{1}}^{e})^{-1}( \gamma _{\tilde{1} 1\tilde{1}}^{e})^{-1}( \gamma _{\tilde{2} 3\tilde{2}}^{c})^{-1}( \gamma _{\tilde{1} 1\tilde{1}}^{d})^{-1}M^a_{5\tilde0}M^b_{6\tilde1}M^c_{7\tilde3}    M^{a}_{\tilde0\tilde1}M^b_{\tilde1\tilde2}(S^{ab}_{e,\mu})_{\tilde2\tilde1}M^c_{\tilde3\tilde2}(S^{ec}_{d,\nu'})_{\tilde2\tilde1}
    M^d_{\tilde10}\ket{d_0,d_{0'}}.
    \esp
\end{equation*}
This step is graphically represented as the follows.
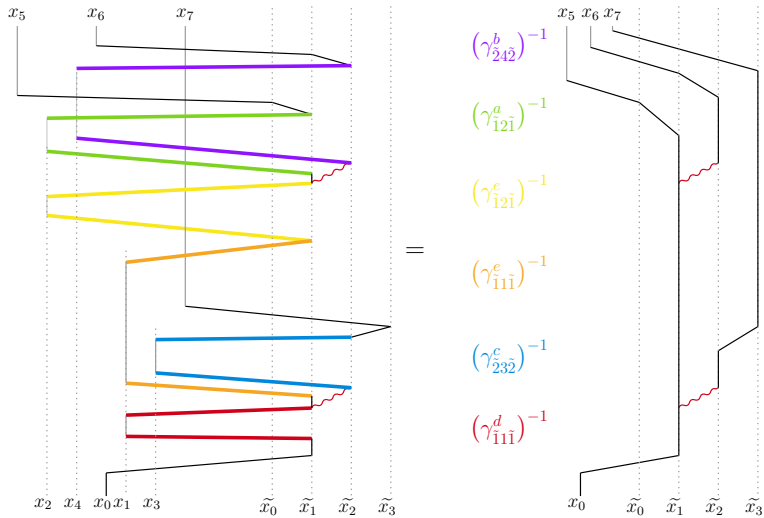
\begin{equation*}
    \scalebox{0.6}{\compareFd}
\end{equation*}
Note that in the above calculation we have used the fact that moving and splitting operators with different endpoints commute. Summarizing the above relations, we have
\beq \label{eq: relating F states 1}
\bsp
&\ket{e,\mu,\nu}=\sum_{\mu'\nu'}\gamma^b_{6\tilde1\tilde24}\gamma^a_{5\tilde0\tilde12}[(\Gamma^{ab}_e)_{\tilde2\tilde142}]_{\mu\mu'}\gamma^e_{2\tilde11}\gamma^c_{7\tilde3\tilde23}[(\Gamma^{ec}_d)_{\tilde2\tilde131}]_{\nu\nu'}\gamma^d_{1\tilde10}\times\\
&( \gamma _{\tilde{2} 4\tilde{2}}^{b})^{-1}( \gamma _{\tilde{1} 2\tilde{1}}^{a})^{-1}( \gamma _{\tilde{1} 2\tilde{1}}^{e})^{-1}( \gamma _{\tilde{1} 1\tilde{1}}^{e})^{-1}( \gamma _{\tilde{2} 3\tilde{2}}^{c})^{-1}( \gamma _{\tilde{1} 1\tilde{1}}^{d})^{-1}M^a_{5\tilde0}M^b_{6\tilde1}M^c_{7\tilde3}\widetilde{\ket{e,\mu',\nu'}}.
\esp
\eeq

We can do similar analysis for the state $\ket{f,\rho,\sigma}$:
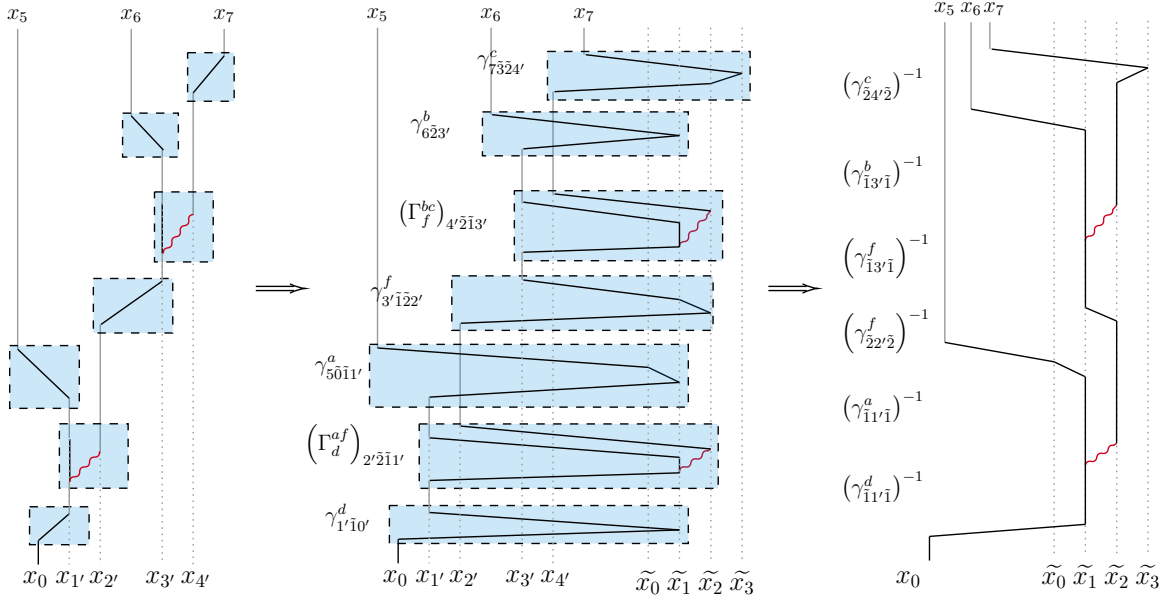
\begin{equation*}
    \scalebox{0.7}{\compareFe}
\end{equation*}
In the above figure and in the following, we use the double-line arrow `$\Rightarrow$' to indicate that the diagram on the tail side of the arrow, excluding the factors, is equal to the diagram on the tip side multiplying the factors on the tip side of the arrow. One multiplies together the factors appearing in the diagram and finds the correct relation/equation. The above diagram is summarized as
\beq \label{eq: relating F states 2}
\bsp
&\ket{f,\rho,\sigma}=\sum_{\rho'\sigma'}\gamma _{7\tilde{3}\tilde{2} 4'}^{c}\gamma _{6\tilde{1} 3'}^{b}[(\Gamma _{f}^{bc})_{\tilde{2}\tilde{1} 4'3'}]_{\rho\rho'}\gamma _{3'\tilde{1}\tilde{2} 2'}^{f}\gamma _{5\tilde{0}\tilde{1} 1'}^{a}[( \Gamma _{d}^{af})_{\tilde{2}\tilde{1} 2'1'}]_{\sigma\sigma'}\gamma _{1'\tilde{1} 0}^{d}\times\\
&( \gamma _{\tilde{2} 4'\tilde{2}}^{c})^{-1}( \gamma _{\tilde{1} 3'\tilde{1}}^{b})^{-1}( \gamma _{\tilde{1} 3'\tilde{1}}^{f})^{-1}( \gamma _{\tilde{2} 2'\tilde{2}}^{f})^{-1}( \gamma _{\tilde{1} 1'\tilde{1}}^{a})^{-1}( \gamma _{\tilde{1} 1'\tilde{1}}^{d})^{-1}M^a_{5\tilde0}M^b_{6\tilde1}M^c_{7\tilde3}\widetilde{\ket{f,\rho',\sigma'}}.
\esp
\eeq

Combining the relations in Eqs. \eqref{eq: relating F states 1} and \eqref{eq: relating F states 2}, and using the fact that $F=\ew{f,\rho,\sigma|e,\mu,\nu}$ and $\tilde F=\widetilde{\bra{f,\rho,\sigma}}\widetilde{e,\mu,\nu\rangle}$, we find the relation between the microscopically defined $F$-matrix and the categorical $\tilde F$-matrix:
\begin{equation}\label{Eq:generalFtocategoryF}
\bsp
    &[(F^{abc}_d)_{\{x\}}]_{\{e,\mu,\nu\},\{f,\rho,\sigma\}}(\{\ket{\psi}\},\{M\},\{S\})=\sum_{\mu'\nu'\rho'\sigma'}
    \gamma^b_{6\tilde1\tilde24}\gamma^a_{5\tilde0\tilde12}[(\Gamma^{ab}_e)_{\tilde2\tilde142}]_{\mu\mu'}\gamma^e_{2\tilde11}\gamma^c_{7\tilde3\tilde23}[(\Gamma^{ec}_d)_{\tilde2\tilde131}]_{\nu\nu'}\gamma^d_{1\tilde10}\cdot\\
    &[(\tilde F^{abc}_{d})_{\{\tilde 0\tilde1\tilde2\tilde3\}}]_{\{e,\mu',\nu'\},\{f,\rho',\sigma'\}}(\{\ket{\psi}\},\{M\},\{S\})\cdot\\
    &(\gamma _{7\tilde{3}\tilde{2} 4'}^{c})^{-1}(\gamma _{6\tilde{1} 3'}^{b})^{-1}[(\Gamma _{f}^{bc})_{\tilde{2}\tilde{1} 4'3'}]_{\rho'\rho}^{-1}(\gamma _{3'\tilde{1}\tilde{2} 2'}^{f})^{-1}(\gamma _{5\tilde{0}\tilde{1} 1'}^{a})^{-1}[(\Gamma _{d}^{af})_{\tilde{2}\tilde{1} 2'1'}]_{\sigma'\sigma}^{-1}(\gamma _{1'\tilde{1} 0}^{d})^{-1} \times
    \frac{\gamma _{\tilde{2} 4'\tilde{2}}^{c}\gamma _{\tilde{1} 3'\tilde{1}}^{b}\gamma _{\tilde{1} 3'\tilde{1}}^{f}\gamma _{\tilde{2} 2'\tilde{2}}^{f}\gamma _{\tilde{1} 1'\tilde{1}}^{a}\gamma _{\tilde{1} 1'\tilde{1}}^{d}}{\gamma _{\tilde{2} 4\tilde{2}}^{b}\gamma _{\tilde{1} 2\tilde{1}}^{a}\gamma _{\tilde{1} 2\tilde{1}}^{e}\gamma _{\tilde{1} 1\tilde{1}}^{e}\gamma _{\tilde{2} 3\tilde{2}}^{c}\gamma _{\tilde{1} 1\tilde{1}}^{d}}.
\esp
\end{equation}
If Eqs. \eqref{eq: seamless} and \eqref{eq: simplifying condition 1} are satisfied, which means that $\gamma=1$ and $\Gamma=\mathbf{1}$, then the microscopic $F$-matrix becomes the categorical $F$-matrix.

An important feature of the relation Eq. \eqref{Eq:generalFtocategoryF} is that it {\it does not} depend on the choice of the moving operators $M_{x\tilde x}$ connecting $x_0,x_1,x_2,x_3,x_4,x_{1'},x_{2'},x_{3'},x_{4'},x_5,x_6,x_7$ and $\tilde x_0,\tilde x_1,\tilde x_2,\tilde x_3$, which is expected because the explicit definitions of the $F$-matrix in Eq. \eqref{Eq:generalF} and the categorical $\tilde F$-symbol in Eq. \eqref{eq: kawagoe levin F matrix app} do not involve those moving operators. To verify this property, we just need to verify that Eq. \eqref{Eq:generalFtocategoryF} is invariant even if we redefine these moving operators. Under this redefinition, according to Eq. \eqref{eq: new moving and splitting operators}, the $\gamma$ phases and $\Gamma$ matrices transform as
\begin{equation*}
    \bsp
    &\gamma^{a}_{2\tilde2\tilde11}\to e^{-i\phi_{\tilde 11}(a)}e^{-i\phi_{2\tilde 2}(a)}\gamma^{a}_{2\tilde2\tilde11},
    \quad 
    \gamma^{a}_{2\tilde21}\to e^{-i\phi_{\tilde 21}(a)}e^{-i\phi_{2\tilde 2}(a)}\gamma^{a}_{2\tilde21},\\
    &[(\Gamma^{ab}_c)_{\tilde2\tilde121}]_{\mu\nu}\to e^{-i\phi_{\tilde 11}(c)}e^{-i\phi_{1\tilde 1}(a)}e^{-i\phi_{2\tilde 2}(b)}[(\Gamma^{ab}_c)_{\tilde2\tilde121}]_{\mu\nu}.
    \esp
\end{equation*}
Substituting the above re-definition into Eq. \eqref{Eq:generalFtocategoryF}, it is straightforward to show that Eq. \eqref{Eq:generalFtocategoryF} is invariant.

Equipped with Eq. \eqref{Eq:generalFtocategoryF}, we can further write down the consistency equations of the microscopic $F$-matrices based on the pentagon identity of categorical $\tilde F$-symbols. We temporarily suppress the notation and make implicit the dependence on the anyon states and the moving and splitting operators for neatness. We also use the abbreviated notation $\{x_i\}$ to represent $\{0_i1_i2_i3_i4_i{1'}_i{2'}_i{3'}_i{4'}_i5_i6_i7_i\}$. Then the equations satisfied by the $F$-matrices defined in Eq. \eqref{Eq:generalF}, which are the analogs of the pentagon equation Eq. \eqref{eq: pentagon equation}, are
\begin{equation} \label{eq: general pentagon}
    \bsp
    &\sum_{\beta'\chi'\nu'\epsilon'\alpha'\mu'\lambda'\kappa'}\sum_{\epsilon}\\
    &[(\Gamma^{fc}_g)_{\tilde2\tilde14_12_1}]_{\beta\beta'}^{-1}[(\Gamma^{gd}_e)_{\tilde2\tilde13_11_1}]_{\chi\chi'}^{-1}[(F^{fcd}_e)_{\{x_1\}}]_{\{g,\beta',\chi'\},\{l,\nu',\epsilon'\}}[(\Gamma _{l}^{cd})_{\tilde{2}\tilde{1} 4_1'3_1'}]_{\nu'\nu}[(\Gamma _{e}^{fl})_{\tilde{2}\tilde{1} 2_1'1_1'}]_{\epsilon'\epsilon}\times\\
    &\frac{\gamma _{7_1\tilde{3}\tilde{2} 4_1'}^{d}\gamma _{6b_1\tilde{1} 3_1'}^{c}\gamma _{3_1'\tilde{1}\tilde{2} 2_1'}^{l}\gamma _{5_1\tilde{0}\tilde{1} 1_1'}^{f}\gamma _{1_1'\tilde{1} 0_1}^{e}}{\gamma^c_{6_1\tilde1\tilde24_1}\gamma^f_{5_1\tilde0\tilde12_1}\gamma^g_{2_1\tilde11_1}\gamma^d_{7_1\tilde3\tilde23_1}\gamma^e_{1_1\tilde10_1}}\times \frac{\gamma _{\tilde{2} 4_1\tilde{2}}^{c}\gamma _{\tilde{1} 2_1\tilde{1}}^{f}\gamma _{\tilde{1} 2_1\tilde{1}}^{g}\gamma _{\tilde{1} 1_1\tilde{1}}^{g}\gamma _{\tilde{2} 3_1\tilde{2}}^{d}\gamma _{\tilde{1} 1_1\tilde{1}}^{e}}{\gamma _{\tilde{2} 4_1'\tilde{2}}^{d}\gamma _{\tilde{1} 3_1'\tilde{1}}^{c}\gamma _{\tilde{1} 3_1'\tilde{1}}^{l}\gamma _{\tilde{2} 2_1'\tilde{2}}^{l}\gamma _{\tilde{1} 1_1'\tilde{1}}^{f}\gamma _{\tilde{1} 1_1'\tilde{1}}^{e}}\times\\
    &[(\Gamma^{ab}_f)_{\tilde2\tilde14_22_2}]_{\alpha\alpha'}^{-1}[(\Gamma^{fl}_e)_{\tilde2\tilde13_21_2}]_{\epsilon\kappa'}^{-1}[(F^{abl}_e)_{\{x_2\}}]_{\{f,\alpha',\kappa'\},\{k,\mu',\lambda'\}}[(\Gamma _{k}^{bl})_{\tilde{2}\tilde{1} 4_2'3_2'}]_{\mu'\mu}[(\Gamma _{e}^{ak})_{\tilde{2}\tilde{1} 2_2'1_2'}]_{\lambda'\lambda}\times\\
    &\frac{\gamma _{7_2\tilde{3}\tilde{2} 4_2'}^{l}\gamma _{6_2\tilde{1} 3_2'}^{b}\gamma _{3_2'\tilde{1}\tilde{2} 2_2'}^{k}\gamma _{5_1\tilde{0}\tilde{1} 1_2'}^{a}\gamma _{1_2'\tilde{1} 0_2}^{e}}{\gamma^b_{6_2\tilde1\tilde24_2}\gamma^a_{5_2\tilde0\tilde12_2}\gamma^f_{2_2\tilde11_2}\gamma^l_{7_2\tilde3\tilde23_2}\gamma^e_{1_2\tilde10_2}}\times \frac{\gamma _{\tilde{2} 4_2\tilde{2}}^{b}\gamma _{\tilde{1} 2_2\tilde{1}}^{a}\gamma _{\tilde{1} 2_2\tilde{1}}^{f}\gamma _{\tilde{1} 1_2\tilde{1}}^{f}\gamma _{\tilde{2} 3_2\tilde{2}}^{l}\gamma _{\tilde{1} 1_2\tilde{1}}^{e}}{\gamma _{\tilde{2} 4_2'\tilde{2}}^{l}\gamma _{\tilde{1} 3_2'\tilde{1}}^{b}\gamma _{\tilde{1} 3_2'\tilde{1}}^{k}\gamma _{\tilde{2} 2_2'\tilde{2}}^{k}\gamma _{\tilde{1} 1_2'\tilde{1}}^{a}\gamma _{\tilde{1} 1_2'\tilde{1}}^{e}}=\\
    &\sum_{\beta'\varphi'\sigma'\chi'\rho'\nu'\alpha'\mu'\lambda'\xi'\iota'\zeta'}\sum_{h,\sigma,\varphi,\rho}\\
    &[(\Gamma^{ab}_f)_{\tilde2\tilde14_32_3}]_{\alpha\alpha'}^{-1}[(\Gamma^{fc}_g)_{\tilde2\tilde13_31_3}]_{\beta\beta'}^{-1}[(F^{abc}_g)_{\{x_3\}}]_{\{f,\alpha',\beta'\},\{h,\varphi',\sigma'\}}[(\Gamma _{h}^{bc})_{\tilde{2}\tilde{1} 4_3'3_3'}]_{\varphi'\varphi}[(\Gamma _{g}^{fh})_{\tilde{2}\tilde{1} 2_3'1_3'}]_{\sigma'\sigma}\times\\
    &\frac{\gamma _{7_3\tilde{3}\tilde{2} 4_3'}^{c}\gamma _{6_3\tilde{1} 3_3'}^{b}\gamma _{3_3'\tilde{1}\tilde{2} 2_3'}^{h}\gamma _{5_3\tilde{0}\tilde{1} 1_3'}^{a}\gamma _{1_3'\tilde{1} 0_3}^{g}}{\gamma^b_{6_3\tilde1\tilde24_3}\gamma^a_{5_3\tilde0\tilde12_3}\gamma^f_{2_3\tilde11_3}\gamma^c_{7_3\tilde3\tilde23_3}\gamma^g_{1_3\tilde10_3}}\times \frac{\gamma _{\tilde{2} 4_3\tilde{2}}^{c}\gamma _{\tilde{1} 2_3\tilde{1}}^{a}\gamma _{\tilde{1} 2_3\tilde{1}}^{f}\gamma _{\tilde{1} 1_3\tilde{1}}^{f}\gamma _{\tilde{2} 3_3\tilde{2}}^{c}\gamma _{\tilde{1} 1_3\tilde{1}}^{g}}{\gamma _{\tilde{2} 4_3'\tilde{2}}^{c}\gamma _{\tilde{1} 3_3'\tilde{1}}^{c}\gamma _{\tilde{1} 3_3'\tilde{1}}^{h}\gamma _{\tilde{2} 2_3'\tilde{2}}^{h}\gamma _{\tilde{1} 1_3'\tilde{1}}^{a}\gamma _{\tilde{1} 1_3'\tilde{1}}^{g}}\times\\
    &[(\Gamma^{ah}_g)_{\tilde2\tilde14_42_4}]_{\sigma\xi'}^{-1}[(\Gamma^{gd}_e)_{\tilde2\tilde13_41_4}]_{\chi\chi'}^{-1}[(F^{ahd}_e)_{\{x_4\}}]_{\{g,\xi',\chi'\},\{k,\rho',\lambda'\}}[(\Gamma _{k}^{hd})_{\tilde{2}\tilde{1} 4_4'3_4'}]_{\rho'\rho}[(\Gamma _{e}^{gk})_{\tilde{2}\tilde{1} 2_4'1_4'}]_{\lambda'\lambda}\times\\
    &\frac{\gamma _{7_4\tilde{3}\tilde{2} 4_4'}^{d}\gamma _{6_4\tilde{1} 3_4'}^{h}\gamma _{3_4'\tilde{1}\tilde{2} 2_4'}^{k}\gamma _{5_4\tilde{0}\tilde{1} 1_4'}^{a}\gamma _{1_4'\tilde{1} 0_4}^{e}}{\gamma^h_{6_4\tilde1\tilde24_4}\gamma^a_{5_4\tilde0\tilde12_4}\gamma^g_{2_4\tilde11_4}\gamma^d_{7_4\tilde3\tilde23_4}\gamma^g_{1_4\tilde10_4}}\times \frac{\gamma _{\tilde{2} 4_4\tilde{2}}^{d}\gamma _{\tilde{1} 2_4\tilde{1}}^{a}\gamma _{\tilde{1} 2_4\tilde{1}}^{g}\gamma _{\tilde{1} 1_4\tilde{1}}^{g}\gamma _{\tilde{2} 3_4\tilde{2}}^{d}\gamma _{\tilde{1} 1_4\tilde{1}}^{e}}{\gamma _{\tilde{2} 4_4'\tilde{2}}^{d}\gamma _{\tilde{1} 3_4'\tilde{1}}^{d}\gamma _{\tilde{1} 3_4'\tilde{1}}^{k}\gamma _{\tilde{2} 2_4'\tilde{2}}^{k}\gamma _{\tilde{1} 1_4'\tilde{1}}^{a}\gamma _{\tilde{1} 1_4'\tilde{1}}^{e}}\times\\
    &[(\Gamma^{bc}_h)_{\tilde2\tilde14_52_5}]_{\varphi\iota'}^{-1}[(\Gamma^{hd}_k)_{\tilde2\tilde13_51_5}]_{\rho\zeta'}^{-1}[(F^{bcd}_k)_{\{x_5\}}]_{\{h,\iota',\zeta'\},\{l,\nu',\mu'\}}[(\Gamma _{l}^{cd})_{\tilde{2}\tilde{1} 4_5'3_5'}]_{\nu'\nu}[(\Gamma _{k}^{hl})_{\tilde{2}\tilde{1} 2_5'1_5'}]_{\mu'\mu}\times\\
    &\frac{\gamma _{7_5\tilde{3}\tilde{2} 4_5'}^{d}\gamma _{6_5\tilde{1} 3_5'}^{c}\gamma _{3_5'\tilde{1}\tilde{2} 2_5'}^{l}\gamma _{5_5\tilde{0}\tilde{1} 1_5'}^{b}\gamma _{1_5'\tilde{1} 0_5}^{k}}{\gamma^c_{6_5\tilde1\tilde24_5}\gamma^h_{5_5\tilde0\tilde12_5}\gamma^b_{2_5\tilde11_5}\gamma^d_{7_5\tilde3\tilde23_5}\gamma^h_{1_5\tilde10_5}}\times \frac{\gamma _{\tilde{2} 4_5\tilde{2}}^{d}\gamma _{\tilde{1} 2_5\tilde{1}}^{b}\gamma _{\tilde{1} 2_5\tilde{1}}^{h}\gamma _{\tilde{1} 1_5\tilde{1}}^{h}\gamma _{\tilde{2} 3_5\tilde{2}}^{d}\gamma _{\tilde{1} 1_5\tilde{1}}^{k}}{\gamma _{\tilde{2} 4_5'\tilde{2}}^{d}\gamma _{\tilde{1} 3_5'\tilde{1}}^{d}\gamma _{\tilde{1} 3_5'\tilde{1}}^{l}\gamma _{\tilde{2} 2_5'\tilde{2}}^{l}\gamma _{\tilde{1} 1_5'\tilde{1}}^{b}\gamma _{\tilde{1} 1_5'\tilde{1}}^{k}}.
    \esp
\end{equation}

The above equation involves $F$-matrices defined at five different sets of positions, with each set containing $12$ positions. When the $F$-matrices coincide with category $F$-symbols, which is the case when Eqs. \eqref{eq: seamless} and \eqref{eq: simplifying condition 1} hold, it is straightforward to check that the above equation reduces to the pentagon identity in Eq. \eqref{eq: pentagon equation}.

\subsubsection{The microscopic transformations of $F$-matrices}

Now we answer the second question: how does the $F$-matrix in Eq. \eqref{Eq:generalF} transform when $\{|\psi\rangle\}$, $M$ and $S$ change? Understanding this question amounts to understanding how $F$-matrices vary under the change of the following choices: the choice of moving and splitting operators, the choice of the local information of the initial anyon states, and choice of the positions of the anyons used to define $F$-matrix. Below we analyze the effects of these three types of changes in turn.

\begin{itemize}

    \item Changing the moving and splitting operators used to define the $F$-matrix.
    
    Under this change, Eq. \eqref{eq: new moving and splitting operators} gives the allowed redefinition of the moving and splitting operators. Under such transformations, the two states $\ket{e,\mu,\nu}$ and $\ket{f,\rho,\sigma}$ change as
    \begin{equation*}
    \bsp
        &\ket{e, \mu, \nu}\to \sum_{\mu'\nu'}e^{i\phi_{52}(a)}e^{i\phi_{64}(b)}e^{i\phi_{73}(c)}[\Omega_0(a,b;e)_{42}]_{\mu\mu'}e^{i\phi_{21}(e)}[\Omega_0(e,c;d)_{31}]_{\nu\nu'}e^{i\phi_{10}(d)}\ket{e, \mu', \nu'};\\
        &\ket{f, \rho, \sigma}\to \sum_{\rho'\sigma'}e^{i\phi_{52'}(a)}e^{i\phi_{64'}(b)}e^{i\phi_{73'}(c)}[\Omega_0(b,c;f)_{4'3'}]_{\rho\rho'}e^{i\phi_{3'2'}(f)}[\Omega_0(a,f;d)_{2'1'}]_{\sigma\sigma'}e^{i\phi_{1'0}(d)}\ket{f, \rho, \sigma}.
    \esp
    \end{equation*}
    As such, the $F$-matrix in Eq. \eqref{Eq:generalF} transforms as
    \begin{equation}\label{Eq:generalFVBGT}
    \bsp
    &[(F^{abc}_d)_{\{x\}}]_{\{e,\mu,\nu\},\{f,\rho,\sigma\}}(\{\ket{\psi}\},\{M\},\{S\})\\
    &\to\frac{e^{i\phi_{52}(a)}e^{i\phi_{64}(b)}e^{i\phi_{73}(c)}e^{i\phi_{21}(e)}e^{i\phi_{10}(d)}}{e^{i\phi_{51'}(a)}e^{i\phi_{63'}(b)}e^{i\phi_{74'}(c)}e^{i\phi_{3'2'}(f)}e^{i\phi_{1'0}(d)}}\cdot\sum_{\mu'\nu'\rho'\sigma'}\\
    &[\Omega_0(a,b;e)_{42}]_{\mu\mu'}[\Omega_0(e,c;d)_{31}]_{\nu\nu'}[(F^{abc}_d)_{\{x\}}]_{\{e,\mu,\nu'\},\{f,\rho',\sigma'\}}(\{\ket{\psi}\},\{M\},\{S\})[\Omega_0(b,c;f)_{4'3'}]_{\rho'\rho}^{-1}[\Omega_0(a,f;d)_{2'1'}]_{\sigma'\sigma}^{-1}.
    \esp
    \end{equation}
    Notably, when $[(F^{abc}_d)_{012341'2'3'4'567}]_{\{e,\mu,\nu\},\{f,\rho,\sigma\}}$ is simplified to the Kawagoe-Levin definition, \ie when $\{x_0, x_1, x_2, x_6, x_{1'}, x_{3'}\}$ are identified with $\tilde x_1$, $\{x_3, x_4, x_{2'}, x_{4'}\}$ are identified with $\tilde x_2$, $x_5$ is identified with $\tilde x_{0}$, and $x_7$ is identified with $\tilde x_{3}$, the above transformation is simplified to
    \begin{equation*}
    \bsp
    &[(F^{abc}_d)_{\tilde0\tilde1\tilde2\tilde3}]_{\{e,\mu,\nu\},\{f,\rho,\sigma\}}(\{\ket{\psi}\},\{M\},\{S\})\to\sum_{\mu'\nu'\rho'\sigma'}\left\{e^{i\phi_{\tilde1\tilde2}(b)}[\Omega_0(a,b;e)_{\tilde2\tilde1}]_{\mu\mu'}\right\}\left\{e^{i\phi_{\tilde1\tilde2}(c)}[\Omega_0(e,c;d)_{\tilde2\tilde1}]_{\nu\nu'}\right\}\times\\
    &[(F^{abc}_d)_{\tilde0\tilde1\tilde2\tilde3}]_{\{e,\mu,\nu'\},\{f,\rho',\sigma'\}}(\{\ket{\psi}\},\{M\},\{S\})\left\{e^{i\phi_{\tilde1\tilde2}(c)}[\Omega_0(b,c;f)_{\tilde2\tilde1}]_{\rho'\rho}\right\}^{-1}\left\{e^{i\phi_{\tilde1\tilde2}(f)}[\Omega_0(a,f;d)_{\tilde2\tilde1}]_{\sigma'\sigma}\right\}^{-1}.
    \esp
    \end{equation*}
    This simplified transformation is also introduced in Ref. \cite{Kawagoe2019}, which takes precisely the same form as the vertex basis gauge transformation of the categorical $\tilde F$-symbols. 

    \item Changing the local information of the chosen anyon states.
    
    Suppose we change the definition of anyon states, such that the anyons have different local information. According to our discussions in Sec. \ref{sec:setup}, the new anyon states can be obtained by applying unitary operators localized around the anyons, \ie the anyon states transform as
    \begin{equation*}
        \ket{a_1,b_2,\cdots}\to W(x_1)W({x_2})\cdots\ket{a_1,b_2,\cdots}=W(x_1, x_2, \cdots)|a_1, b_2, \cdots\rangle,
    \end{equation*}
    where $W(x_i)$ is a unitary operator localized around the position $x_i$, and $W(x_1, x_2, \cdots)=\prod_iW(x_i)$. Correspondingly, the moving and splitting operators should also change according to their definitions in Eq. \eqref{eq: moving operator} and Eq. \eqref{eq: splitting operator}. In particular, we can always make the following choice:
    \begin{equation}\label{Eq:StringChangingLocalInfoa}
    \bsp
        &M^{a}_{21}\to W(x_1, x_2, \cdots)M^{a}_{21}W(x_1, x_2, \cdots)^{\dagger},\\
        &(S^{ab}_{c,\mu})_{21}\to W(x_1, x_2, \cdots)(S^{ab}_{c,\mu})_{21}W(x_1, x_2, \cdots)^{\dagger}.
    \esp 
    \end{equation}
    Note that the new moving and splitting operators still have their supports around their corresponding old operators. As such, the final states $\ket{e,\mu,\nu}$ and $\ket{f,\rho,\sigma}$ are changed to $W(x_1, x_2, \cdots)\ket{e,\mu,\nu}$ and $W(x_1, x_2, \cdots)\ket{f,\rho,\sigma}$, and their inner product does not change. Hence, the microscopic $F$-matrix stays invariant when we change the local information of the anyon states and choose these particular moving and splitting operators. Certainly, we can also make other choices of moving and splitting operators. Then the $F$-matrix will undergo a transformation triggered by changing the definition of moving and splitting operators, which is given in Eq. \eqref{Eq:generalFVBGT}.

    \item Changing the position of anyons.
    
    Since the $F$-matrices have explicit dependence on the positions of anyons, they are also expected to change when we calculate them at different positions. In particular, $[(F^{abc}_d)_{0_11_12_13_14_1{1_1'}{2_1'}3_1'4_1'5_16_17_1}]_{\{e,\mu,\nu\},\{f,\rho,\sigma\}}$ will be different from $[(F^{abc}_d)_{0_21_22_23_24_2{1_2'}{2_2'}3_2'4_2'5_26_27_2}]_{\{e,\mu,\nu\},\{f,\rho,\sigma\}}$, where the positions with subscripts 1 and 2 are two sets of positions, respectively. The strategy to find the relation between them is to relate them to a set of common categorical $F$-symbols extracted at fixed positions $\tilde x_0$, $\tilde x_1$, $\tilde x_2$ and $\tilde x_3$. According to Eq. \eqref{Eq:generalFtocategoryF}, we have (in the following equation, we have made the position dependence explicit and hide the dependence on the anyon states and the moving and splitting operators.)
    \begin{equation*}
    \bsp
    &\sum_{\mu\nu\rho\sigma}[(\Gamma^{ab}_e)_{\tilde2\tilde14_12_1}]_{\mu'\mu}^{-1}[(\Gamma^{ec}_d)_{\tilde2\tilde13_11_1}]_{\nu'\nu}^{-1}[(F^{abc}_d)_{0_11_12_13_14_1{1_1'}{2_1'}3_1'4_1'5_16_17_1}]_{\{e,\mu,\nu\},\{f,\rho,\sigma\}}
    \; [(\Gamma _{f}^{bc})_{\tilde{2}\tilde{1} 4_1'3_1'}]_{\rho\rho'}[(\Gamma _{d}^{af})_{\tilde{2}\tilde{1} 2_1'1_1'}]_{\sigma\sigma'}\times\\
    &\frac{\gamma _{5_1\tilde{3}\tilde{2} 4_1'}^{c}\gamma _{6_1\tilde{1} 3_1'}^{b}\gamma _{3_1'\tilde{1}\tilde{2} 2_1'}^{f}\gamma _{5_1\tilde{0}\tilde{1} 1_1'}^{a}\gamma _{1_1'\tilde{1} 0_1}^{d}}{\gamma^b_{6_1\tilde1\tilde24_1}\gamma^a_{5_1\tilde0\tilde12_1}\gamma^e_{2_1\tilde11_1}\gamma^c_{7_1\tilde3\tilde23_1}\gamma^d_{1_1\tilde10}}\times\frac{\gamma _{\tilde{2} 4_1\tilde{2}}^{b}\gamma _{\tilde{1} 2_1\tilde{1}}^{a}\gamma _{\tilde{1} 2_1\tilde{1}}^{e}\gamma _{\tilde{1} 1_1\tilde{1}}^{e}\gamma _{\tilde{2} 3_1\tilde{2}}^{c}\gamma _{\tilde{1} 1_1\tilde{1}}^{d}}{\gamma _{\tilde{2} 4_1'\tilde{2}}^{c}\gamma _{\tilde{1} 3_1'\tilde{1}}^{b}\gamma _{\tilde{1} 3_1'\tilde{1}}^{f}\gamma _{\tilde{2} 2_1'\tilde{2}}^{f}\gamma _{\tilde{1} 1_1'\tilde{1}}^{a}\gamma _{\tilde{1} 1_1'\tilde{1}}^{d}}=[(\tilde F^{abc}_{d})_{\tilde0\tilde1\tilde2\tilde3}]_{\{e,\mu',\nu'\},\{f,\rho',\sigma'\}}=\\
    &\sum_{\alpha\beta\lambda\kappa}[(\Gamma^{ab}_e)_{\tilde2\tilde14_22_2}]_{\mu'\alpha}^{-1}[(\Gamma^{ec}_d)_{\tilde2\tilde13_21_2}]_{\nu'\beta}^{-1}[(F^{abc}_d)_{0_21_22_23_24_2{1_2'}{2_2'}3_2'4_2'5_26_27_2}]_{\{e,\alpha,\beta\},\{f,\lambda,\kappa\}}
    \; [(\Gamma _{f}^{bc})_{\tilde{2}\tilde{1} 4_2'3_2'}]_{\rho\rho'}[(\Gamma _{d}^{af})_{\tilde{2}\tilde{1} 2_2'1_2'}]_{\sigma\sigma'}\times\\
    &\frac{\gamma _{5_2\tilde{3}\tilde{2} 4_2'}^{c}\gamma _{6_2\tilde{1} 3_2'}^{b}\gamma _{3_2'\tilde{1}\tilde{2} 2_2'}^{f}\gamma _{5_2\tilde{0}\tilde{1} 1_2'}^{a}\gamma _{1_2'\tilde{1} 0_2}^{d}}{\gamma^b_{6_2\tilde1\tilde24_2}\gamma^a_{5_2\tilde0\tilde12_2}\gamma^e_{2_2\tilde11_2}\gamma^c_{7_2\tilde3\tilde23_2}\gamma^d_{1_2\tilde10}}\times\frac{\gamma _{\tilde{2} 4_2\tilde{2}}^{b}\gamma _{\tilde{1} 2_2\tilde{1}}^{a}\gamma _{\tilde{1} 2_2\tilde{1}}^{e}\gamma _{\tilde{1} 1_2\tilde{1}}^{e}\gamma _{\tilde{2} 3_2\tilde{2}}^{c}\gamma _{\tilde{1} 1_2\tilde{1}}^{d}}{\gamma _{\tilde{2} 4_2'\tilde{2}}^{c}\gamma _{\tilde{1} 3_2'\tilde{1}}^{b}\gamma _{\tilde{1} 3_2'\tilde{1}}^{f}\gamma _{\tilde{2} 2_2'\tilde{2}}^{f}\gamma _{\tilde{1} 1_2'\tilde{1}}^{a}\gamma _{\tilde{1} 1_2'\tilde{1}}^{d}}.
\esp
\end{equation*}
Consequently, the $F$-matrices defined at two different sets of positions are related by
\begin{equation}\label{Eq:generalFtransformation}
\bsp
    &[(F^{abc}_d)_{0_11_12_13_14_1{1_1'}{2_1'}3_1'4_1'5_16_17_1}]_{\{e,\mu,\nu\},\{f,\rho,\sigma\}}=\sum_{\mu'\nu'\rho'\sigma'}\sum_{\alpha\beta\lambda\kappa}
    \frac{\gamma^b_{6_1\tilde1\tilde24_1}}{\gamma^b_{6_2\tilde1\tilde24_2}}\frac{\gamma^a_{5_1\tilde0\tilde12_1}}{\gamma^a_{5_2\tilde0\tilde12_2}}\left\{[(\Gamma^{ab}_e)_{\tilde2\tilde14_12_1}]_{\mu\mu'}[(\Gamma^{ab}_e)_{\tilde2\tilde14_22_2}]_{\mu'\alpha}^{-1}\right\}\frac{\gamma^e_{2_1\tilde11_1}}{\gamma^e_{2_2\tilde11_2}}\frac{\gamma^c_{7_1\tilde3\tilde23_1}}{\gamma^c_{7_2\tilde3\tilde23_2}}\\
    &\left\{[(\Gamma^{ec}_d)_{\tilde2\tilde13_11_1}]_{\nu\beta}[(\Gamma^{ec}_d)_{\tilde2\tilde13_21_2}]_{\nu'\beta}^{-1}\right\}\frac{\gamma^d_{1_1\tilde10_1}}{\gamma^d_{1_2\tilde10_2}}\times[(F^{abc}_d)_{0_21_22_23_24_2{1_2'}{2_2'}3_2'4_2'5_26_27_2}]_{\{e,\alpha,\beta\},\{f,\lambda,\kappa\}}\times\\
    &\frac{\gamma _{7_2\tilde{3}\tilde{2} 4_2'}^{c}}{\gamma _{7_1\tilde{3}\tilde{2} 4_1'}^{c}}\frac{\gamma _{6_2\tilde{1} 3_2'}^{b}}{\gamma _{6_1\tilde{1} 3_1'}^{b}}\left\{[(\Gamma _{f}^{bc})_{\tilde{2}\tilde{1} 4_1'3_1'}]_{\lambda\rho}^{-1}[(\Gamma _{f}^{bc})_{\tilde{2}\tilde{1} 4_2'3_2'}]_{\rho\rho'}\right\}\frac{\gamma _{3_2'\tilde{1}\tilde{2} 2_2'}^{f}}{\gamma _{3_1'\tilde{1}\tilde{2} 2_1'}^{f}}\frac{\gamma _{5_2\tilde{0}\tilde{1} 1_2'}^{a}}{\gamma _{5_1\tilde{0}\tilde{1} 1_1'}^{a}}\left\{[(\Gamma _{d}^{af})_{\tilde{2}\tilde{1} 2_1'1_1'}]_{\kappa\sigma}^{-1}[(\Gamma _{d}^{af})_{\tilde{2}\tilde{1} 2_2'1_2'}]_{\sigma\sigma'}\right\}\frac{\gamma _{1_2'\tilde{1} 0_2}^{d}}{\gamma _{1_1'\tilde{1} 0_1}^{d}} \times\\
    &\frac{\gamma _{\tilde{2} 4_1'\tilde{2}}^{c}\gamma _{\tilde{1} 3_1'\tilde{1}}^{b}\gamma _{\tilde{1} 3_1'\tilde{1}}^{f}\gamma _{\tilde{2} 2_1'\tilde{2}}^{f}\gamma _{\tilde{1} 1_1'\tilde{1}}^{a}\gamma _{\tilde{1} 1_1'\tilde{1}}^{d}}{\gamma _{\tilde{2} 4_1\tilde{2}}^{b}\gamma _{\tilde{1} 2_1\tilde{1}}^{a}\gamma _{\tilde{1} 2_1\tilde{1}}^{e}\gamma _{\tilde{1} 1_1\tilde{1}}^{e}\gamma _{\tilde{2} 3_1\tilde{2}}^{c}\gamma _{\tilde{1} 1_1\tilde{1}}^{d}}\times \frac{\gamma _{\tilde{2} 4_2\tilde{2}}^{b}\gamma _{\tilde{1} 2_2\tilde{1}}^{a}\gamma _{\tilde{1} 2_2\tilde{1}}^{e}\gamma _{\tilde{1} 1_2\tilde{1}}^{e}\gamma _{\tilde{2} 3_2\tilde{2}}^{c}\gamma _{\tilde{1} 1_2\tilde{1}}^{d}}{\gamma _{\tilde{2} 4_2'\tilde{2}}^{c}\gamma _{\tilde{1} 3_2'\tilde{1}}^{b}\gamma _{\tilde{1} 3_2'\tilde{1}}^{f}\gamma _{\tilde{2} 2_2'\tilde{2}}^{f}\gamma _{\tilde{1} 1_2'\tilde{1}}^{a}\gamma _{\tilde{1} 1_2'\tilde{1}}^{d}}.
    \esp
    \end{equation}
    Although the phase factors and matrices in the above relation have explicit dependence on the choice of the position $\tilde x_0, \tilde x_1, \tilde x_2, \tilde x_3$, the moving and splitting operators involving only $\tilde x_0,\tilde x_1,\tilde x_2,\tilde x_3$, the moving operators between $\tilde x_0,\tilde x_1,\tilde x_2,\tilde x_3$ and $x_{0_1},x_{1_1},x_{2_1},x_{3_1},x_{4_1},x_{1_1'},x_{2_1'},x_{3_1'},x_{4_1'},x_{5_1},x_{6_1},x_{7_1}$, and the moving operators between $\tilde x_0,\tilde x_1,\tilde x_2,\tilde x_3$ and $x_{0_2},x_{1_2},x_{2_2},x_{3_2},x_{4_2},x_{1_2'},x_{2_2'},x_{3_2'},x_{4_2'},x_{5_2},x_{6_2},x_{7_2}$, the combination of these phases and matrices is independent of these choices, which is expected because $[(F^{abc}_d)_{0_11_12_13_14_1{1_1'}{2_1'}3_1'4_1'5_16_17_1}]_{\{e,\mu,\nu\},\{f,\rho,\sigma\}}$ being different from $[(F^{abc}_d)_{0_21_22_23_24_2{1_2'}{2_2'}3_2'4_2'5_26_27_2}]_{\{e,\mu,\nu\},\{f,\rho,\sigma\}}$ does not depend on them. This independence can be seen by picking up other $\{\tilde x\}$ and re-defining these moving and splitting operators, which will result in precisely the same expression. Again, when both of $[(F^{abc}_d)_{0_11_12_13_14_1{1_1'}{2_1'}3_1'4_1'5_16_17_1}]_{\{e,\mu,\nu\},\{f,\rho,\sigma\}}$ and $[(F^{abc}_d)_{0_21_22_23_24_2{1_2'}{2_2'}3_2'4_2'5_26_27_2}]_{\{e,\mu,\nu\},\{f,\rho,\sigma\}}$ are simplified to the Kawagoe-Levin definition, which gives $[(\tilde F^{abc}_d)_{\{0_11_12_13_1\}}]_{\{e,\mu,\nu\},\{f,\rho,\sigma\}}$ and $[(\tilde F^{abc}_d)_{\{0_21_22_23_2\}}]_{\{e,\mu,\nu\},\{f,\rho,\sigma\}}$, respectively, the above transformation is greatly simplified into
    \beq
    \bsp
    &[(\tilde F^{abc}_d)_{0_11_12_13_1}]_{\{e,\mu,\nu\},\{f,\rho,\sigma\}}\\
    =&\sum_{\alpha\beta\lambda\kappa}[(\tilde\Gamma^{ab}_e)_{2_21_22_11_1}]_{\mu\alpha}[(\tilde\Gamma^{ec}_d)_{2_21_22_11_1}]_{\nu\beta}[(\tilde F^{abc}_d)_{0_21_22_23_2}]_{\{e,\alpha,\beta\},\{f,\lambda,\kappa\}}[(\tilde\Gamma^{bc}_f)_{2_21_22_11_1}]_{\lambda\rho}[(\tilde\Gamma^{af}_d)_{2_21_22_11_1}]_{\kappa\sigma},    
    \esp
    \eeq
    where
    \begin{equation*}
    [(\tilde\Gamma^{ab}_c)_{2_21_22_11_1}]_{\mu\mu'}=\sum_{\mu'}\left\{[(\Gamma^{ab}_c)_{\tilde2\tilde12_11_1}]_{\mu\mu'}[(\Gamma^{ab}_c)_{\tilde2\tilde12_21_1}]_{\mu'\alpha}^{-1}\right\} \frac{\gamma _{\tilde{1} 1_2\tilde{1}}^{a}}{\gamma _{\tilde{1} 1_1\tilde{1}}^{a}}\frac{\gamma _{1_1\tilde{1} 1_1}^{a}}{\gamma _{1_2\tilde{1} 1_2}^{a}}\frac{\gamma^b_{1_1\tilde1\tilde22_1}}{\gamma^b_{1_2\tilde1\tilde22_2}}\frac{\gamma _{\tilde22_2\tilde2}^{b}}{\gamma _{\tilde22_1\tilde2}^{b}}\frac{\gamma _{\tilde{1} 1_2\tilde{1}}^{e}}{\gamma _{\tilde{1} 1_1\tilde{1}}^{e}}.
    \end{equation*}
    This transformation is again the vertex basis gauge transformation of the categorical $F$-symbols.
    \end{itemize}

To summarize, when we change the moving and splitting operators, and/or change the local information of the anyons, the $F$-matrix changes according to Eq. \eqref{Eq:generalFVBGT}. When we change the positions of anyons used to define the $F$-matrix, the $F$-matrix undergoes a transformation given by Eq. \eqref{Eq:generalFtransformation}. When Eqs. \eqref{eq: seamless} and \eqref{eq: simplifying condition 1} hold, Eqs. \eqref{Eq:generalFVBGT} and \eqref{Eq:generalFtransformation} take the form of Eq. \eqref{eq: vertex basis gauge transformation}.

\subsection{The braiding of anyons: $R$-matrices}

\begin{figure*}
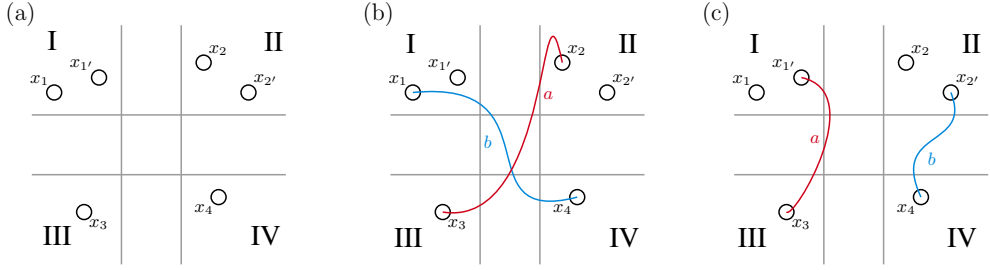

    \centering
    \scalebox{0.75}{\generalRillustrate}
    \caption{(a) Two strips divide the space into $4$ regions: I, II, III, and IV. The four regions {II}-{I}-{III}-{IV} should wind around the overlapping region of the strip in a counter-clockwise manner.
    (b) and (c): The two processes differ by a braiding.}
    \label{fig:generalRillustrate}
\end{figure*}

The second characteristic feature of topological order is the braiding of anyons. Roughly speaking, the braiding property of two anyons is contained in the process of exchanging the positions of the two anyons $a$ and $b$. In real experiments, the exchange process may not be precise in general, namely the anyon $a$ may not be moved to exactly the original position of $b$, but some position close to the original position of $b$ instead. Therefore, we define the general process of braiding as follows. We use two orthogonal strips to divide the space into four regions, as depicted in Fig. \ref{fig:generalRillustrate}(a). The strips must be wide enough to contain the support of anyon moving and splitting operators, and there is no upper limit for the width. The four regions {II}-{I}-{III}-{IV} should wind around the overlapping region of the strip in a counter-clockwise manner. We find two points $x_1$ and $x_{1'}$ in region {I}, two points $x_2$ and $x_{2'}$ in region {II}, one point $x_3$ in region {III}, one point $x_4$ in region {IV}, and one point $x_0$ without any constraint on its position. There is no constraint on the relative position of $x_1$ and $x_{1'}$ ($x_2$ and $x_{2'}$), either. The $R$-matrices that encodes the braiding is defined by
\begin{equation}\label{eq: general R}
    [(R^{ab}_c)_{0121'2'34}]_{\mu\nu}(\{\ket{\psi}\},\{M\},\{S\})=\ew{\nu|\mu},
\end{equation}
where
\begin{equation}
    \bsp
    &\ket{\mu}=M^a_{32}M^b_{41}(S^{ba}_{c,\mu})_{21}M^c_{10}\ket{c_0,\bar c},\\
    &\ket{\nu}=M^b_{42'}M^a_{31'}(S^{ab}_{c,\nu})_{2'1'}M^c_{1'0}\ket{c_0,\bar c}.
\esp
\end{equation}

There are also some requirements on the moving operators. The support of the splitting operator $(S^{ba}_c)_{21}$ and $(S^{ab}_c)_{2'1'}$ should not overlap with regions {III} and {IV}; the support of the moving operator $M^b_{41}$ ($M^a_{32}$) is required to not overlap with regions {II} and {III} (regions {I} and {IV}); the support of the moving operator $M^b_{42'}$ ($M^a_{31'}$) should not overlap with regions {I} and {III} (regions {II} and {IV}). As such, the two processes in Eq. \eqref{eq: general R} differ by an effect of changing the anyon positions in a counter-clockwise direction in a rough sense. 

Using the graphic presentations, the position exchange is more obvious:
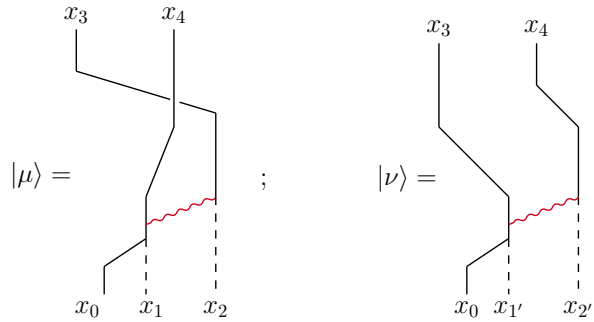
\begin{equation*}
    \scalebox{0.7}{\generalR}
\end{equation*}
Similar to the processes that define the $F$-matrix in Eq. \eqref{Eq:generalF}, the processes considered here are the simplest among the most general braiding processes. Namely, they are general since the splitting processes can happen at different locations, and the anyons can move before and after the splitting processes. At the same time, they are simple since the motion of anyons before and after each splitting process is achieved by applying a single moving operator, rather than multiple moving operators.

Below we discuss the properties of the $R$-matrices. We again answer the two questions: 1) What is the relation between this microscopic $R$-matrix in Eq. \eqref{eq: general R} and the Kawagoe-Levin $R$-symbol in Eq. \eqref{eq: kawagoe levin R matrix app}? And 2) how does the $R$-matrix transform when its microscopic input in Eq. \eqref{eq: general R}, including the moving and splitting operators, and the local information and positions of anyons, changes?

\subsubsection{The relation between microscopic $R$-matrices and categorical $R$-symbols} \label{subsubapp: R relations}

\begin{figure}
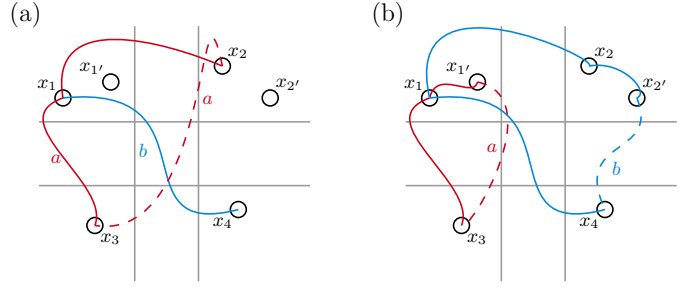

    \centering
    \scalebox{0.8}{\compareRillustrate}
    \caption{(a) and (b) Identifying $M^a_{32}$ with $M^a_{31}M^a_{12}$ and identifying $M^b_{42'}$ with $M^b_{41}M^b_{12'}$.}
    \label{fig:compareRillustrate}
\end{figure}

To understand the relation between the microscopic $R$-matrix in Eq. \eqref{eq: general R} and the categorical $R$-symbol in Eq. \eqref{eq: kawagoe levin R matrix app}, which is denoted by $\tilde R$ in this appendix, we first consider a simplified situation, where $x_1$ and $x_{1'}$ are the same position denoted by $x_1$, $x_2$ and $x_{2'}$ are the same position denoted by $x_2$, identifying $M^a_{32}$ with $M^a_{31}M^a_{12}$, identifying $M^c_{1'0}$ with $M^c_{1'1}M^c_{10}$, identifying $M^b_{42'}$ with $M^b_{41}M^b_{12}M^b_{22'}$, and identifying $M^a_{31'}$ with $M^a_{31}M^a_{11'}$. Figure \ref{fig:compareRillustrate} shows the identification of the moving operators and the support of the new moving operators. The first two identifications amount to making specific choices of anyon positions, and the last three identifications fix $\gamma^a_{312}=\gamma^c_{1'10}=\gamma^b_{4122'}=\gamma^a_{311'}=1$. Also, the support of the moving operator $M^a_{31}$ ($M^a_{12}$) is required to not overlap with regions II and IV (regions III and IV), the support of $M^b_{12}$ is required to not overlap with regions III and IV, the support of $M^a_{11'}$ ($M^b_{22'}$) is required to not overlap with regions II, III, and IV (regions I, III, and IV), and there is no constraint on the support of $M^c_{1'1}$. After the simplification, the definition of $R$-matrices becomes identical to the Kawagoe-Levin definition based on the initial state $M^c_{30}\ket{c_0,\bar c}$. Thus, we can use Eqs. \eqref{Eq:unseamless} and \eqref{Eq:generalMS} to find the relation between $R$ and $\tilde R$. In particular,
\begin{equation}
\bsp
    &\ket{\mu}=\gamma^a_{312}M^a_{31}M^a_{12}M^b_{41}(S^{ba}_{c,\mu})_{21}M^c_{10}\ket{c_0,\bar c},\\
    &\ket{\nu}=\gamma^b_{4122'}\gamma^c_{1'10}\gamma^a_{311'}M^b_{41}M^b_{12}M^b_{22'}M^a_{31}M^a_{11'}(S^{ab}_{c,\nu})_{2'1'}M^c_{1'1}M^c_{10}\ket{c_0,\bar c}\\
    &=\gamma^b_{4122'}\gamma^c_{1'10}\gamma^a_{311'}[(\Gamma^{ab}_c)_{212'1'}]_{\nu\nu'}^{-1}M^b_{41}M^b_{12}M^a_{31}(S^{ab}_{c,\nu})_{21}M^c_{10}\ket{c_0,\bar c}.
\esp
\end{equation}
Note that we have used the fact that $M^a_{11'}$ and $M^b_{22'}$ commute since we have specified their support. In terms of graphic presentations, the above equations are represented as follows. 

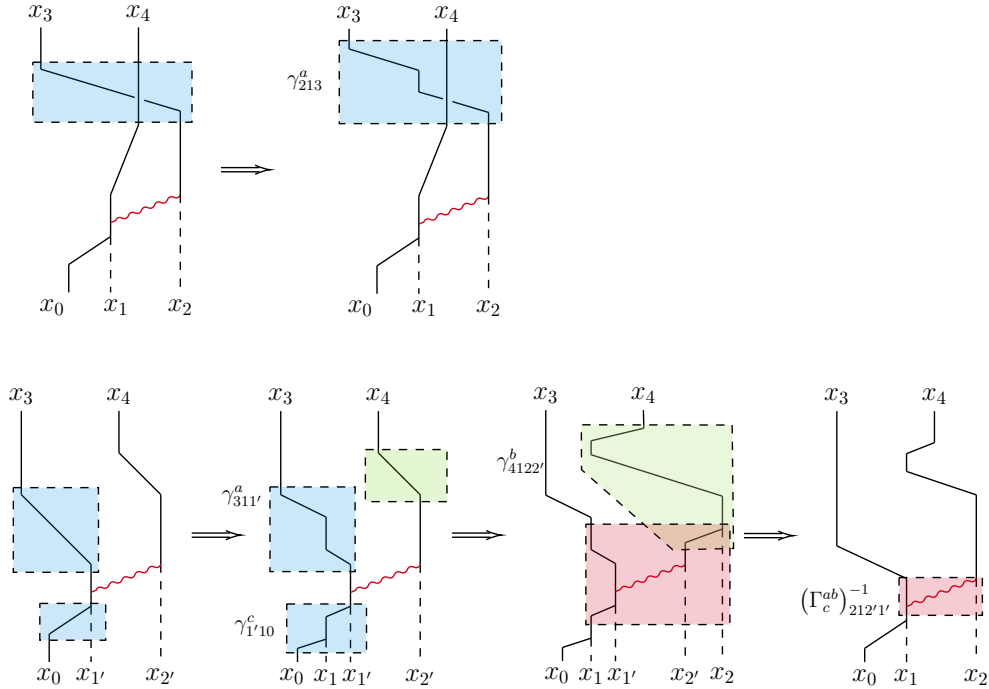
\begin{align*}
    \scalebox{0.7}{\compareR}
\end{align*}

It is then straightforward to see that the microscopic $R$-matrices are related to $\tilde R$-symbols by 
\begin{equation}\label{Eq:generalRtocategoryR}
    [(R^{ab}_c)_{0121'2'34}]_{\mu\nu}(\{\ket{\psi}\},\{M\},\{S\})=\sum_{\nu'}[(\tilde R^{ab}_c)_{3124}]_{\mu\nu'}(\{\ket{\psi}\},\{M\},\{S\})[(\Gamma^{ab}_c)_{212'1'}]_{\nu'\nu}\frac{\gamma^a_{312}}{\gamma^b_{4122'}\gamma^c_{1'10}\gamma^a_{311'}}.
\end{equation}
One finds that when relating $R$ to $\tilde R$, we have used a strategy different from the case of $F$ matrices, where we compare microscopic $F$ matrices to categorical $F$-symbols extracted at different positions. This is because defining $R$-matrices requires special conditions on moving operators, and hence relating $R$-matrices to $R$-symbols extracted at other positions requires special treatments. We will discuss this point in the next section. Also, it must be noted that the above requirements on the moving operators naturally indicate that $M^a_{31}$, $M^b_{41}$, and $M^a_{12}$ ($M^b_{12}$) sweep around $x_1$ in a counter-clockwise manner. If one insist on having the other orientation, e.g. having all our set up mirror-imaged, the extracted matrices via Eq. \eqref{eq: general R} (denoted as $\bar R$) are related to $\tilde R$-symbols by
\begin{equation}\label{Eq:generalRtocategoryRinv}
    [(R^{ab}_c)_{0121'2'34}]_{\mu\nu}(\{\ket{\psi}\},\{M\},\{S\})=\sum_{\nu'}[(\tilde R^{ba}_c)_{3124}]^{-1}_{\mu\nu'}(\{\ket{\psi}\},\{M\},\{S\})[(\Gamma^{ab}_c)_{212'1'}]_{\nu'\nu}\frac{\gamma^a_{312}}{\gamma^b_{4122'}\gamma^c_{1'10}\gamma^a_{311'}}.
\end{equation}

The microscopic $R$-matrices also yield two consistency equations with microscopic $F$-matrices, analogous to the hexagon equations Eq. \eqref{eq: hexagon equations app}. Again, we temporarily suppress the notation and make implicit the dependence on the anyon states and string operators for notational simplicity. In the following equations, all three microscopic $(R^{ab}_c)_{0_i12{1'}_i{2'}_i34}$-matrices share the same points $x_1$, $x_2$, $x_3$, and $x_4$, but the splitting processes of $F$ and $R$ can happen at various positions. There is also a point $x_{\tilde 3}$ in the expression, on which the explicit expression of the equation does not depend. (This can be seen from changing the position of $x_{\tilde 3}$ and verify that the explicit form does not change.) We make explicit the position dependence of all $R$-matrices, and the abbreviated notation $\{x_i\}$ again represents the position dependence of $F$ symbols $\{0_i1_i2_i3_i4_i{1'}_i{2'}_i{3'}_i{4'}_i5_i6_i7_i\}$. Be aware that $x_{k_i}$'s are different points from $x_k$ for $k=1,2,3,4$. The first equation is
\begin{equation} \label{eq: general hexagon 1}
\bsp
    &\sum_{\lambda'\zeta'\xi'}\sum_{\lambda\gamma}
    [(R^{ac}_e)_{0121_1'2_1'34}]_{\alpha\lambda'}[(\Gamma^{ac}_e)_{212_1'1_1'}]_{\lambda'\lambda}^{-1}\frac{\gamma^c_{4122_1'}\gamma^e_{1_1'10}\gamma^a_{311_1'}}{\gamma^a_{312}}\times\\
    &[(\Gamma^{ac}_e)_{214_22_2}]_{\lambda\zeta'}^{-1}[(\Gamma^{eb}_d)_{213_21_2}]_{\beta\beta'}^{-1}[(F^{acb}_d)_{\{x_2\}}]_{\{e,\zeta',\beta'\},\{g,\xi',\nu'\}}[(\Gamma _{g}^{cb})_{21 4_2'3_2'}]_{\xi'\gamma}[(\Gamma _{d}^{ag})_{21 2_2'1_2'}]_{\nu'\nu}\times\\
    &\frac{\gamma _{7_2\tilde{3}2 4_2'}^{b}\gamma _{6_21 3_2'}^{c}\gamma _{3_2'12 2_2'}^{g}\gamma _{5_231 1_2'}^{a}\gamma _{1_2'1 0_2}^{d}}{\gamma^c_{6_2124_2}\gamma^a_{5_2312_2}\gamma^e_{2_211_2}\gamma^d_{7_2\tilde323_2}\gamma^d_{1_210_2}}\times \frac{\gamma _{2 4_22}^{c}\gamma _{1 2_21}^{a}\gamma _{1 2_21}^{e}\gamma _{1 1_21}^{e}\gamma _{2 3_22}^{b}\gamma _{1 1_21}^{d}}{\gamma _{2 4_2'2}^{b}\gamma _{1 3_2'1}^{c}\gamma _{1 3_2'1}^{g}\gamma _{2 2_2'2}^{g}\gamma _{1 1_2'1}^{a}\gamma _{1 1_2'1}^{d}}\times\\
    &[(R^{bc}_g)_{0121_3'2_3'34}]_{\kappa\nu'}[(\Gamma^{ac}_e)_{212_3'1_3'}]_{\nu'\nu}^{-1}\frac{\gamma^c_{4122_3'}\gamma^g_{1_3'10}\gamma^b_{311_3'}}{\gamma^b_{312}}=\\
    &\sum_{\alpha'\beta'\sigma'\iota'\psi'}\sum_{f\sigma\epsilon\phi}[(\Gamma^{ca}_e)_{434_42_4}]_{\alpha\alpha'}^{-1}[(\Gamma^{eb}_d)_{433_41_4}]_{\beta\beta'}^{-1}[(F^{cab}_d)_{\{x_4\}}]_{\{e,\alpha',\beta'\},\{f,\epsilon',\sigma'\}}[(\Gamma _{f}^{ab})_{21 4_4'3_4'}]_{\epsilon'\epsilon}[(\Gamma _{d}^{cf})_{21 2_4'1_4'}]_{\sigma'\sigma}\times\\
    &\frac{\gamma _{7_4\tilde{3}2 4_4'}^{b}\gamma _{6_41 3_4'}^{a}\gamma _{3_4'12 2_4'}^{f}\gamma _{5_431 1_4'}^{c}\gamma _{1_4'1 0_4}^{d}}{\gamma^a_{6_4124_4}\gamma^c_{5_4312_4}\gamma^e_{2_411_4}\gamma^d_{7_4\tilde323_4}\gamma^d_{1_410_4}}\times \frac{\gamma _{2 4_42}^{a}\gamma _{1 2_41}^{c}\gamma _{1 2_41}^{e}\gamma _{1 1_41}^{e}\gamma _{2 3_42}^{b}\gamma _{1 1_41}^{d}}{\gamma _{2 4_4'2}^{b}\gamma _{1 3_4'1}^{a}\gamma _{1 3_4'1}^{f}\gamma _{2 2_4'2}^{f}\gamma _{1 1_4'1}^{c}\gamma _{1 1_4'1}^{d}}\times\\
    &[(R^{fc}_d)_{\{0121_5'2_5'34\}}]_{\iota\psi'}[(\Gamma^{fc}_d)_{212_5'1_5'}]_{\psi'\psi}^{-1}\frac{\gamma^c_{4122_5'}\gamma^d_{1_5'10}\gamma^f_{311_5'}}{\gamma^f_{312}}\times\\
    &[(\Gamma^{ab}_f)_{434_62_6}]_{\epsilon\phi'}^{-1}[(\Gamma^{fc}_d)_{433_61_6}]_{\psi\varphi'}^{-1}[(F^{abc}_d)_{\{x_6\}}]_{\{f,\phi',\varphi'\},\{g,\mu',\nu'\}}[(\Gamma _{g}^{bc})_{21 4_6'3_6'}]_{\mu'\mu}[(\Gamma _{d}^{ag})_{21 2_6'1_6'}]_{\nu'\nu}\times\\
    &\frac{\gamma _{7_6\tilde{3}2 4_6'}^{c}\gamma _{6_61 3_6'}^{b}\gamma _{3_6'12 2_6'}^{g}\gamma _{5_631 1_6'}^{a}\gamma _{1_6'1 0_6}^{d}}{\gamma^b_{6_6124_6}\gamma^a_{5_6312_6}\gamma^f_{2_611_6}\gamma^d_{7_6\tilde323_6}\gamma^d_{1_610_6}}\times \frac{\gamma _{2 4_62}^{b}\gamma _{1 2_61}^{a}\gamma _{1 2_61}^{f}\gamma _{1 1_61}^{f}\gamma _{2 3_62}^{c}\gamma _{1 1_61}^{d}}{\gamma _{2 4_6'2}^{c}\gamma _{1 3_6'1}^{b}\gamma _{1 3_6'1}^{g}\gamma _{2 2_6'2}^{g}\gamma _{1 1_6'1}^{a}\gamma _{1 1_6'1}^{d}}.
\esp
\end{equation}
The second equation is
\begin{equation} \label{eq: general hexagon 2}
\bsp
    &\sum_{\lambda'\zeta'\xi'}\sum_{\lambda\gamma}
    [(R^{ca}_e)_{0121_1'2_1'34}]^{-1}_{\alpha\lambda'}[(\Gamma^{ca}_e)_{212_1'1_1'}]_{\lambda'\lambda}\frac{\gamma^a_{312}}{\gamma^c_{4122_1'}\gamma^e_{1_1'10}\gamma^a_{311_1'}}\times\\
    &[(\Gamma^{ac}_e)_{214_22_2}]_{\lambda\zeta'}^{-1}[(\Gamma^{eb}_d)_{213_21_2}]_{\beta\beta'}^{-1}[(F^{acb}_d)_{\{x_2\}}]_{\{e,\zeta',\beta'\},\{g,\xi',\nu'\}}[(\Gamma _{g}^{cb})_{21 4_2'3_2'}]_{\xi'\gamma}[(\Gamma _{d}^{ag})_{21 2_2'1_2'}]_{\nu'\nu}\times\\
    &\frac{\gamma _{7_2\tilde{3}2 4_2'}^{b}\gamma _{6_21 3_2'}^{c}\gamma _{3_2'12 2_2'}^{g}\gamma _{5_231 1_2'}^{a}\gamma _{1_2'1 0_2}^{d}}{\gamma^c_{6_2124_2}\gamma^a_{5_2312_2}\gamma^e_{2_211_2}\gamma^d_{7_2\tilde323_2}\gamma^d_{1_210_2}}\times \frac{\gamma _{2 4_22}^{c}\gamma _{1 2_21}^{a}\gamma _{1 2_21}^{e}\gamma _{1 1_21}^{e}\gamma _{2 3_22}^{b}\gamma _{1 1_21}^{d}}{\gamma _{2 4_2'2}^{b}\gamma _{1 3_2'1}^{c}\gamma _{1 3_2'1}^{g}\gamma _{2 2_2'2}^{g}\gamma _{1 1_2'1}^{a}\gamma _{1 1_2'1}^{d}}\times\\
    &[(R^{cb}_g)_{0121_3'2_3'34}]^{-1}_{\kappa\nu'}[(\Gamma^{cb}_e)_{212_3'1_3'}]_{\nu'\nu}\frac{\gamma^b_{312}}{\gamma^c_{4122_3'}\gamma^g_{1_3'10}\gamma^b_{311_3'}}=\\
    &\sum_{\alpha'\beta'\sigma'\iota'\psi'}\sum_{f\sigma\epsilon\phi}[(\Gamma^{ca}_e)_{434_42_4}]_{\alpha\alpha'}^{-1}[(\Gamma^{eb}_d)_{433_41_4}]_{\beta\beta'}^{-1}[(F^{cab}_d)_{\{x_4\}}]_{\{e,\alpha',\beta'\},\{f,\epsilon',\sigma'\}}[(\Gamma _{f}^{ab})_{21 4_4'3_4'}]_{\epsilon'\epsilon}[(\Gamma _{d}^{cf})_{21 2_4'1_4'}]_{\sigma'\sigma}\times\\
    &\frac{\gamma _{7_4\tilde{3}2 4_4'}^{b}\gamma _{6_41 3_4'}^{a}\gamma _{3_4'12 2_4'}^{f}\gamma _{5_431 1_4'}^{c}\gamma _{1_4'1 0_4}^{d}}{\gamma^a_{6_4124_4}\gamma^c_{5_4312_4}\gamma^e_{2_411_4}\gamma^d_{7_4\tilde323_4}\gamma^d_{1_410_4}}\times \frac{\gamma _{2 4_42}^{a}\gamma _{1 2_41}^{c}\gamma _{1 2_41}^{e}\gamma _{1 1_41}^{e}\gamma _{2 3_42}^{b}\gamma _{1 1_41}^{d}}{\gamma _{2 4_4'2}^{b}\gamma _{1 3_4'1}^{a}\gamma _{1 3_4'1}^{f}\gamma _{2 2_4'2}^{f}\gamma _{1 1_4'1}^{c}\gamma _{1 1_4'1}^{d}}\times\\
    &[(R^{cf}_d)_{\{0121_5'2_5'34\}}]^{-1}_{\iota\psi'}[(\Gamma^{cf}_d)_{212_5'1_5'}]_{\psi'\psi}^{-1}\frac{\gamma^f_{312}}{\gamma^c_{4122_5'}\gamma^d_{1_5'10}\gamma^f_{311_5'}}\times\\
    &[(\Gamma^{ab}_f)_{434_62_6}]_{\epsilon\phi'}^{-1}[(\Gamma^{fc}_d)_{433_61_6}]_{\psi\varphi'}^{-1}[(F^{abc}_d)_{\{x_6\}}]_{\{f,\phi',\varphi'\},\{g,\mu',\nu'\}}[(\Gamma _{g}^{bc})_{21 4_6'3_6'}]_{\mu'\mu}[(\Gamma _{d}^{ag})_{21 2_6'1_6'}]_{\nu'\nu}\times\\
    &\frac{\gamma _{7_6\tilde{3}2 4_6'}^{c}\gamma _{6_61 3_6'}^{b}\gamma _{3_6'12 2_6'}^{g}\gamma _{5_631 1_6'}^{a}\gamma _{1_6'1 0_6}^{d}}{\gamma^b_{6_6124_6}\gamma^a_{5_6312_6}\gamma^f_{2_611_6}\gamma^d_{7_6\tilde323_6}\gamma^d_{1_610_6}}\times \frac{\gamma _{2 4_62}^{b}\gamma _{1 2_61}^{a}\gamma _{1 2_61}^{f}\gamma _{1 1_61}^{f}\gamma _{2 3_62}^{c}\gamma _{1 1_61}^{d}}{\gamma _{2 4_6'2}^{c}\gamma _{1 3_6'1}^{b}\gamma _{1 3_6'1}^{g}\gamma _{2 2_6'2}^{g}\gamma _{1 1_6'1}^{a}\gamma _{1 1_6'1}^{d}}.
\esp
\end{equation}
When the $F$- and $R$-matrices coincide with categorical $F$- and $R$-symbols, which is the case when Eqs. \eqref{eq: seamless}, \eqref{eq: simplifying condition 1} and \eqref{eq: simplifying condition 2} hold, it is straightforward to check that the above equations reduce to the hexagon equations Eq. \eqref{eq: hexagon equations app}.

\subsubsection{The microscopic transformation of $R$-matrices}

Now we discuss how $R$-matrices vary when changing the three conditions we discussed in the previous section. 

\begin{itemize}
    \item Changing the moving and splitting operators used to define $R$-matrices.

    Under the transformation shown in Eq. \eqref{eq: new moving and splitting operators}, the two states $\ket{\mu}$ and $\ket{\nu}$ transform as
    \begin{equation*}
    \bsp
    &\ket{\mu}\to \sum_{\mu'}e^{i\phi_{32}(a)}e^{i\phi_{41}(b)}[\Omega_0(b,a;c)_{21}]_{\mu\mu'}e^{i\phi_{10}(c)}\ket{\mu'}\\
    &\ket{\nu}\to \sum_{\nu'}e^{i\phi_{42'}(b)}e^{i\phi_{31'}(a)}[\Omega_0(a,b;c)_{2'1'}]_{\nu\nu'}e^{i\phi_{1'0}(c)}\ket{\nu'}.
    \esp        
    \end{equation*}
    As such, the $R$-matrices transform as
    \begin{equation}\label{Eq:generalRVBGT}
    \bsp
        [(R^{ab}_c)_{\{x\}}]_{\mu\nu}(\{\ket{\psi}\},\{M\},\{S\})\to &\sum_{\mu'\nu'}[\Omega_0(b,a;c)_{21}]_{\mu\mu'}([R^{ab}_c]_{\mu\nu})_{\{x\}}(\{\ket{\psi}\},\{M\},\{S\}')[\Omega_0(a,b;c)_{2'1'}]_{\nu'\nu}^{-1}\\
        &\times\frac{e^{i\phi_{32}(a)}e^{i\phi_{41}(b)}e^{i\phi_{10}(c)}}{e^{i\phi_{42'}(b)}e^{i\phi_{31'}(a)}e^{i\phi_{1'0}(c)}}.
        \esp
    \end{equation}

    When the $R$-matrices coincide with the categorical $\tilde R$ symbols, the above transformation is simplified to
    \begin{equation*}
        [(R^{ab}_c)_{\{x\}}]_{\mu\nu}(\{\ket{\psi}\},\{M\},\{S\})\to \sum_{\mu'\nu'}\left(e^{i\phi_{12}(a)}[\Omega_0(b,a;c)_{21}]_{\mu\mu'}\right)[(R^{ab}_c)_{\{x\}}]_{\mu\nu}(\{\ket{\psi}\},\{M\},\{S\})\left(e^{i\phi_{12}(b)}[\Omega_0(a,b;c)_{21}]_{\nu\nu'}\right)^{-1}.
    \end{equation*}
    This simplified transformation is also introduced in Ref. \cite{Kawagoe2019}, which is precisely the same form as the vertex basis gauge transformation of $\tilde R$ symbols in MTC.

    \item Changing the local information of anyons.
    When changing the local information of the chosen anyon states, one can use the same argument as in the case of $F$-matrices to show that the $R$-matrices can be invariant when we choose the moving and splitting operators in Eq. \eqref{Eq:StringChangingLocalInfoa}, and the $R$-matrices transform as Eq. \eqref{Eq:generalRVBGT} when we choose other moving and splitting operators.
    
    \item Changing the position of anyons.

    Since the definition of $R$-matrices has special requirements on moving operators, comparing two microscopic $R$-matrices defined at different positions is a bit more complicated than the case of $F$-matrices. Denote these two $R$-matrices as $(R^{ab}_c)_{0_11_12_11_1'2_1'3_14_1}$ and $(R^{ab}_c)_{0_21_22_21_2'2_2'3_24_2}$. These $R$-matrices should be extracted with the setup of positions and moving operators introduced in Fig. \ref{fig:generalRillustrate} (repeated in Fig. \ref{fig:relateRillustrate}(a)), or be extracted with the mirror-imaged setup of the original setup, as shown in Fig. \ref{fig:relateRillustrate}(b). The strategy to establish their relations is the following. we pick $5$ points $\tilde x_0\tilde x_1\tilde x_2\tilde x_4\tilde x_5$, with $M_{\tilde1\tilde0}$-$M_{\tilde4\tilde0}$-$M_{\tilde2\tilde0}$ organized in a counter-clockwise direction and $M_{\tilde1\tilde0}$-$M_{\tilde5\tilde0}$-$M_{\tilde2\tilde0}$ organized in a clockwise direction. As such, we can extract categorical $\tilde R$-symbols $(\tilde R^{ab}_c)_{\tilde x_0\tilde x_1\tilde x_2\tilde x_4}$ and $(\tilde R^{ab}_c)_{\tilde x_0\tilde x_1\tilde x_2\tilde x_5}=(\tilde R^{ba}_c)^{-1}_{\tilde x_0\tilde x_1\tilde x_2\tilde x_4}$ (see Appendix \ref{app: review Kawagoe-Levin} for details of this equation) from this $5$ points. Then, we compare the microscopic $R$-matrices defined at each position with the $(\tilde R^{ab}_c)_{\tilde x_0\tilde x_1\tilde x_2\tilde x_4}$ or $(\tilde R^{ab}_c)_{\tilde x_0\tilde x_1\tilde x_2\tilde x_5}$. In particular, if the support of $M^a_{3_i1_i}$, $M^b_{4_i1_i}$, and $M^{a}_{1_i2_i}$ ($M^a_{3_i1_i}$, $M^b_{4_i1_i}$, and $M^{b}_{1_i2_i}$) sweeps around $x_{1_i}$ in a counter-clockwise direction, then we can only compare $(R^{ab}_c)_{0_i1_i2_i1_i'2_i'3_i4_i}$ to $(\tilde R^{ab}_c)_{\tilde x_0\tilde x_1\tilde x_2\tilde x_5}$ because otherwise the moving operators will enclose non-trivial anyons and cause uncontrollable phase factors; the support of $M^a_{3_i1_i}$, $M^b_{4_i1_i}$, and $M^{a}_{1_i2_i}$ ($M^a_{3_i1_i}$, $M^b_{4_i1_i}$, and $M^{b}_{1_i2_i}$) sweeps around $x_{1_i}$ in a clockwise direction, then we can only relate $(R^{ab}_c)_{0_i1_i2_i1_i'2_i'3_i4_i}$ to $(\tilde R^{ab}_c)_{\tilde x_0\tilde x_1\tilde x_2\tilde x_4}$ for the same reason. Given the relation between $(R^{ab}_c)_{0_i1_i2_i1_i'2_i'3_i4_i}$ and $(\tilde R^{ab}_c)_{\tilde x_0\tilde x_1\tilde x_2\tilde x_4}$ or $(\tilde R^{ab}_c)_{\tilde x_0\tilde x_1\tilde x_2\tilde x_5}$, we can finally find the relation between $(R^{ab}_c)_{0_11_12_11_1'2_1'3_14_1}$ and $(R^{ab}_c)_{0_21_22_21_2'2_2'3_24_2}$.

    \begin{figure}
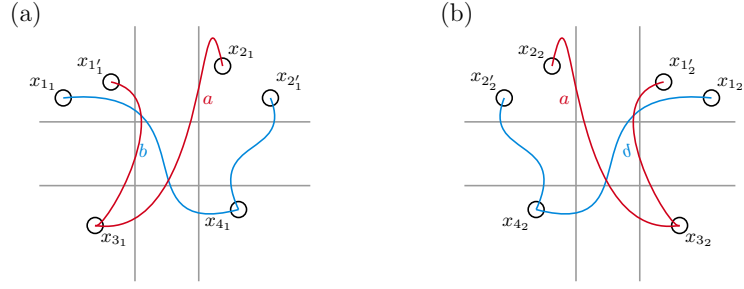

        \centering
        \scalebox{0.8}{\relateRillustrate}
        \caption{(a) The setup of positions and moving operators in Fig. \ref{fig:generalRillustrate}. (b) The setup of positions and moving operators that is a mirror-image of that in Fig. \ref{fig:generalRillustrate}.}
        \label{fig:relateRillustrate}
    \end{figure}

    Suppose $M^a_{3_i1_i}$, $M^b_{4_i1_i}$, and $M^{a}_{1_i2_i}$ ($M^a_{3_i1_i}$, $M^b_{4_i1_i}$, and $M^{b}_{1_i2_i}$) sweeps around $x_{1_i}$ in a counter-clockwise direction for both $i=1,2$. It has been established that the microscopic $R$-matrices are related to the categorical $\tilde R$-symbols defined at $x_{5_i}x_{3_i}x_{4_i}x_{6_i}$ by Eq. \eqref{Eq:generalRtocategoryR}, we only need to find how the two categorical $\tilde R$-symbols defined at different positions are related. To do this, we find the moving operators that connect $x_{1_i}$ to $\tilde x_1$, $x_{2_i}$ to $\tilde x_2$, $x_{3_i}$ to $\tilde x_0$, and $x_{4_i}$ to $\tilde x_5$, as shown in Fig. \ref{fig:relateRs}. As such, we can relate $ (\tilde R^{ab}_c)_{3_i1_i2_i4_i}$ to $(\tilde R^{ab}_c)_{\tilde0\tilde1\tilde2\tilde5}= (\tilde R^{ba}_c)^{-1}_{\tilde0\tilde1\tilde2\tilde4}$ via
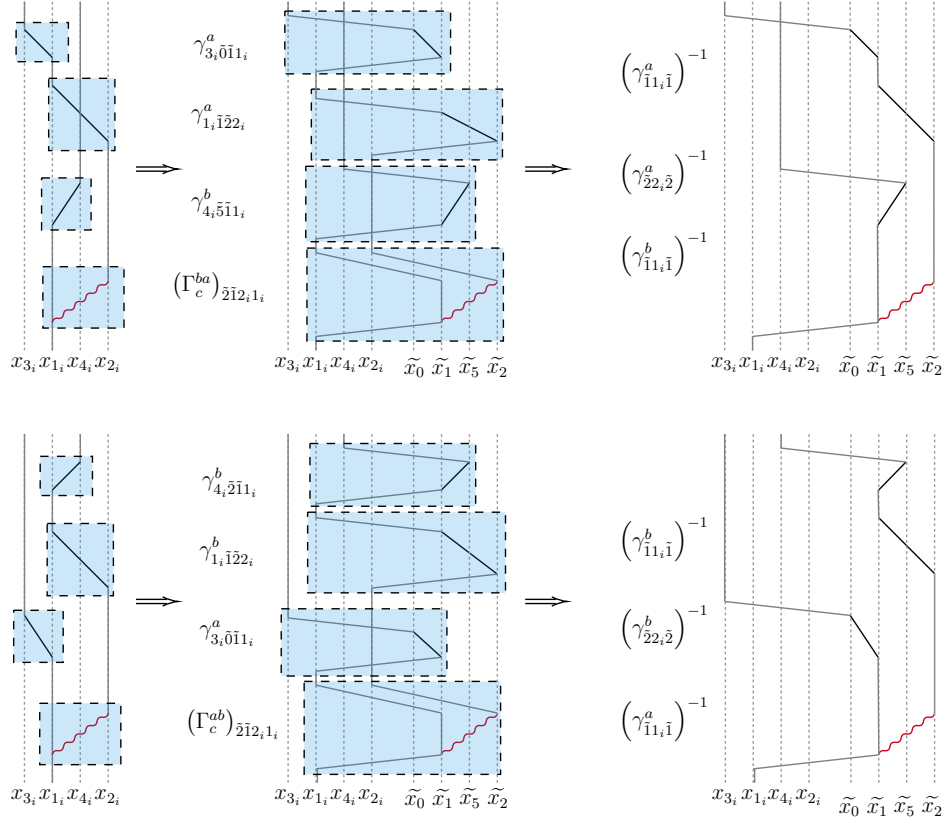
    \begin{equation*}
    \scalebox{0.7}{\relatecategoryR}
    \end{equation*}
    One can easily check that, using the moving operators supported as in Fig. \ref{fig:relateRs}, no non-trivial anyons will be enclosed during the two processes. Consequently, 
    \begin{equation*}
    \bsp
        &[(\tilde R^{ab}_c)_{3_i1_i2_i4_i}]_{\mu\nu}=\sum_{\mu'\nu'}\left(\frac{\gamma^a_{1_i\tilde1\tilde22_i}}{\gamma^a_{\tilde22_i\tilde2}}[(\Gamma^{ba}_c)_{\tilde2\tilde12_i1_i}]_{\mu\mu'}\right)[(\tilde R^{ba}_c)_{\tilde0\tilde1\tilde2\tilde4}]_{\mu'\nu'}^{-1}\left(\frac{\gamma^b_{1_i\tilde1\tilde22_i}}{\gamma^b_{\tilde22_i\tilde2}}[(\Gamma^{ab}_c)_{\tilde2\tilde12_i1_i}]_{\nu'\nu}\right)^{-1}.
        \esp
    \end{equation*}
    It is straightforward to see that two categorical $\tilde R$-symbols defined at different positions are related in the same form as a vertex basis gauge transformation. Combining the relation between $\tilde R$'s and Eq. \eqref{Eq:generalRtocategoryR}, we can find the relation between microscopic $R$-matrices defined at different positions,
    \begin{equation}\label{Eq:generalRtransformation1}
    \bsp    
    &[(R^{ab}_c)_{0_11_12_11_1'2_1'3_14_1}]_{\mu\nu}(\{\ket{\psi}\},\{M\},\{S\})=\sum_{\mu'\nu'\rho\sigma\kappa'\lambda'}[(\Gamma^{ba}_c)_{\tilde2\tilde12_11_1}]_{\mu'\rho}[(\Gamma^{ba}_c)_{\tilde2\tilde12_21_2}]_{\rho\kappa'}\times\\        
    &[(R^{ab}_c)_{0_21_22_21_2'2_2'3_24_2}]_{\kappa\lambda}(\{\ket{\psi}\},\{M\},\{S\})\times[(\Gamma^{ab}_c)_{2_11_12'_11'_1}]_{\nu'\nu}[(\Gamma^{ab}_c)_{\tilde2\tilde12_11_1}]_{\nu\sigma}^{-1}[(\Gamma^{ab}_c)_{\tilde2\tilde12_21_2}]_{\sigma\lambda'}^{-1}[(\Gamma^{ab}_c)_{2_21_22'_21'_2}]_{\lambda'\lambda}\times\\
    &\frac{\gamma^a_{1_1\tilde1\tilde22_1}\gamma^b_{\tilde22_1\tilde2}}{\gamma^b_{1_1\tilde1\tilde22_1}\gamma^a_{\tilde22_1\tilde2}}\frac{\gamma^b_{1_2\tilde1\tilde22_2}\gamma^a_{\tilde22_2\tilde2}}{\gamma^a_{1_2\tilde1\tilde22_2}\gamma^b_{\tilde22_2\tilde2}}\frac{\gamma^a_{3_11_12_1}}{\gamma^b_{4_11_12_12_1'}\gamma^c_{1_1'1_10_1}\gamma^a_{3_11_11_1'}}\frac{\gamma^b_{4_21_22_22_2'}\gamma^c_{1_2'1_20_2}\gamma^a_{3_21_21_2'}}{\gamma^a_{3_21_22_2}}
    \esp
    \end{equation}
    Suppose $M^a_{3_i1_i}$, $M^b_{4_i1_i}$, and $M^{a}_{1_i2_i}$ ($M^a_{3_i1_i}$, $M^b_{4_i1_i}$, and $M^{b}_{1_i2_i}$) sweeps around $x_{1_i}$ in a clockwise direction for both $i=1,2$, the relation between $(R^{ab}_c)_{0_11_12_11_1'2_1'3_14_1}$ and $(R^{ab}_c)_{0_21_22_21_2'2_2'3_24_2}$ is also given by the above equation. When one of the $i$'s gives $M^a_{3_i1_i}$, $M^b_{4_i1_i}$, and $M^{a}_{1_i2_i}$ sweeping around $x_{1_i}$ in a counter-clockwise direction, while the other gives the corresponding moving operators sweeps in a clockwise direction, the relation between $(R^{ab}_c)_{0_11_12_11_1'2_1'3_14_1}$ and $(R^{ab}_c)_{0_21_22_21_2'2_2'3_24_2}$ are correspondingly modified to
    \begin{equation}\label{Eq:generalRtransformation2}
    \bsp
        &[(R^{ab}_c)_{0_11_12_11_1'2_1'3_14_1}]_{\mu\nu}(\{\ket{\psi}\},\{M\},\{S\})=\sum_{\mu'\nu'\rho\sigma\kappa'\lambda'}[(\Gamma^{ba}_c)_{\tilde2\tilde14_13_1}]_{\mu'\rho}[(\Gamma^{ba}_c)_{\tilde2\tilde14_23_2}]_{\rho\kappa'}\times\\        
        &[(R^{ba}_c)_{0_21_22_21_2'2_2'3_24_2}]_{\kappa\lambda}^{-1}(\{\ket{\psi}\},\{M\},\{S\})\times[(\Gamma^{ab}_c)_{4_13_12_1'1_1'}]_{\nu'\nu}[(\Gamma^{ab}_c)_{\tilde2\tilde14_13_1}]_{\nu\sigma}^{-1}[(\Gamma^{ab}_c)_{\tilde2\tilde14_23_2}]_{\sigma\lambda'}^{-1}[(\Gamma^{ab}_c)_{4_23_22_21_2}]_{\lambda'\lambda}\times\\
        &\frac{\gamma^a_{1_1\tilde1\tilde22_1}\gamma^b_{\tilde22_1\tilde2}}{\gamma^b_{1_1\tilde1\tilde22_1}\gamma^a_{\tilde22_1\tilde2}}        \frac{\gamma^a_{1_2\tilde1\tilde22_2}\gamma^b_{\tilde22_2\tilde2}}{\gamma^b_{1_2\tilde1\tilde22_2}\gamma^a_{\tilde22_2\tilde2}}  \frac{\gamma^a_{3_11_12_1}}{\gamma^b_{4_11_12_12_1'}\gamma^c_{1_1'1_10_1}\gamma^a_{3_11_11_1'}}\frac{\gamma^a_{3_21_22_2}}{\gamma^b_{4_21_22_22_2'}\gamma^c_{1_2'1_20_2}\gamma^a_{3_21_21_2'}} 
    \esp
    \end{equation}
    
    \begin{figure}
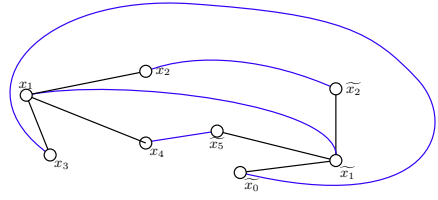

        \centering
        \scalebox{0.6}{\relateRs}
        \caption{The blue lines represents for the moving operators connecting $x$ and $\tilde x$. This choice guarantees that no non-trivial anyon will be enclosed by moving operators.}
        \label{fig:relateRs}
    \end{figure}
    
    Again, the above relations are independent of the choice of the position $\tilde 0\tilde1\tilde2\tilde4\tilde5$, the moving and splitting operators within $\tilde 0\tilde1\tilde2\tilde4\tilde5$, the moving operators between $\tilde 0\tilde1\tilde2\tilde4\tilde5$ and $x_{0_1}x_{1_1}x_{2_1}x_{1_1'}x_{2_1'}x_{3_1}x_{4_1}$, and the moving operators between $\tilde 0\tilde1\tilde2\tilde4\tilde5$ and $x_{0_2}x_{1_2}x_{2_2}x_{1_2'}x_{2_2'}x_{3_2}x_{4_2}$. This independence can be seen by picking other positions as $\{\tilde x\}$ and re-defining the moving and splitting operators, which will result in precisely the same expression. 
\end{itemize}

To summarize, when we change the moving and splitting operators, the $R$-matrices change according to Eq.\eqref{Eq:generalRVBGT}; when we change anyon local information, the $R$-matrices can be invariant or transform according to Eq.\eqref{Eq:generalRVBGT} depending on the choices of the corresponding moving and splitting operators. When we change the anyon positions we used to define them, the $R$-matrices undergo a transformation given by Eqs. \eqref{Eq:generalRtransformation1} and \eqref{Eq:generalRtransformation2} depending on the orientation of the moving operators. When Eqs. \eqref{eq: seamless}, \eqref{eq: simplifying condition 1} and \eqref{eq: simplifying condition 2} hold, these transformation take the form of Eq. \eqref{eq: vertex basis gauge transformation}.

\subsection{The change of vertex basis under symmetry transformation: $U$-matrices}

As we have discussed in the main text, the symmetry properties of a TQSL includes the change of anyon types and fusion vertex basis states under symmetry transformations. These two properties are encoded in the $U$-matrices. The definition of the $U$-matrices is introduced in Eq. \eqref{eq: symmetry localization} of Sec. \ref{subsubsec:symmetrysetup}, and also reformulated in Eq. \eqref{eq: U-matrix} of Sec. \ref{subsec: including symmetries}. For the convenience of the reader, we repeat the definition in Eq. \eqref{eq: U-matrix} here:
\begin{equation} \label{eq: U-matrix app}
 [U_g(a,b;c)_{12}]_{\mu\nu}(\{\ket{\psi}\},\{S\},\{V\})=\ew{\nu|\mu},
\end{equation}
where
\begin{equation}
\bsp
    &\ket{\mu}=R_g(S^{\lsupsc{\bar g}a\lsupsc{\bar g}b}_{\lsupsc{\bar g}c,\mu})_{\lsupsc{\bar g}2\lsupsc{\bar g}1}R_g^{-1}V^{\lsupsc{\bar g}c}_g(x_1)\ket{c_0,\bar c},\\
    &\ket{\nu}=V^{\lsupsc{\bar g}a}_g(x_1)V^{\lsupsc{\bar g}b}_g(x_2)(S^{ab}_{c,\nu})_{21}\ket{c_0,\bar c}.
\esp
\end{equation}

In the main text, we focus on the case where Eqs. \eqref{eq: seamless}, \eqref{eq: simplifying condition 1} and \eqref{eq: simplifying condition 2} hold. However, in general, these conditions do not need to hold. In the following, we will discuss the properties of the $U$-matrices defined above, under the general conditions given by Eqs. \eqref{Eq:unseamless}, \eqref{Eq:generalMS} and \eqref{Eq:generalMV}. Again, we focus on two questions: 1) When the choices of $\{|\psi\rangle\}$, $\{S\}$ and $\{V\}$ change, how do the $U$-matrices defined in Eq. \eqref{eq: U-matrix app} transform? And 2) What consistency equations should the $U$-matrices defined in Eq. \eqref{eq: U-matrix app} satisfy?

\subsubsection{The microscopic transformation of $U$-matrices}

We start with the transformation of the $U$-matrices defined in Eq. \eqref{eq: U-matrix app} when $\{|\psi\rangle\}$, $\{S\}$ and $\{V\}$ change.

\begin{itemize}
    \item Changing the moving and splitting operators used to define $U$-matrices.

    Under the transformation shown in Eq. \eqref{eq: new moving and splitting operators}, the two states $\ket{\mu}$ and $\ket{\nu}$ transform as
    \begin{equation*}
    \bsp
    &\ket{\mu}\to \sum_{\mu'}K^{q(g)}[\Omega_0(\lsupsc{\bar g}a,\lsupsc{\bar g}b;\lsupsc{\bar g}c)_{\lsupsc{\bar g}2\lsupsc{\bar g}1}]_{\mu\mu'}K^{q(g)}\ket{\mu'},\\
    &\ket{\nu}\to \sum_{\nu'}[\Omega_0(a,b;c)_{2'1'}]_{\nu\nu'}\ket{\nu'}.
    \esp        
    \end{equation*}
    As such, the $U$-matrices transform as
    \begin{equation}\label{Eq:generalUVBGT}
    \bsp
        &[U_g(a,b;c)_{12}]_{\mu\nu}(\{\ket{\psi}\},\{M\},\{S\},\{V\})\to\\
        &\sum_{\mu'\nu'}K^{q(g)}[\Omega_0(\lsupsc{\bar g}a,\lsupsc{\bar g}b;\lsupsc{\bar g}c)_{\lsupsc{\bar g}2\lsupsc{\bar g}1}]_{\mu\mu'}K^{q(g)}[U_g(a,b;c)_{12}]_{\mu'\nu'}(\{\ket{\psi}\},\{S\},\{V\})[\Omega_0(a,b;c)_{21}]^{-1}_{\nu'\nu}.
        \esp
    \end{equation}
    When the symmetry is internal, the $U$-matrices are simplified to the $U$-symbols in a $G$-crossed braided tensor category. Moreover, $^{\bar g}x_{1,2}=x_{1,2}$, and the above transformation is simplified to
    \begin{equation*}
    \bsp
        &[U_g(a,b;c)_{12}]_{\mu\nu}(\{\ket{\psi}\},\{M\},\{S\},\{V\})\to\\
        &\sum_{\mu'\nu'}[\Omega_0(\lsupsc{\bar g}a,\lsupsc{\bar g}b;\lsupsc{\bar g}c)_{21}]_{\mu\mu'}[U_g(a,b;c)_{12}]_{\mu'\nu'}(\{\ket{\psi}\},\{M\},\{S\},\{V\})[\Omega_0(a,b;c)_{21}]^{-1}_{\nu'\nu},
    \esp
    \end{equation*}
    which is precisely the vertex basis gauge transformation of the $U$-symbols in a $G$-crossed braided tensor category for TQSLs with an internal symmetry $G$.

    \item Changing the $V$ operators used to define $U$-matrices.

    In defining the $U$-matrices, we need to insert $V$ operators to compensate the change of anyon local information. These $V$ operators, determined according to Eq. \eqref{eq: symmetry localization}, are not unique either. In particular, according to the discussion around Eq. \eqref{eq: symmetry action gauge transformation of states}, re-defining these $V$ operator yields
    \begin{equation}\label{Eq:newVoperators}
        V^{\lsupsc{\bar g}a}_g(x_1)\ket{a,b,c;\mu}\to (\delta^a_g(x_1))^{-1}V^{\lsupsc{\bar g}a}_g(x_1)\ket{a,b,c;\mu}.
    \end{equation}

    Under the re-definition in Eq. \eqref{Eq:newVoperators}, two states $\ket{\mu}$ and $\ket{\nu}$ transform as
    \begin{equation*}
        \ket{\mu}\to (\delta^c_g(x_1))^{-1}\ket{\mu};\qquad\ket{\nu}\to (\delta^a_g(x_1))^{-1}(\delta^b_g(x_2))^{-1}\ket{\nu}.
    \end{equation*}
    As such, the $U$-matrices transform as
    \begin{equation}\label{Eq:generalUSAGT}
        [U_g(a,b;c)_{12}]_{\mu\nu}(\{\ket{\psi}\},\{S\},\{V\})\to [U_g(a,b;c)_{12}]_{\mu\nu})(\ket{\psi},\{S\},\{V\})\frac{\delta^a_g(x_1)\delta^b_g(x_2)}{\delta^c_g(x_1)}.
    \end{equation}

    At the same time, under the re-definition in Eq. \eqref{Eq:newVoperators}, the $\omega$ phases in Eq. \eqref{Eq:generalMV} also change correspondingly,
    \begin{equation*}
        (\omega^a_g)_{12}\to (\omega^a_g)_{12}\delta^a_g(x_1)(\delta^a_g(x_2))^{-1}.
    \end{equation*}
    So when Eq. \eqref{eq: simplifying condition 2} holds, $\delta^a_g(x_1)(\delta^a_g(x_2))^{-1}=1$ for all anyons $a$ and all positions $x_1$ and $x_2$. Then the above transformation is simplified to
    \begin{equation}
        [U_g(a,b;c)_{12}]_{\mu\nu}(\{\ket{\psi}\},\{S\},\{V\})\to \frac{\delta^a_g\delta^b_g}{\delta^c_g}[U_g(a,b;c)_{12}]_{\mu\nu}(\{\ket{\psi}\},\{S\},\{V\}),
    \end{equation}
    which is precisely the symmetry action gauge transformation of $U$-symbols in a $G$-crossed braided tensor category, no matter whether the TQSL under discussion has lattice symmetry or not. We have addressed this transformation in Sec. \ref{subsubsec: equivalence internal}.

    \item Changing the local information of the chosen anyon states.

    When the local information around the anyons is changed, similar to the previous discussions on the $F$- and $R$-matrices, the transformation of the $U$-matrices is already captured by their transformations induced by changing the moving, splitting and $V$ operators. Namely, the $U$-matrices transform according to Eqs. \eqref{Eq:generalUVBGT} and \eqref{Eq:generalUSAGT}.
    
    \item Changing the position of defining $U$-matrices.

    We calculate two $U$-matrices at $x_1x_2$ and $x_3x_4$, respectively, and find their relation. It is simple to directly compare them using a method similar to that in the previous discussions. Denote the two $U$-matrices by $[U_g(a,b;c)_{12}]_{\mu\nu}(\{\ket{\psi}\},\{S\},\{V\})$ and $[U_g(a,b;c)_{34}]_{\mu\nu}(\{\ket{\psi}\},\{S\},\{V\})$, respectively. Starting from the upper left state in the following diagram, we can arrive at the bottom right state via two different ways, where one of them involves $U_g(a, b; c)_{12}$ and the other involves $U_g(a, b; c)_{34}$, which leads to a relation between $U_g(a, b; c)_{12}$ and $U_g(a, b; c)_{34}$.
    \begin{equation*}
        \scalebox{0.9}{\compareU}
    \end{equation*}
    In the above figure, each arrow indicates a substitution using Eqs. \eqref{Eq:generalMS} or \eqref{Eq:generalMV}. Summarizing the above processes, we have   
    \begin{equation}\label{Eq:generalUtransformation}
    \bsp
        &[U_{g}(a,b;c)_{12}]_{\mu\nu}=\sum_{\mu'\nu'}K^{q( g)}\left[(\Gamma _{\lsupsc{\bar g}c}^{\lsupsc{\bar g}a\lsupsc{\bar g}b})_{\lsupsc{\bar g}4\lsupsc{\bar g}3\lsupsc{\bar g}2\lsupsc{\bar g}1}\right]_{\mu\mu'} K^{q( g)}\times [U_{g}(a,b;c)_{34}]_{\mu'\nu'}\times\\
        &\left[( \Gamma _{c}^{ab})_{4321}\right]^{-1}_{\nu'
        \nu}\times(\omega_{g}^{c})_{13}(\omega_{g}^{a})_{31} (\omega^{b}_g)_{24}.
    \esp
    \end{equation}
    When the symmetry action $g$ is internal (indicating that $\lsupsc{\bar g}x_i=x_i$) and all the $\omega$ phases are $1$, this transformation is simplified to the vertex basis gauge transformation of the $U$-symbols in a $G$-crossed braided tensor category for TQSLs with an internal symmetry $G$.
\end{itemize}

\subsubsection{The consistency equations of microscopic $U$-matrices}

The $U$-matrices need to obey some consistency equations that involve the $F$- and $R$-matrices. In particular, the consistency equation between the $U$- and $F$-matrices comes from the ``commutativity" between the symmetry transformation and the fusion, and the equation between the $U$- and $R$-matrices comes from the ``commutativity" between the symmetry transformation and the braiding of anyons.

The ``commutativity" between the symmetry transformation and fusion is reflected in the following diagrams. 
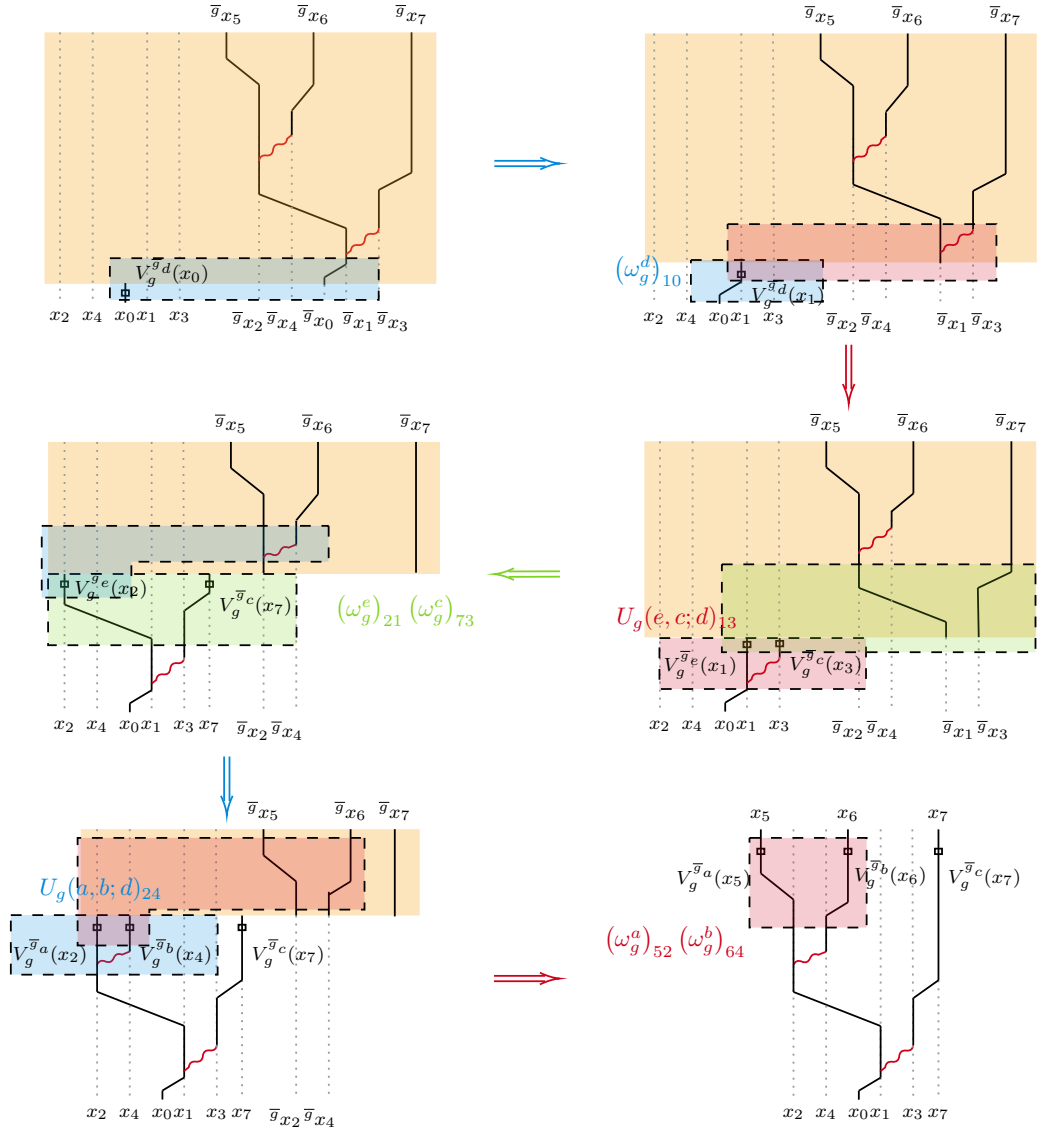
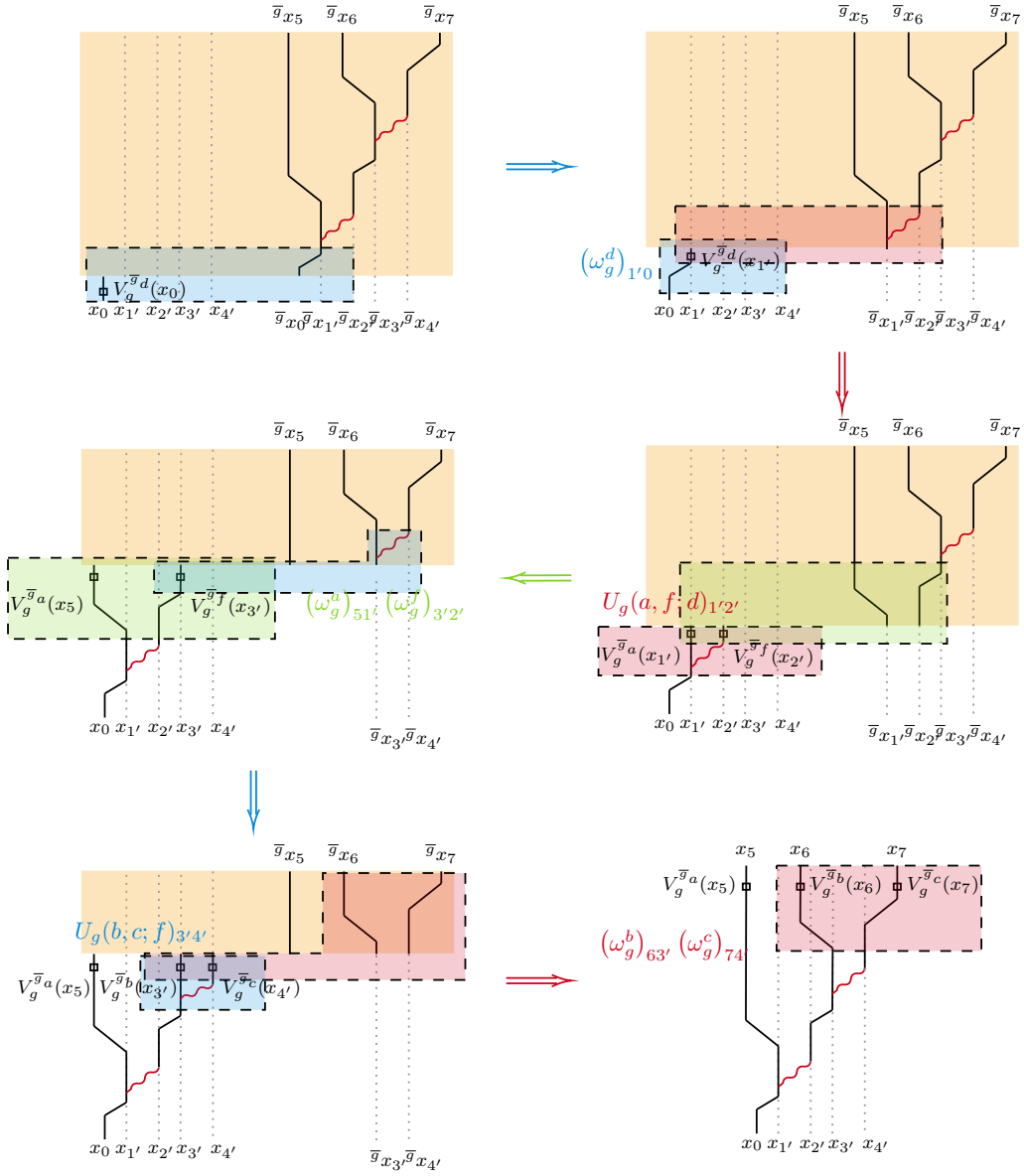
\begin{equation*}
    \scalebox{0.9}{\relationFUa}
\end{equation*}
\begin{equation*}
    \scalebox{0.9}{\relationFUb}
\end{equation*}
In the above two diagrams, each arrow is obtained by either a substitution using \eqref{Eq:generalMS} or Eq. \eqref{Eq:generalMV}, or a substitution using the definition of $U$-matrices. Besides the diagrams shown above, the first and the last figures of each diagram are also related by $F$-matrices.
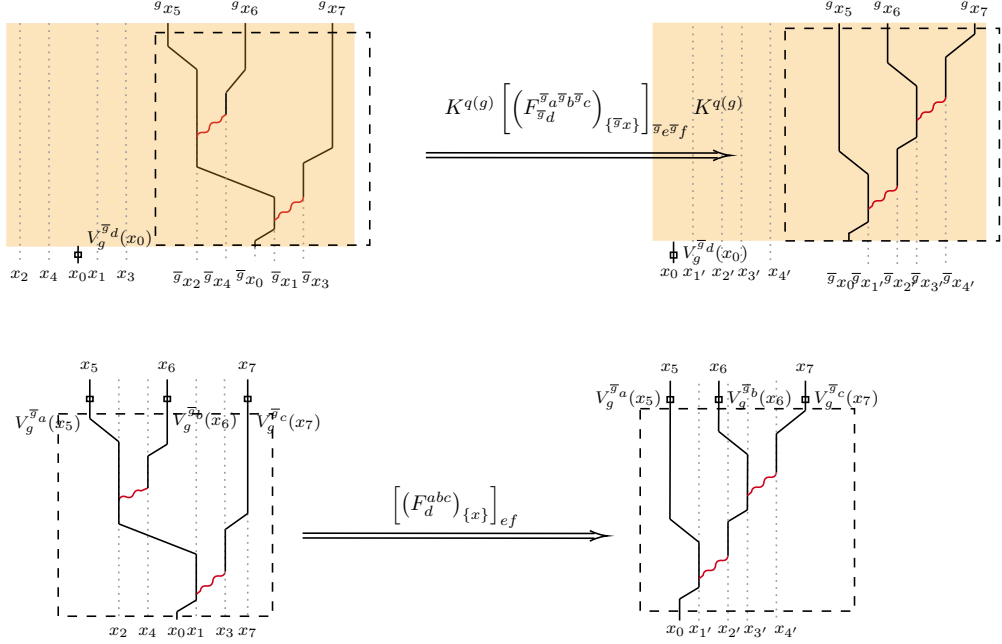
\begin{equation*}
    \scalebox{0.8}{\relationFUc}
\end{equation*}
Summarizing the above processes, the consistency equation between $F$- and $U$-matrices is written as
\begin{equation} \label{eq: general FU equation}
\bsp
    &K^{q( g)}\left[\left( F_{\lsupsc{\bar g}d}^{\lsupsc{\bar g}a\lsupsc{\bar g}b\lsupsc{\bar g}c}\right)_{\lsupsc{\bar g}0\lsupsc{\bar g}1\lsupsc{\bar g}2\lsupsc{\bar g}3\lsupsc{\bar g}4\lsupsc{\bar g}{1'}\lsupsc{\bar g}{2'}\lsupsc{\bar g}{3'}\lsupsc{\bar g}{4'}\lsupsc{\bar g}5\lsupsc{\bar g}6\lsupsc{\bar g}7}\right]_{\{\lsupsc{\bar g}e,\mu ,\nu \} ,\{\lsupsc{\bar g}f,\rho ,\sigma \}} K^{q( g)}=[U_{g}(a,b;e)_{24}]_{\mu\mu'}[U_{g}(e,c;d)_{13}]_{\nu\nu'}\times\\
    &\left[\left( F_{d}^{abc}\right)_{012341'2'3'4'567}\right]_{\{e,\mu ',\nu '\} ,\{f,\rho ',\sigma '\}}[U_g(b,c;f)_{1'2'}]_{\rho'\rho}^{-1}[U_g(a,f;d)_{3'4'}]_{\sigma'\sigma}^{-1}\times\frac{(\omega^d_g)_{10}(\omega^e_g)_{21}(\omega^c_g)_{73}(\omega^a_g)_{52}(\omega^b_g)_{64}}{(\omega^d_g)_{1'0}(\omega^a_g)_{51'}(\omega^f_g)_{3'2'}(\omega^b_g)_{63'}(\omega^c_g)_{74'}}
\esp
\end{equation}

The ``commutativity" between the symmetry transformation and anyon braiding is verified in the following diagram. 
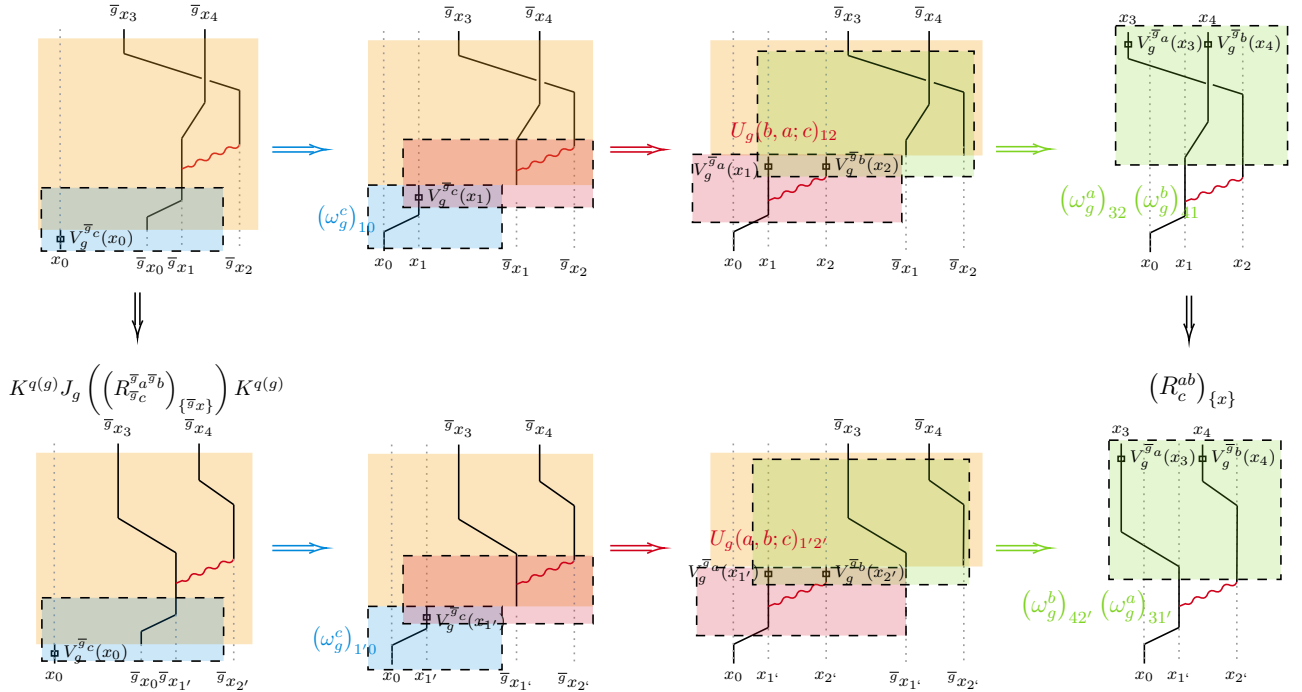
\begin{equation*}
    \scalebox{0.8}{\relationRU}
\end{equation*}
In the above figure, each horizontal arrow indicates either a substitution using \eqref{Eq:generalMS} or Eq. \eqref{Eq:generalMV}, or a substitution using the definition of $U$-matrices. The two figures in the first column differ by a $R$-matrix defined at symmetry-transformed positions $\{\lsupsc{\bar g}x\}$. When the symmetry reverses orientation, we must keep in mind that this $R$-matrix defined at $\{\lsupsc{\bar g}x\}$ is not the originally defined $R$, because the orientation of the moving operators are reversed; instead, this is the $\bar R$-matrix that relates to $(\tilde R^{ba}_c)^{-1}$ via Eq. \eqref{Eq:generalRtocategoryRinv}. We introduce a notation $J_g$ to unify these two cases: when $g$ preserves spatial orientation, $J_g(R^{ab}_c)$ gives $R^{ab}_{c}$; when $g$ reverses spatial orientation, $J_g(R^{ab}_c)$ gives $\bar R^{ab}_{c}$. In particular, for categorical $\tilde R$ symbols, $J_g(\tilde R^{ab}_c)=\tilde R^{ab}_c$ for orientation preserving $g$ and $J_g(\tilde R^{ab}_c)=(\tilde R^{ba}_c)^{-1}$ for orientation reversing $g$. Summarizing the above process, the consistency equation between $R$- and $U$-matrices is written as
\begin{equation} \label{eq: general RU equation}
\bsp
    &K^{q( g)} J_{g}\left[\left[ \left(R_{\lsupsc{\bar g}c}^{\lsupsc{\bar g}a\lsupsc{\bar g}b}\right)_{0{\lsupsc{\bar g}1} {\lsupsc{\bar g}2} {\lsupsc{\bar g}1'} {\lsupsc{\bar g}2'} \lsupsc{\bar g}3\lsupsc{\bar g}4}\right]_{\mu \nu }\right] K^{q( g)}=\\
    &[U_{g}(b,a;c)_{12}]_{\mu\mu'}\left[\left(R_{c}^{ab}\right)_{0121'2'34}\right]_{\mu' \nu' }[U_{g}(a,b;c)_{1'2'}]^{-1}_{\nu'\nu}\frac{(\omega^c_{g})_{10}(\omega^a_{g})_{32}(\omega^b_{g})_{41}}{(\omega^c_{g})_{1'0}(\omega^a_{g})_{31'}(\omega^b_{g})_{42'}}.
\esp
\end{equation}

When $F$- and $R$-matrices are simplified to the categorical $F$, and all $\omega$-phases are $1$, the above two equations simplify to the consistency equations we introduced in the main text.

\subsection{Symmetry fractionalization: $\eta$-phases} \label{subapp: eta}

The last piece of universal data is the symmetry fractionalization of anyons. Physically, non-trivial symmetry fractionalization of an anyon means that an anyon fails to feel the global symmetry acting linearly. This deviation from linearity is characterized by a phase factor $\eta$, which is precisely the $\eta$ phase we defined in the main text. We repeat the definition here.
\begin{equation*}
    \eta_a(g, h)_1(\{\ket{\psi}\},\{V\})=\langle 1|2\rangle,
\end{equation*}
where
\begin{equation*}
    \bsp
&|1\rangle=V_{gh}^{^{\overline{gh}}a}(x_1)|a_1, \bar a\rangle,\\
&|2\rangle=R_gV_h^{^{\overline{gh}}a}(^{\bar g}x_1)R_g^{-1}V_g^{^{\bar g}a}(x_1)|a_1, \bar a\rangle.
\esp
\end{equation*}

The subscript $1$ in $\eta_a(g, h)_1$ is the position label, as before. We have proved in the main text that the two states $\ket{1}$ and $\ket{2}$ differ by merely a phase factor. Below we discuss the properties of the $\eta$-phases.

\subsubsection{The microscopic transformation of the $\eta$-phases}

Here, we discuss how $\eta$-phases change when the setup and relevant operators change. Since the definition of $\eta$-phases does not involve moving or splitting operators, we only need to discuss how $\eta$-phases when changing the local information of anyon states, changing the $V$ operators, and changing the position.

\begin{itemize}
    \item Changing the $V$ operators.

    When the $V$ operators transform as in Eq. \eqref{Eq:newVoperators}, the two states transform as
    \begin{equation*}
    \bsp
        &\ket{1}\to (\delta^a_{gh}(x_1))^{-1}\ket{1},\\
        &\ket{2}\to \left(K^{q(g)}\delta^{\lsupsc{\bar g}a}_{h}(\lsupsc{\bar g}x_1)K^{q(g)}\delta^a_{g}(x_1)\right)^{-1}\ket{2}.
    \esp
    \end{equation*}
    Consequently, the $\eta$-phases change as
    \begin{equation}\label{Eq:generaletaSAGT}
        \eta_a(g,h))_1(\{\ket{\psi}\},\{V\})\to\frac{\delta^a_{gh}(x_1)}{K^{q(g)}\delta^{\lsupsc{\bar g}a}_{h}(\lsupsc{\bar g}x_1)K^{q(g)}\delta^a_{g}(x_1)}\eta_a(g,h)_1(\{\ket{\psi}\},\{V\}).
    \end{equation}
    When the symmetry is internal, \ie $\lsupsc{\bar g}x=x$, the above transformation is simplified to the symmetry action gauge transformation of $\eta$-symbols in the $G$-crossed braided tensor category.

    \item Changing the local information of anyon states.

    When the local information around the anyons is changed, similar to the previous discussions on the $U$-matrices, the transformation of the $\eta$-phases is already captured by their transformations induced by changing the $V$ operators. Namely, the $\eta$-phases transform according to Eq. \eqref{Eq:generaletaSAGT}.
    
    \item Changing the position of defining the $\eta$-phases.

    We find two $\eta_{a}(g,h)$ symbols defined at two different positions $x_1$ and $x_2$. To find the relation between them, we apply the following process.
    \begin{equation*}
        \scalebox{0.85}{\compareeta}
    \end{equation*}
    Each arrow in the above figure indicates a substitution using \eqref{Eq:unseamless} or \eqref{Eq:generalMV}, or a substitution using the definition of $\eta$. The above process is summarized as
    \begin{equation}\label{Eq:generaletatransformation}
        \eta_{a}(g,h)_1=\frac{K^{q( g)} \gamma _{\lsupsc{\bar g}1\lsupsc{\bar g}2\lsupsc{\bar g}1}^{\lsupsc{\bar g}a} K^{q( g)}}{\gamma _{121}^{a}}\times\frac{\left( \omega _{g}^{a}\right)_{21}\left( \omega _{gh}^{a}\right)_{12}}{K^{q( g)}\left( \omega _{h}^{\lsupsc{\bar g}a}\right)_{\lsupsc{\bar g}1\lsupsc{\bar g}2}K^{q( g)}}\eta _{a}( g,h)_{2}.
    \end{equation}
    When the symmetry is internal and all $\gamma$ and $\omega$ phases are $1$, the above transformation is simplified to $\eta_{a}(g,h)_1=\eta_{a}(g,h)_2$, which means $\eta$ symbols are equal everywhere.     
\end{itemize}

\subsubsection{The consistency equation of $\eta$-phases} 

The $\eta$-phases also yield two consistency equations. The first consistency equation comes from the group multiplication of symmetry actions. In particular, successively applying $R_g$ and $R_h$ must be equivalent to applying $R_{gh}$ (up to an irrelevant global phase factor, which is assumed to be $1$ here). This consistency relation is derived as follows. Again we have made implicit the fact that the symmetry may transform the anyon positions. Keep noted that the operators in the shaded region act on the transformed positions.
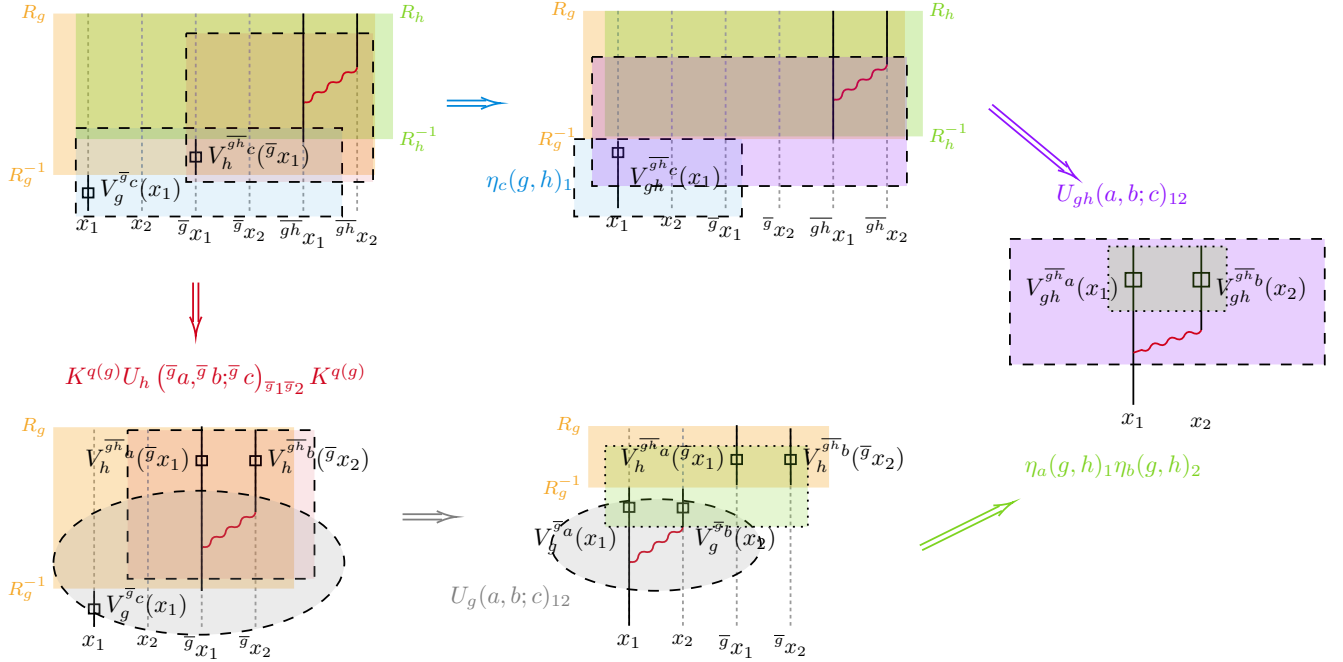
\begin{equation*}
    \scalebox{0.9}{\relationUeta}
\end{equation*}
There are two different ways to verify the diagram in the upper-left corner. In the first step, we can either extract the $\eta_c$-phase defined at $x_1$ or extract the $U_h$-matrix defined at $\lsupsc{\bar g}x_1\lsupsc{\bar g}x_2$. If we first extract the $\eta_c$-phase, it then follows to extract a $U_{gh}$-matrix defined at $x_1x_2$; if we first extract the $U_h$-matrix, we can then extract a $U_g$-matrix defined at $x_1x_2$ and two $\eta$-phases defined at $x_1$ and $x_2$ respectively. Summarizing the above processes, we have
\begin{equation}\label{Eq:relationUeta}
\bsp
    &\sum_\nu K^{q}( g) [U_{h}\left(\lsupsc{\bar g} a,\lsupsc{\bar g} b;\lsupsc{\bar g} c\right)_{\lsupsc{\bar g} 1\lsupsc{\bar g} 2}]_{\mu\nu} K^{q}( g)[U_{g}( a,b;c)_{12}]_{\nu\rho}=\frac{\eta _{c}( g,h)_{1}}{\eta _{a}( g,h)_{1} \eta _{b}( g,h)_{2}}[U_{gh}( a,b;c)_{12}]_{\mu\rho}.
\esp
\end{equation}
Equation \eqref{Eq:relationUeta} is the consistency equation between $U$-matrices and $\eta$-phases. When all phases $\gamma=\omega=1$ (no matter what $\Gamma$ is), the consistency equation is simplified to Eq. \eqref{Eq:internalUeta}. Apart from this equation, there is one more equation that $\eta$-phases themselves should satisfy. This second equation comes from the `associativity' of symmetry actions onto a single anyon.
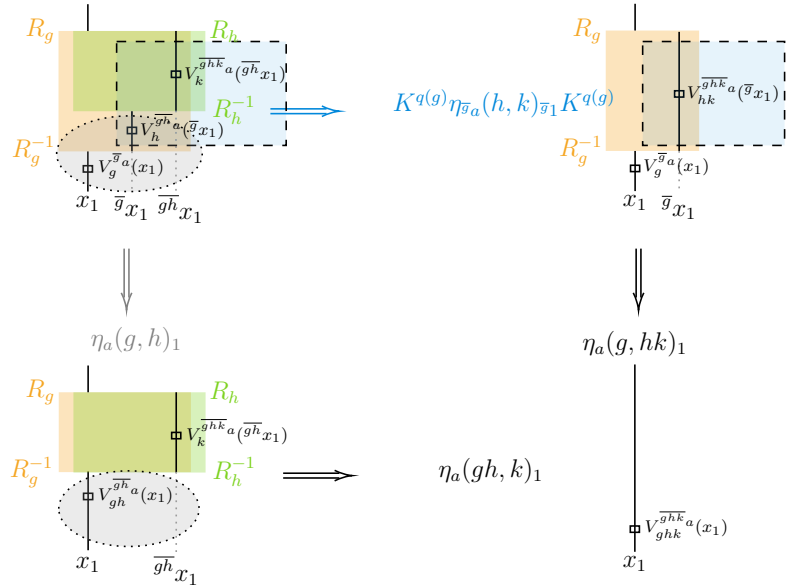
\begin{equation*}
    \scalebox{0.8}{\relationetaeta}
\end{equation*}
The above figure shows two different ways that both start from the upper-left figure and end in the bottom-right figure, and the two ways must be equal. The equivalence results in the consistency of $\eta$-phases,
\begin{equation} \label{eq: general eta eta equation}
    K^{q( g)} \eta _{^{\overline{g}} a}( h,k)_{^{\overline{g}} 1} K^{q( g)}\eta _{a}( g,h)_{1}=\eta _{a}( g,hk)_{1}\eta _{a}( gh,k)_{1}.
\end{equation}
When all phases $\gamma=\omega=1$ (no matter what $\Gamma$ is), this equation is precisely Eq. \eqref{Eq:internaletaeta} in the main text.

\subsection{The graded structure of the microscopic universal data} \label{subapp: graded structure}

Having stated the general version of the universal data, we now discuss the structure of the general data. As we have seen, the general data is quite complicated and has explicit dependence on various subjective choices. In fact, this universal data admits a graded structure parameterized by $\gamma$, $\Gamma$, and $\omega$. Denote the full collection of the universal data as $\mathcal{C}$, which includes all possible data defined at various positions to put anyons, for different anyon states, and for different moving and splitting operators and $V$ operators. Then
\begin{equation}
    \mathcal{C}=\bigoplus_{\gamma,\Gamma,\omega}\mathcal{C}_{\gamma,\Gamma,\omega},
\end{equation}
where each $\mathcal{C}_{\gamma,\Gamma,\omega}$ is the collection of data defined with fixed $\gamma$, $\Gamma$, and $\omega$ that are introduced in Eqs. \eqref{eq: gamma main}, \eqref{eq: Gamma main} and \eqref{eq: omega main}. 

Notice that for fixed $\gamma$, $\Gamma$, and $\omega$, we still have some residual freedom to change the moving and splitting operators and the $V$ operators. In particular, if we change the operators as
\begin{equation}\label{Eq:changeoperators}
\bsp
    &M^a_{21}\ket{a_1,\bar a_{0'}}\to e^{i\phi^a_{21}}M^a_{21}\ket{a_1,\bar a_0},\\
    &(S^{ab}_{c;\mu})_{21}\ket{c_1,\bar c_{0'}}\to \sum_{\mu'}[(\Omega_0(a,b;c)_{21}]_{\mu\mu'}(S^{ab}_{c;\mu})_{21}\ket{c_1,\bar c_{0'}},\\
    &V^a_g(x_1)\ket{\lsupsc{g}a_1,\lsupsc{g}\bar a_{0'}}\to e^{i\varphi^a_g(x_1)}V^a_g(x_1)\ket{\lsupsc{g}a_1,\lsupsc{g}\bar a_{0'}},
\esp
\end{equation}
with the condition
\begin{equation}\label{Eq:VBGTcondition}
\bsp
    &e^{i\phi^a_{21}}e^{i\phi^a_{32}}=e^{i\phi^a_{31}},\quad e^{i\phi^a_{21}}=e^{-i\phi^a_{12}};\\
    &[\Omega_0(a,b;c)_{21}]_{\mu\mu'}=[\Omega_0(a,b;c)_{2'1'}]_{\mu\mu'}\frac{e^{i\phi^a_{11'}}e^{i\phi^b_{22'}}}{e^{i\phi^c_{11'}}};\\
    &e^{i\varphi^a_g(x_1)}e^{i\phi^a_{12}}=K^{q(g)}e^{i\phi^{\lsupsc{\bar g}a}_{\lsupsc{\bar g}1\lsupsc{\bar g}2}}K^{q(g)}e^{i\varphi^a_g(x_2)},
\esp
\end{equation}
one can verify that $\gamma$, $\Gamma$, and $\omega$ are invariant, while the universal data undergo a transformation defined in Eqs. \eqref{Eq:generalFVBGT}, \eqref{Eq:generalRVBGT}, \eqref{Eq:generalUVBGT}, \eqref{Eq:generalUSAGT}, and \eqref{Eq:generaletaSAGT}. We repeat the transformations here and use the condition Eq. \eqref{Eq:VBGTcondition} to simplify them. For $F$- and $R$-matrices, the transformations are
\begin{equation}\label{Eq:generalFRequivalence}
    \bsp
    &[(F^{abc}_d)_{\{x\}}]_{\{e,\mu,\nu\},\{f,\rho,\sigma\}}(\{\ket{\psi}\},\{M\},\{S\})\to\sum_{\mu'\nu'\rho'\sigma'}\left(e^{i\phi^b_{12}}[\Omega_0(a,b;e)_{21}]_{\mu\mu'}\right)\left(e^{i\phi^c_{12}}[\Omega_0(e,c;d)_{21}]_{\nu\nu'}\right)\\
    &[(F^{abc}_d)_{\{x\}}]_{\{e,\mu,\nu'\},\{f,\rho',\sigma'\}}(\{\ket{\psi}\},\{M\},\{S\})\left(e^{i\phi^c_{12}}[\Omega_0(b,c;f)_{21}]_{\rho'\rho}\right)^{-1}\left(e^{i\phi^f_{12}}[\Omega_0(a,f;d)_{21}]_{\sigma'\sigma}\right)^{-1},\\
    &[(R^{ab}_c)_{\{x\}}]_{\mu\nu}(\{\ket{\psi}\},\{M\},\{S\})\to \sum_{\mu'\nu'}\left(e^{i\phi^a_{12}}[\Omega_0(b,a;c)_{21}]_{\mu\mu'}\right)[(R^{ab}_c)_{\{x\}}]_{\mu\nu}(\{\ket{\psi}\},\{M\},\{S\})\left(e^{i\phi^b_{12}}[\Omega_0(a,b;c)_{21}]_{\nu'\nu}\right)^{-1}.
    \esp
\end{equation}
These transformations are in the same form as the vertex basis gauge transformation of categorical $\tilde F$ and $\tilde R$ symbols, although the definition of $F$- and $R$- matrices are general. For $U$-matrices, the transformation is
\begin{equation}\label{Eq:generalUequivalence}
\bsp
    &[U_g(a,b;c)_{12}]_{\mu\nu}(\{\ket{\psi}\},\{M\},\{S\},\{V\})\to \sum_{\mu'\nu'}K^{q(g)}\left(e^{i\phi^{\lsupsc{\bar g}b}_{12}}[\Omega_0(\lsupsc{\bar g}a,\lsupsc{\bar g}b;\lsupsc{\bar g}c)_{21}]_{\mu\mu'}\right)K^{q(g)}\times\\
    &[U_g(a,b;c)_{12}]_{\mu'\nu'}(\{\ket{\psi}\},\{M\},\{S\},\{V\})\left(e^{i\phi^b_{12}}[\Omega_0(a,b;c)_{21}]_{\nu'\nu}\right)^{-1}\times
    K^{q(g)}\frac{e^{i\phi^{\lsupsc{\bar g}a}_{\lsupsc{\bar g}11}}e^{i\phi^{\lsupsc{\bar g}b}_{\lsupsc{\bar g}11}}}{e^{i\phi^{\lsupsc{\bar g}c}_{\lsupsc{\bar g}11}}}K^{q(g)}\frac{e^{i\varphi^c_g(x_1)}}{e^{i\varphi^a_g(x_1)}e^{i\varphi^b_g(x_1)}},
\esp
\end{equation}
which is a hybrid of the vertex basis gauge transformation and the symmetry action gauge transformation (as defined in Sec. \ref{subsubsec: equivalence internal}). Finally, the $\eta$-phases transform as
\begin{equation}\label{Eq:generaletaequivalence}
\bsp
    \eta_a(g,h)_1(\{\ket{\psi}\},\{V\})&\to\frac{K^{q(g)}e^{i\varphi_{h}^{\lsupsc{\bar g}a}(\lsupsc{\bar g}x_1)}K^{q(g)}e^{i\varphi_{g}^a(x_1)}}{e^{i\varphi_{gh}^a(x_1)}}\eta_a(g,h)_1(\{\ket{\psi}\},\{V\})\\
    &=\frac{K^{q(g)}e^{i\varphi_{h}^{\lsupsc{\bar g}a}(x_1)}K^{q(g)}e^{i\varphi_{g}^a(x_1)}}{e^{i\varphi_{gh}^a(x_1)}}\frac{K^{q(gh)}e^{i\phi^{\lsupsc{\overline{gh}}a}_{\lsupsc{\overline{gh}}11}}K^{q(gh)}}{K^{q(g)}e^{i\phi^{\lsupsc{\bar g}a}_{\lsupsc{\overline{g}}11}}K^{q(h)}e^{i\phi^{\lsupsc{\overline{gh}}a}_{\lsupsc{\overline{h}}11}}K^{q(gh)}}\eta_a(g,h)(\{\ket{\psi}\},\{V\}),
\esp
\end{equation}
where we have used Eqs. \eqref{Eq:VBGTcondition} to eliminate $e^{i\varphi^a_h(\lsupsc{\bar g}x_1)}$.

The transformations Eq. \eqref{Eq:generalFRequivalence}-\eqref{Eq:generaletaequivalence} define the equivalence relation within a certain $\mathcal{C}_{\gamma,\Gamma,\omega}$, and the full universal data is therefore graded by $\gamma$, $\Gamma$, and $\omega$.

We note  that the set of data $\mathcal{C}_{1, \mathbbm{1}, 1}$ is not special from a microscopic perspective. However, it is the simplest among all $\mathcal{C}_{\gamma, \Gamma, \omega}$'s. In particular, the data in $\mathcal{C}_{1, \mathbbm{1}, 1}$ is the closest to the topological quantum field theory. For example, the $F$- and $R$-matrices in $\mathcal{C}_{1, \mathbbm{1}, 1}$ can be viewed as the $F$- and $R$-symbols in a unitary modular tensor category, as one can easily check, and their equivalence relations are precisely the vertex basis gauge transformations in such a category; the $U$-matrices and $\eta$-phases in $\mathcal{C}_{1, \mathbbm{1}, 1}$, when restricted to the internal symmetries, are precisely the $U$- and $\eta$-symbols in a $G$-crossed braided tensor category, and their equivalence relations match those in category theory correspondingly. Also, the condition $\gamma=1$ and $\Gamma=\mathbbm{1}$ correspond to the fact that the anyon worldlines and vertices can move freely in topological quantum field theory, and the condition $\omega=1$ corresponds to that the group representation matrices can be smoothly translated via the Wilson lines. Therefore, $\mathcal{C}_{1, \mathbbm{1}, 1}$ is special in the sense that it is potentially the data that has the nicest mathematical description. Another reason for $\mathcal{C}_{1, \mathbbm{1}, 1}$ to be special is that the consistency equations of the data in $\mathcal{C}_{1, \mathbbm{1}, 1}$ are position-independent. By inspecting these consistency equations, we are able to write down an exact crystalline equivalence principle.

\section{The crystalline equivalence principle}\label{appd:CEP}

In this appendix, we explain the details of our precise crystalline equivalence principle discussed in Sec. \ref{sec:equivalence}, which is an explicit bijective map between the universal data characterizing SET phases with a general symmetry $G$ and SET phases with a purely internal symmetry $G$, given by Eq. \eqref{Eq:correspondence}. According to Sec. \ref{sec:framework} and Appendix \ref{app: full structure}, the universal properties of an SET phase can be characterized by $\mc{C}_{1, \mathbbm{1}, 1}$, which consists of the anyon types, fusion rules in Eq. \eqref{eq: fusion}, $F$-symbols in Eq. \eqref{Eq:generalF main} and $R$-symbols in Eq. \eqref{eq: general R main}, the transformation of the anyon types $\rho_g(a)$ in Eq. \eqref{eq: anyon permutation}, the $U$-symbols in Eq. \eqref{eq: U-matrix} and the $\eta$-symbols in Eq. \eqref{eq: eta}. We will achieve the following goals in this appendix. First, in Appendix \ref{subapp: details of CEP} we will give more details of the map in Eq. \eqref{Eq:correspondence}, including the meaning of $A$-matrices there and the inverse of the map Eq. \eqref{Eq:correspondence}. Second, in Appendix \ref{subapp: checking consistency equations} we will apply the map Eq. \eqref{Eq:correspondence} to the data $\mathcal{C}_{1, \mathbbm{1}, 1}$ for an SET phase with a general symmetry $G$, which satisfies the consistency equations given by Eqs. \eqref{Eq:latticeFU}, \eqref{Eq:latticeRU}, \eqref{Eq:latticeUeta} and \eqref{Eq:latticeetaeta}, and show that the resulting data satisfies the consistency equations of the data $\mathcal{C}'_{1, \mathbbm{1}, 1}$ for an SET phase with an internal symmetry $G$, given by Eqs. \eqref{eq: internal F Uprime}, \eqref{eq: internal R Uprime}, \eqref{eq: internal Uprime etaprime} and \eqref{eq: internal etaprime etaprime}. Third, in Appendix \ref{appd:VBGTaftercorrespondence} we will see that if a ``gauge transformation" given by Eqs. \eqref{eq: vertex basis gauge transformation} and \eqref{eq: general U eta transformation} is applied to a set of data $\mathcal{C}_{1, \mathbbm{1}, 1}$ for an SET phase with a general symmetry $G$, the mapped data $\mathcal{C}'_{1, \mathbbm{1}, 1}$ for an SET phase with an internal symmetry $G$ receives a ``gauge transformation" given by Eqs. \eqref{Eq:gaugetransformationUp} and \eqref{Eq:gaugetransformationetap}. Finally, in Appendix \ref{subapp: not RT} we give an intuitive explanation for why the emergent CRT symmetry should be related to the crystalline equivalence principle, instead of the emergent RT symmetry.

Note that we have already shown that the data in $\mc{C}_{1, \mathbbm{1}, 1}$ can be organized as some generalization of a category, so in this appendix we will employ many tools in category theory.

\subsection{Details of the map Eq. \eqref{Eq:correspondence}} \label{subapp: details of CEP}

We begin with explaining more details of the map in Eq. \eqref{Eq:correspondence}. For the readers' convenience, Eq. \eqref{Eq:correspondence} is copied here:

\begin{equation}\label{Eq:correspondence app}
	\bsp
		&\rho'_g(a)=\left\{ \begin{array}{ll}
			\rho_g(a)=^ga, &\textrm{if $g$ preserves spatial orientation,}\\
			\rho_g(\bar a)=^g\bar{a}, &\textrm{if $g$ reverses spatial orientation.}
		\end{array}\right.\\
		&U'_{g}(a,b;c)= \left\{ \begin{array}{ll}
			U_g(a,b;c),&\textrm{if $g$ preserves spatial orientation,}\\
			K^{q(g)}(A^{\lsupsc{\overline g}b\lsupsc{\overline g}a}_{\lsupsc{\overline g}c}R^{\lsupsc{\overline g}a\lsupsc{\overline g}b}_{\lsupsc{\overline g}c})K^{q(g)}U_g(a,b;c),&\textrm{if $g$ reverses spatial orientation.}
		\end{array}\right.\\
		& \eta'_a(g_1,g_2)=\left\{\begin{array}{ll}
			\eta_a(g_1,g_2),&\textrm{if $g_1$ preserves spatial orientation,}\\
			\eta_a(g_1,g_2)K^{q(g_1)}U_{g_2}(\lsupsc{\overline{g_1}} a,\lsupsc{\overline{g_1}}\bar a;0)K^{q(g_1)},&\textrm{if $g_1$ reverses spatial orientation but $g_2$ preserves it,}\\
			\eta_a(g_1,g_2)K^{q(g_1)}U_{g_2}(\lsupsc{\overline{g_1}}\bar a,\lsupsc{\overline{g_1}}a;0)\theta_{\lsupsc{\overline{g_1}}a}\varkappa^{*}_{\lsupsc{\overline{g_1}}a}K^{q(g_1)},&\textrm{if both $g_1$ and $g_2$ reverse spatial orientation.}
		\end{array}\right.
	\esp
\end{equation}
Here the phase factor $\varkappa_a$ is the Frobenius-Schur indicator $\varkappa_a=d_a[F^{a\bar aa}_a]_{00}$, where $d_a$ is called the quantum dimension of the anyon $a$ (see, for example, equation (8.10) in Ref. \cite{Simonbook2023} for its definition). Below we explain the meaning of the unitary matrix $A$ in Appendix \ref{subsubapp: CRT}, which is related to the emergent CRT symmetry of the underlying topological quantum field theory that describes the TQSL. Then we describe the inverse of the map Eq. \eqref{Eq:correspondence} in Appendix \ref{subsubapp: inverse CEP}.

\subsubsection{The CRT action} \label{subsubapp: CRT}

In the categorical description of anyons (see Refs. \cite{Kitaev2006} and \cite{Simonbook2023} for more introduction), the CRT symmetry naturally induces an {\it anti-unitary} map $\mathcal{A}$ between the vertex spaces
\begin{equation*}
    \mathcal{A}: \quad V^{ab}_{c}\rightarrow V^{\bar b\bar a}_{\bar c}.
\end{equation*}
Here, the space $V^{ab}_{c}$ is the vertex space of an anyon $c$ splitting into $a$ and $b$, which can be viewed as a vector space consisting of the morphisms whose source object is $c$ and target object is $a\otimes b$, \ie $V^{ab}_{c}=$Hom$(c,a\otimes b)$; we will interchangeably call them ``morphisms" or ``vectors". Also, we use the following graphic representation to denote the basis of $V^{ab}_{c}$, \footnote{From now on, we will use diagrammatic representations of the anyons to illustrate and determine the properties of the $A$ map. The reader is recommended to refer to Ref. \cite{Simonbook2023} and appendix E of Ref. \cite{Kitaev2006} for details of the diagrammatic calculations. The primary rule of the calculation is that the anyon worldlines can freely move without bending them or changing their order.}
\begin{equation*}
\ket{\mu}=\bmm\scalebox{0.7}{\anyontree{a}{b}{c}{\mu}}\emm.
\end{equation*}
As we will describe in Eq. \eqref{Eq:defineA}, the unitary matrix $A$ is the action of $\mc{A}$ in a basis (note that $\mc{A}$ can be viewed as a composition of a unitary map and a complex conjugation operation).

The map $\mathcal{A}$ is constructed as follows. For a vector $\ket{\mu}=V^{ab}_c$, we first find its Hermitian conjugate $\bra{\mu}$, which is a vector in $V^c_{ab}$,
\begin{equation*}
   H.c.:\quad \bmm\scalebox{0.7}{\anyontree{a}{b}{c}{\mu}}\emm\rightarrow\bmm\scalebox{0.7}{\antianyontree{a}{b}{c}{\mu^\dagger}}\emm.
\end{equation*}
Then we choose three morphisms $\ket{\eta_a}\in V^{\bar a a}_{0}$, $\ket{\eta_b}\in V^{\bar b b}_{0}$, and $\bra{\eta_{\bar c}}\in V^{0}_{c\bar c}$. These three morphisms can be arbitrary but are fixed for $\mathcal{A}$. Composing the above three morphisms with $\bra{\mu^\dagger}$ in the following way gives a morphism $\mathcal{A}(\ket{\mu})\in V^{\bar b\bar a}_{\bar c}$, \footnote{The conjugated version of this construction is mentioned in Appendix E of Ref. \cite{Kitaev2006} and was denoted as $U_l$ in equation (200) there.} and we can expand $\mathcal{A}(\ket{\mu})$ in terms of the basis vectors of $V^{\bar b\bar a}_{\bar c}$.
\begin{equation}\label{Eq:defineA}
\bmm\scalebox{0.7}{\chimap}\emm\equiv \varkappa_{\bar c}\sum_\nu[A^{\bar b\bar a}_{\bar c}]_{\mu\nu}\bmm\scalebox{0.7}{\anyontree{\bar b}{\bar a}{\bar c}{\nu}}\emm.
\end{equation}
This equation defines the matrix $[A^{\bar b\bar a}_{\bar c}]_{\mu\nu}$, which encodes the expansion coefficients of the output morphism $\mathcal{A}(\ket{\mu})$ in terms of the basis of $V^{\bar b\bar a}_{\bar c}$, which depends on the bases of $V^{ab}_c$, $V^{\bar b\bar a}_{\bar c}$, $V^{\bar a a}_{0}$, $V^{\bar bb}_{0}$, and $V^0_{c\bar c}$. The factor $\varkappa_{a}$ is a phase defined as
\begin{equation}
    \bmm\scalebox{0.9}{\FrobeniusSchur}\emm.
\end{equation}
The factor $\varkappa_{a}$ can be derived from the $F$-moves $\varkappa_a=d_aF^{a\bar aa}_d$ (see, for example, equation (14.2) in Ref. \cite{Simonbook2023}) and satisfies $\varkappa_{a}=\varkappa_{\bar a}^*$ (see equation (194) in Ref. \cite{Kitaev2006}). For specific anyons $\bar a_0=a_0$, $\varkappa_{a_0}$ is gauge invariant and coincides with the Frobenius-Schur indicator. We take out this factor in the definition Eq. \eqref{Eq:defineA} to ensure the $A$-matrices to have some nice normalization properties, as we will see below. We can see that the map $\mathcal{A}$ is anti-unitary, because the input vector $\ket{\mu}$ undergoes a Hermitian conjugate in the construction, and multiplying a constant to the vector $c_1\ket{\mu_1}$ gives $c_1^*\mathcal{A}(\ket{\mu_1})$.

In the rest of Appendix \ref{subsubapp: CRT}, we discuss some properties of the matrices $A$ that will be used later in Appendix \ref{appd:CEP}.

\begin{theorem} \label{thm: CRT}
    The unitary matrices $A^{ab}_c$ satisfy the following equations.
    \begin{align}
    &(A^{\bar b\bar a}_{\bar c})^* A^{ab}_c=\frac{\varkappa_a\varkappa_b}{\varkappa_c},\label{Eq:CEPAAcondition}\\
    &A^{\bar b \bar a}_{\bar e}A^{\bar c \bar e}_{\bar d}(F^{\bar c\bar b\bar a}_{\bar d})_{\bar e\bar f}^*=(F^{abc}_{d})_{ef}^*A^{\bar c\bar b}_{\bar f}A^{\bar f\bar a}_{\bar d},\label{Eq:CEPAFcondition}\\
    &(R^{\bar a\bar b}_{\bar c})^{T}=A^{ba}_c R^{ab}_c (A^{ab}_c)^{-1},\label{Eq:CEPARcondition}\\
    &K^{q(g)}A^{\bar b\bar a}_{\bar c}K^{q(g)}U_g(\lsupsc{g} \bar b,\lsupsc{g} \bar a;\lsupsc{g} \bar c)=\frac{U_g(\lsupsc{g} \bar a,\lsupsc{g} a;0)U_g(\lsupsc{g} \bar b,\lsupsc{g} b;0)}{U_g(\lsupsc{g} \bar c,\lsupsc{g} c;0)}U_g^*(\lsupsc{g} a,\lsupsc{g} b;\lsupsc{g} c)A^{\lsupsc{g} \bar b\lsupsc{g} \bar a}_{\lsupsc{g} \bar c}.\label{Eq:CEPAUcondition}
    \end{align}
    Here the $F$- and $R$-matrices are the $F$- and $R$-symbols in the unitary modular tensor category, and the $U_g$-matrices represent the action of certain automorphism $\Phi_g$ of the anyon types. $\Phi_g$ is sometimes called the topological symmetry of the TQSL \cite{barkeshli2014}, and it may microscopiclly originate from an internal or lattice symmetry, \ie $U_g$ here can be viewed as $U_g$ in the data $\mc{C}_{1, \mathbbm{1}, 1}$.
\end{theorem}

\begin{proof}
    \begin{enumerate}
        \item Proof of Eq. \eqref{Eq:CEPAAcondition}.
        
        Applying two $\mc{A}$ maps to a vector $\ket{\mu}$ should yield the following condition.
        \begin{equation*}
        \scalebox{0.8}{\CEPAAcondition}
        \end{equation*}
        The figure on the bottom-left corner differs from the upper-left figure by merely bending the anyon lines, leading to a phase factor $\varkappa_a\varkappa_b\varkappa_c$ (see, \eg chapter 14 of Ref. \cite{Simonbook2023}). Then, using our definition of $A$-matrices Eq. \eqref{Eq:defineA}, we find the resultant symbolic expression of this condition is Eq. \eqref{Eq:CEPAAcondition}.

        \item Proof of Eq. \eqref{Eq:CEPAFcondition}.
        
        We consider the following morphism,
        \begin{equation*}
            \bmm\scalebox{0.7}{\CEPFB}\emm,
        \end{equation*}
        which is constructed by picking up a morphism in $\oplus_eV^{ab}_e\otimes V^{ec}_d$ and then composing it with four morphisms $\ket{\eta_a}\in V^{\bar aa}_0$, $\ket{\eta_b}\in V^{\bar bb}_0$, $\ket{\eta_c}\in V^{\bar cc}_0$, and $\bra{\eta_{\bar d}}\in V^0_{d\bar d}$. These four morphisms are those that we have fixed in defining the map $\mathcal{A}$. Now we notice that we can either first perform a basis transformation using $F$-symbols and then simplify the diagram using $A$-matrices, or first simplify the diagram using $A$-matrices and then perform a basis transformation using $F$-symbols. These two processes are shown in the following diagram.
        \begin{equation*}
        \scalebox{0.8}{\CEPAFcondition}
        \end{equation*}            
        The above two processes should be equal because we are merely performing basis transformations or re-phrasing the vector, and hence all the diagrams in the above represent the same vector. Again, in the above figure and in the following, we use the double-line arrow ``$\Rightarrow$" to indicate that the diagram on the tail side of the arrow, excluding the factors, is equal to the diagram on the tip side multiplying the factors on the tip side or around the arrow. One multiplies together the factors appearing in the diagram and finds the correct relation or equation. Expressing the equation in symbolic form, we get Eq. \eqref{Eq:CEPAFcondition}.

        \item Proof of Eq. \eqref{Eq:CEPARcondition}.

        The proof of this equation is similar to the $F$-$A$ relation Eq. \eqref{Eq:CEPAFcondition}. Since the $R^{ab}_c$-symbols can be viewed as basis transformations in $V^{ab}_c$, we can express the map $\mathcal{A}$ in two different bases connected by $R$, as shown in the following diagram.
        \begin{equation*}
        \scalebox{0.8}{\CEPARcondition}
        \end{equation*}
        The upper layer and the lower layer in the above diagram are exactly the same vectors in different bases, and hence the final result after the $\mathcal{A}$ map should equal. The equality in the dashed line is due to the naturalness of braiding, and hence the cross can freely move to the left. Expressing the equation in symbolic form, we get Eq. \eqref{Eq:CEPARcondition}.

        \item Proof of Eq. \eqref{Eq:CEPAUcondition}.

        The $A$-symbols also yield certain consistency equations with the $U$-matrices. As we have mentioned, these $U$-matrices can be viewed as basis transformation matrices of certain automorphism $\Phi_g$, where $\Phi_g$ might correspond to internal or lattice symmetry in the physical system. Then we notice that the definition of $A$-matrices yields 
        \begin{equation*}
        \scalebox{0.8}{\CEPAUcondition}
        \end{equation*}
        Note that the diagram in the upper-left corner consists of four vectors, and hence applying $\Phi_g$ changes the basis of all four vectors. The symbolic expression of the diagram reads
        \begin{equation*}
            K^{q(g)}\varkappa_{\bar c}A^{\bar b\bar a}_{\bar c}K^{q(g)}U_g(\lsupsc{g} \bar b,\lsupsc{g} \bar a;\lsupsc{g} \bar c)=\frac{U_g(\lsupsc{g} \bar a,\lsupsc{g} a;0)U_g(\lsupsc{g} \bar b,\lsupsc{g} b;0)}{U_g(\lsupsc{g} c,\lsupsc{g} \bar c;0)}U_g^*(\lsupsc{g} a,\lsupsc{g} b;\lsupsc{g} c)\varkappa_{\lsupsc{g}\bar c}A^{\lsupsc{g} \bar b\lsupsc{g} \bar a}_{\lsupsc{g} \bar c}.
        \end{equation*}
        We use the explicit expression of $\varkappa_a=d_aF^{a\bar aa}_a$ to find its transformation under $\Phi_g$, which is
        \begin{equation}\label{Eq:CEPkappaUrelation}
            K^{q(g)}\varkappa_{\lsupsc{\bar g}a}K^{q(g)}=U_g(a,\bar a;0)\varkappa_a U_g(\bar a, a;0)^{-1}.
        \end{equation}        
        Substituting $\frac{\varkappa_{\lsupsc{g}\bar c}}{U_g(\lsupsc{g} c,\lsupsc{g} \bar c;0)}$ with $\frac{K^{q(g)}\varkappa_{c}K^{q(g)}}{U_g(\lsupsc{g} \bar c,\lsupsc{g} c;0)}$, we find the desired equation.
    \end{enumerate}
\end{proof}

It can be seen that the $A$-symbols essentially change anyons to their conjugates. Note that this conjugation cannot be viewed as a microscopic charge conjugation symmetry, because they neither obey the consistency equation for a charge conjugation symmetry (in particular, Eq. \eqref{Eq:CEPAFcondition} is not the equation satisfied by the data characterizing an SET phase with an internal symmetry), nor do they have a guaranteed microscopic realization in a given TQSL. This $\mathcal{A}$ map is interpreted as a CRT action at the level of the low-energy effective topological quantum field theory. We use this $\mathcal{A}$ map, or the $A$-matrices, in Eq. \eqref{Eq:correspondence} to construct the crystalline equivalence principle.

\begin{theorem}
    \begin{equation}\label{Eq:CEPconditionRAtheta}
        A^{\bar a a}_0R^{a \bar a}_0=A^{a\bar a}_0R^{\bar aa}_0=\theta^*_a,
    \end{equation}
    where $\theta_a$ is the topological spin of the anyon $a$ defined by
    \begin{equation}\label{Eq:topologicalspin}
        \bmm\scalebox{0.8}{\topologicalspin}\emm.
    \end{equation}
    See, for example, Eq. (211) in Ref. \cite{Kitaev2006} for this definition.
\end{theorem}
\begin{proof}
    We consider the following equalities.
    \begin{equation}
        \scalebox{0.9}{\CEPconditionRAtheta}
    \end{equation}
    The equality between the first two figures on the upper line is due to our definition of $A$-matrices and the condition Eq. \eqref{Eq:CEPARcondition}, and the second equality is the definition of the quantum dimension $d_a$. See, for example, Eq. (191) in Ref. \cite{Kitaev2006} for this definition. The two figures on the first row are equal because of the naturality of braiding, which allows us to freely move the cross, and the equality in the last line is the definition of topological spin Eq. \eqref{Eq:topologicalspin}. Combining all the equalities and using the fact that $R^{a\bar a}_0$ are phases, we get Eq. \eqref{Eq:CEPconditionRAtheta}.
\end{proof}

Apart from the equations above, there is also another useful equation we will use below:
\begin{equation}\label{Eq:CEPFRcondition}
    R^{ab}_e R^{ec}_d(F^{abc}_d)_{ef}=(F^{cba}_d)_{ef}^* R^{bc}_fR^{af}_d.
\end{equation}
This equation is derived from the following commutative diagram. 
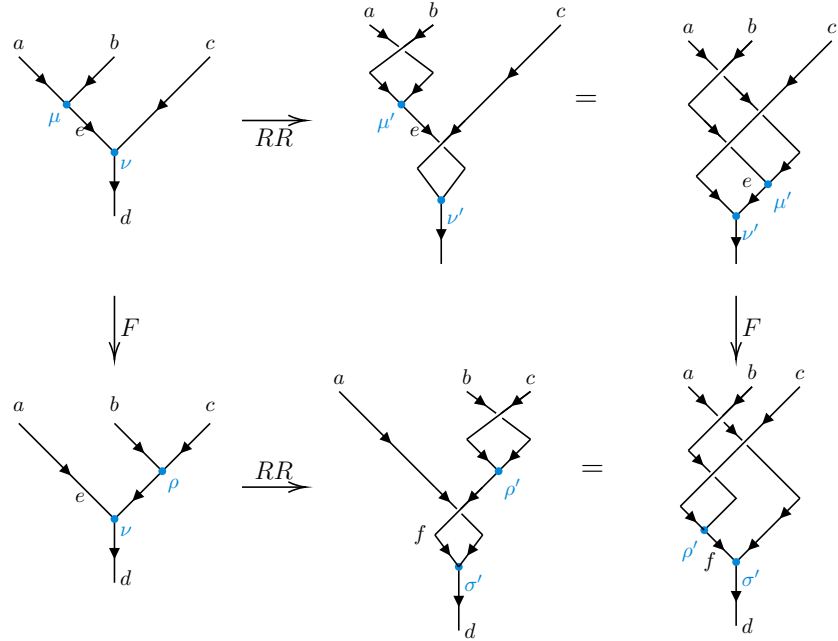
\begin{equation*}
    \scalebox{0.8}{\CEPFRcondition}
\end{equation*}
The diagram commutes because it consists of merely $F$- and $R$- moves. The two equalities in the above diagram both come from the fact that the left hand sides and right hand sides differ merely by moving the worldlines without bending them.

\subsubsection{The inverse map of Eq. \eqref{Eq:correspondence}} \label{subsubapp: inverse CEP}

The above discussion fully specifies Eq. \eqref{Eq:correspondence}. Now we construct the inverse map of Eq. \eqref{Eq:correspondence}, which shows that Eq. \eqref{Eq:correspondence} is bijective.

Suppose that we start from an SET phase with an internal symmetry $G$, and we know whether, after the inverse correspondence, each $g\in G$ becomes a lattice symmetry action (which may preserve or reverse the spatial orientation) or remains internal. Then the explicit form of the inverse map of Eq. \eqref{Eq:correspondence} is as follows.
\begin{equation}\label{Eq:inverse correspondence app}
	\bsp
		&\rho_g(a)=\left\{ \begin{array}{ll}
			\rho'_g(a), &\textrm{if $g$ preserves spatial orientation or remains internal,}\\
			\rho'_g(\bar a), &\textrm{if $g$ reverses spatial orientation.}
		\end{array}\right.\\
		&U_{g}(a,b;c)= \left\{ \begin{array}{ll}
			U'_g(a,b;c),&\begin{array}{l}\textrm{if $g$ preserves spatial orientation}\\ \textrm{    or remains internal,}\end{array}\\
			K^{q'(g)}([R^{\rho'_{\bar g}(\bar a)\rho'_{\bar g}(\bar b)}_{\rho'_{\bar g}(\bar c)}]^{T}[A^{\rho'_{\bar g}(\bar b)\rho'_{\bar g}(\bar a)}_{\rho'_{\bar g}(\bar c)}]^{T})K^{q'(g)}U'_g(a,b;c),&\begin{array}{l}\textrm{if $g$ reverses spatial orientation.}\end{array}
		\end{array}\right.\\
		& \eta_a(g_1,g_2)=\left\{\begin{array}{ll}
			\eta'_a(g_1,g_2),&\begin{array}{l}\textrm{if $g_1$ preserves spatial orientation}\\ \textrm{    or remains internal,}\end{array}\\
			\eta'_a(g_1,g_2)K^{q'(g_1)}U'_{g_2}(\rho'_{\overline{g_1}}(\bar a),\rho'_{\overline{g_1}}(a);0)K^{q'(g_1)},&\begin{array}{l}\textrm{if $g_1$ reverses spatial orientation}\\ \textrm{    but $g_2$ preserves it or remains internal,}\end{array}\\
			\eta'_a(g_1,g_2)K^{q'(g_1)}U'_{g_2}(\rho'_{\overline{g_1}}(a),\rho'_{\overline{g_1}}(\bar a);0)\varkappa^{*}_{\rho'_{\overline{g_1}}(\bar a)}K^{q'(g_1)},&\begin{array}{l}\textrm{if both $g_1$ and $g_2$ reverse spatial orientation.}\end{array}
		\end{array}\right.
	\esp
\end{equation}
Here, as before, the primed data is from the SET phase with an internal symmetry, and the unprimed data is for the SET phase with a general symmetry.

Now we derive this inverse map. For later convenience, it is useful to introduce the notation
\begin{equation} \label{eq: reflection grading}
    p(g)=\left\{\begin{array}{ll}0,&\text{if $g$ preserves spatial orientation,}\\1,&\textrm{if $g$ reverses spatial orientation,}\end{array}\right.
\end{equation}
where $g$ is a symmetry in the TQSL with a general symmetry.

First, $\rho_g(a)$ is simply the inverse function of $\rho'_g(a)$ in Eq. \eqref{Eq:correspondence app}.

For $U_g(a,b;c)$, when $g$ preserves spatial orientation or remains internal, it is clear that $U_g(a,b;c)=U'_g(a,b;c)$. When $g$ reverses spatial orientation after the inverse map is applied, we have, according to the original map,
\begin{equation}
    U'_g(a,b;c)=K^{q(g)}(A^{\lsupsc{\overline g}b\lsupsc{\overline g}a}_{\lsupsc{\overline g}c}R^{\lsupsc{\overline g}a\lsupsc{\overline g}b}_{\lsupsc{\overline g}c})K^{q(g)}U_g(a,b;c).
\end{equation}
We reverse this map by multiplying the inverse of the factor consisting of $A$ and $R$ on both sides and re-expressing the symmetry action on anyons using $\rho'_a(g)$. Note that when $g$ reverses spatial orientation, its unitarity is reversed under the map, so $K^{q(g)}OK^{q(g)}=K^{q'(g)}O^*K^{q'(g)}$ where $q'(g)=q(g)+p(g)$. Consequently, we have 
\begin{equation}
    U_g(a,b;c)=K^{q'(g)}([R^{\rho'_{\bar g}(\bar a)\rho'_{\bar g}(\bar b)}_{\rho'_{\bar g}(\bar c)}]^{T}[A^{\rho'_{\bar g}(\bar b)\rho'_{\bar g}(\bar a)}_{\rho'_{\bar g}(\bar c)}]^{T})K^{q'(g)}U'_g(a,b;c).
\end{equation}

For $\eta_a(g_1,g_2)$, we use the same treatment to derive the inverse map. When $g$ preserves spatial orientation or remains internal, it is clear that $\eta_a(g_1,g_2)=\eta'_a(g_1,g_2)$. When $g_1$ preserves spatial orientation or remains internal while $g_2$ reverses spatial orientation, it is straightforward that we have 
\begin{equation}
\bsp
    &\eta_a(g_1,g_2)=\eta'_a(g_1,g_2)K^{q(g_1)}U^*_{g_2}(\lsupsc{\overline{g_1}} a,\lsupsc{\overline{g_1}}\bar a;0)K^{q(g_1)}\\
    &=\eta'_a(g_1,g_2)K^{q'(g_1)}U'_{g_2}(\rho'_{\overline{g_1}}(\bar a),\rho'_{\overline{g_1}}(a);0)K^{q'(g_1)}.
\esp
\end{equation}
When both $g_1$ and $g_2$ reverse spatial orientation, we have
\begin{equation}
\bsp
    &\eta_a(g_1,g_2)=\eta'_a(g_1,g_2)K^{q(g_1)}U^*_{g_2}(\lsupsc{\overline{g_1}}\bar a,\lsupsc{\overline{g_1}}a;0)\theta^*_{\lsupsc{\overline{g_1}}a}\varkappa_{\lsupsc{\overline{g_1}}a}K^{q(g_1)}\\
    &=\eta'_a(g_1,g_2)K^{q'(g_1)}U_{g_2}(\lsupsc{\overline{g_1}}\bar a,\lsupsc{\overline{g_1}}a;0)\theta_{\lsupsc{\overline{g_1}}a}\varkappa^*_{\lsupsc{\overline{g_1}}a}K^{q'(g_1)}\\
    &=\eta'_a(g_1,g_2)K^{q'(g_1)}K^{q'(g_2)}([R^{\rho'_{\overline{g_2}}(\lsupsc{\overline{g_1}}a)\rho'_{\overline{g_2}}(\lsupsc{\overline{g_1}}\bar a)}_{0}][A^{\rho'_{\overline{g_2}}(\lsupsc{\overline{g_1}}\bar a)\rho'_{\overline{g_2}}(\lsupsc{\overline{g_1}}a)}_{0}])K^{q'(g_2)}U'_{g_2}(\lsupsc{\overline{g_1}}\bar a,\lsupsc{\overline{g_1}}a;0)\theta_{\lsupsc{\overline{g_1}} a}\varkappa^*_{\rho'_{\overline{g_1}}(\bar a)}K^{q'(g_1)}\\
    &=\eta'_a(g_1,g_2)K^{q'(g_1)}U'_{g_2}(\lsupsc{\overline{g_1}}\bar a,\lsupsc{\overline{g_1}}a;0)\left[K^{q'(g_2)}\theta^*_{\rho'_{\overline{g_2}}(\lsupsc{\overline{g_1}}a)}K^{q'(g_2)}\theta_{\lsupsc{\overline{g_1}}a}\right]\varkappa^*_{\rho'_{\overline{g_1}}(\bar a)}K^{q'(g_1)}\\
    &=\eta'_a(g_1,g_2)K^{q'(g_1)}U'_{g_2}(\rho'_{\overline{g_1}}(a),\rho'_{\overline{g_1}}(\bar a);0)\varkappa^*_{\rho'_{\overline{g_1}}(\bar a)}K^{q'(g_1)}.
\esp
\end{equation}
In the last equality above, we have used the property of topological spin $\theta_{\rho'_g(a)}=K^{p'(g)+q'(g)}\theta_a K^{p'(g)+q'(g)}$ and the fact that $p'(g_2)=0$ for any internal symmetry.

Combining all the above, we reach Eq. \eqref{Eq:inverse correspondence app}, which shows that Eq. \eqref{Eq:correspondence} is bijective.

\subsection{Checking the consistency equations} \label{subapp: checking consistency equations}

After describing the map Eq. \eqref{Eq:correspondence app} and its inverse Eq. \eqref{Eq:inverse correspondence app} in Appendix \ref{subapp: details of CEP}, now we check that data obtained from the map Eq. \eqref{Eq:correspondence app} indeed satisfies the correct set of consistency equations.

We copy the consistency equations satisfied by the data $\mc{C}_{1, \mathbbm{1}, 1}$ for an SET phase with a general symmetry $G$:
\begin{align}
    &K^{q(g)}[F^{\lsupsc{\overline{g}}a\lsupsc{\overline{g}}b\lsupsc{\overline{g}}c}_{\lsupsc{\overline{g}}d}]_{(\lsupsc{\overline{g}}e,\mu',\nu')(\lsupsc{\overline{g}}f,\rho',\sigma')}K^{q(g)}\\
    &=\sum_{\mu,\nu,\rho,\sigma}[U_g(a,b;e)]_{\mu'\mu}[U_g(e,c;d)]_{\nu'\nu}[F^{abc}_{d}]_{(e,\mu,\nu)(f,\rho,\sigma)}[(U_g(b,c;f)^{-1})]_{\rho\rho'}[(U_g(a,f;d))^{-1}]_{\sigma\sigma'},\label{Eq:latticeFU}\\
    &K^{q(g)}[J_g(R^{\lsupsc{\overline{g}}a\lsupsc{\overline{g}}b}_{\lsupsc{\overline{g}}c})]_{\mu'\nu'}K^{q(g)}=\sum_{\mu\nu}[U_g(b,a;c)]_{\mu'\mu}[R^{ab}_c]_{\mu\nu}[U_g(a,b;c)]_{\nu\nu'}^{-1},\label{Eq:latticeRU}\\
    &\sum_{\nu}K^{q(g)}[U_h(\lsupsc{\overline{g}}a,\lsupsc{\overline{g}}b;\lsupsc{\overline{g}}c)]_{\mu\nu}K^{q(g)}[U_g(a,b;c)]_{\nu\rho}=\frac{\eta_c(g,h)}{\eta_a(g,h)\eta_b(g,h)}[U_{gh}(a,b;c)]_{\mu\rho},\label{Eq:latticeUeta}\\
    &K^{q(g)}\eta_{\lsupsc{\overline{g}}a}(h,k)K^{q(g)}\eta_a(g,hk)=\eta_a(gh,k)\eta_a(g,h).\label{Eq:latticeetaeta}
\end{align}

Our goal is to prove that, if the unprimed data in Eq. \eqref{Eq:correspondence app} satisfies Eqs. \eqref{Eq:latticeFU}, \eqref{Eq:latticeRU}, \eqref{Eq:latticeUeta} and \eqref{Eq:latticeetaeta}, then the primed data in Eq. \eqref{Eq:correspondence app} satisfies the following equations:
\begin{align}
    &K^{q'(g)}(F^{\rho'_g(a)\rho'_g(b)\rho'_g(c)}_{\rho'_g(d))})_{\rho'_g(e)\rho'_g(f)}K^{q'(g)}=U'_{g}(a,b;e)U'_{g}(e,c;d)(F^{abc}_{d})_{ef}U'_{g}(b,c;f)^{-1}U'_{g}(a,f;d)^{-1},\label{eq: internal F Uprime}\\
    &U'_{g}(b,a;c)R^{ab}_{c}U'_{g}(a,b;c)^{-1}=K^{q'(g)}(R^{\rho'_g(a)\rho'_g(b)}_{\rho'_g(c)})K^{q'(g)},\label{eq: internal R Uprime}\\
    &K^{q'(g)}U'_{h}(\rho'_g(a),\rho'_g(b);\rho'_g(c))K^{q'(g)}U'_{g}(a,b;c)=\frac{\eta_c'(g,h)}{\eta_a'(g,h)\eta_b'(g,h)}U'_{gh}(a,b;c),\label{eq: internal Uprime etaprime}\\
    &K^{q'(g)}\eta'_{\rho'_g(a)}(h,k)K^{q'(g)}\eta'_a(g,h k)=\eta'_a(g h,k)\eta'_a(g,h)\label{eq: internal etaprime etaprime}.
\end{align}
These equations are just the version of Eqs. \eqref{Eq:latticeFU}, \eqref{Eq:latticeRU}, \eqref{Eq:latticeUeta} and \eqref{Eq:latticeetaeta} when there is only internal symmetry, and they must be satisfied by the data $\mc{C}_{1, \mathbbm{1}, 1}$ for an SET phase with a purely internal symmetry $G$.

We first prove that $U'_g(a,b;c)$ satisfies Eq. \eqref{eq: internal F Uprime}. When $g$ preserves spatial orientation, since $U'_g(a,b;c)=U_g(a,b;c)$, it is straightforward to see that Eq. \eqref{eq: internal F Uprime} holds. When $g$ reverses spatial orientation,
\begin{equation*}
\bsp
    &U'_{g}(a,b;e)U'_{g}(e,c;d)(F^{abc}_d)_{ef}U'_{g}(b,c;f)^{-1}U'_{g}(a,f;d)^{-1}\\
    =&K^{q(g)}(A^{\lsupsc{\overline g}b\lsupsc{\overline g}a}_{\lsupsc{\overline g}e}R^{\lsupsc{\overline g}a\lsupsc{\overline g}b}_{\lsupsc{\overline g}e})(A^{\lsupsc{\overline g}c\lsupsc{\overline g}e}_{\lsupsc{\overline g}d}R^{\lsupsc{\overline g}e\lsupsc{\overline g}c}_{\lsupsc{\overline g}d})(F^{\lsupsc{\overline g}a\lsupsc{\overline g}b\lsupsc{\overline g}c}_{\lsupsc{\overline g}d})_{\lsupsc{\overline g}e\lsupsc{\overline g}f}(A^{\lsupsc{\overline g}c\lsupsc{\overline g}b}_{\lsupsc{\overline g}f}R^{\lsupsc{\overline g}b\lsupsc{\overline g}c}_{\lsupsc{\overline g}f})^{-1}(A^{\lsupsc{\overline g}f\lsupsc{\overline g}a}_{\lsupsc{\overline g}d}R^{\lsupsc{\overline g}a\lsupsc{\overline g}f}_{\lsupsc{\overline g}d})^{-1}K^{q(g)}\\
    =&K^{q(g)}A^{\lsupsc{\overline g}b\lsupsc{\overline g}a}_{\lsupsc{\overline g}e}A^{\lsupsc{\overline g}c\lsupsc{\overline g}e}_{\lsupsc{\overline g}d}(F^{\lsupsc{\overline g}c\lsupsc{\overline g}b\lsupsc{\overline g}a}_{\lsupsc{\overline g}d})_{\lsupsc{\overline g}e\lsupsc{\overline g}f}^*(A^{\lsupsc{\overline g}c\lsupsc{\overline g}b}_{\lsupsc{\overline g}f})^{-1}(A^{\lsupsc{\overline g}f\lsupsc{\overline g}a}_{\lsupsc{\overline g}d})^{-1}K^{q(g)}\\
    =&K^{q(g)}(F^{\lsupsc{\overline g}\bar a\lsupsc{\overline g}\bar b\lsupsc{\overline g}\bar c}_{\lsupsc{\overline g}\bar d})_{\lsupsc{\overline g}\bar e\lsupsc{\overline g}\bar f}^*K^{q(g)}.
\esp
\end{equation*}
The first equality is achieved by using Eq. \eqref{Eq:latticeFU}, the second equality is achieved by using Eq. \eqref{Eq:CEPFRcondition}, and the last equality is achieved by using Eq. \eqref{Eq:CEPAFcondition}. Summarizing the two cases, we conclude that
\begin{equation}
    K^{q'(g)}(F^{\rho'_g(a)\rho'_g(b)\rho'_g(c)}_{\rho'_g(d))})_{\rho'_g(e)\rho'_g(f)}K^{q'(g)}=U'_{g}(a,b;e)U'_{g}(e,c;d)(F^{abc}_{d})_{ef}U'_{g}(b,c;f)^{-1}U'_{g}(a,f;d)^{-1},
\end{equation}
which is Eq. \eqref{eq: internal F Uprime}.

We then check Eq. \eqref{eq: internal R Uprime}. When $g$ preserves spatial orientation, it is again straightforward to see that Eq. \eqref{eq: internal R Uprime} holds. When $g$ reverses spatial orientation,
\begin{equation*}
\bsp
    &U'_{g}(b,a;c)R^{ab}_{c}U'_{g}(a,b;c)^{-1}\\
    =&K^{q(g)}(A^{\lsupsc{\overline g}a\lsupsc{\overline g}b}_{\lsupsc{\overline g}c}R^{\lsupsc{\overline g}b\lsupsc{\overline g}a}_{\lsupsc{\overline g}c})K^{q(g)}U_g(b,a;c)R^{ab}_{c}U^{-1}_g(a,b;c)K^{q(g)}(R^{\lsupsc{\overline g}a\lsupsc{\overline g}b}_{\lsupsc{\overline g}c})^{-1}(A^{\lsupsc{\overline g}b\lsupsc{\overline g}a}_{\lsupsc{\overline g}c})^{-1}K^{q(g)}\\
    =&K^{q(g)}(A^{\lsupsc{\overline g}a\lsupsc{\overline g}b}_{\lsupsc{\overline g}c}R^{\lsupsc{\overline g}b\lsupsc{\overline g}a}_{\lsupsc{\overline g}c})(R^{\lsupsc{\bar g}b\lsupsc{\bar g}a}_{\lsupsc{\bar g}c})^{-1}(R^{\lsupsc{\overline g}a\lsupsc{\overline g}b}_{\lsupsc{\overline g}c})^{-1}(A^{\lsupsc{\overline g}b\lsupsc{\overline g}a}_{\lsupsc{\overline g}c})^{-1}K^{q(g)}\\
    =&K^{q(g)}A^{\lsupsc{\overline g}a\lsupsc{\overline g}b}_{\lsupsc{\overline g}c}(R^{\lsupsc{\overline g}a\lsupsc{\overline g}b}_{\lsupsc{\overline g}c})^{-1}(A^{\lsupsc{\overline g}b\lsupsc{\overline g}a}_{\lsupsc{\overline g}c})^{-1}K^{q(g)}\\
    =&K^{q(g)}(R^{\lsupsc{\overline g}\bar a\lsupsc{\overline g}\bar b}_{\lsupsc{\overline g}\bar c})^{*}K^{q(g)}\\
\esp
\end{equation*}
The first equality is obtained by directly inserting the definition of $U'_{g}$, the second equality is obtained by using Eq. \eqref{Eq:latticeRU} for an orientation-reversing symmetry, the third equality is a direct evaluation of the third line, and the last equality is obtained by using Eq. \eqref{Eq:CEPAUcondition}. Summarizing the two cases, we conclude that
\begin{equation}
    K^{q'(g)}(R^{\rho'_g(a)\rho'_g(b)}_{\rho'_g(c)})K^{q'(g)}=U'_{g}(b,a;c)R^{ab}_{c}U'_{g}(a,b;c)^{-1},
\end{equation}
which is Eq. \eqref{eq: internal R Uprime}.

We now check Eq. \eqref{eq: internal Uprime etaprime}. When multiplying $U'_{g}$ and $U'_{h}$ together, there are four cases to be considered: 1) both $g$ and $h$ reverse spatial orientation; 2) $g$ reverses spatial orientation while $h$ does not; 3) $h$ reverses spatial orientation while $g$ does not; 4) both $g$ and $h$ preserves spatial orientation. For the first case, we find that
\begin{equation*}
\bsp
    &K^{q(g)}U'_{h}(\lsupsc{\bar{g}}\bar a,\lsupsc{\bar{g}}\bar b;\lsupsc{\bar{g}}\bar c)^{*}K^{q(g)}U'_{g}(a,b;c)\\
    =&K^{q(g)}K^{q(h)}(A^{\lsupsc{\overline {gh}}\bar b\lsupsc{\overline {gh}}\bar a}_{\lsupsc{\overline {gh}}\bar c}R^{\lsupsc{\overline {gh}}\bar a\lsupsc{\overline {gh}}\bar b}_{\lsupsc{\overline {gh}}\bar c})^*K^{q(h)}U_h^*(\lsupsc{\overline g}\bar a,\lsupsc{\overline g}\bar b;\lsupsc{\overline g}\bar c)(A^{\lsupsc{\overline g}b\lsupsc{\overline g}a}_{\lsupsc{\overline g}c}R^{\lsupsc{\overline g}a\lsupsc{\overline g}b}_{\lsupsc{\overline g}c})K^{q(g)}U_g(a,b;c)\\
    =&K^{q(gh)}(A^{\lsupsc{\overline {gh}}\bar b\lsupsc{\overline {gh}}\bar a}_{\lsupsc{\overline {gh}}\bar c})^*K^{q(h)}U^*_h(\lsupsc{\overline g}\bar b,\lsupsc{\overline g}\bar a;\lsupsc{\overline g}\bar c)(R^{\lsupsc{\overline g}\bar b\lsupsc{\overline g}\bar a}_{\lsupsc{\overline g}\bar c})^T
    (A^{\lsupsc{\overline g}b\lsupsc{\overline g}a}_{\lsupsc{\overline g}c}R^{\lsupsc{\overline g}a\lsupsc{\overline g}b}_{\lsupsc{\overline g}c})K^{q(g)}U_g(a,b;c)\\
    =&K^{q(gh)}(A^{\lsupsc{\overline {gh}}\bar b\lsupsc{\overline {gh}}\bar a}_{\lsupsc{\overline {gh}}\bar c})^*K^{q(h)}U^*_h(\lsupsc{\overline g}\bar b,\lsupsc{\overline g}\bar a;\lsupsc{\overline g}\bar c)(R^{\lsupsc{\overline g}\bar b\lsupsc{\overline g}\bar a}_{\lsupsc{\overline g}\bar c}R^{\lsupsc{\overline g}\bar a\lsupsc{\overline g}\bar b}_{\lsupsc{\overline g}\bar c})^T
    A^{\lsupsc{\overline g}a\lsupsc{\overline g}b}_{\lsupsc{\overline g}c}K^{q(g)}U_g(a,b;c)\\
    =&K^{q(g)}U_h(\lsupsc{\overline g}a,\lsupsc{\overline g}b;\lsupsc{\overline g} c)\frac{U_h(\lsupsc{\overline g}\bar c,\lsupsc{\overline g}c;0)}{U_h(\lsupsc{\overline g}\bar b,\lsupsc{\overline g}b;0)U_h(\lsupsc{\overline g}\bar b,\lsupsc{\overline g}b;0)}\frac{\theta_{\lsupsc{\overline g}c}}{\theta_{\lsupsc{\overline g}a}\theta_{\lsupsc{\overline g}b}
    }(A^{\lsupsc{\overline g}\bar b\lsupsc{\overline g}\bar a}_{\lsupsc{\overline g}\bar c})^*A^{\lsupsc{\overline g}a\lsupsc{\overline g}b}_{\lsupsc{\overline g}c}K^{q(g)}U_g(a,b;c)\\
    =&K^{q(g)}\frac{U_h(\lsupsc{\overline g}\bar c,\lsupsc{\overline g}c;0)\theta_{\lsupsc{\overline g}c}\varkappa^*_{\lsupsc{\overline g}c}}{U_h(\lsupsc{\overline g}\bar a,\lsupsc{\overline g}a;0)\theta_{\lsupsc{\overline g}a}\varkappa^*_{\lsupsc{\overline g}a}U_h(\lsupsc{\overline g}\bar b,\lsupsc{\overline g}b;0)\theta_{\lsupsc{\overline g}b}\varkappa^*_{\lsupsc{\overline g}b}}K^{q(g)}\frac{\eta_c(g,h)}{\eta_a(g,h)\eta_b(g,h)}U_{gh}(a,b;c)\\
    =&\frac{\eta_c'(g,h)}{\eta_a'(g,h)\eta_b'(g,h)}U'_{gh}(a,b;c).
\esp
\end{equation*}
The first equality is achieved by directly inserting the definition of $U'_{g}$ and $U'_{h}$, and $\overline{gh}=\bar h\bar g$ is the inverse group element of $gh$; the second equality is achieved by using Eq. \eqref{Eq:latticeRU} for an orientation-reversing symmetry; the third equality is achieved by using Eq. \eqref{Eq:CEPARcondition}; the fourth equality is obtained by using Eq. \eqref{Eq:CEPAUcondition} and the ribbon identity $R^{ab}_c R^{ba}_c=\theta_c/(\theta_a\theta_b)$; the fifth equality is achieved by using Eq. \eqref{Eq:latticeUeta}; and the last equality uses the definition of $U'_g(a,b;c)$ and $\eta'_a(g,h)$ in Eq. \eqref{Eq:correspondence app}. 

When $g$ is orientation-reversing while $h$ is not, we have
\begin{equation*}
\bsp
    &K^{q(g)}U'_{h}(\lsupsc{\bar{g}}\bar a,\lsupsc{\bar{g}}\bar b;\lsupsc{\bar{g}}\bar c)^{*}K^{q(g)}U'_{g}(a,b;c)\\
    =&K^{q(g)}U_h^*(\lsupsc{\overline g}\bar a,\lsupsc{\overline g}\bar b;\lsupsc{\overline g}\bar c)(A^{\lsupsc{\overline g}b\lsupsc{\overline g}a}_{\lsupsc{\overline g}c}R^{\lsupsc{\overline g}a\lsupsc{\overline g}b}_{\lsupsc{\overline g}c})K^{q(g)}U_g(a,b;c)\\
    =&K^{q(g)}K^{q(h)}A^{\lsupsc{\overline{gh}}b\lsupsc{\overline{gh}}a}_{\lsupsc{\overline{gh}}c}K^{q(h)}U_h(\lsupsc{\overline g}b,\lsupsc{\overline g}a;\lsupsc{\overline g}c)\frac{U_h(\lsupsc{\overline g} c,\lsupsc{\overline g}\bar c;0)}{U_h(\lsupsc{\overline g} a,\lsupsc{\overline g}\bar a;0)U_h(\lsupsc{\overline g} b,\lsupsc{\overline g}\bar b;0)}(R^{\lsupsc{\overline g}a\lsupsc{\overline g}b}_{\lsupsc{\overline g}c})K^{q(g)}U_g(a,b;c)\\
    =&K^{q(gh)}A^{\lsupsc{\overline{gh}}b\lsupsc{\overline{gh}}a}_{\lsupsc{\overline{gh}}c}(R^{\lsupsc{\overline{gh}}a\lsupsc{\overline{gh}}b}_{\lsupsc{\overline{gh}}c})K^{q(h)}\frac{U_h(\lsupsc{\overline{g}} c,\lsupsc{\overline{g}}\bar c;0)}{U_h(\lsupsc{\overline g} a,\lsupsc{\overline g}\bar a;0)U_h(\lsupsc{\overline g} b,\lsupsc{\overline g}\bar b;0)}U_h(\lsupsc{\overline g}a,\lsupsc{\overline g}b;\lsupsc{\overline g}c)K^{q(g)}U_g(a,b;c)\\
    =&K^{q(gh)}A^{\lsupsc{\overline{gh}}b\lsupsc{\overline{gh}}a}_{\lsupsc{\overline{gh}}c}(R^{\lsupsc{\overline{gh}}a\lsupsc{\overline{gh}}b}_{\lsupsc{\overline{gh}}c})K^{q(gh)}\times K^{q(g)}\frac{U_h(\lsupsc{\overline{g}} c,\lsupsc{\overline{g}}\bar c;0)}{U_h(\lsupsc{\overline g} a,\lsupsc{\overline g}\bar a;0)U_h(\lsupsc{\overline g} b,\lsupsc{\overline g}\bar b;0)}K^{q(g)}\frac{\eta_c(g,h)}{\eta_a(g,h)\eta_b(g,h)}U_{gh}(a,b;c)\\
    =&K^{q(g)}\frac{U_h(\lsupsc{\overline{g}} c,\lsupsc{\overline{g}}\bar c;0)}{U_h(\lsupsc{\overline g} a,\lsupsc{\overline g}\bar a;0)U_h(\lsupsc{\overline g} b,\lsupsc{\overline g}\bar b;0)}K^{q(g)}\frac{\eta_c(g,h)}{\eta_a(g,h)\eta_b(g,h)}U'_{gh}(a,b;c)\\
    =&\frac{\eta_c'(g,h)}{\eta_a'(g,h)\eta_b'(g,h)}U'_{gh}(a,b;c).
\esp
\end{equation*}
The first equality is achieved by directly inserting the definition of $U'_{g}$; the second equality is achieved by using Eq. \eqref{Eq:CEPAUcondition}; the third equality is achieved by using Eq. \eqref{Eq:latticeRU} for an orientation preserving symmetry (note that $h$ is orientation preserving here); the fourth equality is obtained by using Eq. \eqref{Eq:latticeUeta}; the fifth equality is achieved by using the definitions of $U'_{gh}(a,b;c)$ in Eq. \eqref{Eq:correspondence app}; and the last equality uses the definition of $\eta'_a(g,h)$ in Eq. \eqref{Eq:correspondence app}. 

When $h$ is orientation-reversing while $g$ is not, we find that
\begin{equation*}
\bsp
    &K^{q(g)}U'_{h}(\lsupsc{\bar{g}}a,\lsupsc{\bar{g}}b;\lsupsc{\bar{g}}c)^{*}K^{q(g)}U'_{g}(a,b;c)\\
    =&K^{q(g)}K^{q(h)}(A^{\lsupsc{\overline {gh}}b\lsupsc{\overline {gh}}a}_{\lsupsc{\overline {gh}}c}R^{\lsupsc{\overline {gh}} a\lsupsc{\overline {gh}}b}_{\lsupsc{\overline {gh}}c})K^{q(h)}U_h(\lsupsc{\overline g}a,\lsupsc{\overline g}b;\lsupsc{\overline g}c)K^{q(g)}U_g(a,b;c)\\
    =&K^{q(gh)}(A^{\lsupsc{\overline {gh}}b\lsupsc{\overline {gh}}a}_{\lsupsc{\overline {gh}}c}R^{\lsupsc{\overline {gh}} a\lsupsc{\overline {gh}}b}_{\lsupsc{\overline {gh}}c})K^{q(gh)}\frac{\eta_c(g,h)}{\eta_a(g,h)\eta_b(g,h)}U_{gh}(a,b;c)\\
    =&\frac{\eta_c'(g,h)}{\eta_a'(g,h)\eta_b'(g,h)}U'_{gh}(a,b;c),  
\esp
\end{equation*}
where we have used Eq. \eqref{Eq:latticeUeta}, and the definition of $U'_{gh}(a,b;c)$ and $\eta'_a(g,h)$ in Eq. \eqref{Eq:correspondence app}. 

When $g$ and $h$ both preserves spatial orientation, the relation between $U_g'$'s are trivially identical to the relation between $U_g$'s. Summarizing the four cases, we have Eq. \eqref{eq: internal Uprime etaprime}:
\begin{equation}
    K^{q(g)+p(g)}U'_{h}(\rho'_g(a),\rho'_g(b);\rho'_g(c))K^{q(g)+p(g)}U'_{g}(a,b;c)=\frac{\eta_c'(g,h)}{\eta_a'(g,h)\eta_b'(g,h)}U'_{gh}(a,b;c),
\end{equation}

The final task is to check Eq. \eqref{eq: internal etaprime etaprime},
where $g$, $h$, and $k$ can be either orientation-preserving or orientation-reversing; hence, there are $8$ cases to be verified. We first show the cases where all three actions reverse spatial orientation. Note that $hk$ and $gh$ preserve spatial orientation. Then, 
\begin{equation*}
\bsp
    &K^{q(g)+p(g)}\eta'_{\rho'_g(a)}(h,k)K^{q(g)+p(g)}\eta'_a(g,h k)\\
    =&K^{q(g)}\left(\eta_{\lsupsc{\overline{g}}\bar a}(h,k)K^{q(h)}U_{k}(\lsupsc{\overline{gh}} a,\lsupsc{\overline{gh}}\bar a;0)(\theta_{\lsupsc{\overline{gh}}\bar a}\varkappa^{*}_{\lsupsc{\overline{gh}}\bar a})\right)^*K^{q(gh)}\left(\eta_a(g,hk)K^{q(g)}U_{hk}(\lsupsc{\overline{g}}a,\lsupsc{\overline{g}}\bar a;0)K^{q(g)}\right)\\
\esp
\end{equation*}
Using the relation,
\begin{equation}\label{Eq:Uaabarrelation}
\bsp
    &K^{q(h)}U_{k}(\lsupsc{\overline{gh}}a,\lsupsc{\overline{gh}}\bar a;0)K^{q(h)}U_{h}(\lsupsc{\overline{g}}a,\lsupsc{\overline{g}}\bar a;0)=\frac{1}{\eta_{\lsupsc{\overline{g}}a}(h,k)\eta_{\lsupsc{\overline{g}}\bar a}(h,k)}U_{hk}(\lsupsc{\overline{g}}a,\lsupsc{\overline{g}}\bar a;0),
\esp
\end{equation}
we substitute $\eta^*_{\lsupsc{\overline{g}}\bar a}(h,k)U_{hk}(\lsupsc{\overline{g}}a,\lsupsc{\overline{g}}\bar a;0)$ with $K^{q(h)}U_{k}(\lsupsc{\overline{gh}}a,\lsupsc{\overline{gh}}\bar a;0)K^{q(h)}U_{h}(\lsupsc{\overline{g}}a,\lsupsc{\overline{g}}\bar a;0)\eta_{\lsupsc{\overline{g}}a}(h,k)$ and get 
\begin{equation*}
\bsp
    &K^{q(g)+p(g)}\eta'_{\rho'_g(a)}(h,k)K^{q(g)+p(g)}\eta'_a(g,h k)\\
    =&\eta_a(g,hk)K^{q(g)}\left(K^{q(h)}U_{k}(\lsupsc{\overline{gh}} a,\lsupsc{\overline{gh}}\bar a;0)\theta_{\lsupsc{\overline{gh}}\bar a}\varkappa^{*}_{\lsupsc{\overline{gh}}\bar a}\right)^*U_{k}(\lsupsc{\overline{gh}}a,\lsupsc{\overline{gh}}\bar a;0)K^{q(h)}U_{h}(\lsupsc{\overline{g}}a,\lsupsc{\overline{g}}\bar a;0)\eta_{\lsupsc{\overline{g}}a}(h,k)K^{q(g)}\\
    =&K^{q(g)}\eta_{\lsupsc{\overline{g}}a}(h,k)K^{q(g)}\eta_a(g,hk)K^{q(gh)}\theta^*_{\lsupsc{\overline{gh}}a}\varkappa_{\lsupsc{\overline{gh}}\bar a}K^{q(h)}U_{h}(\lsupsc{\overline{g}}a,\lsupsc{\overline{g}}\bar a;0)K^{q(g)}\\
    =&\eta_a(gh,k)\eta_a(g,h)\times K^{q(g)}U_h(\lsupsc{\overline{g}}\bar a,\lsupsc{\overline{g}} a;0)\theta_{\lsupsc{\overline{g}}a}\varkappa^{*}_{\lsupsc{\overline{g}}a}K^{q(g)}\\
    =&\eta'_a(gh,k)\eta'_a(g,h). 
    \esp
\end{equation*}    
The second equality is obtained by massaging the factors; the third equality is achieved by using the properties of the topological spin $\theta_{\lsupsc{g}a}=K^{p(g)+q(g)}\theta_a K^{p(g)+q(g)}$ and the $\varkappa$-symbols Eq. \eqref{Eq:CEPkappaUrelation}; the last equality uses the definition of $\eta'_a(g h,k)$ and $\eta'_a(g, h)$. As such, when $g$, $h$, and $k$ are all orientation-reversing, the corresponding $\eta'$ phases satisfy the expected equation. 

When $g$ and $h$ reverse spatial orientation while $k$ preserves spatial orientation, note that $gh$ preserves spatial orientation while $hk$ reverses spatial orientation. We confirm that
\begin{align*}
    &K^{q(g)+p(g)}\eta'_{\rho'_g(a)}(h,k)K^{q(g)+p(g)}\eta'_a(g,h k)\\
    =&K^{q(g)}\eta^*_{\lsupsc{\overline{g}}\bar a}(g,h)K^{q(h)}U_{k}(\lsupsc{\overline{gh}}\bar a,\lsupsc{\overline{gh}}a;0)^*K^{q(gh)}\times \eta_a(g,hk)K^{q(g)}U_{hk}(\lsupsc{\overline{g}}\bar a,\lsupsc{\overline{g}}a;0)\theta_{\lsupsc{\bar g}a}\varkappa^{*}_{\lsupsc{\bar g}a}K^{q(g)}\\
    =&\eta_a(g,hk)K^{q(gh)}U_{k}(\lsupsc{\overline{gh}}\bar a,\lsupsc{\overline{gh}}a;0)K^{q(h)}U_{h}(\lsupsc{\overline{g}}\bar a,\lsupsc{\overline{g}}a;0)\eta_{\lsupsc{\overline{g}}a}(h,k)K^{q(h)}U_{k}(\lsupsc{\overline{gh}}\bar a,\lsupsc{\overline{gh}}a;0)^*K^{q(h)}\theta_{\lsupsc{\bar g}a}\varkappa^{*}_{\lsupsc{\bar g}a}K^{q(g)}\\
    =&K^{q(g)}\eta_{\lsupsc{\overline{g}}a}(h,k)K^{q(g)}\eta_a(g,hk)\times K^{q(g)}U_{h}(\lsupsc{\overline{g}}\bar a,\lsupsc{\overline{g}}a;0)\theta_{\lsupsc{\bar g}a}\varkappa^{*}_{\lsupsc{\bar g}a}K^{q(g)}\\
    =&\eta_a(gh,k)\eta_a(g,h)\times K^{q(g)}U_{h}(\lsupsc{\overline{g}}\bar a,\lsupsc{\overline{g}}a;0)\theta_{\lsupsc{\bar g}a}\varkappa^{*}_{\lsupsc{\bar g}a}K^{q(g)}\\
    =&\eta'_a(gh,k)\eta'_a(g,h).
\end{align*}
The first equality is obtained by directly plugging in the definition of $\eta'$; the second equality uses the relation Eq. \eqref{Eq:Uaabarrelation}; the third equality is obtained by merely massaging the factors; the fourth equality uses the relation between $\eta$'s; the last equality again uses the definition of $\eta'$. As such, when $g$ and $h$ reverse spatial orientation while $k$ preserves spatial orientation, the corresponding $\eta'$ phases satisfy the expected equation.

When $g$ and $k$ reverse spatial orientation while $h$ preserves spatial orientation, note that $gh$ and $hk$ both reverse spatial orientation. We verify that
\begin{align*}
    &K^{q(g)+p(g)}\eta'_{\rho'_g(a)}(h,k)K^{q(g)+p(g)}\eta'_a(g,h k)\\
    =&K^{q(g)}\eta^*_{\lsupsc{\bar g}\bar a}(h,k)K^{q(g)}\eta_a(g,hk)K^{q(g)}U_{hk}(\lsupsc{\overline{g}}\bar a,\lsupsc{\overline{g}}a;0)\theta_{\lsupsc{\bar g}a}\varkappa^{*}_{\lsupsc{\bar g}a}K^{q(g)}\\
    =&\eta_a(g,hk)K^{q(gh)}U_{k}(\lsupsc{\overline{gh}}\bar a,\lsupsc{\overline{gh}}a;0)K^{q(h)}U_{h}(\lsupsc{\overline{g}}\bar a,\lsupsc{\overline{g}}a;0)\eta_{\lsupsc{\overline{g}}a}(h,k)\theta_{\lsupsc{\bar g}a}\varkappa^{*}_{\lsupsc{\bar g}a}K^{q(g)}\\
    =&\eta_a(gh,k)\eta_a(g,h)\times K^{q(gh)}U_{k}(\lsupsc{\overline{gh}}\bar a,\lsupsc{\overline{gh}}a;0)K^{q(gh)}\times K^{q(g)}U_{h}(\lsupsc{\overline{g}}\bar a,\lsupsc{\overline{g}}a;0)\theta_{\lsupsc{\bar g}a}\varkappa^{*}_{\lsupsc{\bar g}a}K^{q(g)}\\
    =&\eta_a(gh,k)\eta_a(g,h)\times K^{q(gh)}U_{k}(\lsupsc{\overline{gh}}\bar a,\lsupsc{\overline{gh}}a;0)\theta_{\lsupsc{\overline{gh}}a}\varkappa^{*}_{\lsupsc{\overline{gh}}a}K^{q(gh)}\times K^{q(g)}U_{h}(\lsupsc{\overline{g}}a,\lsupsc{\overline{g}}\bar a;0)K^{q(g)}\\
    =&\eta'_a(gh,k)\eta'_a(g,h).
\end{align*}
The first equality is obtained by directly plugging in the definition of $\eta'$; the second equality uses the relation Eq. \eqref{Eq:Uaabarrelation}; the third equality uses the relation between $\eta$'s; the fourth equality uses the property of $\theta$ and $\kappa$; the last equality again uses the definition of $\eta'$. As such, when $g$ and $k$ reverse spatial orientation while $h$ preserves spatial orientation, the corresponding $\eta'$ phases satisfy the expected equation.

When $g$ reverses spatial orientation while $h$ and $k$ preserve spatial orientation, note that $gh$ reverses spatial orientation while $hk$ preserves spatial orientation. We verify that
\begin{align*}
    &K^{q(g)+p(g)}\eta'_{\rho'_g(a)}(h,k)K^{q(g)+p(g)}\eta'_a(g,h k)\\
    =&K^{q(g)}\eta^*_{\lsupsc{\bar g}\bar a}(h,k)K^{q(g)}\eta_a(g,hk)K^{q(g)}U_{hk}(\lsupsc{\overline{g}} a,\lsupsc{\overline{g}}\bar a;0)K^{q(g)}\\
    =&\eta_a(g,hk)K^{q(gh)}U_{k}(\lsupsc{\overline{gh}}a,\lsupsc{\overline{gh}}\bar a;0)K^{q(h)}U_{h}(\lsupsc{\overline{g}}a,\lsupsc{\overline{g}}\bar a;0)\eta_{\lsupsc{\overline{g}}a}(h,k)K^{q(g)}\\
    =&K^{q(g)}\eta_{\lsupsc{\overline{g}}a}(h,k)K^{q(g)}\eta_a(g,hk)\times K^{q(gh)}U_{k}(\lsupsc{\overline{gh}}a,\lsupsc{\overline{gh}}\bar a;0)K^{q(h)}U_{h}(\lsupsc{\overline{g}}a,\lsupsc{\overline{g}}\bar a;0)K^{q(g)}\\
    =&\eta_a(gh,k)\eta_a(g,h)\times K^{q(gh)}U_{k}(\lsupsc{\overline{gh}}a,\lsupsc{\overline{gh}}\bar a;0)K^{q(gh)} \times K^{q(g)}U_{h}(\lsupsc{\overline{g}}a,\lsupsc{\overline{g}}\bar a;0)K^{q(g)}\\
    =&\eta'_a(gh,k)\eta'_a(g,h).
\end{align*}
The verification is essentially the same as the previous verifications. We thereby confirm that when $g$ reverses spatial orientation while $h$ and $k$ preserve spatial orientation, the corresponding $\eta'$ phases satisfy the expected equation.

When $g$ preserves spatial orientation while $h$ and $k$ both reverse spatial orientation, note that $gh$ reverses spatial orientation while $hk$ preserves spatial orientation. We check that
\begin{align*}
    &K^{q(g)+p(g)}\eta'_{\rho'_g(a)}(h,k)K^{q(g)+p(g)}\eta'_a(g,h k)\\
    =&K^{q(g)}\eta_{\lsupsc{\bar g}a}(h,k)K^{q(h)}U_{k}(\lsupsc{\overline{gh}}\bar a,\lsupsc{\overline{gh}}a;0)\theta_{\lsupsc{\overline{gh}}a}\varkappa^{*}_{\lsupsc{\overline{gh}}a}K^{q(gh)}\eta_a(g,hk)\\
    =&\eta_{a}(g,h)\eta_a(gh,k)\times K^{q(gh)}U_{k}(\lsupsc{\overline{gh}}\bar a,\lsupsc{\overline{gh}}a;0)\theta_{\lsupsc{\overline{gh}}a}\varkappa^{*}_{\lsupsc{\overline{gh}}a}K^{q(gh)}\\
    =&\eta'_a(gh,k)\eta'_a(g,h).
\end{align*}
When $g$ and $k$ preserve spatial orientation while $h$ reverses spatial orientation, note that $gh$ and $hk$ both reverse spatial orientation. We check that
\begin{align*}
    &K^{q(g)+p(g)}\eta'_{\rho'_g(a)}(h,k)K^{q(g)+p(g)}\eta'_a(g,h k)\\
    =&K^{q(g)}\eta_{\lsupsc{\bar g}a}(h,k)K^{q(h)}U_{k}(\lsupsc{\overline{gh}}a,\lsupsc{\overline{gh}}\bar a;0)K^{q(gh)}\eta_a(g,hk)\\
    =&\eta_{a}(g,h)\eta_a(gh,k)\times K^{q(gh)}U_{k}(\lsupsc{\overline{gh}} a,\lsupsc{\overline{gh}}\bar a;0)K^{q(gh)}\\
    =&\eta'_a(gh,k)\eta'_a(g,h).
\end{align*}
When $g$ and $h$ preserve spatial orientation, no matter $k$ reverses or preserves spatial orientation, we always have
\begin{align*}
    &K^{q(g)+p(g)}\eta'_{\rho'_g(a)}(h,k)K^{q(g)+p(g)}\eta'_a(g,h k)\\
    =&K^{q(g)}\eta_{\lsupsc{\bar g}a}(h,k)K^{q(h)}\eta_a(g,hk)\\
    =&\eta_{a}(g,h)\eta_a(gh,k)\\
    =&\eta'_a(gh,k)\eta'_a(g,h).
\end{align*}
Summarizing the eight cases, we have Eq. \eqref{eq: internal etaprime etaprime}:
\begin{equation}
    K^{q'(g)}\eta'_{\rho'_g(a)}(h,k)K^{q'(g)}\eta'_a(g,h k)=\eta'_a(g h,k)\eta'_a(g,h).
\end{equation}

The above discussion shows that the data obtained from the map Eq. \eqref{Eq:correspondence app} satisfy the correct set of consistency equations.

\subsection{The correspondence of the gauge transformation}\label{appd:VBGTaftercorrespondence}

In the above, we have explicitly shown that the primed data $U_g'(a,b;c)$ and $\eta'_a(g,h)$ satisfy Eqs. \eqref{eq: internal F Uprime}-\eqref{eq: internal etaprime etaprime}. Now we show that when the original data $U_g(a,b;c)$ and $\eta_a(g,h)$ undergo gauge transformation Eqs. \eqref{Eq:generalFRequivalence}-\eqref{Eq:generaletaequivalence}, which are copied below:

\begin{equation}\label{Eq:generalequivalence}
    \bsp
    &[(F^{abc}_d)_{\{x\}}]_{\{e,\mu,\nu\},\{f,\rho,\sigma\}}\to\sum_{\mu'\nu'\rho'\sigma'}[\Omega^{ab}_e]_{\mu\mu'}[\Omega^{ec}_d]_{\nu\nu'}[(F^{abc}_d)_{\{x\}}]_{\{e,\mu,\nu'\},\{f,\rho',\sigma'\}}[\Omega^{bc}_f]_{\rho'\rho}^{-1}[\Omega^{af}_d]_{\sigma'\sigma}^{-1},\\
    &[(R^{ab}_c)_{\{x\}}]_{\mu\nu}\to \sum_{\mu'\nu'}[\Omega^{ba}_c]_{\mu\mu'}[(R^{ab}_c)_{\{x\}}]_{\mu\nu}[\Omega^{ab}_c]_{\nu'\nu}^{-1},\\
    &[U_g(a,b;c)_{12}]_{\mu\nu}\to \sum_{\mu'\nu'}K^{q(g)}[\Omega^{\lsupsc{\bar g}a\lsupsc{\bar g}b}_{\lsupsc{\bar g}c}]_{\mu\mu'}K^{q(g)}[U_g(a,b;c)_{12}]_{\mu'\nu'}[\Omega^{ab}_c]_{\nu'\nu}^{-1}\times
    \frac{\delta^a_g \delta^b_g}{\delta^c_g},\\
    &\eta_a(g,h)_1\to\frac{\delta^a_{gh}}{\delta^a_g K^{q(g)}\delta^a_hK^{q(g)}}\eta_a(g,h),
    \esp
\end{equation}
where $\Omega^{ab}_c=e^{i\phi^b_{12}}\Omega_0(a,b;c)_{21}$ and $\delta^a_g=\frac{K^{q(g)}e^{i\phi^{\lsupsc{\bar g}a}_{\lsupsc{\bar g}11}}K^{q(g)}}{e^{i\varphi^a_g(x_1)}}$, then the primed data $U_g'(a,b;c)$ and $\eta'_a(g,h)$ transform in the same form as the gauge transformation in $G$-BTC:
\begin{equation}
\bsp
    &U'_g(a,b;c)\to K^{q'(g)}\left(\Omega^{\rho'_{\bar g}(a)\rho'_{\bar g}(b)}_{\rho'_{\bar g}(c)}\right)K^{q'(g)} U'_g(a,b;c)\left(\Omega^{ab}_c\right)^{-1}\times \frac{\Delta^a_g\Delta^b_g}{\Delta^c_g},\\
    &\eta'_a(g,h)\to\eta'_a(g,h)\times\frac{\Delta^a_{gh}}{\Delta^a_g K^{q'(g)}(\Delta_h^{\rho'_{\bar g}(a)})K^{q'(g)}},
\esp
\end{equation}
where
\begin{equation}
    \bsp
    &\Omega^{ab}_c=e^{i\phi^b_{12}}\Omega_0(a,b;c)_{21},\\
    &\Delta^a_g=\left\{\begin{array}{ll} K^{q(g)}e^{i\phi^{\lsupsc{\bar g}a}_{\lsupsc{\bar g}11}}K^{q(g)}e^{-i\varphi^a_g(x_1)} & \textrm{when $g$ preserves spatial orientation,}\\K^{q(g)}e^{i\phi^{\lsupsc{\bar g}a}_{\lsupsc{\bar g}11}}\left(e_{12}^{i\phi^{\lsupsc{\overline g} \bar a}}\Omega_0(\lsupsc{\overline g}{a},\lsupsc{\overline g}\bar a;0)_{21}\right)K^{q(g)}e^{-i\varphi^a_g(x_1)} & \textrm{when $g$ reverses spatial orientation.}\end{array}\right.
    \esp
\end{equation}

To prove the statement, we first note that, according to the definition of $A$ matrices, they should transform under the vertex basis gauge transformation as,
\begin{equation}\label{Eq:CEP A VBGT}
    A^{\bar b\bar a}_{\bar c}\to \frac{\Omega^{\bar a a}_0\Omega^{\bar b b}_0}{\Omega^{\bar c c}_0}\left(\Omega^{ab}_c\right)^*A^{\bar b\bar a}_{\bar c}\left(\Omega^{\bar b\bar a}_{\bar c}\right)^{-1},
\end{equation}
with $\Omega^{ab}_c=e^{i\phi^b_{12}}[\Omega_0(a,b;c)_{21}]$. It is straightforward to show that the transformation of $A$, together with Eq. \eqref{Eq:generalequivalence} keeps the equations Eqs. \eqref{Eq:CEPAAcondition}-\eqref{Eq:CEPFRcondition} invariant. 

When the symmetry $g$ preserves spatial orientation, the CEP map is an identity map, and it is straightforward that
\begin{equation*}
\bsp
    &U'_g(a,b;c)=U_g(a,b;c)\to K^{q(g)}\left(e^{i\phi^{\lsupsc{\bar g}b}_{12}}[\Omega_0(\lsupsc{\bar g}a,\lsupsc{\bar g}b;\lsupsc{\bar g}c)_{21}]\right)K^{q(g)}\times\\
    &[U'_g(a,b;c)]\left(e^{i\phi^b_{12}}[\Omega_0(a,b;c)_{21}]\right)^{-1}\times
    K^{q(g)}\frac{e^{i\phi^{\lsupsc{\bar g}a}_{\lsupsc{\bar g}11}}e^{i\phi^{\lsupsc{\bar g}b}_{\lsupsc{\bar g}11}}}{e^{i\phi^{\lsupsc{\bar g}c}_{\lsupsc{\bar g}11}}}K^{q(g)}\frac{e^{i\varphi^c_g(x_1)}}{e^{i\varphi^a_g(x_1)}e^{i\varphi^b_g(x_1)}},
\esp
\end{equation*}
When the symmetry action $g$ reverses spatial orientation, we find that
\begin{equation*}
\bsp
    &U'_g(a,b;c)=K^{q(g)}(A^{\lsupsc{\overline g}b\lsupsc{\overline g}a}_{\lsupsc{\overline g}c}R^{\lsupsc{\overline g}a\lsupsc{\overline g}b}_{\lsupsc{\overline g}c})K^{q(g)}U_g(a,b;c)\to\\
    &K^{q(g)}\frac{\left(e_{12}^{i\phi^{\lsupsc{\overline g} \bar a}}\Omega_0(\lsupsc{\overline g}{a},\lsupsc{\overline g}\bar a;0)_{21}\right)\left(e_{12}^{i\phi^{\lsupsc{\overline g}\bar b}}\Omega_0(\lsupsc{\overline g}{b},\lsupsc{\overline g}\bar b;0)_{21}\right)}{\left(e_{12}^{i\phi^{\lsupsc{\overline g}\bar c}}\Omega_0(\lsupsc{\overline g}{c},\lsupsc{\overline g}\bar c;0)_{21}\right)}\left(e^{i\phi_{12}^{\lsupsc{\overline g}\bar b}}\Omega_0(\lsupsc{\overline g}\bar a,\lsupsc{\overline g}\bar b;\lsupsc{\overline g}\bar c)_{21}\right)^*A^{\lsupsc{\overline g}b\lsupsc{\overline g}a}_{\lsupsc{\overline g}c}\left(e^{i\phi_{12}^{\lsupsc{\overline g}a}}\Omega_0(\lsupsc{\overline g}b,\lsupsc{\overline g}a;\lsupsc{\overline g}c)_{21}\right)^{-1}\\
    &\left(e^{i\phi^{\lsupsc{\overline g}a}_{12}}[\Omega_0(\lsupsc{\overline g}b,\lsupsc{\overline g}a;\lsupsc{\overline g}c)_{21}]\right)([R^{\lsupsc{\overline g}a\lsupsc{\overline g}b}_{\lsupsc{\overline g}c}])\left(e^{i\phi^{\lsupsc{\overline g}b}_{12}}[\Omega_0(\lsupsc{\overline g}a,\lsupsc{\overline g}b;\lsupsc{\overline g}c)_{21}]\right)^{-1}\left(e^{i\phi^{\lsupsc{\bar g}b}_{12}}[\Omega_0(\lsupsc{\bar g}a,\lsupsc{\bar g}b;\lsupsc{\bar g}c)_{21}]\right)K^{q(g)}\times\\
    &\quad[U_g(a,b;c)]\left(e^{i\phi^b_{12}}[\Omega_0(a,b;c)_{21}]^{-1}\right)\times K^{q(g)}\frac{e^{i\phi^{\lsupsc{\bar g}a}_{\lsupsc{\bar g}11}}e^{i\phi^{\lsupsc{\bar g}b}_{\lsupsc{\bar g}11}}}{e^{i\phi^{\lsupsc{\bar g}c}_{\lsupsc{\bar g}11}}}K^{q(g)}
    \frac{e^{i\varphi^c_g(x_1)}}{e^{i\varphi^a_g(x_1)}e^{i\varphi^b_g(x_1)}}\\
    =&K^{q(g)}\left(e^{i\phi_{12}^{\lsupsc{\overline g}\bar b}}\Omega_0(\lsupsc{\overline g}\bar a,\lsupsc{\overline g}\bar b;\lsupsc{\overline g}\bar c)_{21}\right)^*K^{q(g)}\times K^{q(g)}(A^{\lsupsc{\overline g}b\lsupsc{\overline g}a}_{\lsupsc{\overline g}c}R^{\lsupsc{\overline g}a\lsupsc{\overline g}b}_{\lsupsc{\overline g}c})K^{q(g)}U_g(a,b;c)\times\left(e^{i\phi^b_{12}}[\Omega_0(a,b;c)_{21}]^{-1}\right)\\
    &\times K^{q(g)}\frac{e^{i\phi^{\lsupsc{\bar g}a}_{\lsupsc{\bar g}11}}e^{i\phi^{\lsupsc{\bar g}b}_{\lsupsc{\bar g}11}}}{e^{i\phi^{\lsupsc{\bar g}c}_{\lsupsc{\bar g}11}}}\frac{\left(e_{12}^{i\phi^{\lsupsc{\overline g} \bar a}}\Omega_0(\lsupsc{\overline g}{a},\lsupsc{\overline g}\bar a;0)_{21}\right)\left(e_{12}^{i\phi^{\lsupsc{\overline g}\bar b}}\Omega_0(\lsupsc{\overline g}{b},\lsupsc{\overline g}\bar b;0)_{21}\right)}{\left(e_{12}^{i\phi^{\lsupsc{\overline g}\bar c}}\Omega_0(\lsupsc{\overline g}{c},\lsupsc{\overline g}\bar c;0)_{21}\right)}K^{q(g)}
    \frac{e^{i\varphi^c_g(x_1)}}{e^{i\varphi^a_g(x_1)}e^{i\varphi^b_g(x_1)}}\\
    =&K^{q(g)}\left(e^{i\phi_{12}^{\lsupsc{\overline g}\bar b}}\Omega_0(\lsupsc{\overline g}\bar a,\lsupsc{\overline g}\bar b;\lsupsc{\overline g}\bar c)_{21}\right)^*K^{q(g)} U'_g(a,b;c) \left(e^{i\phi^b_{12}}[\Omega_0(a,b;c)_{21}]^{-1}\right) \\
    &\times K^{q(g)}\frac{e^{i\phi^{\lsupsc{\bar g}a}_{\lsupsc{\bar g}11}}e^{i\phi^{\lsupsc{\bar g}b}_{\lsupsc{\bar g}11}}}{e^{i\phi^{\lsupsc{\bar g}c}_{\lsupsc{\bar g}11}}}\frac{\left(e_{12}^{i\phi^{\lsupsc{\overline g} \bar a}}\Omega_0(\lsupsc{\overline g}{a},\lsupsc{\overline g}\bar a;0)_{21}\right)\left(e_{12}^{i\phi^{\lsupsc{\overline g}\bar b}}\Omega_0(\lsupsc{\overline g}{b},\lsupsc{\overline g}\bar b;0)_{21}\right)}{\left(e_{12}^{i\phi^{\lsupsc{\overline g}\bar c}}\Omega_0(\lsupsc{\overline g}{c},\lsupsc{\overline g}\bar c;0)_{21}\right)}K^{q(g)}\frac{e^{i\varphi^c_g(x_1)}}{e^{i\varphi^a_g(x_1)}e^{i\varphi^b_g(x_1)}}.
\esp
\end{equation*}
Summarizing the above two cases, we find that, under the transformations in Eq. \eqref{Eq:generalequivalence},
\begin{equation}\label{Eq:gaugetransformationUp}
    U'_g(a,b;c)\to K^{q(g)+p(g)}\left(e^{i\phi_{12}^{\rho'_{\bar g}(b)}}\Omega_0(\rho'_{\bar g}(a),\rho'_{\bar g}(b);\rho'_{\bar g}( c))_{21}\right)K^{q(g)+p(g)} U'_g(a,b;c) \left(e^{i\phi^b_{12}}[\Omega_0(a,b;c)_{21}]^{-1}\right)\times \frac{\Delta^a_g\Delta^b_g}{\Delta^c_g},
\end{equation}
where
\begin{equation*}
    \Delta^a_g=\left\{\begin{array}{ll} K^{q(g)}e^{i\phi^{\lsupsc{\bar g}a}_{\lsupsc{\bar g}11}}K^{q(g)}e^{-i\varphi^a_g(x_1)} & \textrm{when $g$ preserves spatial orientation,}\\K^{q(g)}e^{i\phi^{\lsupsc{\bar g}a}_{\lsupsc{\bar g}11}}\left(e_{12}^{i\phi^{\lsupsc{\overline g} \bar a}}\Omega_0(\lsupsc{\overline g}{a},\lsupsc{\overline g}\bar a;0)_{21}\right)K^{q(g)}e^{-i\varphi^a_g(x_1)} & \textrm{when $g$ reverses spatial orientation.}\end{array}\right.
\end{equation*}
This transformation is precisely the gauge transformation for $U$-symbols in $G$-BTC.

For the $\eta'_a(g,h)$-phases, when both $g$ and $h$ reverse spatial orientation, we check that 
\begin{equation*}
\bsp
    &\eta'_a(g,h)=\eta_a(g,h)K^{q(g)}U_{h}(\lsupsc{\overline{g}}\bar a,\lsupsc{\overline{g}}a;0)(R^{\lsupsc{\overline{g}}\bar a\lsupsc{\overline{g}}a}_{0})^{-1}K^{q(g)}\to \\
    &\frac{K^{q(g)}e^{i\varphi_{h}^{\lsupsc{\bar g}a}(x_1)}K^{q(g)}e^{i\varphi_{g}^a(x_1)}}{e^{i\varphi_{gh}^a(x_1)}}\frac{K^{q(gh)}e^{i\phi^{\lsupsc{\overline{gh}}a}_{\lsupsc{\overline{gh}}11}}K^{q(gh)}}{K^{q(g)}e^{i\phi^{\lsupsc{\bar g}a}_{\lsupsc{\overline{g}}11}}K^{q(h)}e^{i\phi^{\lsupsc{\overline{gh}}a}_{\lsupsc{\overline{h}}11}}K^{q(gh)}}\eta_a(g,h)\times\\
    &K^{q(gh)}\left(e^{i\phi^{\lsupsc{\overline{gh}}a}_{12}}[\Omega_0(\lsupsc{\overline{gh}}\bar a,\lsupsc{\overline{gh}}a;0)_{21}]\right)K^{q(h)}U_h(\lsupsc{\bar g}\bar a,\lsupsc{\bar g}a;0)\left(e^{i\phi^{\lsupsc{\bar g}a}_{12}}[\Omega_0(\lsupsc{\bar g}\bar a,\lsupsc{\bar g}a;0)_{21}]\right)^{-1}\times
    \frac{K^{q(h)}e^{i\phi^{\lsupsc{\overline{gh}}\bar a}_{\lsupsc{\overline{h}}11}}e^{i\phi^{\lsupsc{\overline{gh}}a}_{\lsupsc{\overline{h}}11}}K^{q(h)}}{e^{i\varphi^{\lsupsc{\bar g}\bar a}_h(x_1)}e^{i\varphi^{\lsupsc{\bar g}a}_h(x_1)}}\times\\
    &\left(e^{i\phi^{\lsupsc{\overline{g}}\bar a}_{12}}[\Omega_0(\lsupsc{\overline{g}} a,\lsupsc{\overline{g}}\bar a;0)_{21}]\right)^{-1}(R^{\lsupsc{\overline{g}}\bar a\lsupsc{\overline{g}}a}_{0})^{-1}\left(e^{i\phi^{\lsupsc{\overline{g}}a}_{12}}[\Omega_0(\lsupsc{\overline{g}}\bar a,\lsupsc{\overline{g}}a;0)_{21}]\right) K^{q(g)}\\
    &=\frac{K^{q(g)}e^{-i\varphi_{h}^{\lsupsc{\bar g}\bar a}(x_1)}K^{q(g)}e^{i\varphi_{g}^a(x_1)}}{e^{i\varphi_{gh}^a(x_1)}}\frac{K^{q(gh)}e^{i\phi^{\lsupsc{\overline{gh}}a}_{\lsupsc{\overline{gh}}11}}K^{q(gh)}}{K^{q(g)}e^{i\phi^{\lsupsc{\bar g}a}_{\lsupsc{\overline{g}}11}}K^{q(h)}e^{-i\phi^{\lsupsc{\overline{gh}}\bar a}_{\lsupsc{\overline{h}}11}}K^{q(gh)}}\eta_a(g,h)\times\\
    &K^{q(gh)}\left(e^{i\phi^{\lsupsc{\overline{gh}}a}_{12}}[\Omega_0(\lsupsc{\overline{gh}}\bar a,\lsupsc{\overline{gh}}a;0)_{21}]\right)K^{q(h)}U_h(\lsupsc{\bar g}\bar a,\lsupsc{\bar g}a;0)\left(e^{i\phi^{\lsupsc{\overline{g}}\bar a}_{12}}[\Omega_0(\lsupsc{\overline{g}} a,\lsupsc{\overline{g}}\bar a;0)_{21}]\right)^{-1}(R^{\lsupsc{\overline{g}}\bar a\lsupsc{\overline{g}}a}_{0})^{-1}K^{q(g)}\\
    &=\eta_a(g,h)K^{q(g)}U_h(\lsupsc{\bar g}\bar a,\lsupsc{\bar g}a;0)(R^{\lsupsc{\overline{g}}\bar a\lsupsc{\overline{g}}a}_{0})^{-1}K^{q(g)}\times\frac{K^{q(gh)}e^{i\phi^{\lsupsc{\overline{gh}}a}_{\lsupsc{\overline{gh}}11}}K^{q(gh)}e^{-i\varphi_{gh}^a(x_1)}}{K^{q(g)}e^{i\phi^{\lsupsc{\overline{g}}a}_{\lsupsc{\overline{g}}11}}\left(e^{i\phi^{\lsupsc{\overline{g}}\bar a}_{12}}[\Omega_0(\lsupsc{\overline{g}} a,\lsupsc{\overline{g}}\bar a;0)_{21}]\right)K^{q(g)}e^{-i\varphi_{g}^a(x_1)}}\times\\
    &\frac{1}{K^{q(g)}\left(K^{q(h)}e^{i\phi^{\lsupsc{\overline{gh}}\bar a}_{\lsupsc{\overline{h}}11}}\left(e^{i\phi^{\lsupsc{\overline{gh}}a}_{12}}[\Omega_0(\lsupsc{\overline{gh}}\bar a,\lsupsc{\overline{gh}}a;0)_{21}]\right)K^{q(h)}e^{-i\varphi_{h}^{\lsupsc{\bar g}\bar a}(x_1)}\right)^*K^{q(g)}}\\
    &=\eta'_a(g,h)\times\frac{\Delta^a_{gh}}{\Delta^a_g K^{q(g)}(\Delta_h^{\rho'_{\bar g}(a)})^*K^{q(g)}}.
\esp
\end{equation*}
When $g$ reverses spatial orientation while $h$ preserves spatial orientation (keep in mind that $gh$ reverses spatial orientation),
\begin{equation*}
\bsp
    &\eta'_a(g,h)=\eta_a(g,h)K^{q(g)}U_{h}(\lsupsc{\overline{g}} a,\lsupsc{\overline{g}}\bar a;0)K^{q(g)}\to\\
    &\frac{K^{q(g)}e^{i\varphi_{h}^{\lsupsc{\bar g}a}(x_1)}K^{q(g)}e^{i\varphi_{g}^a(x_1)}}{e^{i\varphi_{gh}^a(x_1)}}\frac{K^{q(gh)}e^{i\phi^{\lsupsc{\overline{gh}}a}_{\lsupsc{\overline{gh}}11}}K^{q(gh)}}{K^{q(g)}e^{i\phi^{\lsupsc{\bar g}a}_{\lsupsc{\overline{g}}11}}K^{q(h)}e^{i\phi^{\lsupsc{\overline{gh}}a}_{\lsupsc{\overline{h}}11}}K^{q(gh)}}\eta_a(g,h)\times\\
    &K^{q(gh)}\left(e^{i\phi^{\lsupsc{\overline{gh}}\bar a}_{12}}[\Omega_0(\lsupsc{\overline{gh}}a,\lsupsc{\overline{gh}}\bar a;0)_{21}]\right)K^{q(h)}U_h(\lsupsc{\bar g}a,\lsupsc{\bar g}\bar a;0)\left(e^{i\phi^{\lsupsc{\bar g}\bar a}_{12}}[\Omega_0(\lsupsc{\bar g}a,\lsupsc{\bar g}\bar a;0)_{21}]\right)^{-1}\times    \frac{K^{q(h)}e^{i\phi^{\lsupsc{\overline{gh}}a}_{\lsupsc{\overline{h}}11}}e^{i\phi^{\lsupsc{\overline{gh}}\bar a}_{\lsupsc{\overline{h}}11}}K^{q(h)}}{e^{i\varphi^{\lsupsc{\bar g}a}_h(x_1)}e^{i\varphi^{\lsupsc{\bar g}\bar a}_h(x_1)}}K^{q(g)}\\
    &=\eta_a(g,h)K^{q(g)}U_h(\lsupsc{\bar g}a,\lsupsc{\bar g}\bar a;0)K^{q(g)}\times\frac{K^{q(gh)}e^{i\phi^{\lsupsc{\overline{gh}}a}_{\lsupsc{\overline{gh}}11}}    
    \left(e^{i\phi^{\lsupsc{\overline{gh}}\bar a}_{12}}[\Omega_0(\lsupsc{\overline{gh}}a,\lsupsc{\overline{gh}}\bar a;0)_{21}]\right)K^{q(gh)}e^{-i\varphi_{gh}^a(x_1)}}{K^{q(g)}e^{i\phi^{\lsupsc{\overline{g}}a}_{\lsupsc{\overline{g}}11}}\left(e^{i\phi^{\lsupsc{\overline{g}}\bar a}_{12}}[\Omega_0(\lsupsc{\overline{g}} a,\lsupsc{\overline{g}}\bar a;0)_{21}]\right)K^{q(g)}e^{-i\varphi_{g}^a(x_1)}}\times\\
    &\frac{1}{K^{q(g)}\left(K^{q(h)}e^{i\phi^{\lsupsc{\overline{gh}}\bar a}_{\lsupsc{\overline{h}}11}}K^{q(h)}e^{-i\varphi_{h}^{\lsupsc{\bar g}\bar a}(x_1)}\right)^*K^{q(g)}}\\
    &=\eta'_a(g,h)\times\frac{\Delta^a_{gh}}{\Delta^a_g K^{q(g)}(\Delta_h^{\rho'_{\bar g}(a)})^*K^{q(g)}}.
\esp
\end{equation*}
When $g$ preserves spatial orientation while $h$ reverses spatial orientation,
\begin{equation*}
\bsp
    &\eta'_a(g,h)=\eta_a(g,h)\to\\
    &\frac{K^{q(g)}e^{i\varphi_{h}^{\lsupsc{\bar g}a}(x_1)}K^{q(g)}e^{i\varphi_{g}^a(x_1)}}{e^{i\varphi_{gh}^a(x_1)}}\frac{K^{q(gh)}e^{i\phi^{\lsupsc{\overline{gh}}a}_{\lsupsc{\overline{gh}}11}}K^{q(gh)}}{K^{q(g)}e^{i\phi^{\lsupsc{\bar g}a}_{\lsupsc{\overline{g}}11}}K^{q(h)}e^{i\phi^{\lsupsc{\overline{gh}}a}_{\lsupsc{\overline{h}}11}}K^{q(gh)}}\eta_a(g,h)\\
    &=\eta_a(g,h)\times\frac{K^{q(gh)}e^{i\phi^{\lsupsc{\overline{gh}}a}_{\lsupsc{\overline{gh}}11}}\left(e^{i\phi^{\lsupsc{\overline{gh}}\bar a}_{12}}[\Omega_0(\lsupsc{\overline{gh}}a,\lsupsc{\overline{gh}}\bar a;0)_{21}]\right)K^{q(gh)}e^{-i\varphi_{gh}^a(x_1)}}{K^{q(g)}\left(K^{q(h)}e^{i\phi^{\lsupsc{\overline{gh}}a}_{\lsupsc{\overline{h}}11}}\left(e^{i\phi^{\lsupsc{\overline{gh}}\bar a}_{12}}[\Omega_0(\lsupsc{\overline{gh}}a,\lsupsc{\overline{gh}}\bar a;0)_{21}]K^{q(h)}\right)e^{i\varphi_{h}^{\lsupsc{\bar g}a}(x_1)}\right){K^{q(g)}}\left(K^{q(g)}e^{i\phi^{\lsupsc{\bar g}a}_{\lsupsc{\bar g}11}}K^{q(g)}e^{-i\varphi^a_g(x_1)}\right)}\\
    &=\eta'_a(g,h)\times\frac{\Delta^a_{gh}}{\Delta^a_g K^{q(g)}(\Delta_h^{\rho'_{\bar g}(a)})K^{q(g)}}.
\esp
\end{equation*}
Finally, when $g$ and $h$ both preserve spatial orientation, it is straightforward that 
\begin{equation*}
\bsp
    &\eta'_a(g,h)=\eta_a(g,h)\to\\
    &\frac{K^{q(g)}e^{i\varphi_{h}^{\lsupsc{\bar g}a}(x_1)}K^{q(g)}e^{i\varphi_{g}^a(x_1)}}{e^{i\varphi_{gh}^a(x_1)}}\frac{K^{q(gh)}e^{i\phi^{\lsupsc{\overline{gh}}a}_{\lsupsc{\overline{gh}}11}}K^{q(gh)}}{K^{q(g)}e^{i\phi^{\lsupsc{\bar g}a}_{\lsupsc{\overline{g}}11}}K^{q(h)}e^{i\phi^{\lsupsc{\overline{gh}}a}_{\lsupsc{\overline{h}}11}}K^{q(gh)}}\eta_a(g,h)\\
    &=\eta_a(g,h)\times\frac{K^{q(gh)}e^{i\phi^{\lsupsc{\overline{gh}}a}_{\lsupsc{\overline{gh}}11}}K^{q(gh)}e^{-i\varphi_{gh}^a(x_1)}}{K^{q(g)}\left(K^{q(h)}e^{i\phi^{\lsupsc{\overline{gh}}a}_{\lsupsc{\overline{h}}11}}e^{i\varphi_{h}^{\lsupsc{\bar g}a}(x_1)}\right){K^{q(g)}}\left(K^{q(g)}e^{i\phi^{\lsupsc{\bar g}a}_{\lsupsc{\bar g}11}}K^{q(g)}e^{-i\varphi^a_g(x_1)}\right)}\\
    &=\eta'_a(g,h)\times\frac{\Delta^a_{gh}}{\Delta^a_g K^{q(g)}(\Delta_h^{\rho'_{\bar g}(a)})K^{q(g)}}.
\esp
\end{equation*}

Summarizing the above four cases, under the transformations Eq. \eqref{Eq:generalequivalence}, the $\eta'$-phases transform as
\begin{equation}\label{Eq:gaugetransformationetap}
    \eta'_a(g,h)\to\eta'_a(g,h)\times\frac{\Delta^a_{gh}}{\Delta^a_g K^{q(g)+p(g)}(\Delta_h^{\rho'_{\bar g}(a)})K^{q(g)+p(g)}},
\end{equation}
which is precisely the symmetry action gauge transformation in $G$-BTC.

\subsection{Emergent RT symmetry is not directly related to crystalline equivalence principle} \label{subapp: not RT}

In constructing the map Eq. \eqref{Eq:correspondence}, the emergent CRT symmetry of the TQSLs play a crucial role. As discussed in Sec. \ref{sec:equivalence}, all TQSLs also have an emergent RT symmetry. But unlike the emergent CRT symmetry, this RT symmetry is not directly related to the crystalline equivalence principle. This fact is also why $\rho'_g(a)$ in Eq. \eqref{Eq:correspondence} takes that form. We finish this appendix by giving an intuitive explanation of this point.

Concretely, consider an example of symmetry-protected topological (SPT) phase with a $\z_4$ internal symmetry and a reflection symmetry $R$, which commute with each other. In the effective field theory of this SPT phase, denote by $\psi$ the field that carries a unit $\z_4$ charge, \ie under the $\z_4$ and $R$ symmetries, $\psi$ transforms as
\beq
\bsp
&\z_4: \psi(x)\rightarrow i\psi(x),\\
&R: \psi(x)\rightarrow\psi(^Rx).
\esp
\eeq
Suppose that this SPT phase is related to an SPT phase with a $\z_4$ internal symmetry and a $\z_2^T$ time reversal symmetry, via the crystalline equivalence principle. The crystalline equivalence principle asserts that this $\z_2^T$ symmetry should commute with the $\z_4$ internal symmetry. So if we denote by $\psi'$ the field that carries a unit $\z_4$ charge in this new SPT, then it transforms under the $\z_4$ and $\z_2^T$ symmetry as
\beq
\bsp
&\z_4: \psi'(x)\rightarrow i\psi'(x),\\
&\z_2^T: \psi'(x)\rightarrow\psi'^\dagger(x),
\esp
\eeq
where the hermitian conjugate in the $\z_2^T$ action is needed for the actions of $\z_4$ and $\z_2^T$ to commute. Note that this $\z_2^T$ is often considered as the CT symmetry in the field theoretic context, rather than the T symmetry. This observation already suggests that it is CRT, rather than RT, that is the building block of the crystalline equivalence principle.

Now suppose that we gauge the internal $\z_4$ symmetry, which turns the above two SPT phases into a $\z_4$ topological order with a reflection symmetry $R$ and a $\z_4$ topological order with a time reversal symmetry $\z_2^T$. These two $\z_4$ topological orders should still be related by the crystalline equivalence principle. From the actions of $R$ and $\z_2^T$ in the two SPT phases, we see that the anyon permutation patterns in these two gauged theories should be related as in Eq. \eqref{Eq:correspondence}.

\section{Symmetry-enriched $\z_2$ TQSLs} \label{app: SET Z2}

In this appendix, we give more details on symmetry-enriched $\z_2$ TQSLs discussed in Sec. \ref{sec:application}. We will explain the physical meanings of $\eta'_e(g_1, g_2)$ and $\eta'_m(g_1, g_2)$, and explain how to obtain the expansion coefficients in Eqs. \eqref{eq: coefficients toric code}, \eqref{eq: coefficients flipped toric code} and \eqref{eq: coefficients ruby}.

\subsection{$p4m\times\z_2^T$ symmetric $\z_2$ TQSLs}

We start with $p4m\times\z_2^T$ symmetric $\z_2$ TQSLs, where the symmetries do not permute the anyons. These TQSLs are characterized by the fractionalization pattern of the $p4m\times\z_2^T$ symmetry on the $e$ and $m$ anyons, which, mathematically, are characterized by an element in $\mathcal{H}^2(p4m\times\z_2^T, \mathbb{Z}_2\times \mathbb{Z}_2)$ \cite{Ye2023}. The physical meanings of $\eta'_e(g_1, g_2)$ and $\eta'_m(g_1, g_2)$ can be read off using the following 20 ``cohomology invariants":
\begin{equation} \label{eq: invaraints p4m}
    \bsp
        &\frac{\eta'_e(C_2,C_2)}{\eta'_e(\mathbbm{1}, \mathbbm{1})}, \frac{\eta'_e(T_1T_2C_2,T_1T_2C_2)}{\eta'_e(\mathbbm{1}, \mathbbm{1})}, \frac{\eta'_e(T_1C_2,T_1C_2)}{\eta'_e(\mathbbm{1}, \mathbbm{1})},\frac{\eta'_e(M,M)}{\eta'_e(\mathbbm{1}, \mathbbm{1})},
         \frac{\eta'_e(T_1M,T_1M)}{\eta'_e(\mathbbm{1}, \mathbbm{1})},\\
         &\frac{\eta'_e(C_4M,C_4M)}{\eta'_e(\mathbbm{1}, \mathbbm{1})},\frac{\eta'_e(T_1,\mathcal{T})}{\eta'_e(\mathcal{T},T_1)}, \frac{\eta'_e(C_4,\mathcal{T})}{\eta'_e(\mathcal{T},C_4)},
         \frac{\eta'_e(M,\mathcal{T})}{\eta'_e(\mathcal{T},M)}, \eta'_e(\mathcal{T},\mathcal{T})\eta'_e(\mathbbm{1}, \mathbbm{1}),\\
         &\frac{\eta'_m(C_2,C_2)}{\eta'_m(\mathbbm{1}, \mathbbm{1})}, \frac{\eta'_m(T_1T_2C_2,T_1T_2C_2)}{\eta'_m(\mathbbm{1}, \mathbbm{1})}, \frac{\eta'_m(T_1C_2,T_1C_2)}{\eta'_m(\mathbbm{1}, \mathbbm{1})},\frac{\eta'_m(M,M)}{\eta'_m(\mathbbm{1}, \mathbbm{1})},
         \frac{\eta'_m(T_1M,T_1M)}{\eta'_m(\mathbbm{1}, \mathbbm{1})},\\
         &\frac{\eta'_m(C_4M,C_4M)}{\eta'_m(\mathbbm{1}, \mathbbm{1})},\frac{\eta'_m(T_1,\mathcal{T})}{\eta'_m(\mathcal{T},T_1)}, \frac{\eta'_m(C_4,\mathcal{T})}{\eta'_m(\mathcal{T},C_4)},
         \frac{\eta'_m(M,\mathcal{T})}{\eta'_m(\mathcal{T},M)}, \eta'_m(\mathcal{T},\mathcal{T})\eta'_m(\mathbbm{1}, \mathbbm{1}).
    \esp
\end{equation}
These 20 quantities are invariant under the coboundary transformations relevant to $\mathcal{H}^2(p4m\times\z_2^T, \mathbb{Z}_2\times \mathbb{Z}_2)$, and their values completely characterize an element in this cohomology. They are also listed in Table XXVI of Ref. \cite{Ye2023}. The meaning of each of them can be easily speculated from its definition. For example, roughly speaking, $\eta'_e(C_2, C_2)/\eta'_e(\mathbbm{1}, \mathbbm{1})=1$ and $\eta'_e(C_2, C_2)/\eta'_e(\mathbbm{1}, \mathbbm{1})=-1$ mean that acting $C_2$ twice on $e$ results in a $1$ and $-1$ phase factor, respectively, $\frac{\eta_m(T_1, \mc{T})}{\eta_m(\mc{T}, T_1)}=1$ and $\frac{\eta_m(T_1, \mc{T})}{\eta_m(\mc{T}, T_1)}=-1$ mean that $T_1$ and $\mc{T}$ commute and anti-commute on $m$, respectively, and $\eta'_m(\mathcal{T},\mathcal{T})\eta'_m(\mathbbm{1}, \mathbbm{1})=1$ and $\eta'_m(\mathcal{T},\mathcal{T})\eta'_m(\mathbbm{1}, \mathbbm{1})=-1$ mean that $m$ is a Kramers doublet under $\z_2^T$.

To get Eq. \eqref{eq: coefficients toric code} for the toric code, we first use our crystalline equivalence principle Eq. \eqref{Eq:correspondence} to get $\eta'_e(g_1, g_2)$ and $\eta'_m(g_1, g_2)$, based on which we can calculate the invariants in Eq. \eqref{eq: invaraints p4m}. The results are
\begin{equation} \label{eq: invaraints toric code}
    \bsp
        &\frac{\eta'_e(C_2,C_2)}{\eta'_e(\mathbbm{1}, \mathbbm{1})}=1, \frac{\eta'_e(T_1T_2C_2,T_1T_2C_2)}{\eta'_e(\mathbbm{1}, \mathbbm{1})}=1, \frac{\eta'_e(T_1C_2,T_1C_2)}{\eta'_e(\mathbbm{1}, \mathbbm{1})}=1,\frac{\eta'_e(M,M)}{\eta'_e(\mathbbm{1}, \mathbbm{1})}=1,
         \frac{\eta'_e(T_1M,T_1M)}{\eta'_e(\mathbbm{1}, \mathbbm{1})}=1,\\
         &\frac{\eta'_e(C_4M,C_4M)}{\eta'_e(\mathbbm{1}, \mathbbm{1})}=1,\frac{\eta'_e(T_1,\mathcal{T})}{\eta'_e(\mathcal{T},T_1)}=-1, \frac{\eta'_e(C_4,\mathcal{T})}{\eta'_e(\mathcal{T},C_4)}=1,
         \frac{\eta'_e(M,\mathcal{T})}{\eta'_e(\mathcal{T},M)}=1, \eta'_e(\mathcal{T},\mathcal{T})\eta'_e(\mathbbm{1}, \mathbbm{1})=1,\\
         &\frac{\eta'_m(C_2,C_2)}{\eta'_m(\mathbbm{1}, \mathbbm{1})}=1, \frac{\eta'_m(T_1T_2C_2,T_1T_2C_2)}{\eta'_m(\mathbbm{1}, \mathbbm{1})}=1, \frac{\eta'_m(T_1C_2,T_1C_2)}{\eta'_m(\mathbbm{1}, \mathbbm{1})}=1,\frac{\eta'_m(M,M)}{\eta'_m(\mathbbm{1}, \mathbbm{1})}=1,
         \frac{\eta'_m(T_1M,T_1M)}{\eta'_m(\mathbbm{1}, \mathbbm{1})}=1,\\
         &\frac{\eta'_m(C_4M,C_4M)}{\eta'_m(\mathbbm{1}, \mathbbm{1})}=1,\frac{\eta'_m(T_1,\mathcal{T})}{\eta'_m(\mathcal{T},T_1)}=-1, \frac{\eta'_m(C_4,\mathcal{T})}{\eta'_m(\mathcal{T},C_4)}=-1,
         \frac{\eta'_m(M,\mathcal{T})}{\eta'_m(\mathcal{T},M)}=-1, \eta'_m(\mathcal{T},\mathcal{T})\eta'_m(\mathbbm{1}, \mathbbm{1})=1.
    \esp
\end{equation}

On the other hand, one can calculate the invariants in Eq. \eqref{eq: invaraints p4m} for each entry in Eq. \eqref{Eq:p4mbasis}. By matching the obtained values with Eq. \eqref{eq: invaraints toric code}, we can get Eq. \eqref{eq: coefficients toric code}. To finish this calculation, we need the definitions of each entry in Eq. \eqref{Eq:p4mbasis}, which is reproduced from Appendix B. 4 in Ref. \cite{Ye2023}:
\begin{equation} \label{eq: p4m basis cocycles}
	\bsp
	&B_{xy}(g_1,g_2)=P_c(c_1)y_1x_2+P(c_1)y_2(y_1+x_2),\\
	&B_{c^2}(g_1,g_2)=\frac{1}{4}\left([c_1]_4+(-1)^{m_1}[c_2]_4-[c_1+(-1)^{m_1}c_2]_4\right),\\
	&A^2_{x+y}(g_1,g_2)=(x_1+y_1)(x_2+y_2),\\
	&A_{x+y}A_m(g_1,g_2)=m_1(x_2+y_2),\\
	&A_{c}^2=c_1c_2,\quad A_m^2=m_1m_2, \quad A_ct=c_1t_2,\\
	&A_{x+y}t=(x_1+y_1)t_2,\quad A_mt=m_1t_2,\quad t^2=t_1t_2,
	\esp
\end{equation}
where we have labeled each group element as $g=T_1^xT_2^yC_4^cM^m\mathcal{T}^t$ with $x,y \in \mathbb{Z}$, $c\in \{0, 1, 2, 3\}$, and $m,t\in \{0, 1\}$. The functions $P$, $P_c$, and $[x]_a$ are
\begin{equation}\label{Eq:defineP}
	\bsp
	&P(x)=\left\{\begin{array}{ll}1,&x\text{ is odd,}\\0,&x \text{ is even,}\end{array}\right.\quad P_c(x)=1-P(x),\\
	&[x]_a=\{y=x\;(\text{mod } a)|0\leqslant y<x\}.
	\esp
\end{equation}

Similar analysis also leads to Eq. \eqref{eq: coefficients flipped toric code} for the flipped toric code.

\subsection{$p6m\times\z_2^T$ symmetric $\z_2$ TQSLs}

Next, we turn to $p6m\times\z_2^T$ symmetric $\z_2$ TQSLs, where the symmetries do not permute anyons. These TQSLs are characterized by the fractionalization pattern of the $p6m\times\z_2^T$ symmetry on the $e$ and $m$ anyons, which, mathematically, are characterized by an elment in $\mc{H}^2(p6m\times\z_2^T, \z_2\times\z_2)$ \cite{Ye2023}. The physical meanings of $\eta'_e(g_1, g_2)$ and $\eta'_m(g_1, g_2)$ can be read off from the following 14 ``cohomology invariants":
\beq \label{eq: invariants p6m}
\bsp
&\frac{\eta'_e(C_2, C_2)}{\eta'_e(\mathbbm{1}, \mathbbm{1})}, \frac{\eta'_e(T_1C_2, T_1C_2)}{\eta'_e(\mathbbm{1}, \mathbbm{1})},
\eta'_e(C_2M, C_2M)\eta'_e(\mathbbm{1}, \mathbbm{1}), \eta'_e(M, M)\eta'_e(\mathbbm{1}, \mathbbm{1}),\\
&\frac{\eta'_e(C_6, \mc{T})}{\eta'_e(\mc{T}, C_6)}, \frac{\eta'_e(M, \mc{T})}{\eta'_e(\mc{T}, M)}, \eta'_e(\mc{T}, \mc{T})\eta'_e(\mathbbm{1}, \mathbbm{1}),\\
&\frac{\eta'_m(C_2, C_2)}{\eta'_m(\mathbbm{1}, \mathbbm{1})}, \frac{\eta'_m(T_1C_2, T_1C_2)}{\eta'_m(\mathbbm{1}, \mathbbm{1})},
\eta'_m(C_2M, C_2M)\eta'_m(\mathbbm{1}, \mathbbm{1}), \eta'_m(M, M)\eta'_m(\mathbbm{1}, \mathbbm{1}),\\
&\frac{\eta'_m(C_6, \mc{T})}{\eta'_m(\mc{T}, C_6)}, \frac{\eta'_m(M, \mc{T})}{\eta'_m(\mc{T}, M)}, \eta'_m(\mc{T}, \mc{T})\eta'_m(\mathbbm{1}, \mathbbm{1}).
\esp
\eeq
These 14 quantities are invariant under the coboundary transformations relevant to $\mc{H}^2(p6m\times\z_2^T, \z_2\times\z_2)$, and their values completely characterize an element in this cohomology. The meaning of each of them can be similarly read off, as in the case of $p4m\times\z_2^T$ symmetric $\z_2$ TQSLs.

To get Eq. \eqref{eq: coefficients ruby}, we first use our crystalline equivalence principle Eq. \eqref{Eq:correspondence} to get $\eta'_e(g_1, g_2)$ and $\eta'_m(g_1, g_2)$, based on which we can calculate the invariants in Eq. \eqref{eq: invariants p6m}. We find that all 14 invariants are 1 in this case, which means that we can take $\vec\chi^e=\vec\chi^m=\vec 0$.

For completeness, we reproduce the definition of the 2-cocycles in Eq. \eqref{eq: vector basis p6m} from Appendix B.2 in Ref. \cite{Ye2023}.
\begin{equation} \label{eq: basis p6m}
	\bsp
	&B_{xy}(g_1,g_2)=P_{60}[P_c(m_1)y_1x_2+m_1y_2(x_2+y_1)]\\
	&\quad+P_{61}\left[P_c(m_1)\left(\frac{x_2(x_2-1)}{2}+y_1x_2-y_2(x_2+y_1)\right)\right.\\
	&\quad\left.+m_1\left(\frac{y_2(y_2-1)}{2}+y_1(y_2-x_2)\right)\right]\\
	&\quad+P_{62}\left[P_c(m_1)\left(\frac{y_2(y_2-1)}{2}-x_2-y_2(x_2+y_1)\right)\right.\\
	&\quad\left.+m_1\left(\frac{x_2(x_2-1)}{2}-y_2-y_1x_2\right)\right]\\
	&\quad+P_{63}[P_c(m_1)(y_2-x_2-y_1x_2)+m_1(x_2-y_2+y_2(x_2-y_1)]\\
	&\quad+P_{64}\left[P_c(m_1)\left(\frac{x_2(x_2-1)}{2}+y_2-y_1x_2-y_2(x_2-y_1)\right)\right.\\
	&\quad\left.+m_1\left(\frac{y_2(y_2-1)}{2}+x_2+y_1(x_2-y_2)\right)\right]\\
	&\quad+P_{65}\left[P_c(m_1)\left(\frac{y_2(y_2-1)}{2}-y_2(x_2-y_1)\right)\right.\\
	&\quad\left.+m_1\left(\frac{x_2(x_2-1)}{2}+y_1x_2\right)\right],\\
	&A_{c}A_m(g_1,g_2)=m_1c_2,\quad A_{c}^2=c_1c_2,\\
	&A_m^2=m_1m_2, \quad A_ct=c_1t_2,\\
	&A_mt=m_1t_2,\quad t^2=t_1t_2.
	\esp
\end{equation}
where we have labeled each group element as $g=T_1^xT_2^yC_6^cM^m\mathcal{T}^t$ with $x,y \in \mathbb{Z}$, $c\in\{0, 1, 2, 3, 4, 5\}$, and $m,t\in \{0, 1\}$. The function $P_c$ is defined in Eq. \eqref{Eq:defineP} and $P_{ab}$ is defined by
\begin{equation}
	\bsp
	&P_{ab}(x)=\left\{\begin{array}{ll}1,&x=b\text{ mod }a,\\0,&\text{ otherwise.}\end{array}\right.
	\esp
\end{equation}

\end{appendices}

\bibliography{lib.bib}

\begin{thebibliography}{91}%
\makeatletter
\providecommand \@ifxundefined [1]{%
 \@ifx{#1\undefined}
}%
\providecommand \@ifnum [1]{%
 \ifnum #1\expandafter \@firstoftwo
 \else \expandafter \@secondoftwo
 \fi
}%
\providecommand \@ifx [1]{%
 \ifx #1\expandafter \@firstoftwo
 \else \expandafter \@secondoftwo
 \fi
}%
\providecommand \natexlab [1]{#1}%
\providecommand \enquote  [1]{``#1''}%
\providecommand \bibnamefont  [1]{#1}%
\providecommand \bibfnamefont [1]{#1}%
\providecommand \citenamefont [1]{#1}%
\providecommand \href@noop [0]{\@secondoftwo}%
\providecommand \href [0]{\begingroup \@sanitize@url \@href}%
\providecommand \@href[1]{\@@startlink{#1}\@@href}%
\providecommand \@@href[1]{\endgroup#1\@@endlink}%
\providecommand \@sanitize@url [0]{\catcode `\\12\catcode `\$12\catcode
  `\&12\catcode `\#12\catcode `\^12\catcode `\_12\catcode `\%12\relax}%
\providecommand \@@startlink[1]{}%
\providecommand \@@endlink[0]{}%
\providecommand \url  [0]{\begingroup\@sanitize@url \@url }%
\providecommand \@url [1]{\endgroup\@href {#1}{\urlprefix }}%
\providecommand \urlprefix  [0]{URL }%
\providecommand \Eprint [0]{\href }%
\providecommand \doibase [0]{http://dx.doi.org/}%
\providecommand \selectlanguage [0]{\@gobble}%
\providecommand \bibinfo  [0]{\@secondoftwo}%
\providecommand \bibfield  [0]{\@secondoftwo}%
\providecommand \translation [1]{[#1]}%
\providecommand \BibitemOpen [0]{}%
\providecommand \bibitemStop [0]{}%
\providecommand \bibitemNoStop [0]{.\EOS\space}%
\providecommand \EOS [0]{\spacefactor3000\relax}%
\providecommand \BibitemShut  [1]{\csname bibitem#1\endcsname}%
\let\auto@bib@innerbib\@empty
%</preamble>
\bibitem [{\citenamefont {Wen}(2004)}]{wen2004quantum}%
  \BibitemOpen
  \bibfield  {author} {\bibinfo {author} {\bibfnamefont {X.G.}\ \bibnamefont
  {Wen}},\ }\href {https://books.google.ca/books?id=RYESDAAAQBAJ} {\emph
  {\bibinfo {title} {Quantum Field Theory of Many-Body Systems: From the Origin
  of Sound to an Origin of Light and Electrons}}},\ Oxford Graduate Texts\
  (\bibinfo  {publisher} {OUP Oxford},\ \bibinfo {year} {2004})\BibitemShut
  {NoStop}%
\bibitem [{\citenamefont {{Kitaev}}(2003)}]{Kitaev1997}%
  \BibitemOpen
  \bibfield  {author} {\bibinfo {author} {\bibfnamefont {A.~Yu.}\ \bibnamefont
  {{Kitaev}}},\ }\bibfield  {title} {\enquote {\bibinfo {title}
  {{Fault-tolerant quantum computation by anyons}},}\ }\href {\doibase
  10.1016/S0003-4916(02)00018-0} {\bibfield  {journal} {\bibinfo  {journal}
  {Annals of Physics}\ }\textbf {\bibinfo {volume} {303}},\ \bibinfo {pages}
  {2--30} (\bibinfo {year} {2003})},\ \Eprint
  {http://arxiv.org/abs/quant-ph/9707021} {arXiv:quant-ph/9707021 [quant-ph]}
  \BibitemShut {NoStop}%
\bibitem [{\citenamefont {Freedman}\ \emph {et~al.}(2003)\citenamefont
  {Freedman}, \citenamefont {Kitaev}, \citenamefont {Larsen},\ and\
  \citenamefont {Wang}}]{Freedman2002}%
  \BibitemOpen
  \bibfield  {author} {\bibinfo {author} {\bibfnamefont {Michael~H.}\
  \bibnamefont {Freedman}}, \bibinfo {author} {\bibfnamefont {Alexei}\
  \bibnamefont {Kitaev}}, \bibinfo {author} {\bibfnamefont {Michael~J.}\
  \bibnamefont {Larsen}}, \ and\ \bibinfo {author} {\bibfnamefont {Zhenghan}\
  \bibnamefont {Wang}},\ }\bibfield  {title} {\enquote {\bibinfo {title}
  {Topological quantum computation},}\ }\href {\doibase
  10.1090/S0273-0979-02-00964-3} {\bibfield  {journal} {\bibinfo  {journal}
  {Bulletin of the American Mathematical Society}\ }\textbf {\bibinfo {volume}
  {40}},\ \bibinfo {pages} {31--38} (\bibinfo {year} {2003})},\ \Eprint
  {http://arxiv.org/abs/quant-ph/0101025} {arXiv:quant-ph/0101025 [quant-ph]}
  \BibitemShut {NoStop}%
\bibitem [{\citenamefont {{Nayak}}\ \emph {et~al.}(2008)\citenamefont
  {{Nayak}}, \citenamefont {{Simon}}, \citenamefont {{Stern}}, \citenamefont
  {{Freedman}},\ and\ \citenamefont {{Das Sarma}}}]{Nayak2007}%
  \BibitemOpen
  \bibfield  {author} {\bibinfo {author} {\bibfnamefont {Chetan}\ \bibnamefont
  {{Nayak}}}, \bibinfo {author} {\bibfnamefont {Steven~H.}\ \bibnamefont
  {{Simon}}}, \bibinfo {author} {\bibfnamefont {Ady}\ \bibnamefont {{Stern}}},
  \bibinfo {author} {\bibfnamefont {Michael}\ \bibnamefont {{Freedman}}}, \
  and\ \bibinfo {author} {\bibfnamefont {Sankar}\ \bibnamefont {{Das Sarma}}},\
  }\bibfield  {title} {\enquote {\bibinfo {title} {{Non-Abelian anyons and
  topological quantum computation}},}\ }\href {\doibase
  10.1103/RevModPhys.80.1083} {\bibfield  {journal} {\bibinfo  {journal}
  {Reviews of Modern Physics}\ }\textbf {\bibinfo {volume} {80}},\ \bibinfo
  {pages} {1083--1159} (\bibinfo {year} {2008})},\ \Eprint
  {http://arxiv.org/abs/0707.1889} {arXiv:0707.1889 [cond-mat.str-el]}
  \BibitemShut {NoStop}%
\bibitem [{\citenamefont {Broholm}\ \emph {et~al.}(2020)\citenamefont
  {Broholm}, \citenamefont {Cava}, \citenamefont {Kivelson}, \citenamefont
  {Nocera}, \citenamefont {Norman},\ and\ \citenamefont
  {Senthil}}]{Broholm2019}%
  \BibitemOpen
  \bibfield  {author} {\bibinfo {author} {\bibfnamefont {C.}~\bibnamefont
  {Broholm}}, \bibinfo {author} {\bibfnamefont {R.~J.}\ \bibnamefont {Cava}},
  \bibinfo {author} {\bibfnamefont {S.~A.}\ \bibnamefont {Kivelson}}, \bibinfo
  {author} {\bibfnamefont {D.~G.}\ \bibnamefont {Nocera}}, \bibinfo {author}
  {\bibfnamefont {M.~R.}\ \bibnamefont {Norman}}, \ and\ \bibinfo {author}
  {\bibfnamefont {T.}~\bibnamefont {Senthil}},\ }\bibfield  {title} {\enquote
  {\bibinfo {title} {Quantum spin liquids},}\ }\href {\doibase
  10.1126/science.aay0668} {\bibfield  {journal} {\bibinfo  {journal}
  {Science}\ }\textbf {\bibinfo {volume} {367}},\ \bibinfo {pages} {eaay0668}
  (\bibinfo {year} {2020})},\ \Eprint {http://arxiv.org/abs/1905.07040}
  {arXiv:1905.07040 [cond-mat.str-el]} \BibitemShut {NoStop}%
\bibitem [{\citenamefont {{Levin}}\ and\ \citenamefont
  {{Wen}}(2005)}]{Levin2004}%
  \BibitemOpen
  \bibfield  {author} {\bibinfo {author} {\bibfnamefont {Michael~A.}\
  \bibnamefont {{Levin}}}\ and\ \bibinfo {author} {\bibfnamefont {Xiao-Gang}\
  \bibnamefont {{Wen}}},\ }\bibfield  {title} {\enquote {\bibinfo {title}
  {{String-net condensation:{\quad}A physical mechanism for topological
  phases}},}\ }\href {\doibase 10.1103/PhysRevB.71.045110} {\bibfield
  {journal} {\bibinfo  {journal} {\prb}\ }\textbf {\bibinfo {volume} {71}},\
  \bibinfo {eid} {045110} (\bibinfo {year} {2005})},\ \Eprint
  {http://arxiv.org/abs/cond-mat/0404617} {arXiv:cond-mat/0404617
  [cond-mat.str-el]} \BibitemShut {NoStop}%
\bibitem [{\citenamefont {{Kitaev}}(2006)}]{Kitaev2006}%
  \BibitemOpen
  \bibfield  {author} {\bibinfo {author} {\bibfnamefont {Alexei}\ \bibnamefont
  {{Kitaev}}},\ }\bibfield  {title} {\enquote {\bibinfo {title} {{Anyons in an
  exactly solved model and beyond}},}\ }\href {\doibase
  10.1016/j.aop.2005.10.005} {\bibfield  {journal} {\bibinfo  {journal} {Annals
  of Physics}\ }\textbf {\bibinfo {volume} {321}},\ \bibinfo {pages} {2--111}
  (\bibinfo {year} {2006})},\ \Eprint {http://arxiv.org/abs/cond-mat/0506438}
  {arXiv:cond-mat/0506438 [cond-mat.mes-hall]} \BibitemShut {NoStop}%
\bibitem [{\citenamefont {{Gong}}\ \emph {et~al.}(2014)\citenamefont {{Gong}},
  \citenamefont {{Zhu}},\ and\ \citenamefont {{Sheng}}}]{Gong2013}%
  \BibitemOpen
  \bibfield  {author} {\bibinfo {author} {\bibfnamefont {Shou-Shu}\
  \bibnamefont {{Gong}}}, \bibinfo {author} {\bibfnamefont {Wei}\ \bibnamefont
  {{Zhu}}}, \ and\ \bibinfo {author} {\bibfnamefont {D.~N.}\ \bibnamefont
  {{Sheng}}},\ }\bibfield  {title} {\enquote {\bibinfo {title} {{Emergent
  Chiral Spin Liquid: Fractional Quantum Hall Effect in a Kagome Heisenberg
  Model}},}\ }\href {\doibase 10.1038/srep06317} {\bibfield  {journal}
  {\bibinfo  {journal} {Scientific Reports}\ }\textbf {\bibinfo {volume} {4}},\
  \bibinfo {eid} {6317} (\bibinfo {year} {2014})},\ \Eprint
  {http://arxiv.org/abs/1312.4519} {arXiv:1312.4519 [cond-mat.str-el]}
  \BibitemShut {NoStop}%
\bibitem [{\citenamefont {{Gong}}\ \emph {et~al.}(2015)\citenamefont {{Gong}},
  \citenamefont {{Zhu}}, \citenamefont {{Balents}},\ and\ \citenamefont
  {{Sheng}}}]{Gong2014}%
  \BibitemOpen
  \bibfield  {author} {\bibinfo {author} {\bibfnamefont {Shou-Shu}\
  \bibnamefont {{Gong}}}, \bibinfo {author} {\bibfnamefont {Wei}\ \bibnamefont
  {{Zhu}}}, \bibinfo {author} {\bibfnamefont {Leon}\ \bibnamefont {{Balents}}},
  \ and\ \bibinfo {author} {\bibfnamefont {D.~N.}\ \bibnamefont {{Sheng}}},\
  }\bibfield  {title} {\enquote {\bibinfo {title} {{Global phase diagram of
  competing ordered and quantum spin-liquid phases on the kagome lattice}},}\
  }\href {\doibase 10.1103/PhysRevB.91.075112} {\bibfield  {journal} {\bibinfo
  {journal} {\prb}\ }\textbf {\bibinfo {volume} {91}},\ \bibinfo {eid} {075112}
  (\bibinfo {year} {2015})},\ \Eprint {http://arxiv.org/abs/1412.1571}
  {arXiv:1412.1571 [cond-mat.str-el]} \BibitemShut {NoStop}%
\bibitem [{\citenamefont {{Heinrich}}\ \emph {et~al.}(2016)\citenamefont
  {{Heinrich}}, \citenamefont {{Burnell}}, \citenamefont {{Fidkowski}},\ and\
  \citenamefont {{Levin}}}]{Heinrich2016}%
  \BibitemOpen
  \bibfield  {author} {\bibinfo {author} {\bibfnamefont {Chris}\ \bibnamefont
  {{Heinrich}}}, \bibinfo {author} {\bibfnamefont {Fiona}\ \bibnamefont
  {{Burnell}}}, \bibinfo {author} {\bibfnamefont {Lukasz}\ \bibnamefont
  {{Fidkowski}}}, \ and\ \bibinfo {author} {\bibfnamefont {Michael}\
  \bibnamefont {{Levin}}},\ }\bibfield  {title} {\enquote {\bibinfo {title}
  {{Symmetry-enriched string nets: Exactly solvable models for SET phases}},}\
  }\href {\doibase 10.1103/PhysRevB.94.235136} {\bibfield  {journal} {\bibinfo
  {journal} {\prb}\ }\textbf {\bibinfo {volume} {94}},\ \bibinfo {eid} {235136}
  (\bibinfo {year} {2016})},\ \Eprint {http://arxiv.org/abs/1606.07816}
  {arXiv:1606.07816 [cond-mat.str-el]} \BibitemShut {NoStop}%
\bibitem [{\citenamefont {{Cheng}}\ \emph {et~al.}(2017)\citenamefont
  {{Cheng}}, \citenamefont {{Gu}}, \citenamefont {{Jiang}},\ and\ \citenamefont
  {{Qi}}}]{Cheng2016}%
  \BibitemOpen
  \bibfield  {author} {\bibinfo {author} {\bibfnamefont {Meng}\ \bibnamefont
  {{Cheng}}}, \bibinfo {author} {\bibfnamefont {Zheng-Cheng}\ \bibnamefont
  {{Gu}}}, \bibinfo {author} {\bibfnamefont {Shenghan}\ \bibnamefont
  {{Jiang}}}, \ and\ \bibinfo {author} {\bibfnamefont {Yang}\ \bibnamefont
  {{Qi}}},\ }\bibfield  {title} {\enquote {\bibinfo {title} {{Exactly solvable
  models for symmetry-enriched topological phases}},}\ }\href {\doibase
  10.1103/PhysRevB.96.115107} {\bibfield  {journal} {\bibinfo  {journal}
  {\prb}\ }\textbf {\bibinfo {volume} {96}},\ \bibinfo {eid} {115107} (\bibinfo
  {year} {2017})},\ \Eprint {http://arxiv.org/abs/1606.08482} {arXiv:1606.08482
  [cond-mat.str-el]} \BibitemShut {NoStop}%
\bibitem [{\citenamefont {{Gong}}\ \emph {et~al.}(2017)\citenamefont {{Gong}},
  \citenamefont {{Zhu}}, \citenamefont {{Zhu}}, \citenamefont {{Sheng}},\ and\
  \citenamefont {{Yang}}}]{Gong2017}%
  \BibitemOpen
  \bibfield  {author} {\bibinfo {author} {\bibfnamefont {Shou-Shu}\
  \bibnamefont {{Gong}}}, \bibinfo {author} {\bibfnamefont {W.}~\bibnamefont
  {{Zhu}}}, \bibinfo {author} {\bibfnamefont {J.-X.}\ \bibnamefont {{Zhu}}},
  \bibinfo {author} {\bibfnamefont {D.~N.}\ \bibnamefont {{Sheng}}}, \ and\
  \bibinfo {author} {\bibfnamefont {Kun}\ \bibnamefont {{Yang}}},\ }\bibfield
  {title} {\enquote {\bibinfo {title} {{Global phase diagram and quantum spin
  liquids in a spin-1/2 triangular antiferromagnet}},}\ }\href {\doibase
  10.1103/PhysRevB.96.075116} {\bibfield  {journal} {\bibinfo  {journal}
  {\prb}\ }\textbf {\bibinfo {volume} {96}},\ \bibinfo {eid} {075116} (\bibinfo
  {year} {2017})},\ \Eprint {http://arxiv.org/abs/1705.00510} {arXiv:1705.00510
  [cond-mat.str-el]} \BibitemShut {NoStop}%
\bibitem [{\citenamefont {{Verresen}}\ \emph {et~al.}(2021)\citenamefont
  {{Verresen}}, \citenamefont {{Lukin}},\ and\ \citenamefont
  {{Vishwanath}}}]{Verresen2020}%
  \BibitemOpen
  \bibfield  {author} {\bibinfo {author} {\bibfnamefont {Ruben}\ \bibnamefont
  {{Verresen}}}, \bibinfo {author} {\bibfnamefont {Mikhail~D.}\ \bibnamefont
  {{Lukin}}}, \ and\ \bibinfo {author} {\bibfnamefont {Ashvin}\ \bibnamefont
  {{Vishwanath}}},\ }\bibfield  {title} {\enquote {\bibinfo {title}
  {{Prediction of Toric Code Topological Order from Rydberg Blockade}},}\
  }\href {\doibase 10.1103/PhysRevX.11.031005} {\bibfield  {journal} {\bibinfo
  {journal} {Physical Review X}\ }\textbf {\bibinfo {volume} {11}},\ \bibinfo
  {eid} {031005} (\bibinfo {year} {2021})},\ \Eprint
  {http://arxiv.org/abs/2011.12310} {arXiv:2011.12310 [cond-mat.str-el]}
  \BibitemShut {NoStop}%
\bibitem [{\citenamefont {{Huang}}\ \emph {et~al.}(2022)\citenamefont
  {{Huang}}, \citenamefont {{Zhu}}, \citenamefont {{Gong}}, \citenamefont
  {{Jiang}},\ and\ \citenamefont {{Sheng}}}]{Huang2021}%
  \BibitemOpen
  \bibfield  {author} {\bibinfo {author} {\bibfnamefont {Yixuan}\ \bibnamefont
  {{Huang}}}, \bibinfo {author} {\bibfnamefont {W.}~\bibnamefont {{Zhu}}},
  \bibinfo {author} {\bibfnamefont {Shou-Shu}\ \bibnamefont {{Gong}}}, \bibinfo
  {author} {\bibfnamefont {Hong-Chen}\ \bibnamefont {{Jiang}}}, \ and\ \bibinfo
  {author} {\bibfnamefont {D.~N.}\ \bibnamefont {{Sheng}}},\ }\bibfield
  {title} {\enquote {\bibinfo {title} {{Coexistence of non-Abelian chiral spin
  liquid and magnetic order in a spin-1 antiferromagnet}},}\ }\href {\doibase
  10.1103/PhysRevB.105.155104} {\bibfield  {journal} {\bibinfo  {journal}
  {\prb}\ }\textbf {\bibinfo {volume} {105}},\ \bibinfo {eid} {155104}
  (\bibinfo {year} {2022})},\ \Eprint {http://arxiv.org/abs/2108.06676}
  {arXiv:2108.06676 [cond-mat.str-el]} \BibitemShut {NoStop}%
\bibitem [{\citenamefont {{Luo}}\ \emph {et~al.}(2023)\citenamefont {{Luo}},
  \citenamefont {{Huang}}, \citenamefont {{Sheng}},\ and\ \citenamefont
  {{Zhu}}}]{Luo2022}%
  \BibitemOpen
  \bibfield  {author} {\bibinfo {author} {\bibfnamefont {Wei-Wei}\ \bibnamefont
  {{Luo}}}, \bibinfo {author} {\bibfnamefont {Yixuan}\ \bibnamefont {{Huang}}},
  \bibinfo {author} {\bibfnamefont {D.~N.}\ \bibnamefont {{Sheng}}}, \ and\
  \bibinfo {author} {\bibfnamefont {W.}~\bibnamefont {{Zhu}}},\ }\bibfield
  {title} {\enquote {\bibinfo {title} {{Global quantum phase diagram and
  non-Abelian chiral spin liquid in a spin-3/2 square-lattice
  antiferromagnet}},}\ }\href {\doibase 10.1103/PhysRevB.108.035130} {\bibfield
   {journal} {\bibinfo  {journal} {\prb}\ }\textbf {\bibinfo {volume} {108}},\
  \bibinfo {eid} {035130} (\bibinfo {year} {2023})},\ \Eprint
  {http://arxiv.org/abs/2212.14223} {arXiv:2212.14223 [cond-mat.str-el]}
  \BibitemShut {NoStop}%
\bibitem [{\citenamefont {{Zhang}}\ \emph {et~al.}(2024)\citenamefont
  {{Zhang}}, \citenamefont {{Huang}}, \citenamefont {{Wu}}, \citenamefont
  {{Sheng}},\ and\ \citenamefont {{Gong}}}]{Zhang2024}%
  \BibitemOpen
  \bibfield  {author} {\bibinfo {author} {\bibfnamefont {Xiao-Tian}\
  \bibnamefont {{Zhang}}}, \bibinfo {author} {\bibfnamefont {Yixuan}\
  \bibnamefont {{Huang}}}, \bibinfo {author} {\bibfnamefont {Han-Qing}\
  \bibnamefont {{Wu}}}, \bibinfo {author} {\bibfnamefont {D.~N.}\ \bibnamefont
  {{Sheng}}}, \ and\ \bibinfo {author} {\bibfnamefont {Shou-Shu}\ \bibnamefont
  {{Gong}}},\ }\bibfield  {title} {\enquote {\bibinfo {title} {{Chiral spin
  liquid and quantum phase diagram of spin-12 J1-J2-J{\ensuremath{\chi}} model
  on the square lattice}},}\ }\href {\doibase 10.1103/PhysRevB.109.125146}
  {\bibfield  {journal} {\bibinfo  {journal} {\prb}\ }\textbf {\bibinfo
  {volume} {109}},\ \bibinfo {eid} {125146} (\bibinfo {year} {2024})},\ \Eprint
  {http://arxiv.org/abs/2401.07461} {arXiv:2401.07461 [cond-mat.str-el]}
  \BibitemShut {NoStop}%
\bibitem [{\citenamefont {{Satzinger}}\ \emph {et~al.}(2021)\citenamefont
  {{Satzinger}}, \citenamefont {{Liu}}, \citenamefont {{Smith}}, \citenamefont
  {{Knapp}}, \citenamefont {{Newman}}, \citenamefont {{Jones}}, \citenamefont
  {{Chen}}, \citenamefont {{Quintana}}, \citenamefont {{Mi}}, \citenamefont
  {{Dunsworth}}, \citenamefont {{Gidney}}, \citenamefont {{Aleiner}},
  \citenamefont {{Arute}}, \citenamefont {{Arya}}, \citenamefont {{Atalaya}},
  \citenamefont {{Babbush}}, \citenamefont {{Bardin}}, \citenamefont
  {{Barends}}, \citenamefont {{Basso}}, \citenamefont {{Bengtsson}},
  \citenamefont {{Bilmes}}, \citenamefont {{Broughton}}, \citenamefont
  {{Buckley}}, \citenamefont {{Buell}}, \citenamefont {{Burkett}},
  \citenamefont {{Bushnell}}, \citenamefont {{Chiaro}}, \citenamefont
  {{Collins}}, \citenamefont {{Courtney}}, \citenamefont {{Demura}},
  \citenamefont {{Derk}}, \citenamefont {{Eppens}}, \citenamefont {{Erickson}},
  \citenamefont {{Faoro}}, \citenamefont {{Farhi}}, \citenamefont {{Fowler}},
  \citenamefont {{Foxen}}, \citenamefont {{Giustina}}, \citenamefont
  {{Greene}}, \citenamefont {{Gross}}, \citenamefont {{Harrigan}},
  \citenamefont {{Harrington}}, \citenamefont {{Hilton}}, \citenamefont
  {{Hong}}, \citenamefont {{Huang}}, \citenamefont {{Huggins}}, \citenamefont
  {{Ioffe}}, \citenamefont {{Isakov}}, \citenamefont {{Jeffrey}}, \citenamefont
  {{Jiang}}, \citenamefont {{Kafri}}, \citenamefont {{Kechedzhi}},
  \citenamefont {{Khattar}}, \citenamefont {{Kim}}, \citenamefont {{Klimov}},
  \citenamefont {{Korotkov}}, \citenamefont {{Kostritsa}}, \citenamefont
  {{Landhuis}}, \citenamefont {{Laptev}}, \citenamefont {{Locharla}},
  \citenamefont {{Lucero}}, \citenamefont {{Martin}}, \citenamefont
  {{McClean}}, \citenamefont {{McEwen}}, \citenamefont {{Miao}}, \citenamefont
  {{Mohseni}}, \citenamefont {{Montazeri}}, \citenamefont {{Mruczkiewicz}},
  \citenamefont {{Mutus}}, \citenamefont {{Naaman}}, \citenamefont {{Neeley}},
  \citenamefont {{Neill}}, \citenamefont {{Niu}}, \citenamefont {{O'Brien}},
  \citenamefont {{Opremcak}}, \citenamefont {{Pat{\'o}}}, \citenamefont
  {{Petukhov}}, \citenamefont {{Rubin}}, \citenamefont {{Sank}}, \citenamefont
  {{Shvarts}}, \citenamefont {{Strain}}, \citenamefont {{Szalay}},
  \citenamefont {{Villalonga}}, \citenamefont {{White}}, \citenamefont {{Yao}},
  \citenamefont {{Yeh}}, \citenamefont {{Yoo}}, \citenamefont {{Zalcman}},
  \citenamefont {{Neven}}, \citenamefont {{Boixo}}, \citenamefont {{Megrant}},
  \citenamefont {{Chen}}, \citenamefont {{Kelly}}, \citenamefont
  {{Smelyanskiy}}, \citenamefont {{Kitaev}}, \citenamefont {{Knap}},
  \citenamefont {{Pollmann}},\ and\ \citenamefont {{Roushan}}}]{Satzinger2021}%
  \BibitemOpen
  \bibfield  {author} {\bibinfo {author} {\bibfnamefont {K.~J.}\ \bibnamefont
  {{Satzinger}}}, \bibinfo {author} {\bibfnamefont {Y.-J.}\ \bibnamefont
  {{Liu}}}, \bibinfo {author} {\bibfnamefont {A.}~\bibnamefont {{Smith}}},
  \bibinfo {author} {\bibfnamefont {C.}~\bibnamefont {{Knapp}}}, \bibinfo
  {author} {\bibfnamefont {M.}~\bibnamefont {{Newman}}}, \bibinfo {author}
  {\bibfnamefont {C.}~\bibnamefont {{Jones}}}, \bibinfo {author} {\bibfnamefont
  {Z.}~\bibnamefont {{Chen}}}, \bibinfo {author} {\bibfnamefont
  {C.}~\bibnamefont {{Quintana}}}, \bibinfo {author} {\bibfnamefont
  {X.}~\bibnamefont {{Mi}}}, \bibinfo {author} {\bibfnamefont {A.}~\bibnamefont
  {{Dunsworth}}}, \bibinfo {author} {\bibfnamefont {C.}~\bibnamefont
  {{Gidney}}}, \bibinfo {author} {\bibfnamefont {I.}~\bibnamefont {{Aleiner}}},
  \bibinfo {author} {\bibfnamefont {F.}~\bibnamefont {{Arute}}}, \bibinfo
  {author} {\bibfnamefont {K.}~\bibnamefont {{Arya}}}, \bibinfo {author}
  {\bibfnamefont {J.}~\bibnamefont {{Atalaya}}}, \bibinfo {author}
  {\bibfnamefont {R.}~\bibnamefont {{Babbush}}}, \bibinfo {author}
  {\bibfnamefont {J.~C.}\ \bibnamefont {{Bardin}}}, \bibinfo {author}
  {\bibfnamefont {R.}~\bibnamefont {{Barends}}}, \bibinfo {author}
  {\bibfnamefont {J.}~\bibnamefont {{Basso}}}, \bibinfo {author} {\bibfnamefont
  {A.}~\bibnamefont {{Bengtsson}}}, \bibinfo {author} {\bibfnamefont
  {A.}~\bibnamefont {{Bilmes}}}, \bibinfo {author} {\bibfnamefont
  {M.}~\bibnamefont {{Broughton}}}, \bibinfo {author} {\bibfnamefont {B.~B.}\
  \bibnamefont {{Buckley}}}, \bibinfo {author} {\bibfnamefont {D.~A.}\
  \bibnamefont {{Buell}}}, \bibinfo {author} {\bibfnamefont {B.}~\bibnamefont
  {{Burkett}}}, \bibinfo {author} {\bibfnamefont {N.}~\bibnamefont
  {{Bushnell}}}, \bibinfo {author} {\bibfnamefont {B.}~\bibnamefont
  {{Chiaro}}}, \bibinfo {author} {\bibfnamefont {R.}~\bibnamefont {{Collins}}},
  \bibinfo {author} {\bibfnamefont {W.}~\bibnamefont {{Courtney}}}, \bibinfo
  {author} {\bibfnamefont {S.}~\bibnamefont {{Demura}}}, \bibinfo {author}
  {\bibfnamefont {A.~R.}\ \bibnamefont {{Derk}}}, \bibinfo {author}
  {\bibfnamefont {D.}~\bibnamefont {{Eppens}}}, \bibinfo {author}
  {\bibfnamefont {C.}~\bibnamefont {{Erickson}}}, \bibinfo {author}
  {\bibfnamefont {L.}~\bibnamefont {{Faoro}}}, \bibinfo {author} {\bibfnamefont
  {E.}~\bibnamefont {{Farhi}}}, \bibinfo {author} {\bibfnamefont {A.~G.}\
  \bibnamefont {{Fowler}}}, \bibinfo {author} {\bibfnamefont {B.}~\bibnamefont
  {{Foxen}}}, \bibinfo {author} {\bibfnamefont {M.}~\bibnamefont {{Giustina}}},
  \bibinfo {author} {\bibfnamefont {A.}~\bibnamefont {{Greene}}}, \bibinfo
  {author} {\bibfnamefont {J.~A.}\ \bibnamefont {{Gross}}}, \bibinfo {author}
  {\bibfnamefont {M.~P.}\ \bibnamefont {{Harrigan}}}, \bibinfo {author}
  {\bibfnamefont {S.~D.}\ \bibnamefont {{Harrington}}}, \bibinfo {author}
  {\bibfnamefont {J.}~\bibnamefont {{Hilton}}}, \bibinfo {author}
  {\bibfnamefont {S.}~\bibnamefont {{Hong}}}, \bibinfo {author} {\bibfnamefont
  {T.}~\bibnamefont {{Huang}}}, \bibinfo {author} {\bibfnamefont {W.~J.}\
  \bibnamefont {{Huggins}}}, \bibinfo {author} {\bibfnamefont {L.~B.}\
  \bibnamefont {{Ioffe}}}, \bibinfo {author} {\bibfnamefont {S.~V.}\
  \bibnamefont {{Isakov}}}, \bibinfo {author} {\bibfnamefont {E.}~\bibnamefont
  {{Jeffrey}}}, \bibinfo {author} {\bibfnamefont {Z.}~\bibnamefont {{Jiang}}},
  \bibinfo {author} {\bibfnamefont {D.}~\bibnamefont {{Kafri}}}, \bibinfo
  {author} {\bibfnamefont {K.}~\bibnamefont {{Kechedzhi}}}, \bibinfo {author}
  {\bibfnamefont {T.}~\bibnamefont {{Khattar}}}, \bibinfo {author}
  {\bibfnamefont {S.}~\bibnamefont {{Kim}}}, \bibinfo {author} {\bibfnamefont
  {P.~V.}\ \bibnamefont {{Klimov}}}, \bibinfo {author} {\bibfnamefont {A.~N.}\
  \bibnamefont {{Korotkov}}}, \bibinfo {author} {\bibfnamefont
  {F.}~\bibnamefont {{Kostritsa}}}, \bibinfo {author} {\bibfnamefont
  {D.}~\bibnamefont {{Landhuis}}}, \bibinfo {author} {\bibfnamefont
  {P.}~\bibnamefont {{Laptev}}}, \bibinfo {author} {\bibfnamefont
  {A.}~\bibnamefont {{Locharla}}}, \bibinfo {author} {\bibfnamefont
  {E.}~\bibnamefont {{Lucero}}}, \bibinfo {author} {\bibfnamefont
  {O.}~\bibnamefont {{Martin}}}, \bibinfo {author} {\bibfnamefont {J.~R.}\
  \bibnamefont {{McClean}}}, \bibinfo {author} {\bibfnamefont {M.}~\bibnamefont
  {{McEwen}}}, \bibinfo {author} {\bibfnamefont {K.~C.}\ \bibnamefont
  {{Miao}}}, \bibinfo {author} {\bibfnamefont {M.}~\bibnamefont {{Mohseni}}},
  \bibinfo {author} {\bibfnamefont {S.}~\bibnamefont {{Montazeri}}}, \bibinfo
  {author} {\bibfnamefont {W.}~\bibnamefont {{Mruczkiewicz}}}, \bibinfo
  {author} {\bibfnamefont {J.}~\bibnamefont {{Mutus}}}, \bibinfo {author}
  {\bibfnamefont {O.}~\bibnamefont {{Naaman}}}, \bibinfo {author}
  {\bibfnamefont {M.}~\bibnamefont {{Neeley}}}, \bibinfo {author}
  {\bibfnamefont {C.}~\bibnamefont {{Neill}}}, \bibinfo {author} {\bibfnamefont
  {M.~Y.}\ \bibnamefont {{Niu}}}, \bibinfo {author} {\bibfnamefont {T.~E.}\
  \bibnamefont {{O'Brien}}}, \bibinfo {author} {\bibfnamefont {A.}~\bibnamefont
  {{Opremcak}}}, \bibinfo {author} {\bibfnamefont {B.}~\bibnamefont
  {{Pat{\'o}}}}, \bibinfo {author} {\bibfnamefont {A.}~\bibnamefont
  {{Petukhov}}}, \bibinfo {author} {\bibfnamefont {N.~C.}\ \bibnamefont
  {{Rubin}}}, \bibinfo {author} {\bibfnamefont {D.}~\bibnamefont {{Sank}}},
  \bibinfo {author} {\bibfnamefont {V.}~\bibnamefont {{Shvarts}}}, \bibinfo
  {author} {\bibfnamefont {D.}~\bibnamefont {{Strain}}}, \bibinfo {author}
  {\bibfnamefont {M.}~\bibnamefont {{Szalay}}}, \bibinfo {author}
  {\bibfnamefont {B.}~\bibnamefont {{Villalonga}}}, \bibinfo {author}
  {\bibfnamefont {T.~C.}\ \bibnamefont {{White}}}, \bibinfo {author}
  {\bibfnamefont {Z.}~\bibnamefont {{Yao}}}, \bibinfo {author} {\bibfnamefont
  {P.}~\bibnamefont {{Yeh}}}, \bibinfo {author} {\bibfnamefont
  {J.}~\bibnamefont {{Yoo}}}, \bibinfo {author} {\bibfnamefont
  {A.}~\bibnamefont {{Zalcman}}}, \bibinfo {author} {\bibfnamefont
  {H.}~\bibnamefont {{Neven}}}, \bibinfo {author} {\bibfnamefont
  {S.}~\bibnamefont {{Boixo}}}, \bibinfo {author} {\bibfnamefont
  {A.}~\bibnamefont {{Megrant}}}, \bibinfo {author} {\bibfnamefont
  {Y.}~\bibnamefont {{Chen}}}, \bibinfo {author} {\bibfnamefont
  {J.}~\bibnamefont {{Kelly}}}, \bibinfo {author} {\bibfnamefont
  {V.}~\bibnamefont {{Smelyanskiy}}}, \bibinfo {author} {\bibfnamefont
  {A.}~\bibnamefont {{Kitaev}}}, \bibinfo {author} {\bibfnamefont
  {M.}~\bibnamefont {{Knap}}}, \bibinfo {author} {\bibfnamefont
  {F.}~\bibnamefont {{Pollmann}}}, \ and\ \bibinfo {author} {\bibfnamefont
  {P.}~\bibnamefont {{Roushan}}},\ }\bibfield  {title} {\enquote {\bibinfo
  {title} {{Realizing topologically ordered states on a quantum processor}},}\
  }\href {\doibase 10.1126/science.abi8378} {\bibfield  {journal} {\bibinfo
  {journal} {Science}\ }\textbf {\bibinfo {volume} {374}},\ \bibinfo {pages}
  {1237--1241} (\bibinfo {year} {2021})},\ \Eprint
  {http://arxiv.org/abs/2104.01180} {arXiv:2104.01180 [quant-ph]} \BibitemShut
  {NoStop}%
\bibitem [{\citenamefont {{Semeghini}}\ \emph {et~al.}(2021)\citenamefont
  {{Semeghini}}, \citenamefont {{Levine}}, \citenamefont {{Keesling}},
  \citenamefont {{Ebadi}}, \citenamefont {{Wang}}, \citenamefont {{Bluvstein}},
  \citenamefont {{Verresen}}, \citenamefont {{Pichler}}, \citenamefont
  {{Kalinowski}}, \citenamefont {{Samajdar}}, \citenamefont {{Omran}},
  \citenamefont {{Sachdev}}, \citenamefont {{Vishwanath}}, \citenamefont
  {{Greiner}}, \citenamefont {{Vuleti{\'c}}},\ and\ \citenamefont
  {{Lukin}}}]{Semeghini2021}%
  \BibitemOpen
  \bibfield  {author} {\bibinfo {author} {\bibfnamefont {G.}~\bibnamefont
  {{Semeghini}}}, \bibinfo {author} {\bibfnamefont {H.}~\bibnamefont
  {{Levine}}}, \bibinfo {author} {\bibfnamefont {A.}~\bibnamefont
  {{Keesling}}}, \bibinfo {author} {\bibfnamefont {S.}~\bibnamefont {{Ebadi}}},
  \bibinfo {author} {\bibfnamefont {T.~T.}\ \bibnamefont {{Wang}}}, \bibinfo
  {author} {\bibfnamefont {D.}~\bibnamefont {{Bluvstein}}}, \bibinfo {author}
  {\bibfnamefont {R.}~\bibnamefont {{Verresen}}}, \bibinfo {author}
  {\bibfnamefont {H.}~\bibnamefont {{Pichler}}}, \bibinfo {author}
  {\bibfnamefont {M.}~\bibnamefont {{Kalinowski}}}, \bibinfo {author}
  {\bibfnamefont {R.}~\bibnamefont {{Samajdar}}}, \bibinfo {author}
  {\bibfnamefont {A.}~\bibnamefont {{Omran}}}, \bibinfo {author} {\bibfnamefont
  {S.}~\bibnamefont {{Sachdev}}}, \bibinfo {author} {\bibfnamefont
  {A.}~\bibnamefont {{Vishwanath}}}, \bibinfo {author} {\bibfnamefont
  {M.}~\bibnamefont {{Greiner}}}, \bibinfo {author} {\bibfnamefont
  {V.}~\bibnamefont {{Vuleti{\'c}}}}, \ and\ \bibinfo {author} {\bibfnamefont
  {M.~D.}\ \bibnamefont {{Lukin}}},\ }\bibfield  {title} {\enquote {\bibinfo
  {title} {{Probing topological spin liquids on a programmable quantum
  simulator}},}\ }\href {\doibase 10.1126/science.abi8794} {\bibfield
  {journal} {\bibinfo  {journal} {Science}\ }\textbf {\bibinfo {volume}
  {374}},\ \bibinfo {pages} {1242--1247} (\bibinfo {year} {2021})},\ \Eprint
  {http://arxiv.org/abs/2104.04119} {arXiv:2104.04119 [quant-ph]} \BibitemShut
  {NoStop}%
\bibitem [{\citenamefont {{Iqbal}}\ \emph
  {et~al.}(2024{\natexlab{a}})\citenamefont {{Iqbal}}, \citenamefont
  {{Tantivasadakarn}}, \citenamefont {{Gatterman}}, \citenamefont {{Gerber}},
  \citenamefont {{Gilmore}}, \citenamefont {{Gresh}}, \citenamefont {{Hankin}},
  \citenamefont {{Hewitt}}, \citenamefont {{Horst}}, \citenamefont {{Matheny}},
  \citenamefont {{Mengle}}, \citenamefont {{Neyenhuis}}, \citenamefont
  {{Vishwanath}}, \citenamefont {{Foss-Feig}}, \citenamefont {{Verresen}},\
  and\ \citenamefont {{Dreyer}}}]{Iqbal2023}%
  \BibitemOpen
  \bibfield  {author} {\bibinfo {author} {\bibfnamefont {Mohsin}\ \bibnamefont
  {{Iqbal}}}, \bibinfo {author} {\bibfnamefont {Nathanan}\ \bibnamefont
  {{Tantivasadakarn}}}, \bibinfo {author} {\bibfnamefont {Thomas~M.}\
  \bibnamefont {{Gatterman}}}, \bibinfo {author} {\bibfnamefont {Justin~A.}\
  \bibnamefont {{Gerber}}}, \bibinfo {author} {\bibfnamefont {Kevin}\
  \bibnamefont {{Gilmore}}}, \bibinfo {author} {\bibfnamefont {Dan}\
  \bibnamefont {{Gresh}}}, \bibinfo {author} {\bibfnamefont {Aaron}\
  \bibnamefont {{Hankin}}}, \bibinfo {author} {\bibfnamefont {Nathan}\
  \bibnamefont {{Hewitt}}}, \bibinfo {author} {\bibfnamefont {Chandler~V.}\
  \bibnamefont {{Horst}}}, \bibinfo {author} {\bibfnamefont {Mitchell}\
  \bibnamefont {{Matheny}}}, \bibinfo {author} {\bibfnamefont {Tanner}\
  \bibnamefont {{Mengle}}}, \bibinfo {author} {\bibfnamefont {Brian}\
  \bibnamefont {{Neyenhuis}}}, \bibinfo {author} {\bibfnamefont {Ashvin}\
  \bibnamefont {{Vishwanath}}}, \bibinfo {author} {\bibfnamefont {Michael}\
  \bibnamefont {{Foss-Feig}}}, \bibinfo {author} {\bibfnamefont {Ruben}\
  \bibnamefont {{Verresen}}}, \ and\ \bibinfo {author} {\bibfnamefont {Henrik}\
  \bibnamefont {{Dreyer}}},\ }\bibfield  {title} {\enquote {\bibinfo {title}
  {{Topological order from measurements and feed-forward on a trapped ion
  quantum computer}},}\ }\href {\doibase 10.1038/s42005-024-01698-3} {\bibfield
   {journal} {\bibinfo  {journal} {Communications Physics}\ }\textbf {\bibinfo
  {volume} {7}},\ \bibinfo {eid} {205} (\bibinfo {year}
  {2024}{\natexlab{a}})},\ \Eprint {http://arxiv.org/abs/2302.01917}
  {arXiv:2302.01917 [quant-ph]} \BibitemShut {NoStop}%
\bibitem [{\citenamefont {{Foss-Feig}}\ \emph {et~al.}(2023)\citenamefont
  {{Foss-Feig}}, \citenamefont {{Tikku}}, \citenamefont {{Lu}}, \citenamefont
  {{Mayer}}, \citenamefont {{Iqbal}}, \citenamefont {{Gatterman}},
  \citenamefont {{Gerber}}, \citenamefont {{Gilmore}}, \citenamefont {{Gresh}},
  \citenamefont {{Hankin}}, \citenamefont {{Hewitt}}, \citenamefont {{Horst}},
  \citenamefont {{Matheny}}, \citenamefont {{Mengle}}, \citenamefont
  {{Neyenhuis}}, \citenamefont {{Dreyer}}, \citenamefont {{Hayes}},
  \citenamefont {{Hsieh}},\ and\ \citenamefont {{Kim}}}]{Foss-Feig2023}%
  \BibitemOpen
  \bibfield  {author} {\bibinfo {author} {\bibfnamefont {Michael}\ \bibnamefont
  {{Foss-Feig}}}, \bibinfo {author} {\bibfnamefont {Arkin}\ \bibnamefont
  {{Tikku}}}, \bibinfo {author} {\bibfnamefont {Tsung-Cheng}\ \bibnamefont
  {{Lu}}}, \bibinfo {author} {\bibfnamefont {Karl}\ \bibnamefont {{Mayer}}},
  \bibinfo {author} {\bibfnamefont {Mohsin}\ \bibnamefont {{Iqbal}}}, \bibinfo
  {author} {\bibfnamefont {Thomas~M.}\ \bibnamefont {{Gatterman}}}, \bibinfo
  {author} {\bibfnamefont {Justin~A.}\ \bibnamefont {{Gerber}}}, \bibinfo
  {author} {\bibfnamefont {Kevin}\ \bibnamefont {{Gilmore}}}, \bibinfo {author}
  {\bibfnamefont {Dan}\ \bibnamefont {{Gresh}}}, \bibinfo {author}
  {\bibfnamefont {Aaron}\ \bibnamefont {{Hankin}}}, \bibinfo {author}
  {\bibfnamefont {Nathan}\ \bibnamefont {{Hewitt}}}, \bibinfo {author}
  {\bibfnamefont {Chandler~V.}\ \bibnamefont {{Horst}}}, \bibinfo {author}
  {\bibfnamefont {Mitchell}\ \bibnamefont {{Matheny}}}, \bibinfo {author}
  {\bibfnamefont {Tanner}\ \bibnamefont {{Mengle}}}, \bibinfo {author}
  {\bibfnamefont {Brian}\ \bibnamefont {{Neyenhuis}}}, \bibinfo {author}
  {\bibfnamefont {Henrik}\ \bibnamefont {{Dreyer}}}, \bibinfo {author}
  {\bibfnamefont {David}\ \bibnamefont {{Hayes}}}, \bibinfo {author}
  {\bibfnamefont {Timothy~H.}\ \bibnamefont {{Hsieh}}}, \ and\ \bibinfo
  {author} {\bibfnamefont {Isaac~H.}\ \bibnamefont {{Kim}}},\ }\bibfield
  {title} {\enquote {\bibinfo {title} {{Experimental demonstration of the
  advantage of adaptive quantum circuits}},}\ }\href {\doibase
  10.48550/arXiv.2302.03029} {\bibfield  {journal} {\bibinfo  {journal} {arXiv
  e-prints}\ ,\ \bibinfo {eid} {arXiv:2302.03029}} (\bibinfo {year} {2023})},\
  \Eprint {http://arxiv.org/abs/2302.03029} {arXiv:2302.03029 [quant-ph]}
  \BibitemShut {NoStop}%
\bibitem [{\citenamefont {{Iqbal}}\ \emph
  {et~al.}(2024{\natexlab{b}})\citenamefont {{Iqbal}}, \citenamefont
  {{Tantivasadakarn}}, \citenamefont {{Verresen}}, \citenamefont {{Campbell}},
  \citenamefont {{Dreiling}}, \citenamefont {{Figgatt}}, \citenamefont
  {{Gaebler}}, \citenamefont {{Johansen}}, \citenamefont {{Mills}},
  \citenamefont {{Moses}}, \citenamefont {{Pino}}, \citenamefont {{Ransford}},
  \citenamefont {{Rowe}}, \citenamefont {{Siegfried}}, \citenamefont {{Stutz}},
  \citenamefont {{Foss-Feig}}, \citenamefont {{Vishwanath}},\ and\
  \citenamefont {{Dreyer}}}]{Iqbal2023a}%
  \BibitemOpen
  \bibfield  {author} {\bibinfo {author} {\bibfnamefont {Mohsin}\ \bibnamefont
  {{Iqbal}}}, \bibinfo {author} {\bibfnamefont {Nathanan}\ \bibnamefont
  {{Tantivasadakarn}}}, \bibinfo {author} {\bibfnamefont {Ruben}\ \bibnamefont
  {{Verresen}}}, \bibinfo {author} {\bibfnamefont {Sara~L.}\ \bibnamefont
  {{Campbell}}}, \bibinfo {author} {\bibfnamefont {Joan~M.}\ \bibnamefont
  {{Dreiling}}}, \bibinfo {author} {\bibfnamefont {Caroline}\ \bibnamefont
  {{Figgatt}}}, \bibinfo {author} {\bibfnamefont {John~P.}\ \bibnamefont
  {{Gaebler}}}, \bibinfo {author} {\bibfnamefont {Jacob}\ \bibnamefont
  {{Johansen}}}, \bibinfo {author} {\bibfnamefont {Michael}\ \bibnamefont
  {{Mills}}}, \bibinfo {author} {\bibfnamefont {Steven~A.}\ \bibnamefont
  {{Moses}}}, \bibinfo {author} {\bibfnamefont {Juan~M.}\ \bibnamefont
  {{Pino}}}, \bibinfo {author} {\bibfnamefont {Anthony}\ \bibnamefont
  {{Ransford}}}, \bibinfo {author} {\bibfnamefont {Mary}\ \bibnamefont
  {{Rowe}}}, \bibinfo {author} {\bibfnamefont {Peter}\ \bibnamefont
  {{Siegfried}}}, \bibinfo {author} {\bibfnamefont {Russell~P.}\ \bibnamefont
  {{Stutz}}}, \bibinfo {author} {\bibfnamefont {Michael}\ \bibnamefont
  {{Foss-Feig}}}, \bibinfo {author} {\bibfnamefont {Ashvin}\ \bibnamefont
  {{Vishwanath}}}, \ and\ \bibinfo {author} {\bibfnamefont {Henrik}\
  \bibnamefont {{Dreyer}}},\ }\bibfield  {title} {\enquote {\bibinfo {title}
  {{Non-Abelian topological order and anyons on a trapped-ion processor}},}\
  }\href {\doibase 10.1038/s41586-023-06934-4} {\bibfield  {journal} {\bibinfo
  {journal} {\nat}\ }\textbf {\bibinfo {volume} {626}},\ \bibinfo {pages}
  {505--511} (\bibinfo {year} {2024}{\natexlab{b}})},\ \Eprint
  {http://arxiv.org/abs/2305.03766} {arXiv:2305.03766 [quant-ph]} \BibitemShut
  {NoStop}%
\bibitem [{\citenamefont {Chen}(2017)}]{Chen2016}%
  \BibitemOpen
  \bibfield  {author} {\bibinfo {author} {\bibfnamefont {Xie}\ \bibnamefont
  {Chen}},\ }\bibfield  {title} {\enquote {\bibinfo {title} {Symmetry
  fractionalization in two dimensional topological phases},}\ }\href {\doibase
  https://doi.org/10.1016/j.revip.2017.02.002} {\bibfield  {journal} {\bibinfo
  {journal} {Reviews in Physics}\ }\textbf {\bibinfo {volume} {2}},\ \bibinfo
  {pages} {3--18} (\bibinfo {year} {2017})},\ \Eprint
  {http://arxiv.org/abs/1606.07569} {arXiv:1606.07569 [cond-mat.str-el]}
  \BibitemShut {NoStop}%
\bibitem [{\citenamefont {{Bombin}}(2010)}]{Bombin2010}%
  \BibitemOpen
  \bibfield  {author} {\bibinfo {author} {\bibfnamefont {H.}~\bibnamefont
  {{Bombin}}},\ }\bibfield  {title} {\enquote {\bibinfo {title} {{Topological
  Order with a Twist: Ising Anyons from an Abelian Model}},}\ }\href {\doibase
  10.1103/PhysRevLett.105.030403} {\bibfield  {journal} {\bibinfo  {journal}
  {\prl}\ }\textbf {\bibinfo {volume} {105}},\ \bibinfo {eid} {030403}
  (\bibinfo {year} {2010})},\ \Eprint {http://arxiv.org/abs/1004.1838}
  {arXiv:1004.1838 [cond-mat.str-el]} \BibitemShut {NoStop}%
\bibitem [{\citenamefont {{Lindner}}\ \emph {et~al.}(2012)\citenamefont
  {{Lindner}}, \citenamefont {{Berg}}, \citenamefont {{Refael}},\ and\
  \citenamefont {{Stern}}}]{Lindner2012}%
  \BibitemOpen
  \bibfield  {author} {\bibinfo {author} {\bibfnamefont {Netanel~H.}\
  \bibnamefont {{Lindner}}}, \bibinfo {author} {\bibfnamefont {Erez}\
  \bibnamefont {{Berg}}}, \bibinfo {author} {\bibfnamefont {Gil}\ \bibnamefont
  {{Refael}}}, \ and\ \bibinfo {author} {\bibfnamefont {Ady}\ \bibnamefont
  {{Stern}}},\ }\bibfield  {title} {\enquote {\bibinfo {title}
  {{Fractionalizing Majorana Fermions: Non-Abelian Statistics on the Edges of
  Abelian Quantum Hall States}},}\ }\href {\doibase 10.1103/PhysRevX.2.041002}
  {\bibfield  {journal} {\bibinfo  {journal} {Physical Review X}\ }\textbf
  {\bibinfo {volume} {2}},\ \bibinfo {eid} {041002} (\bibinfo {year} {2012})},\
  \Eprint {http://arxiv.org/abs/1204.5733} {arXiv:1204.5733
  [cond-mat.mes-hall]} \BibitemShut {NoStop}%
\bibitem [{\citenamefont {{Clarke}}\ \emph {et~al.}(2013)\citenamefont
  {{Clarke}}, \citenamefont {{Alicea}},\ and\ \citenamefont
  {{Shtengel}}}]{Clarke2012}%
  \BibitemOpen
  \bibfield  {author} {\bibinfo {author} {\bibfnamefont {David~J.}\
  \bibnamefont {{Clarke}}}, \bibinfo {author} {\bibfnamefont {Jason}\
  \bibnamefont {{Alicea}}}, \ and\ \bibinfo {author} {\bibfnamefont {Kirill}\
  \bibnamefont {{Shtengel}}},\ }\bibfield  {title} {\enquote {\bibinfo {title}
  {{Exotic non-Abelian anyons from conventional fractional quantum Hall
  states}},}\ }\href {\doibase 10.1038/ncomms2340} {\bibfield  {journal}
  {\bibinfo  {journal} {Nature Communications}\ }\textbf {\bibinfo {volume}
  {4}},\ \bibinfo {eid} {1348} (\bibinfo {year} {2013})},\ \Eprint
  {http://arxiv.org/abs/1204.5479} {arXiv:1204.5479 [cond-mat.str-el]}
  \BibitemShut {NoStop}%
\bibitem [{\citenamefont {{Alicea}}\ and\ \citenamefont
  {{Fendley}}(2016)}]{Alicea2015}%
  \BibitemOpen
  \bibfield  {author} {\bibinfo {author} {\bibfnamefont {Jason}\ \bibnamefont
  {{Alicea}}}\ and\ \bibinfo {author} {\bibfnamefont {Paul}\ \bibnamefont
  {{Fendley}}},\ }\bibfield  {title} {\enquote {\bibinfo {title} {{Topological
  Phases with Parafermions: Theory and Blueprints}},}\ }\href {\doibase
  10.1146/annurev-conmatphys-031115-011336} {\bibfield  {journal} {\bibinfo
  {journal} {Annual Review of Condensed Matter Physics}\ }\textbf {\bibinfo
  {volume} {7}},\ \bibinfo {pages} {119--139} (\bibinfo {year} {2016})},\
  \Eprint {http://arxiv.org/abs/1504.02476} {arXiv:1504.02476
  [cond-mat.str-el]} \BibitemShut {NoStop}%
\bibitem [{\citenamefont {{Barkeshli}}\ \emph
  {et~al.}(2019{\natexlab{a}})\citenamefont {{Barkeshli}}, \citenamefont
  {{Bonderson}}, \citenamefont {{Cheng}},\ and\ \citenamefont
  {{Wang}}}]{barkeshli2014}%
  \BibitemOpen
  \bibfield  {author} {\bibinfo {author} {\bibfnamefont {Maissam}\ \bibnamefont
  {{Barkeshli}}}, \bibinfo {author} {\bibfnamefont {Parsa}\ \bibnamefont
  {{Bonderson}}}, \bibinfo {author} {\bibfnamefont {Meng}\ \bibnamefont
  {{Cheng}}}, \ and\ \bibinfo {author} {\bibfnamefont {Zhenghan}\ \bibnamefont
  {{Wang}}},\ }\bibfield  {title} {\enquote {\bibinfo {title} {{Symmetry
  Fractionalization, Defects, and Gauging of Topological Phases}},}\ }\href
  {\doibase 10.1103/physrevb.100.115147} {\bibfield  {journal} {\bibinfo
  {journal} {\prb}\ }\textbf {\bibinfo {volume} {100}},\ \bibinfo {eid}
  {115147} (\bibinfo {year} {2019}{\natexlab{a}})},\ \Eprint
  {http://arxiv.org/abs/1410.4540} {arXiv:1410.4540 [cond-mat.str-el]}
  \BibitemShut {NoStop}%
\bibitem [{\citenamefont {{Tarantino}}\ \emph {et~al.}(2016)\citenamefont
  {{Tarantino}}, \citenamefont {{Lindner}},\ and\ \citenamefont
  {{Fidkowski}}}]{Tarantino2015}%
  \BibitemOpen
  \bibfield  {author} {\bibinfo {author} {\bibfnamefont {Nicolas}\ \bibnamefont
  {{Tarantino}}}, \bibinfo {author} {\bibfnamefont {Netanel~H.}\ \bibnamefont
  {{Lindner}}}, \ and\ \bibinfo {author} {\bibfnamefont {Lukasz}\ \bibnamefont
  {{Fidkowski}}},\ }\bibfield  {title} {\enquote {\bibinfo {title} {{Symmetry
  fractionalization and twist defects}},}\ }\href {\doibase
  10.1088/1367-2630/18/3/035006} {\bibfield  {journal} {\bibinfo  {journal}
  {New Journal of Physics}\ }\textbf {\bibinfo {volume} {18}},\ \bibinfo {eid}
  {035006} (\bibinfo {year} {2016})},\ \Eprint
  {http://arxiv.org/abs/1506.06754} {arXiv:1506.06754 [cond-mat.str-el]}
  \BibitemShut {NoStop}%
\bibitem [{\citenamefont {{Lan}}\ \emph {et~al.}(2024)\citenamefont {{Lan}},
  \citenamefont {{Yue}},\ and\ \citenamefont {{Wang}}}]{Lan2023}%
  \BibitemOpen
  \bibfield  {author} {\bibinfo {author} {\bibfnamefont {Tian}\ \bibnamefont
  {{Lan}}}, \bibinfo {author} {\bibfnamefont {Gen}\ \bibnamefont {{Yue}}}, \
  and\ \bibinfo {author} {\bibfnamefont {Longye}\ \bibnamefont {{Wang}}},\
  }\bibfield  {title} {\enquote {\bibinfo {title} {{Category of SET orders}},}\
  }\href {\doibase 10.1007/JHEP11(2024)111} {\bibfield  {journal} {\bibinfo
  {journal} {Journal of High Energy Physics}\ }\textbf {\bibinfo {volume}
  {2024}},\ \bibinfo {eid} {111} (\bibinfo {year} {2024})},\ \Eprint
  {http://arxiv.org/abs/2312.15958} {arXiv:2312.15958 [cond-mat.str-el]}
  \BibitemShut {NoStop}%
\bibitem [{\citenamefont {{Essin}}\ and\ \citenamefont
  {{Hermele}}(2013)}]{Essin2012}%
  \BibitemOpen
  \bibfield  {author} {\bibinfo {author} {\bibfnamefont {Andrew~M.}\
  \bibnamefont {{Essin}}}\ and\ \bibinfo {author} {\bibfnamefont {Michael}\
  \bibnamefont {{Hermele}}},\ }\bibfield  {title} {\enquote {\bibinfo {title}
  {{Classifying fractionalization: Symmetry classification of gapped Z$_{2}$
  spin liquids in two dimensions}},}\ }\href {\doibase
  10.1103/PhysRevB.87.104406} {\bibfield  {journal} {\bibinfo  {journal}
  {\prb}\ }\textbf {\bibinfo {volume} {87}},\ \bibinfo {eid} {104406} (\bibinfo
  {year} {2013})},\ \Eprint {http://arxiv.org/abs/1212.0593} {arXiv:1212.0593
  [cond-mat.str-el]} \BibitemShut {NoStop}%
\bibitem [{\citenamefont {{Qi}}\ and\ \citenamefont {{Fu}}(2015)}]{Qi2015c}%
  \BibitemOpen
  \bibfield  {author} {\bibinfo {author} {\bibfnamefont {Yang}\ \bibnamefont
  {{Qi}}}\ and\ \bibinfo {author} {\bibfnamefont {Liang}\ \bibnamefont
  {{Fu}}},\ }\bibfield  {title} {\enquote {\bibinfo {title} {{Detecting crystal
  symmetry fractionalization from the ground state: Application to Z$_{2}$ spin
  liquids on the kagome lattice}},}\ }\href {\doibase
  10.1103/PhysRevB.91.100401} {\bibfield  {journal} {\bibinfo  {journal}
  {\prb}\ }\textbf {\bibinfo {volume} {91}},\ \bibinfo {eid} {100401} (\bibinfo
  {year} {2015})},\ \Eprint {http://arxiv.org/abs/1501.00009} {arXiv:1501.00009
  [cond-mat.str-el]} \BibitemShut {NoStop}%
\bibitem [{\citenamefont {Zaletel}\ \emph {et~al.}(2017)\citenamefont
  {Zaletel}, \citenamefont {Lu},\ and\ \citenamefont
  {Vishwanath}}]{Zaletel2015}%
  \BibitemOpen
  \bibfield  {author} {\bibinfo {author} {\bibfnamefont {Michael~P.}\
  \bibnamefont {Zaletel}}, \bibinfo {author} {\bibfnamefont {Yuan-Ming}\
  \bibnamefont {Lu}}, \ and\ \bibinfo {author} {\bibfnamefont {Ashvin}\
  \bibnamefont {Vishwanath}},\ }\bibfield  {title} {\enquote {\bibinfo {title}
  {Measuring space-group symmetry fractionalization in ${\mathbb{z}}_{2}$ spin
  liquids},}\ }\href {\doibase 10.1103/PhysRevB.96.195164} {\bibfield
  {journal} {\bibinfo  {journal} {Phys. Rev. B}\ }\textbf {\bibinfo {volume}
  {96}},\ \bibinfo {pages} {195164} (\bibinfo {year} {2017})},\ \Eprint
  {http://arxiv.org/abs/1501.01395} {arXiv:1501.01395 [cond-mat.str-el]}
  \BibitemShut {NoStop}%
\bibitem [{\citenamefont {{Barkeshli}}\ \emph
  {et~al.}(2019{\natexlab{b}})\citenamefont {{Barkeshli}}, \citenamefont
  {{Bonderson}}, \citenamefont {{Cheng}}, \citenamefont {{Jian}},\ and\
  \citenamefont {{Walker}}}]{Barkeshli2016}%
  \BibitemOpen
  \bibfield  {author} {\bibinfo {author} {\bibfnamefont {Maissam}\ \bibnamefont
  {{Barkeshli}}}, \bibinfo {author} {\bibfnamefont {Parsa}\ \bibnamefont
  {{Bonderson}}}, \bibinfo {author} {\bibfnamefont {Meng}\ \bibnamefont
  {{Cheng}}}, \bibinfo {author} {\bibfnamefont {Chao-Ming}\ \bibnamefont
  {{Jian}}}, \ and\ \bibinfo {author} {\bibfnamefont {Kevin}\ \bibnamefont
  {{Walker}}},\ }\bibfield  {title} {\enquote {\bibinfo {title} {{Reflection
  and Time Reversal Symmetry Enriched Topological Phases of Matter: Path
  Integrals, Non-orientable Manifolds, and Anomalies}},}\ }\href {\doibase
  10.1007/s00220-019-03475-8} {\bibfield  {journal} {\bibinfo  {journal}
  {Communications in Mathematical Physics}\ }\textbf {\bibinfo {volume}
  {374}},\ \bibinfo {pages} {1021--1124} (\bibinfo {year}
  {2019}{\natexlab{b}})},\ \Eprint {http://arxiv.org/abs/1612.07792}
  {arXiv:1612.07792 [cond-mat.str-el]} \BibitemShut {NoStop}%
\bibitem [{\citenamefont {{Qi}}\ \emph {et~al.}(2019)\citenamefont {{Qi}},
  \citenamefont {{Jian}},\ and\ \citenamefont {{Wang}}}]{Qi2017}%
  \BibitemOpen
  \bibfield  {author} {\bibinfo {author} {\bibfnamefont {Yang}\ \bibnamefont
  {{Qi}}}, \bibinfo {author} {\bibfnamefont {Chao-Ming}\ \bibnamefont
  {{Jian}}}, \ and\ \bibinfo {author} {\bibfnamefont {Chenjie}\ \bibnamefont
  {{Wang}}},\ }\bibfield  {title} {\enquote {\bibinfo {title} {{Folding
  approach to topological order enriched by mirror symmetry}},}\ }\href
  {\doibase 10.1103/PhysRevB.99.085128} {\bibfield  {journal} {\bibinfo
  {journal} {\prb}\ }\textbf {\bibinfo {volume} {99}},\ \bibinfo {eid} {085128}
  (\bibinfo {year} {2019})},\ \Eprint {http://arxiv.org/abs/1710.09391}
  {arXiv:1710.09391 [cond-mat.str-el]} \BibitemShut {NoStop}%
\bibitem [{\citenamefont {{Ding}}\ and\ \citenamefont {{Qi}}(2025)}]{Ding2025}%
  \BibitemOpen
  \bibfield  {author} {\bibinfo {author} {\bibfnamefont {Zhaoyang}\
  \bibnamefont {{Ding}}}\ and\ \bibinfo {author} {\bibfnamefont {Yang}\
  \bibnamefont {{Qi}}},\ }\bibfield  {title} {\enquote {\bibinfo {title}
  {{Point-group symmetry enriched topological orders}},}\ }\href {\doibase
  10.1103/ghds-2st8} {\bibfield  {journal} {\bibinfo  {journal} {\prb}\
  }\textbf {\bibinfo {volume} {112}},\ \bibinfo {eid} {115102} (\bibinfo {year}
  {2025})},\ \Eprint {http://arxiv.org/abs/2502.11106} {arXiv:2502.11106
  [cond-mat.str-el]} \BibitemShut {NoStop}%
\bibitem [{\citenamefont {{Kawagoe}}\ and\ \citenamefont
  {{Levin}}(2020)}]{Kawagoe2019}%
  \BibitemOpen
  \bibfield  {author} {\bibinfo {author} {\bibfnamefont {Kyle}\ \bibnamefont
  {{Kawagoe}}}\ and\ \bibinfo {author} {\bibfnamefont {Michael}\ \bibnamefont
  {{Levin}}},\ }\bibfield  {title} {\enquote {\bibinfo {title} {{Microscopic
  definitions of anyon data}},}\ }\href {\doibase 10.1103/PhysRevB.101.115113}
  {\bibfield  {journal} {\bibinfo  {journal} {\prb}\ }\textbf {\bibinfo
  {volume} {101}},\ \bibinfo {eid} {115113} (\bibinfo {year} {2020})},\ \Eprint
  {http://arxiv.org/abs/1910.11353} {arXiv:1910.11353 [cond-mat.str-el]}
  \BibitemShut {NoStop}%
\bibitem [{\citenamefont {{Ye}}\ and\ \citenamefont {{Zou}}(2024)}]{Ye2023}%
  \BibitemOpen
  \bibfield  {author} {\bibinfo {author} {\bibfnamefont {Weicheng}\
  \bibnamefont {{Ye}}}\ and\ \bibinfo {author} {\bibfnamefont {Liujun}\
  \bibnamefont {{Zou}}},\ }\bibfield  {title} {\enquote {\bibinfo {title}
  {{Classification of Symmetry-Enriched Topological Quantum Spin Liquids}},}\
  }\href {\doibase 10.1103/PhysRevX.14.021053} {\bibfield  {journal} {\bibinfo
  {journal} {Physical Review X}\ }\textbf {\bibinfo {volume} {14}},\ \bibinfo
  {eid} {021053} (\bibinfo {year} {2024})},\ \Eprint
  {http://arxiv.org/abs/2309.15118} {arXiv:2309.15118 [cond-mat.str-el]}
  \BibitemShut {NoStop}%
\bibitem [{\citenamefont {{Song}}\ \emph {et~al.}(2017)\citenamefont {{Song}},
  \citenamefont {{Huang}}, \citenamefont {{Fu}},\ and\ \citenamefont
  {{Hermele}}}]{Song2016}%
  \BibitemOpen
  \bibfield  {author} {\bibinfo {author} {\bibfnamefont {Hao}\ \bibnamefont
  {{Song}}}, \bibinfo {author} {\bibfnamefont {Sheng-Jie}\ \bibnamefont
  {{Huang}}}, \bibinfo {author} {\bibfnamefont {Liang}\ \bibnamefont {{Fu}}}, \
  and\ \bibinfo {author} {\bibfnamefont {Michael}\ \bibnamefont {{Hermele}}},\
  }\bibfield  {title} {\enquote {\bibinfo {title} {{Topological Phases
  Protected by Point Group Symmetry}},}\ }\href {\doibase
  10.1103/PhysRevX.7.011020} {\bibfield  {journal} {\bibinfo  {journal}
  {Physical Review X}\ }\textbf {\bibinfo {volume} {7}},\ \bibinfo {eid}
  {011020} (\bibinfo {year} {2017})},\ \Eprint
  {http://arxiv.org/abs/1604.08151} {arXiv:1604.08151 [cond-mat.str-el]}
  \BibitemShut {NoStop}%
\bibitem [{\citenamefont {Thorngren}\ and\ \citenamefont
  {Else}(2018{\natexlab{a}})}]{Thorngren2016}%
  \BibitemOpen
  \bibfield  {author} {\bibinfo {author} {\bibfnamefont {Ryan}\ \bibnamefont
  {Thorngren}}\ and\ \bibinfo {author} {\bibfnamefont {Dominic~V.}\
  \bibnamefont {Else}},\ }\bibfield  {title} {\enquote {\bibinfo {title}
  {Gauging spatial symmetries and the classification of topological crystalline
  phases},}\ }\href {\doibase 10.1103/PhysRevX.8.011040} {\bibfield  {journal}
  {\bibinfo  {journal} {Phys. Rev. X}\ }\textbf {\bibinfo {volume} {8}},\
  \bibinfo {pages} {011040} (\bibinfo {year} {2018}{\natexlab{a}})},\ \Eprint
  {http://arxiv.org/abs/1612.00846} {arXiv:1612.00846 [cond-mat.str-el]}
  \BibitemShut {NoStop}%
\bibitem [{\citenamefont {{Zou}}\ and\ \citenamefont
  {{Cheng}}(2026)}]{Zou2026}%
  \BibitemOpen
  \bibfield  {author} {\bibinfo {author} {\bibfnamefont {Liujun}\ \bibnamefont
  {{Zou}}}\ and\ \bibinfo {author} {\bibfnamefont {Meng}\ \bibnamefont
  {{Cheng}}},\ }\bibfield  {title} {\enquote {\bibinfo {title}
  {{Lieb-Schultz-Mattis Anomalies and Anomaly Matching}},}\ }\href@noop {}
  {\bibfield  {journal} {\bibinfo  {journal} {arXiv e-prints}\ ,\ \bibinfo
  {eid} {arXiv:2604.00347}} (\bibinfo {year} {2026})},\ \Eprint
  {http://arxiv.org/abs/2604.00347} {arXiv:2604.00347 [cond-mat.str-el]}
  \BibitemShut {NoStop}%
\bibitem [{\citenamefont {{Ye}}\ \emph {et~al.}(2022)\citenamefont {{Ye}},
  \citenamefont {{Guo}}, \citenamefont {{He}}, \citenamefont {{Wang}},\ and\
  \citenamefont {{Zou}}}]{Ye2021a}%
  \BibitemOpen
  \bibfield  {author} {\bibinfo {author} {\bibfnamefont {Weicheng}\
  \bibnamefont {{Ye}}}, \bibinfo {author} {\bibfnamefont {Meng}\ \bibnamefont
  {{Guo}}}, \bibinfo {author} {\bibfnamefont {Yin-Chen}\ \bibnamefont {{He}}},
  \bibinfo {author} {\bibfnamefont {Chong}\ \bibnamefont {{Wang}}}, \ and\
  \bibinfo {author} {\bibfnamefont {Liujun}\ \bibnamefont {{Zou}}},\ }\bibfield
   {title} {\enquote {\bibinfo {title} {{Topological characterization of
  Lieb-Schultz-Mattis constraints and applications to symmetry-enriched quantum
  criticality}},}\ }\href {\doibase 10.21468/SciPostPhys.13.3.066} {\bibfield
  {journal} {\bibinfo  {journal} {SciPost Physics}\ }\textbf {\bibinfo {volume}
  {13}},\ \bibinfo {eid} {066} (\bibinfo {year} {2022})},\ \Eprint
  {http://arxiv.org/abs/2111.12097} {arXiv:2111.12097 [cond-mat.str-el]}
  \BibitemShut {NoStop}%
\bibitem [{\citenamefont {{Ye}}\ and\ \citenamefont {{Zou}}(2023)}]{Ye2022}%
  \BibitemOpen
  \bibfield  {author} {\bibinfo {author} {\bibfnamefont {Weicheng}\
  \bibnamefont {{Ye}}}\ and\ \bibinfo {author} {\bibfnamefont {Liujun}\
  \bibnamefont {{Zou}}},\ }\bibfield  {title} {\enquote {\bibinfo {title}
  {{Anomaly of (2+1)-dimensional symmetry-enriched topological order from
  (3+1)-dimensional topological quantum field theory}},}\ }\href {\doibase
  10.21468/SciPostPhys.15.1.004} {\bibfield  {journal} {\bibinfo  {journal}
  {SciPost Physics}\ }\textbf {\bibinfo {volume} {15}},\ \bibinfo {eid} {004}
  (\bibinfo {year} {2023})},\ \Eprint {http://arxiv.org/abs/2210.02444}
  {arXiv:2210.02444 [cond-mat.str-el]} \BibitemShut {NoStop}%
\bibitem [{\citenamefont {{Liu}}\ \emph {et~al.}(2025)\citenamefont {{Liu}},
  \citenamefont {{Yi}}, \citenamefont {{Zhou}},\ and\ \citenamefont
  {{Zou}}}]{Liu2024a}%
  \BibitemOpen
  \bibfield  {author} {\bibinfo {author} {\bibfnamefont {Ruizhi}\ \bibnamefont
  {{Liu}}}, \bibinfo {author} {\bibfnamefont {Jinmin}\ \bibnamefont {{Yi}}},
  \bibinfo {author} {\bibfnamefont {Shiyu}\ \bibnamefont {{Zhou}}}, \ and\
  \bibinfo {author} {\bibfnamefont {Liujun}\ \bibnamefont {{Zou}}},\ }\bibfield
   {title} {\enquote {\bibinfo {title} {{Entanglement area law and
  Lieb-Schultz-Mattis theorem in long-range interacting systems, and
  symmetry-enforced long-range entanglement}},}\ }\href {\doibase
  10.1103/2jgh-nrj1} {\bibfield  {journal} {\bibinfo  {journal} {\prb}\
  }\textbf {\bibinfo {volume} {112}},\ \bibinfo {eid} {214408} (\bibinfo {year}
  {2025})},\ \Eprint {http://arxiv.org/abs/2405.14929} {arXiv:2405.14929
  [cond-mat.str-el]} \BibitemShut {NoStop}%
\bibitem [{\citenamefont {{Li}}\ and\ \citenamefont {{Zou}}(2025)}]{Li2024}%
  \BibitemOpen
  \bibfield  {author} {\bibinfo {author} {\bibfnamefont {Kangle}\ \bibnamefont
  {{Li}}}\ and\ \bibinfo {author} {\bibfnamefont {Liujun}\ \bibnamefont
  {{Zou}}},\ }\bibfield  {title} {\enquote {\bibinfo {title}
  {{Symmetry-enforced minimal entanglement and correlation in quantum spin
  chains}},}\ }\href {\doibase 10.21468/SciPostPhys.19.1.020} {\bibfield
  {journal} {\bibinfo  {journal} {SciPost Physics}\ }\textbf {\bibinfo {volume}
  {19}},\ \bibinfo {eid} {020} (\bibinfo {year} {2025})},\ \Eprint
  {http://arxiv.org/abs/2412.20765} {arXiv:2412.20765 [cond-mat.str-el]}
  \BibitemShut {NoStop}%
\bibitem [{\citenamefont {{Hastings}}\ and\ \citenamefont
  {{Wen}}(2005)}]{Hastings2005}%
  \BibitemOpen
  \bibfield  {author} {\bibinfo {author} {\bibfnamefont {M.~B.}\ \bibnamefont
  {{Hastings}}}\ and\ \bibinfo {author} {\bibfnamefont {Xiao-Gang}\
  \bibnamefont {{Wen}}},\ }\bibfield  {title} {\enquote {\bibinfo {title}
  {{Quasiadiabatic continuation of quantum states: The stability of topological
  ground-state degeneracy and emergent gauge invariance}},}\ }\href {\doibase
  10.1103/PhysRevB.72.045141} {\bibfield  {journal} {\bibinfo  {journal}
  {\prb}\ }\textbf {\bibinfo {volume} {72}},\ \bibinfo {eid} {045141} (\bibinfo
  {year} {2005})},\ \Eprint {http://arxiv.org/abs/cond-mat/0503554}
  {arXiv:cond-mat/0503554 [cond-mat.str-el]} \BibitemShut {NoStop}%
\bibitem [{\citenamefont {{Hastings}}(2010{\natexlab{a}})}]{Hastings2010}%
  \BibitemOpen
  \bibfield  {author} {\bibinfo {author} {\bibfnamefont {M.~B.}\ \bibnamefont
  {{Hastings}}},\ }\bibfield  {title} {\enquote {\bibinfo {title}
  {{Quasi-adiabatic Continuation for Disordered Systems: Applications to
  Correlations, Lieb-Schultz-Mattis, and Hall Conductance}},}\ }\href {\doibase
  10.48550/arXiv.1001.5280} {\bibfield  {journal} {\bibinfo  {journal} {arXiv
  e-prints}\ ,\ \bibinfo {eid} {arXiv:1001.5280}} (\bibinfo {year}
  {2010}{\natexlab{a}})},\ \Eprint {http://arxiv.org/abs/1001.5280}
  {arXiv:1001.5280 [math-ph]} \BibitemShut {NoStop}%
\bibitem [{\citenamefont {{Bachmann}}\ \emph {et~al.}(2012)\citenamefont
  {{Bachmann}}, \citenamefont {{Michalakis}}, \citenamefont {{Nachtergaele}},\
  and\ \citenamefont {{Sims}}}]{Bachmann2012automorphic}%
  \BibitemOpen
  \bibfield  {author} {\bibinfo {author} {\bibfnamefont {Sven}\ \bibnamefont
  {{Bachmann}}}, \bibinfo {author} {\bibfnamefont {Spyridon}\ \bibnamefont
  {{Michalakis}}}, \bibinfo {author} {\bibfnamefont {Bruno}\ \bibnamefont
  {{Nachtergaele}}}, \ and\ \bibinfo {author} {\bibfnamefont {Robert}\
  \bibnamefont {{Sims}}},\ }\bibfield  {title} {\enquote {\bibinfo {title}
  {{Automorphic Equivalence within Gapped Phases of Quantum Lattice
  Systems}},}\ }\href {\doibase 10.1007/s00220-011-1380-0} {\bibfield
  {journal} {\bibinfo  {journal} {Communications in Mathematical Physics}\
  }\textbf {\bibinfo {volume} {309}},\ \bibinfo {pages} {835--871} (\bibinfo
  {year} {2012})},\ \Eprint {http://arxiv.org/abs/1102.0842} {arXiv:1102.0842
  [math-ph]} \BibitemShut {NoStop}%
\bibitem [{\citenamefont {{Kapustin}}\ and\ \citenamefont
  {{Sopenko}}(2022)}]{Kapustin2022Noether}%
  \BibitemOpen
  \bibfield  {author} {\bibinfo {author} {\bibfnamefont {Anton}\ \bibnamefont
  {{Kapustin}}}\ and\ \bibinfo {author} {\bibfnamefont {Nikita}\ \bibnamefont
  {{Sopenko}}},\ }\bibfield  {title} {\enquote {\bibinfo {title} {{Local
  Noether theorem for quantum lattice systems and topological invariants of
  gapped states}},}\ }\href {\doibase 10.1063/5.0085964} {\bibfield  {journal}
  {\bibinfo  {journal} {Journal of Mathematical Physics}\ }\textbf {\bibinfo
  {volume} {63}},\ \bibinfo {eid} {091903} (\bibinfo {year} {2022})},\ \Eprint
  {http://arxiv.org/abs/2201.01327} {arXiv:2201.01327 [math-ph]} \BibitemShut
  {NoStop}%
\bibitem [{\citenamefont {{Shi}}\ \emph {et~al.}(2020)\citenamefont {{Shi}},
  \citenamefont {{Kato}},\ and\ \citenamefont {{Kim}}}]{Shi2019}%
  \BibitemOpen
  \bibfield  {author} {\bibinfo {author} {\bibfnamefont {Bowen}\ \bibnamefont
  {{Shi}}}, \bibinfo {author} {\bibfnamefont {Kohtaro}\ \bibnamefont {{Kato}}},
  \ and\ \bibinfo {author} {\bibfnamefont {Isaac~H.}\ \bibnamefont {{Kim}}},\
  }\bibfield  {title} {\enquote {\bibinfo {title} {{Fusion rules from
  entanglement}},}\ }\href {\doibase 10.1016/j.aop.2020.168164} {\bibfield
  {journal} {\bibinfo  {journal} {Annals of Physics}\ }\textbf {\bibinfo
  {volume} {418}},\ \bibinfo {eid} {168164} (\bibinfo {year} {2020})},\ \Eprint
  {http://arxiv.org/abs/1906.09376} {arXiv:1906.09376 [cond-mat.str-el]}
  \BibitemShut {NoStop}%
\bibitem [{\citenamefont {{Cian}}\ \emph {et~al.}(2022)\citenamefont {{Cian}},
  \citenamefont {{Hafezi}},\ and\ \citenamefont {{Barkeshli}}}]{Cian2022}%
  \BibitemOpen
  \bibfield  {author} {\bibinfo {author} {\bibfnamefont {Ze-Pei}\ \bibnamefont
  {{Cian}}}, \bibinfo {author} {\bibfnamefont {Mohammad}\ \bibnamefont
  {{Hafezi}}}, \ and\ \bibinfo {author} {\bibfnamefont {Maissam}\ \bibnamefont
  {{Barkeshli}}},\ }\bibfield  {title} {\enquote {\bibinfo {title} {{Extracting
  Wilson loop operators and fractional statistics from a single bulk ground
  state}},}\ }\href {\doibase 10.48550/arXiv.2209.14302} {\bibfield  {journal}
  {\bibinfo  {journal} {arXiv e-prints}\ ,\ \bibinfo {eid} {arXiv:2209.14302}}
  (\bibinfo {year} {2022})},\ \Eprint {http://arxiv.org/abs/2209.14302}
  {arXiv:2209.14302 [cond-mat.str-el]} \BibitemShut {NoStop}%
\bibitem [{\citenamefont {{Cheng}}\ and\ \citenamefont
  {{Williamson}}(2020)}]{Cheng2020}%
  \BibitemOpen
  \bibfield  {author} {\bibinfo {author} {\bibfnamefont {Meng}\ \bibnamefont
  {{Cheng}}}\ and\ \bibinfo {author} {\bibfnamefont {Dominic~J.}\ \bibnamefont
  {{Williamson}}},\ }\bibfield  {title} {\enquote {\bibinfo {title} {{Relative
  anomaly in (1 +1 )d rational conformal field theory}},}\ }\href {\doibase
  10.1103/PhysRevResearch.2.043044} {\bibfield  {journal} {\bibinfo  {journal}
  {Physical Review Research}\ }\textbf {\bibinfo {volume} {2}},\ \bibinfo {eid}
  {043044} (\bibinfo {year} {2020})},\ \Eprint
  {http://arxiv.org/abs/2002.02984} {arXiv:2002.02984 [cond-mat.str-el]}
  \BibitemShut {NoStop}%
\bibitem [{\citenamefont {{Chen}}\ \emph {et~al.}(2011)\citenamefont {{Chen}},
  \citenamefont {{Gu}},\ and\ \citenamefont {{Wen}}}]{Chen2010}%
  \BibitemOpen
  \bibfield  {author} {\bibinfo {author} {\bibfnamefont {Xie}\ \bibnamefont
  {{Chen}}}, \bibinfo {author} {\bibfnamefont {Zheng-Cheng}\ \bibnamefont
  {{Gu}}}, \ and\ \bibinfo {author} {\bibfnamefont {Xiao-Gang}\ \bibnamefont
  {{Wen}}},\ }\bibfield  {title} {\enquote {\bibinfo {title} {{Classification
  of gapped symmetric phases in one-dimensional spin systems}},}\ }\href
  {\doibase 10.1103/PhysRevB.83.035107} {\bibfield  {journal} {\bibinfo
  {journal} {\prb}\ }\textbf {\bibinfo {volume} {83}},\ \bibinfo {eid} {035107}
  (\bibinfo {year} {2011})},\ \Eprint {http://arxiv.org/abs/1008.3745}
  {arXiv:1008.3745 [cond-mat.str-el]} \BibitemShut {NoStop}%
\bibitem [{\citenamefont {{Isobe}}\ and\ \citenamefont
  {{Fu}}(2015)}]{Isobe2015}%
  \BibitemOpen
  \bibfield  {author} {\bibinfo {author} {\bibfnamefont {Hiroki}\ \bibnamefont
  {{Isobe}}}\ and\ \bibinfo {author} {\bibfnamefont {Liang}\ \bibnamefont
  {{Fu}}},\ }\bibfield  {title} {\enquote {\bibinfo {title} {{Theory of
  interacting topological crystalline insulators}},}\ }\href {\doibase
  10.1103/PhysRevB.92.081304} {\bibfield  {journal} {\bibinfo  {journal}
  {\prb}\ }\textbf {\bibinfo {volume} {92}},\ \bibinfo {eid} {081304} (\bibinfo
  {year} {2015})},\ \Eprint {http://arxiv.org/abs/1502.06962} {arXiv:1502.06962
  [cond-mat.str-el]} \BibitemShut {NoStop}%
\bibitem [{\citenamefont {{Huang}}\ \emph {et~al.}(2017)\citenamefont
  {{Huang}}, \citenamefont {{Song}}, \citenamefont {{Huang}},\ and\
  \citenamefont {{Hermele}}}]{Huang2017}%
  \BibitemOpen
  \bibfield  {author} {\bibinfo {author} {\bibfnamefont {Sheng-Jie}\
  \bibnamefont {{Huang}}}, \bibinfo {author} {\bibfnamefont {Hao}\ \bibnamefont
  {{Song}}}, \bibinfo {author} {\bibfnamefont {Yi-Ping}\ \bibnamefont
  {{Huang}}}, \ and\ \bibinfo {author} {\bibfnamefont {Michael}\ \bibnamefont
  {{Hermele}}},\ }\bibfield  {title} {\enquote {\bibinfo {title} {{Building
  crystalline topological phases from lower-dimensional states}},}\ }\href
  {\doibase 10.1103/PhysRevB.96.205106} {\bibfield  {journal} {\bibinfo
  {journal} {\prb}\ }\textbf {\bibinfo {volume} {96}},\ \bibinfo {eid} {205106}
  (\bibinfo {year} {2017})},\ \Eprint {http://arxiv.org/abs/1705.09243}
  {arXiv:1705.09243 [cond-mat.str-el]} \BibitemShut {NoStop}%
\bibitem [{\citenamefont {{Cheng}}(2018)}]{Cheng2017a}%
  \BibitemOpen
  \bibfield  {author} {\bibinfo {author} {\bibfnamefont {Meng}\ \bibnamefont
  {{Cheng}}},\ }\bibfield  {title} {\enquote {\bibinfo {title} {{Microscopic
  Theory of Surface Topological Order for Topological Crystalline
  Superconductors}},}\ }\href {\doibase 10.1103/PhysRevLett.120.036801}
  {\bibfield  {journal} {\bibinfo  {journal} {\prl}\ }\textbf {\bibinfo
  {volume} {120}},\ \bibinfo {eid} {036801} (\bibinfo {year} {2018})},\ \Eprint
  {http://arxiv.org/abs/1707.02079} {arXiv:1707.02079 [cond-mat.str-el]}
  \BibitemShut {NoStop}%
\bibitem [{\citenamefont {{Zou}}(2018)}]{Zou2017a}%
  \BibitemOpen
  \bibfield  {author} {\bibinfo {author} {\bibfnamefont {Liujun}\ \bibnamefont
  {{Zou}}},\ }\bibfield  {title} {\enquote {\bibinfo {title} {{Bulk
  characterization of topological crystalline insulators: Stability under
  interactions and relations to symmetry enriched U (1) quantum spin
  liquids}},}\ }\href {\doibase 10.1103/PhysRevB.97.045130} {\bibfield
  {journal} {\bibinfo  {journal} {\prb}\ }\textbf {\bibinfo {volume} {97}},\
  \bibinfo {eid} {045130} (\bibinfo {year} {2018})},\ \Eprint
  {http://arxiv.org/abs/1711.03090} {arXiv:1711.03090 [cond-mat.str-el]}
  \BibitemShut {NoStop}%
\bibitem [{\citenamefont {{Else}}\ and\ \citenamefont
  {{Thorngren}}(2019)}]{Else2018a}%
  \BibitemOpen
  \bibfield  {author} {\bibinfo {author} {\bibfnamefont {Dominic~V.}\
  \bibnamefont {{Else}}}\ and\ \bibinfo {author} {\bibfnamefont {Ryan}\
  \bibnamefont {{Thorngren}}},\ }\bibfield  {title} {\enquote {\bibinfo {title}
  {{Crystalline topological phases as defect networks}},}\ }\href {\doibase
  10.1103/PhysRevB.99.115116} {\bibfield  {journal} {\bibinfo  {journal}
  {\prb}\ }\textbf {\bibinfo {volume} {99}},\ \bibinfo {eid} {115116} (\bibinfo
  {year} {2019})},\ \Eprint {http://arxiv.org/abs/1810.10539} {arXiv:1810.10539
  [cond-mat.str-el]} \BibitemShut {NoStop}%
\bibitem [{\citenamefont {Thorngren}\ and\ \citenamefont
  {Else}(2018{\natexlab{b}})}]{Else2018}%
  \BibitemOpen
  \bibfield  {author} {\bibinfo {author} {\bibfnamefont {Ryan}\ \bibnamefont
  {Thorngren}}\ and\ \bibinfo {author} {\bibfnamefont {Dominic~V.}\
  \bibnamefont {Else}},\ }\bibfield  {title} {\enquote {\bibinfo {title}
  {Gauging spatial symmetries and the classification of topological crystalline
  phases},}\ }\href {\doibase 10.1103/PhysRevX.8.011040} {\bibfield  {journal}
  {\bibinfo  {journal} {Phys. Rev. X}\ }\textbf {\bibinfo {volume} {8}},\
  \bibinfo {eid} {arXiv:1612.00846} (\bibinfo {year} {2018}{\natexlab{b}})},\
  \Eprint {http://arxiv.org/abs/1612.00846} {arXiv:1612.00846
  [cond-mat.str-el]} \BibitemShut {NoStop}%
\bibitem [{\citenamefont {{Else}}\ and\ \citenamefont
  {{Thorngren}}(2020)}]{Else2019}%
  \BibitemOpen
  \bibfield  {author} {\bibinfo {author} {\bibfnamefont {Dominic~V.}\
  \bibnamefont {{Else}}}\ and\ \bibinfo {author} {\bibfnamefont {Ryan}\
  \bibnamefont {{Thorngren}}},\ }\bibfield  {title} {\enquote {\bibinfo {title}
  {{Topological theory of Lieb-Schultz-Mattis theorems in quantum spin
  systems}},}\ }\href {\doibase 10.1103/PhysRevB.101.224437} {\bibfield
  {journal} {\bibinfo  {journal} {\prb}\ }\textbf {\bibinfo {volume} {101}},\
  \bibinfo {eid} {224437} (\bibinfo {year} {2020})},\ \Eprint
  {http://arxiv.org/abs/1907.08204} {arXiv:1907.08204 [cond-mat.str-el]}
  \BibitemShut {NoStop}%
\bibitem [{\citenamefont {Jiang}\ \emph {et~al.}(2021)\citenamefont {Jiang},
  \citenamefont {Cheng}, \citenamefont {Qi},\ and\ \citenamefont
  {Lu}}]{Jiang2019}%
  \BibitemOpen
  \bibfield  {author} {\bibinfo {author} {\bibfnamefont {Shenghan}\
  \bibnamefont {Jiang}}, \bibinfo {author} {\bibfnamefont {Meng}\ \bibnamefont
  {Cheng}}, \bibinfo {author} {\bibfnamefont {Yang}\ \bibnamefont {Qi}}, \ and\
  \bibinfo {author} {\bibfnamefont {Yuan-Ming}\ \bibnamefont {Lu}},\ }\bibfield
   {title} {\enquote {\bibinfo {title} {{Generalized Lieb-Schultz-Mattis
  theorem on bosonic symmetry protected topological phases}},}\ }\href
  {\doibase 10.21468/SciPostPhys.11.2.024} {\bibfield  {journal} {\bibinfo
  {journal} {SciPost Phys.}\ }\textbf {\bibinfo {volume} {11}},\ \bibinfo
  {pages} {024} (\bibinfo {year} {2021})},\ \Eprint
  {http://arxiv.org/abs/1907.08596} {arXiv:1907.08596 [cond-mat.str-el]}
  \BibitemShut {NoStop}%
\bibitem [{\citenamefont {Simon}(2023)}]{Simonbook2023}%
  \BibitemOpen
  \bibfield  {author} {\bibinfo {author} {\bibfnamefont {Steven~H.}\
  \bibnamefont {Simon}},\ }\href {\doibase 10.1093/oso/9780198886723.001.0001}
  {\emph {\bibinfo {title} {Topological Quantum}}}\ (\bibinfo  {publisher}
  {Oxford University Press},\ \bibinfo {year} {2023})\BibitemShut {NoStop}%
\bibitem [{\citenamefont {{Essin}}\ and\ \citenamefont
  {{Hermele}}(2014)}]{Essin2014}%
  \BibitemOpen
  \bibfield  {author} {\bibinfo {author} {\bibfnamefont {Andrew~M.}\
  \bibnamefont {{Essin}}}\ and\ \bibinfo {author} {\bibfnamefont {Michael}\
  \bibnamefont {{Hermele}}},\ }\bibfield  {title} {\enquote {\bibinfo {title}
  {{Spectroscopic signatures of crystal momentum fractionalization}},}\ }\href
  {\doibase 10.1103/PhysRevB.90.121102} {\bibfield  {journal} {\bibinfo
  {journal} {\prb}\ }\textbf {\bibinfo {volume} {90}},\ \bibinfo {eid} {121102}
  (\bibinfo {year} {2014})},\ \Eprint {http://arxiv.org/abs/1401.1846}
  {arXiv:1401.1846 [cond-mat.str-el]} \BibitemShut {NoStop}%
\bibitem [{\citenamefont {{Tarabunga}}\ \emph {et~al.}(2022)\citenamefont
  {{Tarabunga}}, \citenamefont {{Surace}}, \citenamefont {{Andreoni}},
  \citenamefont {{Angelone}},\ and\ \citenamefont
  {{Dalmonte}}}]{Tarabunga2022}%
  \BibitemOpen
  \bibfield  {author} {\bibinfo {author} {\bibfnamefont {P.~S.}\ \bibnamefont
  {{Tarabunga}}}, \bibinfo {author} {\bibfnamefont {F.~M.}\ \bibnamefont
  {{Surace}}}, \bibinfo {author} {\bibfnamefont {R.}~\bibnamefont
  {{Andreoni}}}, \bibinfo {author} {\bibfnamefont {A.}~\bibnamefont
  {{Angelone}}}, \ and\ \bibinfo {author} {\bibfnamefont {M.}~\bibnamefont
  {{Dalmonte}}},\ }\bibfield  {title} {\enquote {\bibinfo {title}
  {{Gauge-Theoretic Origin of Rydberg Quantum Spin Liquids}},}\ }\href
  {\doibase 10.1103/PhysRevLett.129.195301} {\bibfield  {journal} {\bibinfo
  {journal} {\prl}\ }\textbf {\bibinfo {volume} {129}},\ \bibinfo {eid}
  {195301} (\bibinfo {year} {2022})},\ \Eprint
  {http://arxiv.org/abs/2205.13000} {arXiv:2205.13000 [cond-mat.str-el]}
  \BibitemShut {NoStop}%
\bibitem [{\citenamefont {{Samajdar}}\ \emph {et~al.}(2023)\citenamefont
  {{Samajdar}}, \citenamefont {{Joshi}}, \citenamefont {{Teng}},\ and\
  \citenamefont {{Sachdev}}}]{Samajdar2022}%
  \BibitemOpen
  \bibfield  {author} {\bibinfo {author} {\bibfnamefont {Rhine}\ \bibnamefont
  {{Samajdar}}}, \bibinfo {author} {\bibfnamefont {Darshan~G.}\ \bibnamefont
  {{Joshi}}}, \bibinfo {author} {\bibfnamefont {Yanting}\ \bibnamefont
  {{Teng}}}, \ and\ \bibinfo {author} {\bibfnamefont {Subir}\ \bibnamefont
  {{Sachdev}}},\ }\bibfield  {title} {\enquote {\bibinfo {title} {{Emergent
  Z$_{2}$ Gauge Theories and Topological Excitations in Rydberg Atom
  Arrays}},}\ }\href {\doibase 10.1103/PhysRevLett.130.043601} {\bibfield
  {journal} {\bibinfo  {journal} {\prl}\ }\textbf {\bibinfo {volume} {130}},\
  \bibinfo {eid} {043601} (\bibinfo {year} {2023})},\ \Eprint
  {http://arxiv.org/abs/2204.00632} {arXiv:2204.00632 [cond-mat.quant-gas]}
  \BibitemShut {NoStop}%
\bibitem [{\citenamefont {{Verresen}}\ and\ \citenamefont
  {{Vishwanath}}(2022)}]{Verresen2022}%
  \BibitemOpen
  \bibfield  {author} {\bibinfo {author} {\bibfnamefont {Ruben}\ \bibnamefont
  {{Verresen}}}\ and\ \bibinfo {author} {\bibfnamefont {Ashvin}\ \bibnamefont
  {{Vishwanath}}},\ }\bibfield  {title} {\enquote {\bibinfo {title} {{Unifying
  Kitaev Magnets, Kagom{\'e} Dimer Models, and Ruby Rydberg Spin Liquids}},}\
  }\href {\doibase 10.1103/PhysRevX.12.041029} {\bibfield  {journal} {\bibinfo
  {journal} {Physical Review X}\ }\textbf {\bibinfo {volume} {12}},\ \bibinfo
  {eid} {041029} (\bibinfo {year} {2022})},\ \Eprint
  {http://arxiv.org/abs/2205.15302} {arXiv:2205.15302 [cond-mat.str-el]}
  \BibitemShut {NoStop}%
\bibitem [{\citenamefont {{Cha}}\ \emph {et~al.}(2019)\citenamefont {{Cha}},
  \citenamefont {{Naaijkens}},\ and\ \citenamefont {{Nachtergaele}}}]{Cha2018}%
  \BibitemOpen
  \bibfield  {author} {\bibinfo {author} {\bibfnamefont {Matthew}\ \bibnamefont
  {{Cha}}}, \bibinfo {author} {\bibfnamefont {Pieter}\ \bibnamefont
  {{Naaijkens}}}, \ and\ \bibinfo {author} {\bibfnamefont {Bruno}\ \bibnamefont
  {{Nachtergaele}}},\ }\bibfield  {title} {\enquote {\bibinfo {title} {{On the
  Stability of Charges in Infinite Quantum Spin Systems}},}\ }\href {\doibase
  10.1007/s00220-019-03630-1} {\bibfield  {journal} {\bibinfo  {journal}
  {Communications in Mathematical Physics}\ }\textbf {\bibinfo {volume}
  {373}},\ \bibinfo {pages} {219--264} (\bibinfo {year} {2019})},\ \Eprint
  {http://arxiv.org/abs/1804.03203} {arXiv:1804.03203 [math-ph]} \BibitemShut
  {NoStop}%
\bibitem [{\citenamefont {{Ogata}}(2021)}]{Ogata2021a}%
  \BibitemOpen
  \bibfield  {author} {\bibinfo {author} {\bibfnamefont {Yoshiko}\ \bibnamefont
  {{Ogata}}},\ }\bibfield  {title} {\enquote {\bibinfo {title} {{A derivation
  of braided $C^*$-tensor categories from gapped ground states satisfying the
  approximate Haag duality}},}\ }\href {\doibase 10.48550/arXiv.2106.15741}
  {\bibfield  {journal} {\bibinfo  {journal} {arXiv e-prints}\ ,\ \bibinfo
  {eid} {arXiv:2106.15741}} (\bibinfo {year} {2021})},\ \Eprint
  {http://arxiv.org/abs/2106.15741} {arXiv:2106.15741 [math-ph]} \BibitemShut
  {NoStop}%
\bibitem [{\citenamefont {{Kawagoe}}\ \emph {et~al.}(2024)\citenamefont
  {{Kawagoe}}, \citenamefont {{Vadnerkar}},\ and\ \citenamefont
  {{Wallick}}}]{Kawagoe2024}%
  \BibitemOpen
  \bibfield  {author} {\bibinfo {author} {\bibfnamefont {Kyle}\ \bibnamefont
  {{Kawagoe}}}, \bibinfo {author} {\bibfnamefont {Siddharth}\ \bibnamefont
  {{Vadnerkar}}}, \ and\ \bibinfo {author} {\bibfnamefont {Daniel}\
  \bibnamefont {{Wallick}}},\ }\bibfield  {title} {\enquote {\bibinfo {title}
  {{An operator algebraic approach to symmetry defects and
  fractionalization}},}\ }\href {\doibase 10.48550/arXiv.2410.23380} {\bibfield
   {journal} {\bibinfo  {journal} {arXiv e-prints}\ ,\ \bibinfo {eid}
  {arXiv:2410.23380}} (\bibinfo {year} {2024})},\ \Eprint
  {http://arxiv.org/abs/2410.23380} {arXiv:2410.23380 [math-ph]} \BibitemShut
  {NoStop}%
\bibitem [{\citenamefont {{Morampudi}}\ \emph {et~al.}(2017)\citenamefont
  {{Morampudi}}, \citenamefont {{Turner}}, \citenamefont {{Pollmann}},\ and\
  \citenamefont {{Wilczek}}}]{Morampudi2016}%
  \BibitemOpen
  \bibfield  {author} {\bibinfo {author} {\bibfnamefont {Siddhardh~C.}\
  \bibnamefont {{Morampudi}}}, \bibinfo {author} {\bibfnamefont {Ari~M.}\
  \bibnamefont {{Turner}}}, \bibinfo {author} {\bibfnamefont {Frank}\
  \bibnamefont {{Pollmann}}}, \ and\ \bibinfo {author} {\bibfnamefont {Frank}\
  \bibnamefont {{Wilczek}}},\ }\bibfield  {title} {\enquote {\bibinfo {title}
  {{Statistics of Fractionalized Excitations through Threshold
  Spectroscopy}},}\ }\href {\doibase 10.1103/PhysRevLett.118.227201} {\bibfield
   {journal} {\bibinfo  {journal} {\prl}\ }\textbf {\bibinfo {volume} {118}},\
  \bibinfo {eid} {227201} (\bibinfo {year} {2017})},\ \Eprint
  {http://arxiv.org/abs/1608.05700} {arXiv:1608.05700 [cond-mat.str-el]}
  \BibitemShut {NoStop}%
\bibitem [{\citenamefont {{Kirchner}}\ \emph {et~al.}(2025)\citenamefont
  {{Kirchner}}, \citenamefont {{Choi}},\ and\ \citenamefont
  {{Pollmann}}}]{Kirchner2025}%
  \BibitemOpen
  \bibfield  {author} {\bibinfo {author} {\bibfnamefont {Nico}\ \bibnamefont
  {{Kirchner}}}, \bibinfo {author} {\bibfnamefont {Wonjune}\ \bibnamefont
  {{Choi}}}, \ and\ \bibinfo {author} {\bibfnamefont {Frank}\ \bibnamefont
  {{Pollmann}}},\ }\bibfield  {title} {\enquote {\bibinfo {title} {{Measuring
  Anyonic Exchange Phases Using Two-Dimensional Coherent Spectroscopy}},}\
  }\href {\doibase 10.48550/arXiv.2511.17420} {\bibfield  {journal} {\bibinfo
  {journal} {arXiv e-prints}\ ,\ \bibinfo {eid} {arXiv:2511.17420}} (\bibinfo
  {year} {2025})},\ \Eprint {http://arxiv.org/abs/2511.17420} {arXiv:2511.17420
  [cond-mat.str-el]} \BibitemShut {NoStop}%
\bibitem [{\citenamefont {Stormer}\ \emph {et~al.}(1999)\citenamefont
  {Stormer}, \citenamefont {Tsui},\ and\ \citenamefont
  {Gossard}}]{Stormer1999}%
  \BibitemOpen
  \bibfield  {author} {\bibinfo {author} {\bibfnamefont {Horst~L.}\
  \bibnamefont {Stormer}}, \bibinfo {author} {\bibfnamefont {Daniel~C.}\
  \bibnamefont {Tsui}}, \ and\ \bibinfo {author} {\bibfnamefont {Arthur~C.}\
  \bibnamefont {Gossard}},\ }\bibfield  {title} {\enquote {\bibinfo {title}
  {The fractional quantum hall effect},}\ }\href {\doibase
  10.1103/RevModPhys.71.S298} {\bibfield  {journal} {\bibinfo  {journal} {Rev.
  Mod. Phys.}\ }\textbf {\bibinfo {volume} {71}},\ \bibinfo {pages}
  {S298--S305} (\bibinfo {year} {1999})}\BibitemShut {NoStop}%
\bibitem [{\citenamefont {{Aasen}}\ \emph {et~al.}(2021)\citenamefont
  {{Aasen}}, \citenamefont {{Bonderson}},\ and\ \citenamefont
  {{Knapp}}}]{Aasen2021}%
  \BibitemOpen
  \bibfield  {author} {\bibinfo {author} {\bibfnamefont {David}\ \bibnamefont
  {{Aasen}}}, \bibinfo {author} {\bibfnamefont {Parsa}\ \bibnamefont
  {{Bonderson}}}, \ and\ \bibinfo {author} {\bibfnamefont {Christina}\
  \bibnamefont {{Knapp}}},\ }\bibfield  {title} {\enquote {\bibinfo {title}
  {{Characterization and Classification of Fermionic Symmetry Enriched
  Topological Phases}},}\ }\href@noop {} {\bibfield  {journal} {\bibinfo
  {journal} {arXiv e-prints}\ ,\ \bibinfo {eid} {arXiv:2109.10911}} (\bibinfo
  {year} {2021})},\ \Eprint {http://arxiv.org/abs/2109.10911} {arXiv:2109.10911
  [cond-mat.str-el]} \BibitemShut {NoStop}%
\bibitem [{\citenamefont {{Bulmash}}\ and\ \citenamefont
  {{Barkeshli}}(2022{\natexlab{a}})}]{Bulmash2021}%
  \BibitemOpen
  \bibfield  {author} {\bibinfo {author} {\bibfnamefont {Daniel}\ \bibnamefont
  {{Bulmash}}}\ and\ \bibinfo {author} {\bibfnamefont {Maissam}\ \bibnamefont
  {{Barkeshli}}},\ }\bibfield  {title} {\enquote {\bibinfo {title} {{Fermionic
  symmetry fractionalization in (2+1) dimensions}},}\ }\href {\doibase
  10.1103/PhysRevB.105.125114} {\bibfield  {journal} {\bibinfo  {journal}
  {\prb}\ }\textbf {\bibinfo {volume} {105}},\ \bibinfo {eid} {125114}
  (\bibinfo {year} {2022}{\natexlab{a}})},\ \Eprint
  {http://arxiv.org/abs/2109.10913} {arXiv:2109.10913 [cond-mat.str-el]}
  \BibitemShut {NoStop}%
\bibitem [{\citenamefont {{Bulmash}}\ and\ \citenamefont
  {{Barkeshli}}(2022{\natexlab{b}})}]{Bulmash2021b}%
  \BibitemOpen
  \bibfield  {author} {\bibinfo {author} {\bibfnamefont {Daniel}\ \bibnamefont
  {{Bulmash}}}\ and\ \bibinfo {author} {\bibfnamefont {Maissam}\ \bibnamefont
  {{Barkeshli}}},\ }\bibfield  {title} {\enquote {\bibinfo {title} {{Anomaly
  cascade in (2+1)-dimensional fermionic topological phases}},}\ }\href
  {\doibase 10.1103/PhysRevB.105.155126} {\bibfield  {journal} {\bibinfo
  {journal} {\prb}\ }\textbf {\bibinfo {volume} {105}},\ \bibinfo {eid}
  {155126} (\bibinfo {year} {2022}{\natexlab{b}})},\ \Eprint
  {http://arxiv.org/abs/2109.10922} {arXiv:2109.10922 [cond-mat.str-el]}
  \BibitemShut {NoStop}%
\bibitem [{\citenamefont {{Barkeshli}}\ \emph {et~al.}(2022)\citenamefont
  {{Barkeshli}}, \citenamefont {{Chen}}, \citenamefont {{Hsin}},\ and\
  \citenamefont {{Manjunath}}}]{Barkeshli2021}%
  \BibitemOpen
  \bibfield  {author} {\bibinfo {author} {\bibfnamefont {Maissam}\ \bibnamefont
  {{Barkeshli}}}, \bibinfo {author} {\bibfnamefont {Yu-An}\ \bibnamefont
  {{Chen}}}, \bibinfo {author} {\bibfnamefont {Po-Shen}\ \bibnamefont
  {{Hsin}}}, \ and\ \bibinfo {author} {\bibfnamefont {Naren}\ \bibnamefont
  {{Manjunath}}},\ }\bibfield  {title} {\enquote {\bibinfo {title}
  {{Classification of (2 +1 )D invertible fermionic topological phases with
  symmetry}},}\ }\href {\doibase 10.1103/PhysRevB.105.235143} {\bibfield
  {journal} {\bibinfo  {journal} {\prb}\ }\textbf {\bibinfo {volume} {105}},\
  \bibinfo {eid} {235143} (\bibinfo {year} {2022})},\ \Eprint
  {http://arxiv.org/abs/2109.11039} {arXiv:2109.11039 [cond-mat.str-el]}
  \BibitemShut {NoStop}%
\bibitem [{\citenamefont {{Wen}}(2002)}]{Wen2001}%
  \BibitemOpen
  \bibfield  {author} {\bibinfo {author} {\bibfnamefont {Xiao-Gang}\
  \bibnamefont {{Wen}}},\ }\bibfield  {title} {\enquote {\bibinfo {title}
  {{Quantum orders and symmetric spin liquids}},}\ }\href {\doibase
  10.1103/PhysRevB.65.165113} {\bibfield  {journal} {\bibinfo  {journal}
  {\prb}\ }\textbf {\bibinfo {volume} {65}},\ \bibinfo {eid} {165113} (\bibinfo
  {year} {2002})},\ \Eprint {http://arxiv.org/abs/cond-mat/0107071}
  {arXiv:cond-mat/0107071 [cond-mat.str-el]} \BibitemShut {NoStop}%
\bibitem [{\citenamefont {{Ma}}\ and\ \citenamefont {{Wang}}(2023)}]{Ma2022a}%
  \BibitemOpen
  \bibfield  {author} {\bibinfo {author} {\bibfnamefont {Ruochen}\ \bibnamefont
  {{Ma}}}\ and\ \bibinfo {author} {\bibfnamefont {Chong}\ \bibnamefont
  {{Wang}}},\ }\bibfield  {title} {\enquote {\bibinfo {title} {{Average
  Symmetry-Protected Topological Phases}},}\ }\href {\doibase
  10.1103/PhysRevX.13.031016} {\bibfield  {journal} {\bibinfo  {journal}
  {Physical Review X}\ }\textbf {\bibinfo {volume} {13}},\ \bibinfo {eid}
  {031016} (\bibinfo {year} {2023})},\ \Eprint
  {http://arxiv.org/abs/2209.02723} {arXiv:2209.02723 [cond-mat.str-el]}
  \BibitemShut {NoStop}%
\bibitem [{\citenamefont {{Ma}}\ \emph {et~al.}(2025)\citenamefont {{Ma}},
  \citenamefont {{Zhang}}, \citenamefont {{Bi}}, \citenamefont {{Cheng}},\ and\
  \citenamefont {{Wang}}}]{Ma2023}%
  \BibitemOpen
  \bibfield  {author} {\bibinfo {author} {\bibfnamefont {Ruochen}\ \bibnamefont
  {{Ma}}}, \bibinfo {author} {\bibfnamefont {Jian-Hao}\ \bibnamefont
  {{Zhang}}}, \bibinfo {author} {\bibfnamefont {Zhen}\ \bibnamefont {{Bi}}},
  \bibinfo {author} {\bibfnamefont {Meng}\ \bibnamefont {{Cheng}}}, \ and\
  \bibinfo {author} {\bibfnamefont {Chong}\ \bibnamefont {{Wang}}},\ }\bibfield
   {title} {\enquote {\bibinfo {title} {{Topological Phases with Average
  Symmetries: The Decohered, the Disordered, and the Intrinsic}},}\ }\href
  {\doibase 10.1103/PhysRevX.15.021062} {\bibfield  {journal} {\bibinfo
  {journal} {Physical Review X}\ }\textbf {\bibinfo {volume} {15}},\ \bibinfo
  {eid} {021062} (\bibinfo {year} {2025})},\ \Eprint
  {http://arxiv.org/abs/2305.16399} {arXiv:2305.16399 [cond-mat.str-el]}
  \BibitemShut {NoStop}%
\bibitem [{\citenamefont {{Delmastro}}\ \emph {et~al.}(2023)\citenamefont
  {{Delmastro}}, \citenamefont {{Gomis}}, \citenamefont {{Hsin}},\ and\
  \citenamefont {{Komargodski}}}]{Delmastro2022}%
  \BibitemOpen
  \bibfield  {author} {\bibinfo {author} {\bibfnamefont {Diego~Gabriel}\
  \bibnamefont {{Delmastro}}}, \bibinfo {author} {\bibfnamefont {Jaume}\
  \bibnamefont {{Gomis}}}, \bibinfo {author} {\bibfnamefont {Po-Shen}\
  \bibnamefont {{Hsin}}}, \ and\ \bibinfo {author} {\bibfnamefont {Zohar}\
  \bibnamefont {{Komargodski}}},\ }\bibfield  {title} {\enquote {\bibinfo
  {title} {{Anomalies and symmetry fractionalization}},}\ }\href {\doibase
  10.21468/SciPostPhys.15.3.079} {\bibfield  {journal} {\bibinfo  {journal}
  {SciPost Physics}\ }\textbf {\bibinfo {volume} {15}},\ \bibinfo {eid} {079}
  (\bibinfo {year} {2023})},\ \Eprint {http://arxiv.org/abs/2206.15118}
  {arXiv:2206.15118 [hep-th]} \BibitemShut {NoStop}%
\bibitem [{\citenamefont {{Brennan}}\ \emph {et~al.}(2022)\citenamefont
  {{Brennan}}, \citenamefont {{Cordova}},\ and\ \citenamefont
  {{Dumitrescu}}}]{Brennan2022}%
  \BibitemOpen
  \bibfield  {author} {\bibinfo {author} {\bibfnamefont {T.~Daniel}\
  \bibnamefont {{Brennan}}}, \bibinfo {author} {\bibfnamefont {Clay}\
  \bibnamefont {{Cordova}}}, \ and\ \bibinfo {author} {\bibfnamefont
  {Thomas~T.}\ \bibnamefont {{Dumitrescu}}},\ }\bibfield  {title} {\enquote
  {\bibinfo {title} {{Line Defect Quantum Numbers \& Anomalies}},}\ }\href
  {\doibase 10.48550/arXiv.2206.15401} {\bibfield  {journal} {\bibinfo
  {journal} {arXiv e-prints}\ ,\ \bibinfo {eid} {arXiv:2206.15401}} (\bibinfo
  {year} {2022})},\ \Eprint {http://arxiv.org/abs/2206.15401} {arXiv:2206.15401
  [hep-th]} \BibitemShut {NoStop}%
\bibitem [{\citenamefont {Yi}\ \emph {et~al.}(2026)\citenamefont {Yi},
  \citenamefont {Li}, \citenamefont {Liu}, \citenamefont {Li},\ and\
  \citenamefont {Zou}}]{Yi2025}%
  \BibitemOpen
  \bibfield  {author} {\bibinfo {author} {\bibfnamefont {Jinmin}\ \bibnamefont
  {Yi}}, \bibinfo {author} {\bibfnamefont {Kangle}\ \bibnamefont {Li}},
  \bibinfo {author} {\bibfnamefont {Chuan}\ \bibnamefont {Liu}}, \bibinfo
  {author} {\bibfnamefont {Zixuan}\ \bibnamefont {Li}}, \ and\ \bibinfo
  {author} {\bibfnamefont {Liujun}\ \bibnamefont {Zou}},\ }\bibfield  {title}
  {\enquote {\bibinfo {title} {Universal decay of mutual information and
  conditional mutual information in gapped pure- and mixed-state quantum
  matter},}\ }\href {\doibase 10.1103/mqp8-y1m7} {\bibfield  {journal}
  {\bibinfo  {journal} {Phys. Rev. Lett.}\ }\textbf {\bibinfo {volume} {136}},\
  \bibinfo {pages} {116604} (\bibinfo {year} {2026})},\ \Eprint
  {http://arxiv.org/abs/2510.22867} {arXiv:2510.22867 [cond-mat.str-el]}
  \BibitemShut {NoStop}%
\bibitem [{\citenamefont {Muger}(2003)}]{Muger2003}%
  \BibitemOpen
  \bibfield  {author} {\bibinfo {author} {\bibfnamefont {Michael}\ \bibnamefont
  {Muger}},\ }\bibfield  {title} {\enquote {\bibinfo {title} {On the structure
  of modular categories},}\ }\href {\doibase 10.1112/S0024611503014187}
  {\bibfield  {journal} {\bibinfo  {journal} {Proceedings of the London
  Mathematical Society}\ }\textbf {\bibinfo {volume} {87}},\ \bibinfo {pages}
  {291--308} (\bibinfo {year} {2003})}\BibitemShut {NoStop}%
\bibitem [{\citenamefont {{Kirillov}}\ and\ \citenamefont
  {{Balsam}}(2010)}]{Kirillov2010}%
  \BibitemOpen
  \bibfield  {author} {\bibinfo {author} {\bibfnamefont {Alexander}\
  \bibnamefont {{Kirillov}}, \bibfnamefont {Jr.}}\ and\ \bibinfo {author}
  {\bibfnamefont {Benjamin}\ \bibnamefont {{Balsam}}},\ }\bibfield  {title}
  {\enquote {\bibinfo {title} {{Turaev-Viro invariants as an extended TQFT}},}\
  }\href {\doibase 10.48550/arXiv.1004.1533} {\bibfield  {journal} {\bibinfo
  {journal} {arXiv e-prints}\ ,\ \bibinfo {eid} {arXiv:1004.1533}} (\bibinfo
  {year} {2010})},\ \Eprint {http://arxiv.org/abs/1004.1533} {arXiv:1004.1533
  [math.GT]} \BibitemShut {NoStop}%
\bibitem [{\citenamefont {{Davydov}}\ \emph {et~al.}(2010)\citenamefont
  {{Davydov}}, \citenamefont {{Mueger}}, \citenamefont {{Nikshych}},\ and\
  \citenamefont {{Ostrik}}}]{Davydov2010}%
  \BibitemOpen
  \bibfield  {author} {\bibinfo {author} {\bibfnamefont {Alexei}\ \bibnamefont
  {{Davydov}}}, \bibinfo {author} {\bibfnamefont {Michael}\ \bibnamefont
  {{Mueger}}}, \bibinfo {author} {\bibfnamefont {Dmitri}\ \bibnamefont
  {{Nikshych}}}, \ and\ \bibinfo {author} {\bibfnamefont {Victor}\ \bibnamefont
  {{Ostrik}}},\ }\bibfield  {title} {\enquote {\bibinfo {title} {{The Witt
  group of non-degenerate braided fusion categories}},}\ }\href {\doibase
  10.48550/arXiv.1009.2117} {\bibfield  {journal} {\bibinfo  {journal} {arXiv
  e-prints}\ ,\ \bibinfo {eid} {arXiv:1009.2117}} (\bibinfo {year} {2010})},\
  \Eprint {http://arxiv.org/abs/1009.2117} {arXiv:1009.2117 [math.QA]}
  \BibitemShut {NoStop}%
\bibitem [{\citenamefont {{Balsam}}(2010{\natexlab{a}})}]{Balsam2010}%
  \BibitemOpen
  \bibfield  {author} {\bibinfo {author} {\bibfnamefont {Benjamin}\
  \bibnamefont {{Balsam}}},\ }\bibfield  {title} {\enquote {\bibinfo {title}
  {{Turaev-Viro invariants as an extended TQFT II}},}\ }\href {\doibase
  10.48550/arXiv.1010.1222} {\bibfield  {journal} {\bibinfo  {journal} {arXiv
  e-prints}\ ,\ \bibinfo {eid} {arXiv:1010.1222}} (\bibinfo {year}
  {2010}{\natexlab{a}})},\ \Eprint {http://arxiv.org/abs/1010.1222}
  {arXiv:1010.1222 [math.QA]} \BibitemShut {NoStop}%
\bibitem [{\citenamefont {{Balsam}}(2010{\natexlab{b}})}]{Balsam2010a}%
  \BibitemOpen
  \bibfield  {author} {\bibinfo {author} {\bibfnamefont {Benjamin}\
  \bibnamefont {{Balsam}}},\ }\bibfield  {title} {\enquote {\bibinfo {title}
  {{Turaev-Viro invariants as an extended TQFT III}},}\ }\href {\doibase
  10.48550/arXiv.1012.0560} {\bibfield  {journal} {\bibinfo  {journal} {arXiv
  e-prints}\ ,\ \bibinfo {eid} {arXiv:1012.0560}} (\bibinfo {year}
  {2010}{\natexlab{b}})},\ \Eprint {http://arxiv.org/abs/1012.0560}
  {arXiv:1012.0560 [math.QA]} \BibitemShut {NoStop}%
\bibitem [{\citenamefont {{Kirillov}}(2011)}]{Kirillov2011}%
  \BibitemOpen
  \bibfield  {author} {\bibinfo {author} {\bibfnamefont {Alexander}\
  \bibnamefont {{Kirillov}}, \bibfnamefont {Jr}},\ }\bibfield  {title}
  {\enquote {\bibinfo {title} {{String-net model of Turaev-Viro invariants}},}\
  }\href {\doibase 10.48550/arXiv.1106.6033} {\bibfield  {journal} {\bibinfo
  {journal} {arXiv e-prints}\ ,\ \bibinfo {eid} {arXiv:1106.6033}} (\bibinfo
  {year} {2011})},\ \Eprint {http://arxiv.org/abs/1106.6033} {arXiv:1106.6033
  [math.AT]} \BibitemShut {NoStop}%
\bibitem [{\citenamefont {Lieb}\ and\ \citenamefont
  {Robinson}(1972)}]{Lieb:1972wy}%
  \BibitemOpen
  \bibfield  {author} {\bibinfo {author} {\bibfnamefont {E.~H.}\ \bibnamefont
  {Lieb}}\ and\ \bibinfo {author} {\bibfnamefont {D.~W.}\ \bibnamefont
  {Robinson}},\ }\bibfield  {title} {\enquote {\bibinfo {title} {{The finite
  group velocity of quantum spin systems}},}\ }\href {\doibase
  10.1007/BF01645779} {\bibfield  {journal} {\bibinfo  {journal} {Commun. Math.
  Phys.}\ }\textbf {\bibinfo {volume} {28}},\ \bibinfo {pages} {251--257}
  (\bibinfo {year} {1972})}\BibitemShut {NoStop}%
\bibitem [{\citenamefont {{Hastings}}(2010{\natexlab{b}})}]{Hastings2010a}%
  \BibitemOpen
  \bibfield  {author} {\bibinfo {author} {\bibfnamefont {M.~B.}\ \bibnamefont
  {{Hastings}}},\ }\bibfield  {title} {\enquote {\bibinfo {title} {{Locality in
  Quantum Systems}},}\ }\href {\doibase 10.48550/arXiv.1008.5137} {\bibfield
  {journal} {\bibinfo  {journal} {arXiv e-prints}\ ,\ \bibinfo {eid}
  {arXiv:1008.5137}} (\bibinfo {year} {2010}{\natexlab{b}})},\ \Eprint
  {http://arxiv.org/abs/1008.5137} {arXiv:1008.5137 [math-ph]} \BibitemShut
  {NoStop}%
\bibitem [{\citenamefont {{Shi}}(2019)}]{Shi2018}%
  \BibitemOpen
  \bibfield  {author} {\bibinfo {author} {\bibfnamefont {Bowen}\ \bibnamefont
  {{Shi}}},\ }\bibfield  {title} {\enquote {\bibinfo {title} {{Seeing
  topological entanglement through the information convex}},}\ }\href {\doibase
  10.1103/PhysRevResearch.1.033048} {\bibfield  {journal} {\bibinfo  {journal}
  {Physical Review Research}\ }\textbf {\bibinfo {volume} {1}},\ \bibinfo {eid}
  {033048} (\bibinfo {year} {2019})},\ \Eprint
  {http://arxiv.org/abs/1810.01986} {arXiv:1810.01986 [cond-mat.str-el]}
  \BibitemShut {NoStop}%
\bibitem [{\citenamefont {{Wolf}}\ \emph {et~al.}(2008)\citenamefont {{Wolf}},
  \citenamefont {{Verstraete}}, \citenamefont {{Hastings}},\ and\ \citenamefont
  {{Cirac}}}]{Wolf2007}%
  \BibitemOpen
  \bibfield  {author} {\bibinfo {author} {\bibfnamefont {Michael~M.}\
  \bibnamefont {{Wolf}}}, \bibinfo {author} {\bibfnamefont {Frank}\
  \bibnamefont {{Verstraete}}}, \bibinfo {author} {\bibfnamefont {Matthew~B.}\
  \bibnamefont {{Hastings}}}, \ and\ \bibinfo {author} {\bibfnamefont
  {J.~Ignacio}\ \bibnamefont {{Cirac}}},\ }\bibfield  {title} {\enquote
  {\bibinfo {title} {{Area Laws in Quantum Systems: Mutual Information and
  Correlations}},}\ }\href {\doibase 10.1103/PhysRevLett.100.070502} {\bibfield
   {journal} {\bibinfo  {journal} {\prl}\ }\textbf {\bibinfo {volume} {100}},\
  \bibinfo {eid} {070502} (\bibinfo {year} {2008})},\ \Eprint
  {http://arxiv.org/abs/0704.3906} {arXiv:0704.3906 [quant-ph]} \BibitemShut
  {NoStop}%
\end{thebibliography}%

\end{document}